\documentclass[namedreferences]{solarphysics}

\usepackage[hyperref,optionalrh]{spr-sola-addons} 
\usepackage{graphicx}        
\usepackage{epstopdf}
\usepackage{courier}         
\usepackage{color}           
\usepackage{breakurl}        
\usepackage{subfig}

\usepackage{amsmath}
\usepackage[figuresright]{rotating}

\begin{document}

\begin{article}

\begin{opening}

\title{Segmentation of Coronal Holes Using Active Contours Without Edges$^{*}$}

\author{L.~E.~\surname{Boucheron}$^{1}$\sep
        M.~\surname{Valluri}$^{1,2}$\sep
        R.~T.~J.~\surname{McAteer}$^{3}$      
       }
\runningauthor{Valluri et al.}
\runningtitle{Segmentation of Coronal Holes with Active Contours}

   \institute{$^{*}$ Authors' accepted manuscript. The final publication is available at Springer via \url{http://dx.doi.org/10.1007/s11207-016-0985-z}\\
	      $^{1}$ Klipsch School of Electrical and Computer Engineering, New Mexico State University, Las Cruces, 		NM 88003, USA
                     email: \url{lboucher@nmsu.edu} \\
              $^{2}$ Qualcomm, 5775 Morehouse Drive, San Diego, CA 92121, USA
                     email: \url{c_mvallu@qti.qualcomm.com}\\ 
              $^{3}$ Department of Astronomy, New Mexico State University, Las Cruces, NM 88003, USA
                     email: \url{mcateer@nmsu.edu}\\
             }

\begin{abstract}
An application of active contours without edges is presented as an efficient and effective means of extracting and characterizing coronal holes. Coronal holes are regions of low-density plasma on the Sun with open magnetic field lines. As the source of the fast solar wind, the detection and characterization of these regions is important for both testing theories of their formation and evolution and from a space weather perspective. Coronal holes are detected in full disk extreme ultraviolet (EUV) images of the corona obtained with the  Solar Dynamics Observatory Atmospheric Imaging Assembly (SDO/AIA). The proposed method detects coronal boundaries without determining any fixed intensity value in the data. Instead, the active contour segmentation employs an energy-minimization in which coronal holes are assumed to have more homogeneous intensities than surrounding active regions and quiet Sun. The segmented coronal holes tend to correspond to unipolar magnetic regions, are consistent with concurrent solar wind observations, and qualitatively match the coronal holes segmented by other methods.  The means to identify a coronal hole without specification of a final intensity threshold may allow this algorithm to be more robust across multiple datasets, regardless of data type, resolution, and quality.
\end{abstract}
\end{opening}

\section{Introduction}
\label{intro}

Coronal holes (CHs) are large-scale, cold, low density regions on the Sun. As the electron density in CHs is 2-3 times less than that of the quiet Sun~\citep{antonucci2004}, CHs appear as dark regions in X-ray and extreme ultraviolet (EUV) images~\citep{altschuler1972,wang1996}. As regions of magnetically open field lines, CHs are the source of high-speed solar wind~\citep{krieger1973,wang1996,hassler1999,schwadron2003,antonucci2004}. Detection of CHs is thus important for solar wind forecasting and has direct implications for space weather prediction~\citep{robbins2006,vrsnak2007,gopalswamy2009,rotter2015}.

Initial research in the segmentation of CHs involved hand-drawn synoptic maps based upon 10830 \AA\ He I images \citep{harvey2002}. Recently, several automated detection methods have been proposed for segmentation of CHs. \citet{henney2005} present an algorithm which uses morphological image processing, thresholding (determined based on~\citet{harvey2002}), and smoothing to estimate the location, boundaries, and flux of CH regions in 10830 \AA\ He I images.  The results of this algorithm were compared to hand-drawn maps and were found to have approximately 3\% smaller area. \citet{malanushenko2005} use image spectroscopy in the He I line to separate CHs from quiet sun (QS) and results were qualitatively compared to National Solar Observatory (NSO) CH maps.  

Several researchers have used multispectral information to segment CHs from active regions and quiet Sun (QS).  \citet{dudokdewit2006} uses a multispectral feature vector consisting of 171 \AA, 195 \AA, 284 \AA, and 304 \AA~from SOHO/EIT in a Bayesian classification to segment CHs and qualitative results were presented.  \citet{scholl2008} use a combination of thresholding and morphological operations for segmentation of 171 \AA, 193 \AA, and 304 \AA~SOHO/EIT images for detection of CHs and magnetograms for differentiation between filaments and CHs. Results presented were qualitative in nature.  \citet{detoma2011} uses multiple wavelengths from SOHO/EIT and STEREO/EUVI (171 \AA, 195 \AA, 284 \AA, 304 \AA, 10830 \AA~or 6560 \AA) and magnetograms to segment CHs.  Magnetograms and chromosphere images are used to eliminate filaments and filament channels, CHs are segmented from each of the remaining EUV wavelengths using fixed thresholds for each wavelength, the segmented images are morphologically cleaned, and CHs are defined as a weighted average of the segmented regions in the different wavelength images.  Accuracy of the CH segmentation was qualitatively discussed.  \citet{colak2013} use fuzzy rules on 193 \AA, 304 \AA, and 1700 \AA~SDO/AIA images to segment CHs, active regions, and QS.  Filling factors of the three solar structures were found to be highly correlated to irradiance measurements. \citet{verbeeck2014} use a fuzzy c-means clustering to separate CHs and active regions from QS in SDO/AIA images.  Performance of segmentation was evaluated qualitatively and in reference to the relation of the filling factor to the solar cycle.  The method in~\citet{verbeeck2014} is used as part of the SDO image processing pipeline~\citep{martens2011}.

Several researchers have also investigated the use of single wavelength images for segmentation of CHs.  \citet{kirk2009} present a method for segmentation of polar CHs in 171 \AA, 195 \AA, and 304 \AA~SOHO/EIT images using thresholding and perimeter tracing.  The accuracy was validated on a synthetic image sequence, but is specifically designed for detection of polar CHs. \citet{krista2009} use a histogram-based local intensity thresholding technique to delineate CHs in 195 \AA\ images from SOHO/EIT, STEREO/SECCHI, and Hinode/XRT.  \citet{krista2009} use magnetic flux skewness to distinguish CHs from filaments, a threshold consistency test for robustness of their intensity thresholds, and solar wind data from ACE and STEREO to compare to CH locations.  \citet{lowder2014} extend the method of \citet{krista2009} to use data from multiple instruments, study CH persistence, and compare CH magnetic flux to potential field source surface models.  \citet{caplan2016} propose point-spread function decovolution and limb brightening correction to normalize for instrumental variation and use a two-threshold region growing approach for segmentation of CHs and study CH persistence in SDO/AIA and STEREO/EUVI images.

As opposed to the many existing methods which rely on threshold-based techniques~\citep{henney2005,scholl2008,detoma2011,colak2013,kirk2009,krista2009,lowder2014,caplan2016} for detection of CH boundaries, the work presented here uses active contour-based methods for segmentation of coronal holes. This allows independence from reliance on an absolute threshold value to define CH boundaries which may vary for different wavelengths or imaging conditions.  Instead, we use an intensity threshold only to seed the segmentation process and the final CH boundaries are determined by the active contours without edges algorithm.

In Section 2 we briefly introduce active contours without edges.  In Section 3 we describe the datasets, our proposed CH segmentation, and our study of reasonable segmentation parameters.  In Section 4 we present our results and check consistency of those results by a study of the magnetic unipolarity underlying our segmented CHs, a comparison to high-speed solar wind measurements, and a comparison to another CH segmentation method.  Finally, in Section 5 we conclude and briefly summarize some future work.

\section{Active Contours Without Edges}
\label{sec:AC}
Active contours (or ``snakes'') formulate the segmentation of an image into foreground and background as an energy minimization process subject to external constraint forces (e.g., smoothness) and internal image forces (e.g., gradient)~\citep{kass1988}.  A common image force used in segmentation is the intensity gradient which pulls the contour toward image edges.  The use of the gradient can be problematic, however, for segmentation of more complicated images with texture or smoothly varying edges.  

\citet{chan2001} propose a formulation of active contours ``without edges'' as follows.  The energy functional $F$ to be minimized is defined as
\begin{equation}
 F(m_o,m_i,C) = \mu\ell(C)+\lambda_i\int\limits_{C_{i}}|I(x,y)-m_i|^2dxdy + \lambda_o\int\limits_{C_{o}}|I(x,y)-m_o|^2dxdy,
 \label{eq:acwe}
\end{equation}
where $C$ is the contour and $C_{i}$ and $C_{o}$ denote the region inside and outside of $C$, respectively; $m_i$ and $m_o$ are the average intensity values inside and outside of $C$, respectively; $\mu$, $\lambda_i$, and $\lambda_o$ are constant weighting terms; $\ell(C)$ denotes the length of contour $C$; and $I(x,y)$ is the intensity value of the image at pixel $(x,y)$.  The minimization of this energy functional seeks a short contour $C$ (controlled by the first term) with homogeneous intensity (although not the same average value) both inside and outside contour $C$ (controlled by the second and third terms).  The choice of parameters $\mu$, $\lambda_i$, and $\lambda_o$ control the relative strength of each term. An initial curve will propagate in normal directions to iteratively minimize the energy functional, and can be implemented using the Mumford-Shah level set formulation~\citep{mumford1989}. We use the \verb+activecontour+ function provided in the MATLAB image processing toolbox in this work and choose $\mu=0$ to ignore length constraints on the contour.

\section{Segmentation of Coronal Holes}
\label{sec:image_analysis}
\begin{table}[t!]
\begin{tabular}{lccccc}\hline
 CR   & Start Time          & End Time            & Wavelength & \# Files\\\hline
 2099 & 2010 Jul. 13, 09:35 & 2010 Aug. 09, 14:44 & 193 \AA    & 28\\
 2106 & 2011 Jan. 20, 09:15 & 2011 Feb. 16, 17:26 & 193 \AA    & 28\\
 2133 & 2013 Jan. 25, 19:56 & 2013 Feb. 22, 04:06 & 193 \AA    & 28\\
 2150 & 2014 May 04, 14:42  & 2014 May 31, 19:55  & 193 \AA    & 28\\\hline
\end{tabular}
\caption{Datasets used for testing the CH segmentation algorithm.}
\label{tab:Dataused}
\end{table}
\subsection{Datasets}
The data used here are from Carrington rotations (CRs) 2099 and 2106 as used in the SHINE 2014 conference\footnote{\url{http://shinecon.org/PastMeetings/Meetings_past.php}} and an additional two CRs (2133 and 2150) from nearer the peak of the solar cycle.  The data are summarized in Table \ref{tab:Dataused} where we include one image per day from SDO/AIA~\citep{lemen2012} for the duration of each CR.  Along with the AIA datasets, additional datasets are used to check the consistency of the segmented CH boundaries.  Concurrent Helioseismic and Magnetic Imagery (HMI)~\citep{scherrer2012} magnetogram data are used for measuring magnetic flux distribution within the segmented CHs and Advanced Composition Explorer (ACE)~\citep{chiu1998} solar wind measurements are used for comparison with CH location.  The remainder of this section details the image processing steps for segmentation of CHs using active contours without edges (ACWE).  The limb brightening correction as described in~\cite{verbeeck2014} is first applied before all subsequent processing.  Additionally, all processing is applied to reduced resolution images (decimation by $8\times$) to speed computation; all processing described here can be applied without modification to the original high resolution images.  We will briefly discuss the computational requirements and the effects of $8\times$ decimation in Section~\ref{sec:computation}.

\subsection{Initialization of the Contour by Thresholding}
\label{sec:threshold}
An initial contour must be defined which will subsequently evolve according to the ACWE energy functional. This initial contour should to satisfy several constraints, namely that the initial contour should: (1) contain the CH on the interior, (2) contain all regions with strong potential to be part of a CH, and (3) not contain extraneous regions.  As such, the contour is initialized by thresholding at a low intensity which is certain to be within the CH, similar to the seeding threshold of \citet{caplan2016}.  In this work a threshold of $t_0=\alpha I_{QS}$ is chosen where $I_{QS}$ is an estimate of the QS intensity.  First, a solar disk mask $SD$ is defined using the solar radius metadata from the SDO files in order to ignore the off-disk pixels:
\begin{equation}
 SD(x,y) = \begin{cases}
            1,&~\text{if}~\sqrt{(x-x_c)^2+(y-y_c)^2}<R\\
            0,&~\text{else}
           \end{cases},
\end{equation}
where $(x_c,y_c)$ is the center pixel of the image and $R$ is the solar radius in pixels.  Then, the QS intensity $I_{QS}$ is defined as the most frequent intensity within the solar disk~\citep{detoma2011} as
\begin{equation}
 I_{QS} = \lbrace I_k | h(I_k)>h(I_j),~k\ne j \rbrace,
 \label{eq:quiet_sun}
\end{equation}
where $h(\cdot)$ is the histogram of intensities with 100 bins, $I_k$ is the $k$-th intensity bin, and we only consider pixels in the solar disk $\{(x,y)|SD(x,y)=1\}$.

\begin{figure}
  \centering
  \subfloat[13 July 2010, 09:35]{\includegraphics[width=0.33\textwidth]{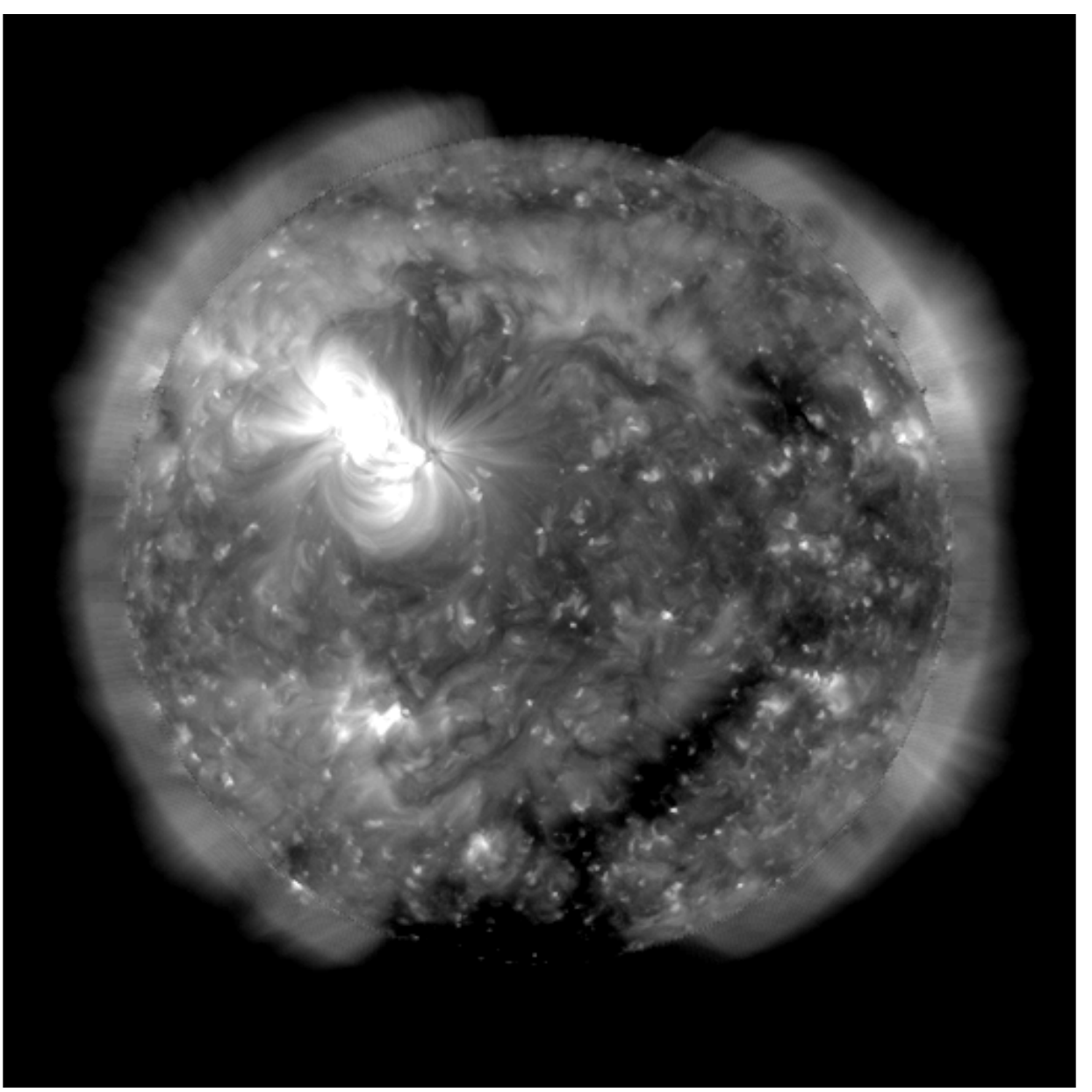}}~~
  \subfloat[Image thresholded at $t_0=0.3 I_{QS}$ (82 for this image).]{\includegraphics[width=0.33\textwidth]{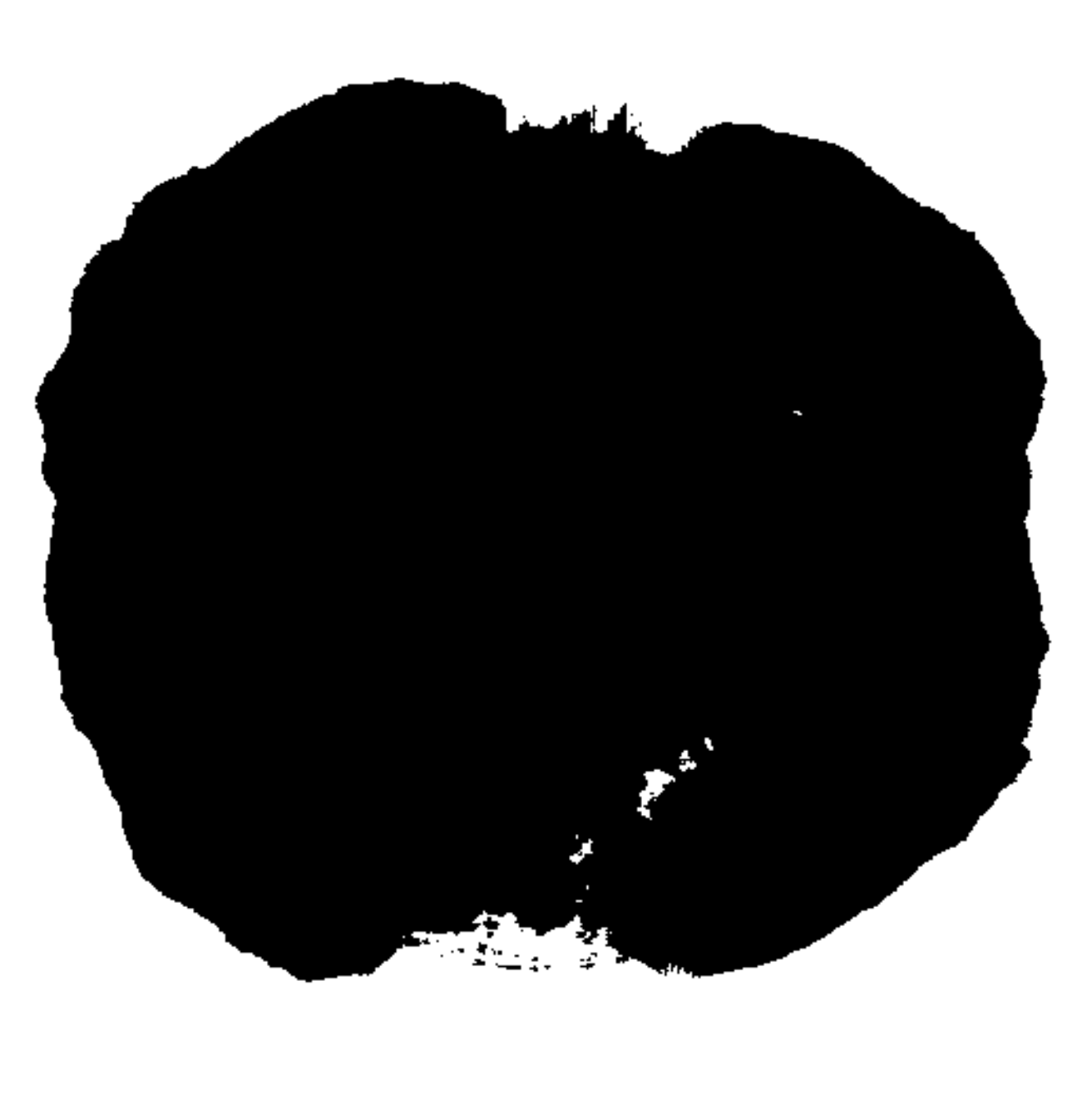}}\\
  \subfloat[Removal of off-disk regions.]{\includegraphics[width=0.33\textwidth]{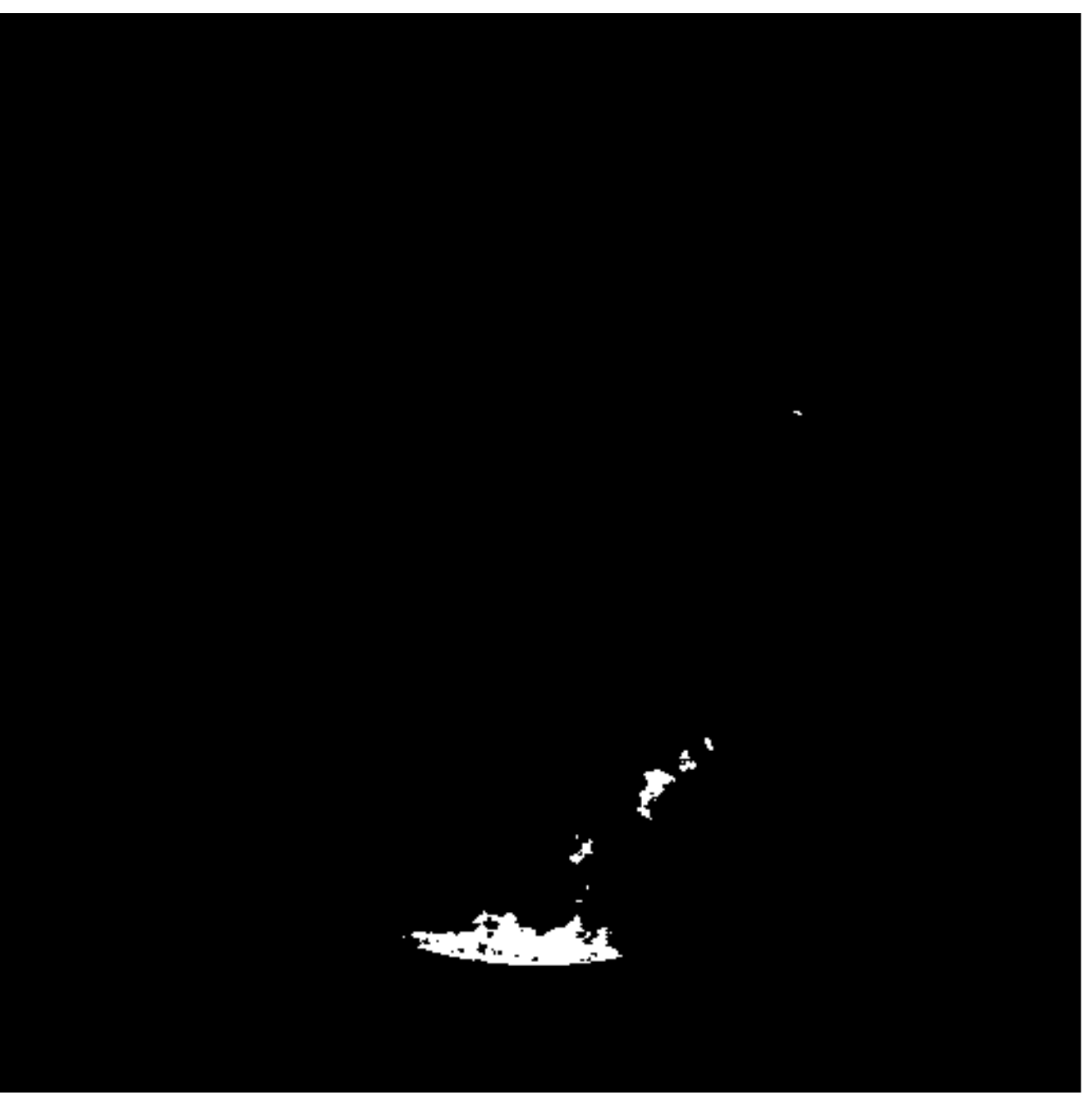}}~~
  \subfloat[Holes filled.]{\includegraphics[width=0.33\textwidth]{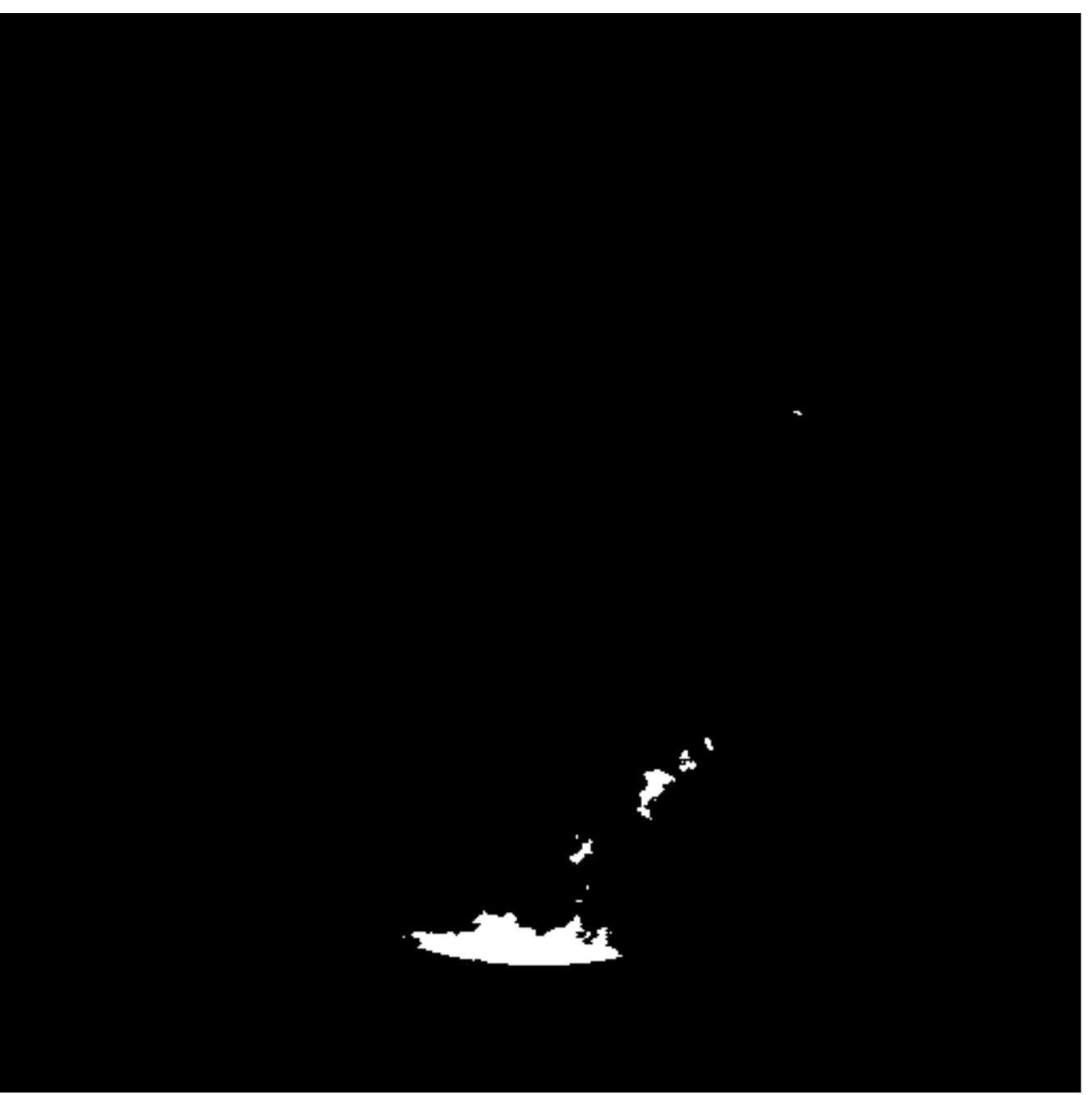}}
  \caption{Initialization of the CH contour by thresholding.  (a) Original image, clipped to $[100,2500]$ and log-scaled, (b) image thresholded to 0.3 of the QS intensity, (c) removal of the off-disk regions, and (d) filling of holes in the initialization.}
  \label{fig:initialize}
\end{figure}

As qualitative illustration Figure~\ref{fig:initialize}(a) shows an image from CR 2099 and Figure~\ref{fig:initialize}(b) shows the regions corresponding to a threshold of $t_0=0.3 I_{QS}$, i.e., $I(x,y)\le t_0$.  Since off-disk regions are also low intensity pixels, they are included in the thresholded image.  The solar disk mask $SD$ is used to remove the off-disk pixels as shown in Figure~\ref{fig:initialize}(c).  As a last step in initialization, holes in the initial mask are filled using a morphological hole filling operation.  This step is required since, as specified in the documentation\footnote{\url{http://www.mathworks.com/help/images/ref/activecontour.html}}, the MATLAB \verb+activecontour+ function may yield unpredictable results if initial masks contain holes.  The final initialization with holes filled is shown in Figure~\ref{fig:initialize}(d); this initialization mask is denoted $M^{(0)}$ in further discussion.  Section~\ref{sec:params} will discuss the effect of different thresholds in initialization of the ACWE contour. Empirically, in limited studies, we have found no significant difference in performance when holes are not filled in the initial mask and additionally find large numbers of holes in the converged CH boundaries.  This may indicate that the hole filling operation is not needed.  Future work will study the effect of removing the hole filling operation on the final converged CH boundaries.

\subsection{Ignoring the Off-Disk Regions in ACWE Evolution}
\begin{figure}
  \centering
  \subfloat[13 July 2010, 09:35]{\includegraphics[width=0.24\textwidth]{figures/CR2099_file1}}~~~~~
  \subfloat[Image with off disk regions set to the mean value of pixels outside the initialization.]{\includegraphics[width=0.24\textwidth]{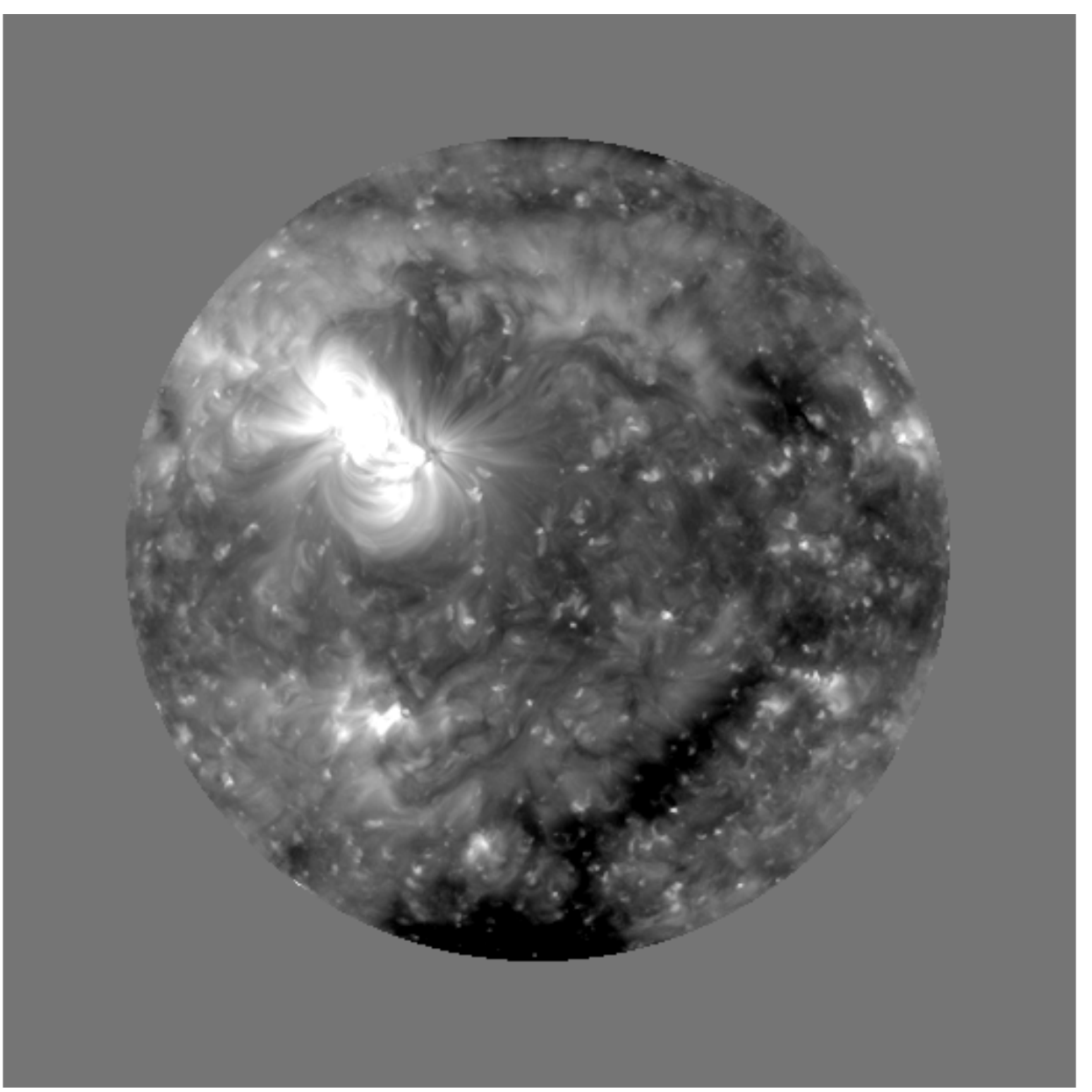}}
  \caption{ACWE can be forced to ignore the off-disk regions by setting all off-disk pixels to the mean value of the non-CH pixels on the Sun.  Both images are clipped to $[100,2500]$ and log scaled for visualization.  The off disk regions are set to 439 in this case, assuming an initialization at $t_0=0.3I_{QS}=82$ in this case.}
  \label{fig:off_disk_mean}
\end{figure}

As seen in Figure~\ref{fig:initialize}(b), pixels located off the solar disk are similarly dark to those pixels of CHs.  It is desired, however, that ACWE evolution ignore these off-disk pixels.  Here we propose a means to ignore the off-disk pixels while still using the out-of-the-box MATLAB \verb+activecontour+ function; future work may find computational advantages in modifying the underlying ACWE evolution to ignore off-disk pixels.  If the off-disk pixels are set to a constant value equal to the mean value of pixels outside the CH, i.e., $m_o$ in Equation (\ref{eq:acwe}), the off-disk pixels will contribute zero energy and will thus be effectively ignored in the ACWE contour evolution.  The following steps are thus iterated until convergence (defined in Section~\ref{sec:stopping}):
\renewcommand{\labelenumi}{Step~\theenumi.}
\begin{enumerate}[indent=0]
 \item Set $M^{(k)}=C_i^{(k-1)}$ where $M^{(k)}$ denotes the initialization mask used for iteration $k$ and $C_i^{(k-1)}$ is the interior of the contour $C$ in the ACWE segmentation from iteration $k-1$.
 \item Set
  \begin{equation}
   I^{(k)}(x,y) = \begin{cases}
                  I(x,y),&~\text{if}~\sqrt{(x-x_c)^2+(y-y_c)^2}<R\\
                  m_o^{(k-1)},&~\text{else}
                 \end{cases},
  \end{equation}
  where $I^{(k)}$ denotes the image used as input for iteration $k$ and $m_o^{(k-1)}$ is the mean value of pixels outside the contour at iteration $k-1$. See Figure~\ref{fig:off_disk_mean} for an example of $I^{(0)}$.
 \item Iterate the ACWE algorithm until convergence with $I^{(k)}$ as the input image and $M^{(k)}$ as the initial mask.
\end{enumerate}
By definition $C_i^{(0)}=M^{(0)}$ and $m_o^{(0)}$ is the mean value of pixels within the solar disk and outside of the initial mask $M^{(0)}$:
\begin{equation}
 m_o^{(0)} = \frac{\sum_{x}\sum_{y}I(x,y)SD(x,y)M^{(0)}(x,y)}{|SD\cap \overline{M^{(0)}}|}
\end{equation}
where $\cap$ is the set intersection operator, $\overline{M^{(0)}}$ is the complement of $M^{(0)}$, and $|\cdot|$ is the cardinality operator.

\subsection{Stopping Criterion}
\label{sec:stopping}
A stopping criterion is defined as a measure of percentage change in area between subsequent ACWE iterations.  The total percentage change in area $\Delta A_{tot}$ is calculated with respect to the previous iteration by  
\begin{equation}
 \Delta A_{tot} = \frac{|C_i^{(k)}\ominus C_i^{(k-1)}|}{|C_i^{(k)}|},
\end{equation}
where $\ominus$ is the symmetric set difference operator (i.e., pixels in $C_i^{(k)}$ or $C_i^{(k-1)}$ but not both).  There are, however, some pixels which will continuously oscillate between the interior $C_i$ and exterior $C_o$ of the segmentation between iterations.  This oscillation is due to the discrete nature of the digital image, since a pixel cannot partially belong to the interior and exterior. The energy minimization process may evolve the contour away from a pixel only to overshoot the minimum; evolving the contour back toward the pixel overshoots in the other direction.

Mitigation of the effect of these oscillating pixels on the stopping criterion is achieved by defining a percentage of the pixels which have changed (either added to or subtracted from the previous segmentation) but are not those pixels which continuously change.  The number of times a pixel $(x,y)$ has changed from interior to exterior is accumulated in matrix $B$:
\begin{equation}
 B^{(k)}(x,y) = \begin{cases}
           B^{(k-1)}(x,y)+1,&~\text{if}~C_i^{(k)}(x,y)\ominus C_i^{(k-1)}(x,y)\\
           B^{(k-1)}(x,y),&~\text{else}
          \end{cases},
\end{equation}
where $B^{(0)}=\mathbf{0}^{N\times N}$ and $N$ is the size of image $I$.  The percentage of newly changed pixels $\Delta A_{new}$ is calculated as
\begin{equation}
 \Delta A_{new} = \frac{|(C_i^{(k)}\ominus C_i^{(k-1)})\cap B_1|}{|(C_i^{(k)}\ominus C_i^{(k-1)})\cap B_n|},
\end{equation}
where $B_1$ are pixels which have changed exactly once and $B_n$ are pixels which have changed more than once. ACWE iteration is stopped if 
$\Delta A_{new}=0$.

\subsection{Choice of ACWE Parameters for AIA Imagery}
\label{sec:params}
ACWE will be applied for segmentation of CHs in a series of AIA images and appropriate parameters for automated application of the ACWE algorithm must therefore be determined.  These parameters include parameter $\alpha$ which controls the threshold used to initialize the algorithm, parameter $\lambda_i$ which weights the energy term associated with the region inside the contour, and parameter $\lambda_o$ which weights the energy term associated with the region outside the contour.  $\lambda_i$ and $\lambda_o$ can be simultaneously changed by varying the ratio $\lambda_i/\lambda_o$, leaving only two parameters to optimize ($\lambda_i/\lambda_o$ and $\alpha$).  We first present a qualitative justification for threshold parameter $\alpha$ in Section~\ref{sec:threshold_qualitative} followed by a qualitative study of the effect of $\lambda_i/\lambda_o$ in Section~\ref{sec:homogeneity_qualitative}.  We will then provide a quantitative analysis of both $\alpha$ and $\lambda_i/\lambda_o$ via grid search in Section~\ref{sec:grid_search}.

\subsubsection{Initialization Threshold Parameter $\alpha$--Qualitative Analysis}
\label{sec:threshold_qualitative}
\begin{figure}
  \centering
  \subfloat[$\alpha=0.30$.]{\includegraphics[width=0.19\textwidth]{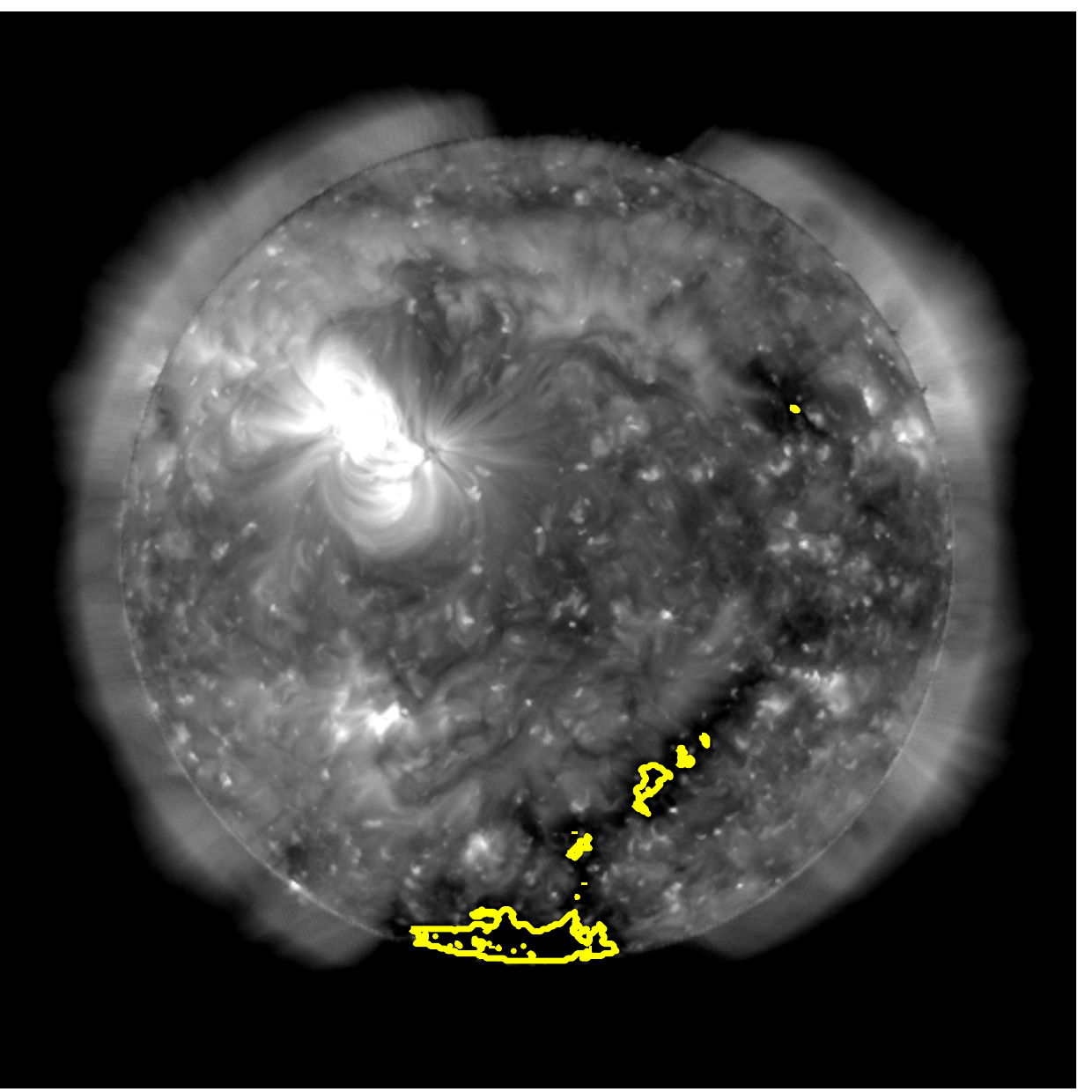}}~
  \subfloat[$\alpha=0.40$.]{\includegraphics[width=0.19\textwidth]{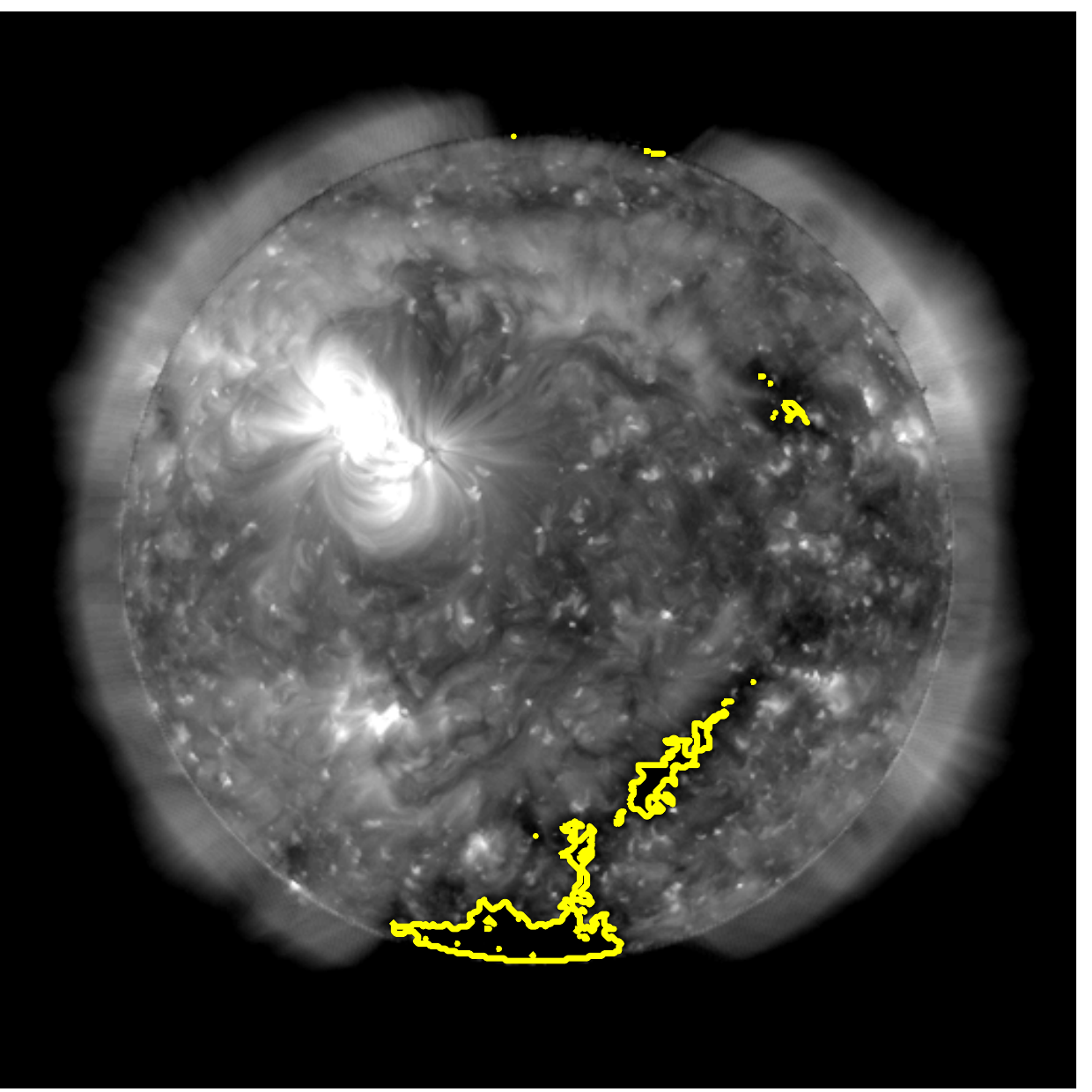}}~
  \subfloat[$\alpha=0.50$.]{\includegraphics[width=0.19\textwidth]{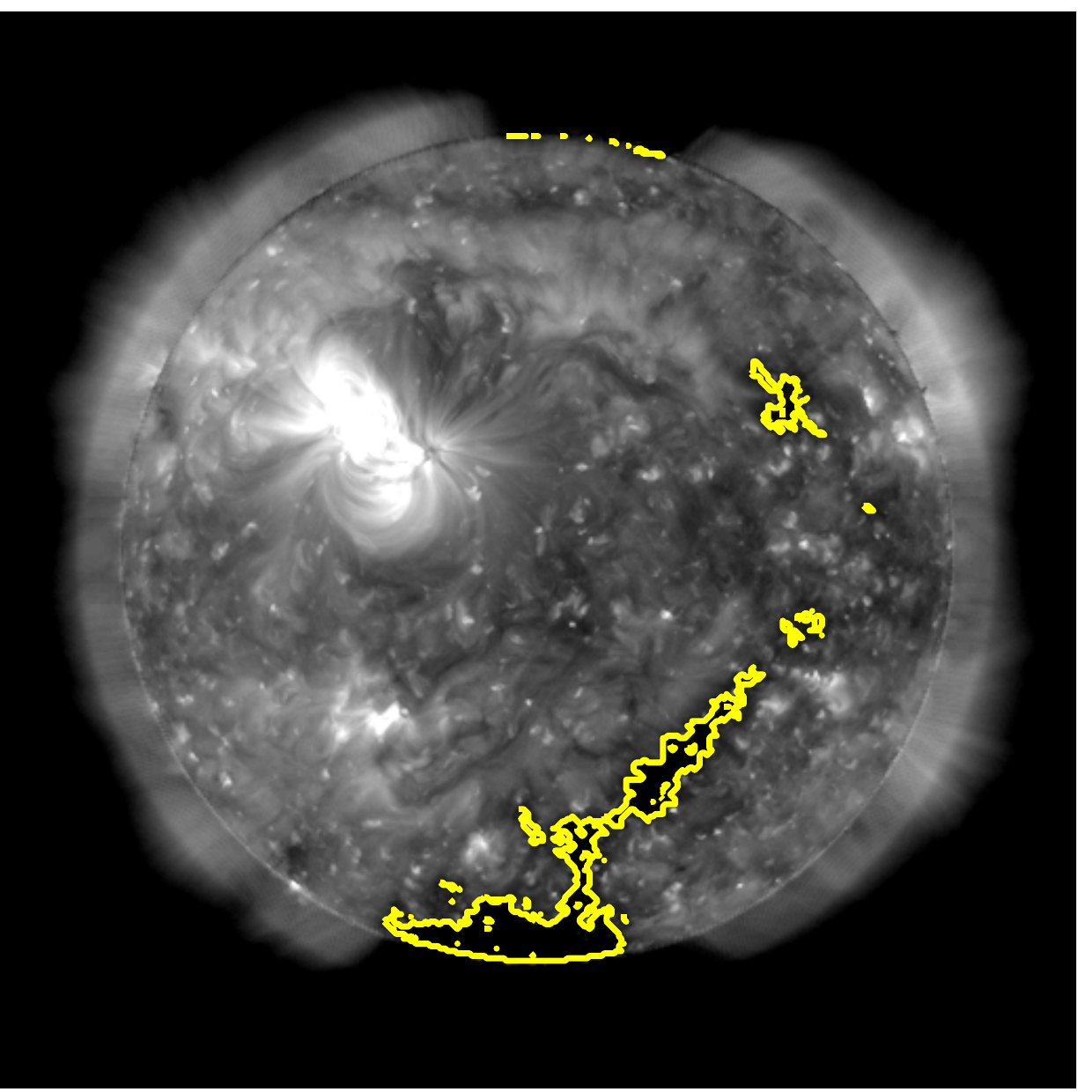}}~
  \subfloat[$\alpha=0.60$.]{\includegraphics[width=0.19\textwidth]{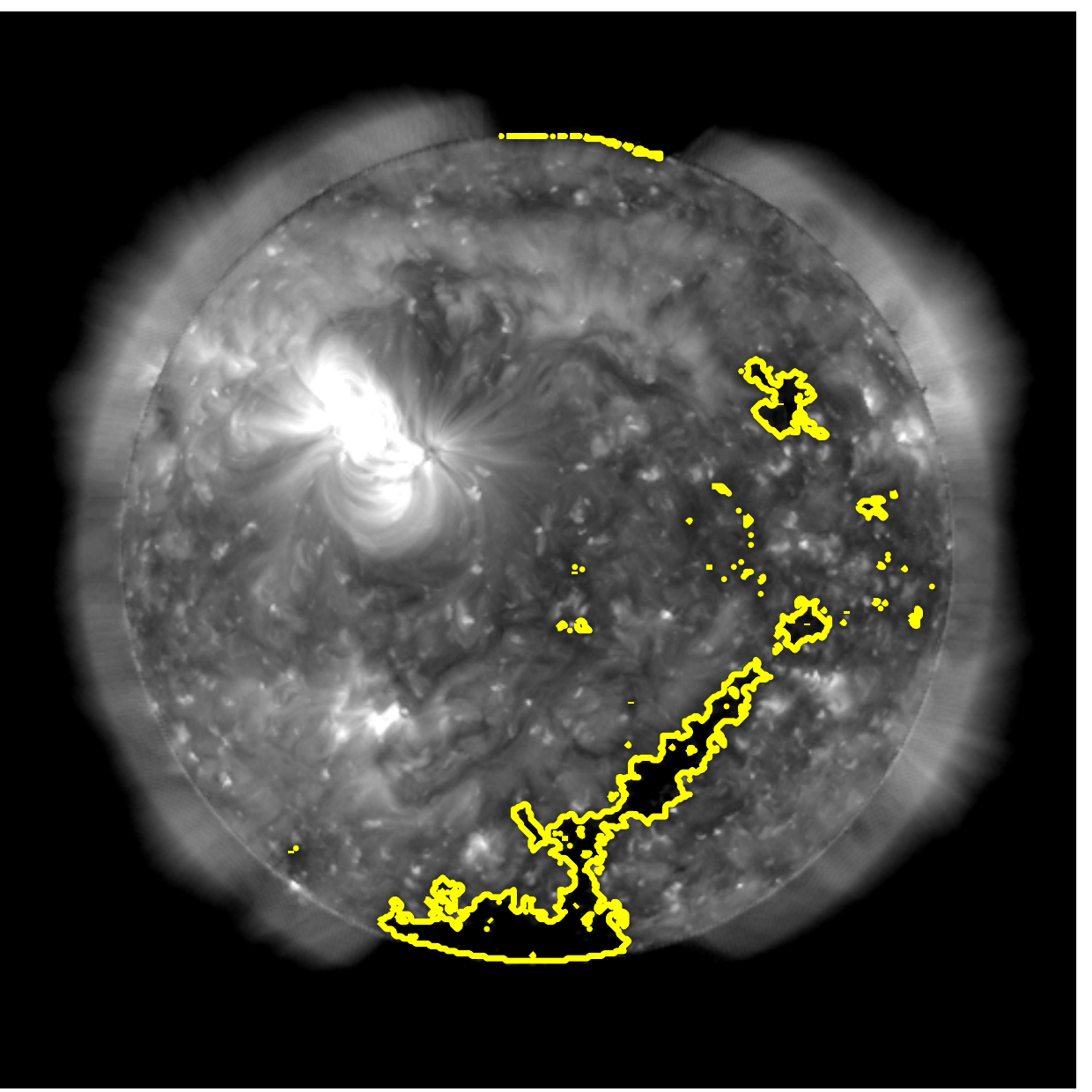}}~
  \subfloat[$\alpha=0.70$.]{\includegraphics[width=0.19\textwidth]{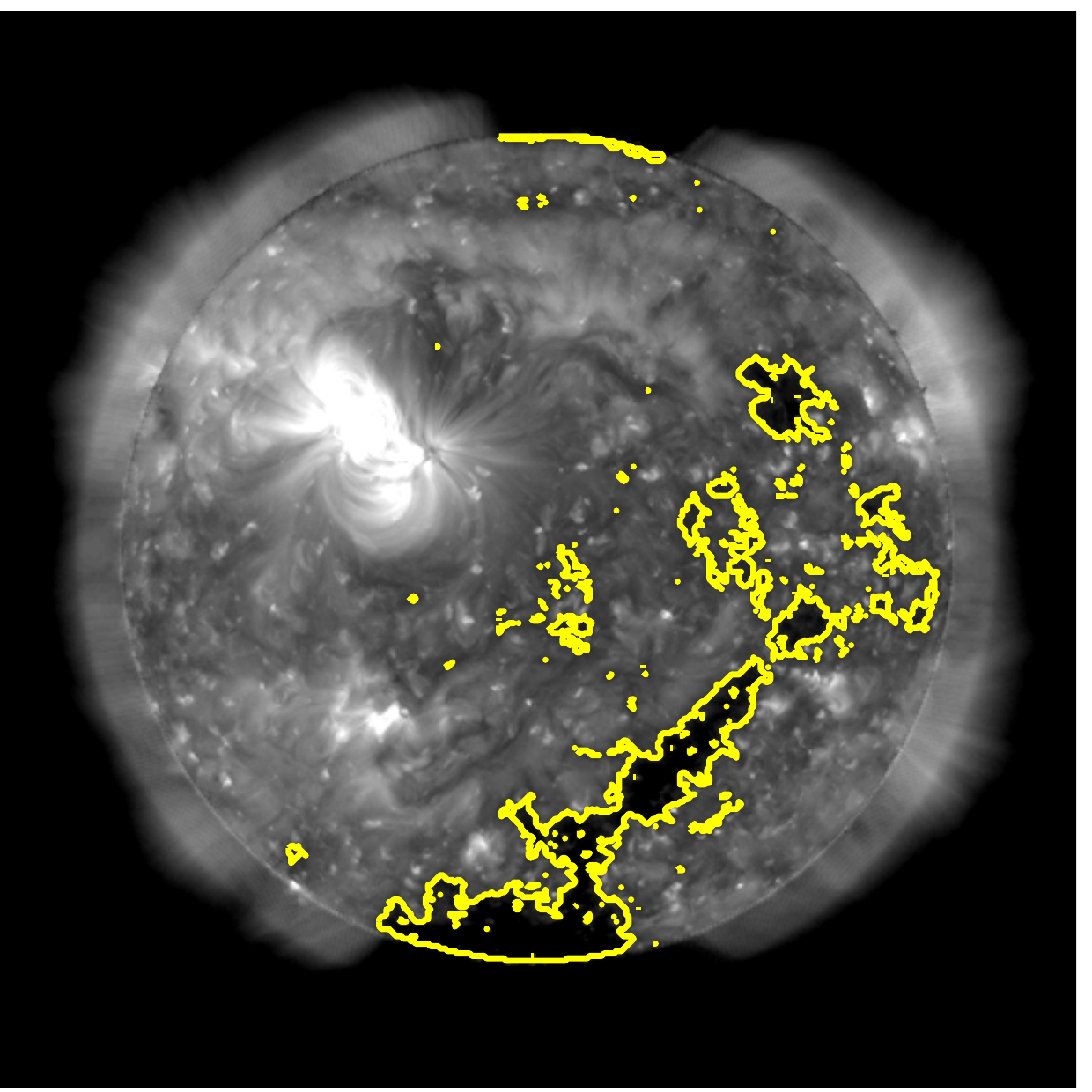}}
  \caption{Initialization for $I(x,y)\le t_0$ for different $t_0=\alpha I_{QS}$, where the yellow outlines indicate the thresholded region $I(x,y)\le t_0$.  Note that $0.3\le\alpha\le0.4$ initializes within the CHs without too many extraneous regions.}
  \label{fig:threshold}
\end{figure}

Figure~\ref{fig:threshold} qualitatively illustrates some results for different initialization thresholds $t_0=\alpha I_{QS}$ for values of $\alpha\in[0.3,0.7]$.  This range of $\alpha$ is chosen on the basis of the discussion in~\citet{krista2009}.  As it is unclear how \citet{krista2009} define QS intensity, we use the definition as in Equation~(\ref{eq:quiet_sun}) and as described by~\citet{detoma2011}.  From Figure~\ref{fig:threshold} we see that $0.3\le\alpha\le0.4$ initializes within the CHs without too many extraneous regions. We will quantitatively assess the effect of the initial threshold parameter $\alpha$ in Section~\ref{sec:grid_search}.

\subsubsection{Homogeneity Parameters $\lambda_i$ and $\lambda_o$--Qualitative Analysis}
\label{sec:homogeneity_qualitative}
\begin{figure}
  \centering
  \subfloat[$\lambda_i/\lambda_o=7$, $276\pm76$ inside, $790\pm750$ outside.]{\includegraphics[width=0.19\textwidth]{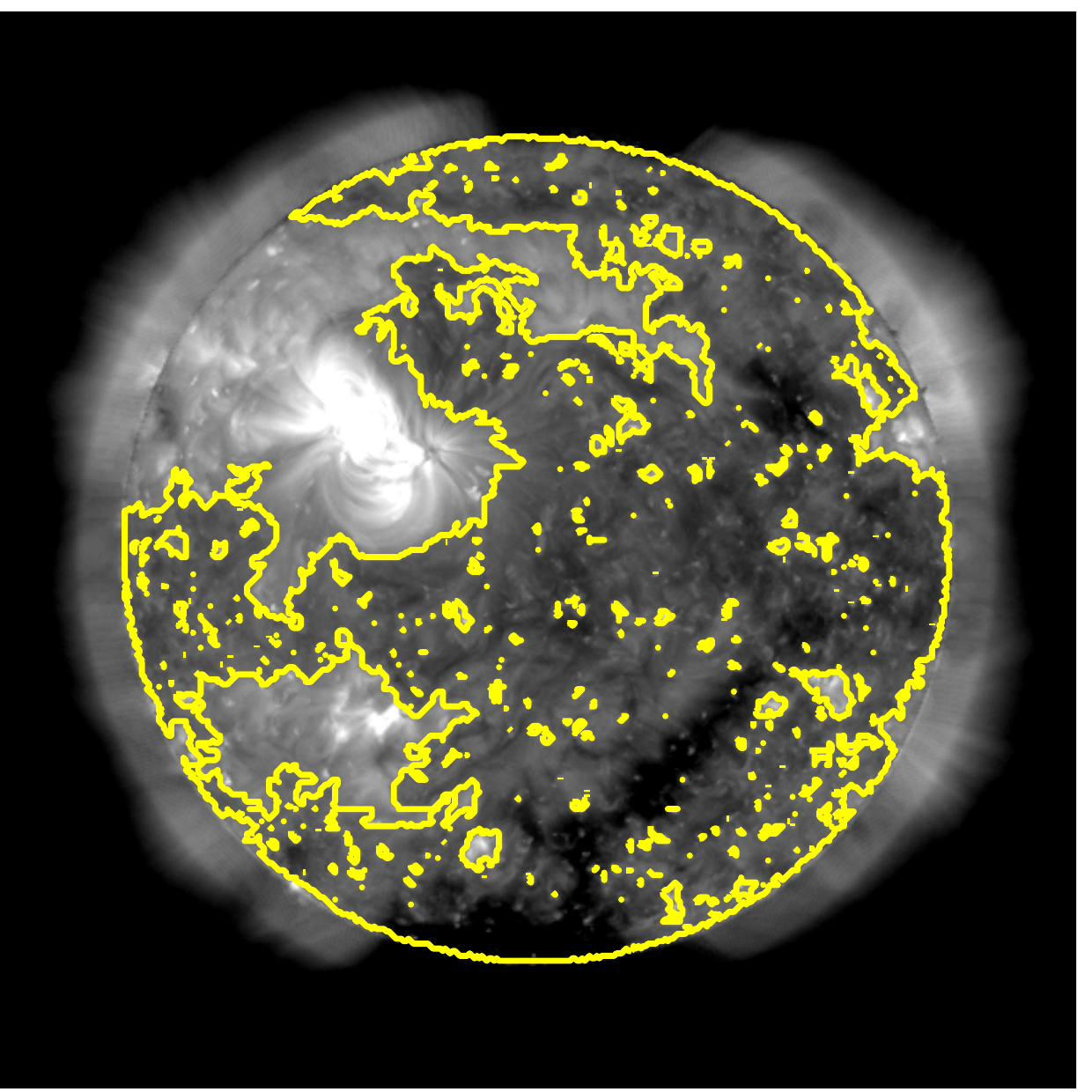}}~
  \subfloat[$\lambda_i/\lambda_o=8$, $139\pm52$ inside, $458\pm494$ outside.]{\includegraphics[width=0.19\textwidth]{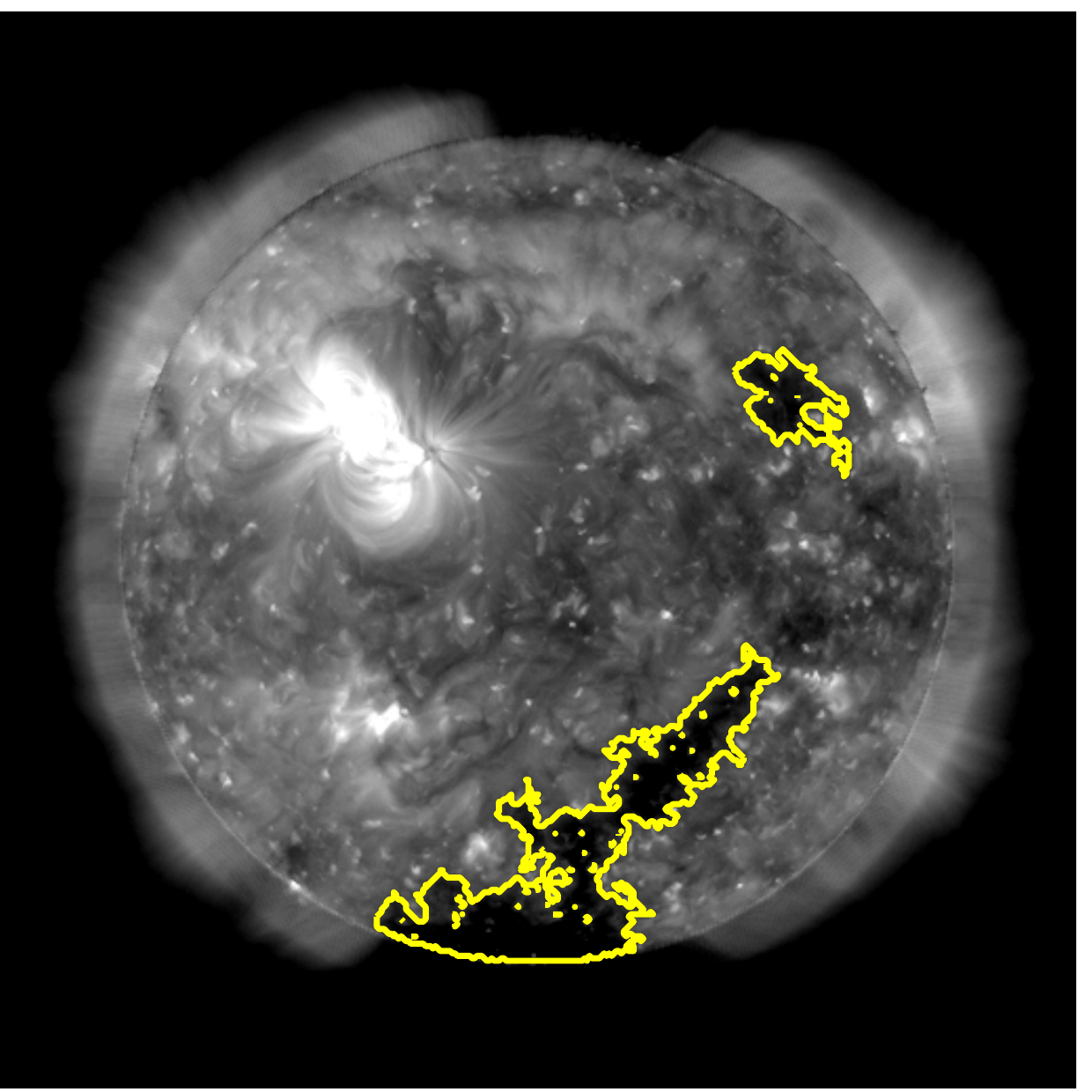}}~
  \subfloat[$\lambda_i/\lambda_o=10$, $129\pm 47$ inside, $456\pm492$ outside.]{\includegraphics[width=0.19\textwidth]{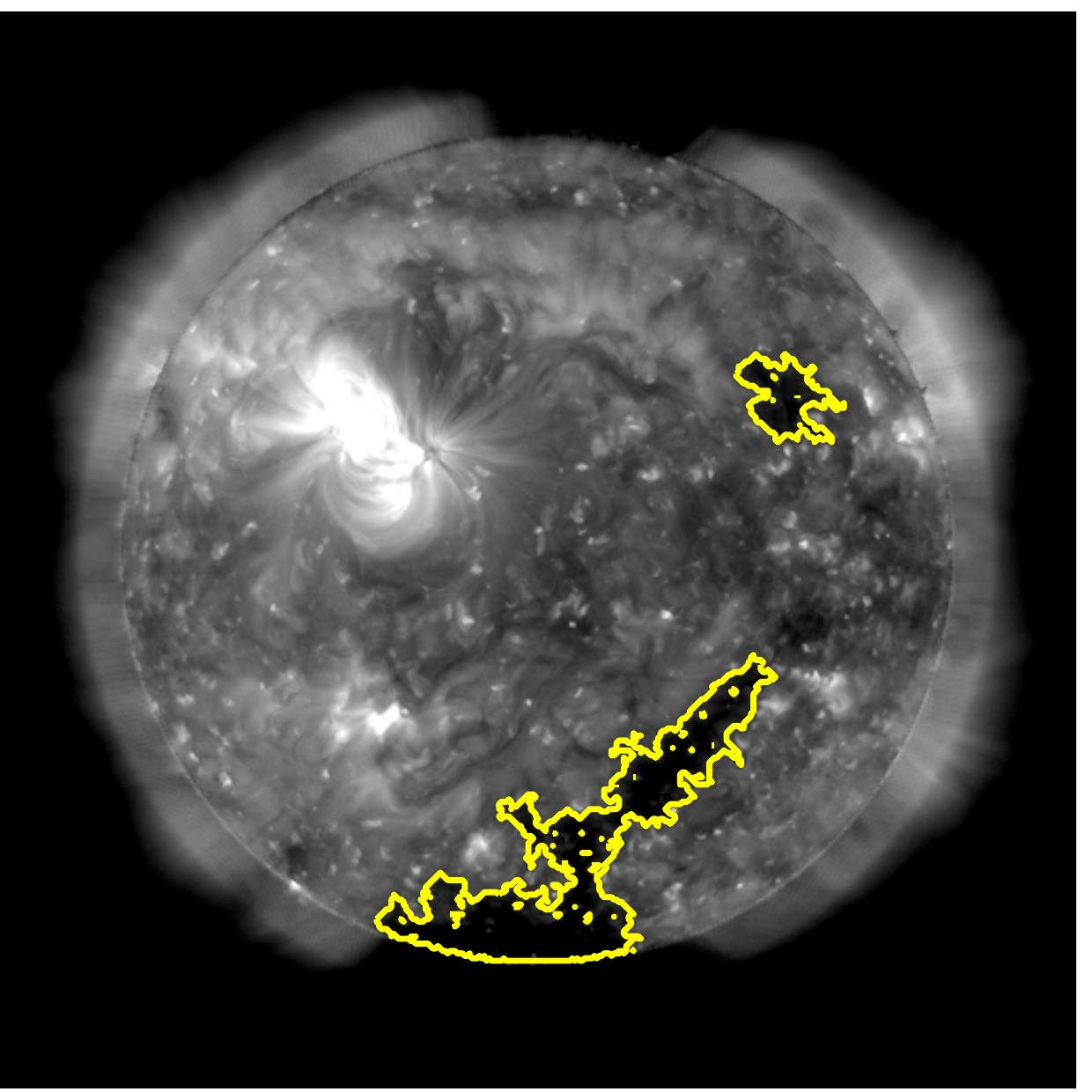}}~
  \subfloat[$\lambda_i/\lambda_o=50$, $80\pm 27$ inside, $443\pm485$ outside.]{\includegraphics[width=0.19\textwidth]{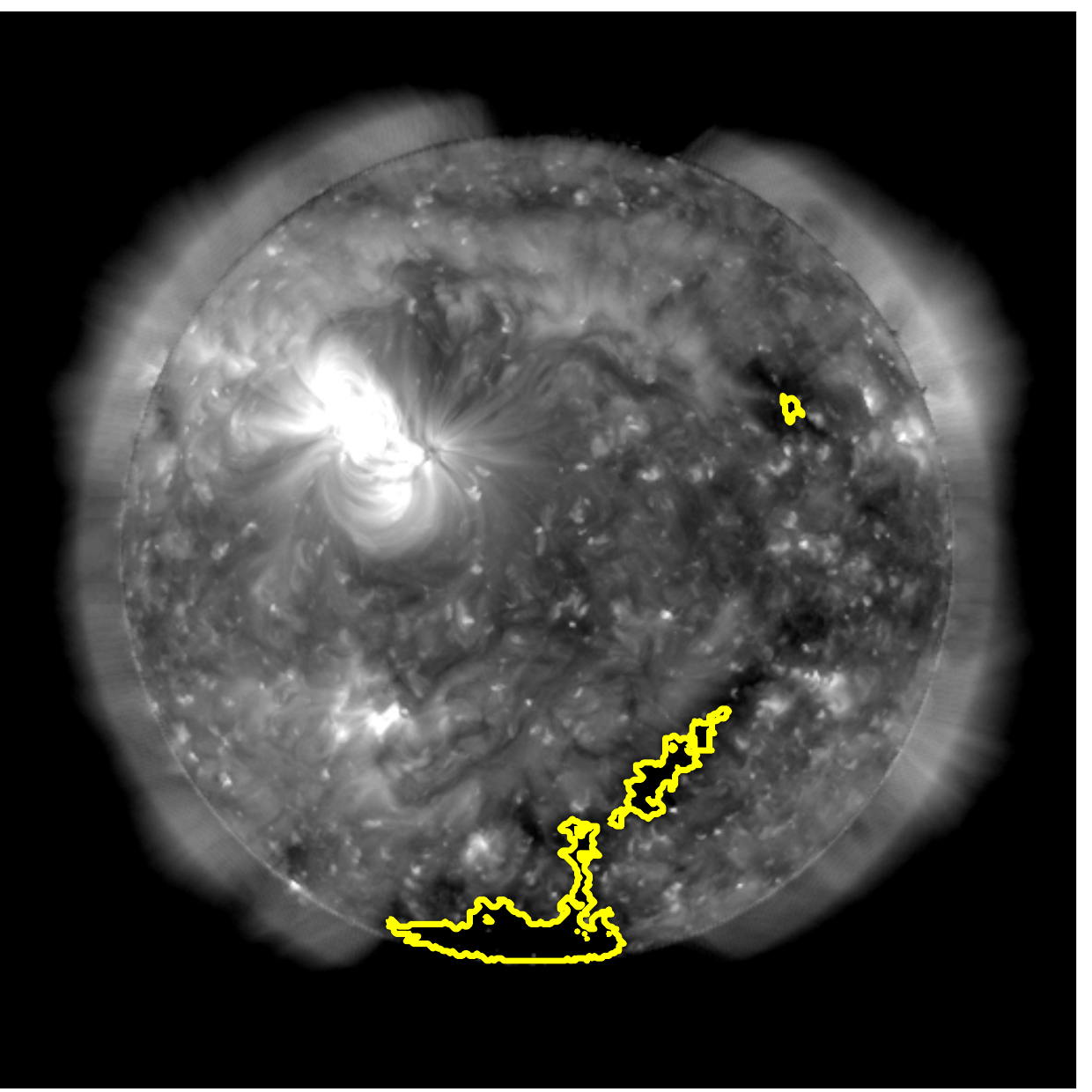}}~
  \subfloat[$\lambda_i/\lambda_o=100$, $69\pm 22$ inside, $441\pm 484$ outside.]{\includegraphics[width=0.19\textwidth]{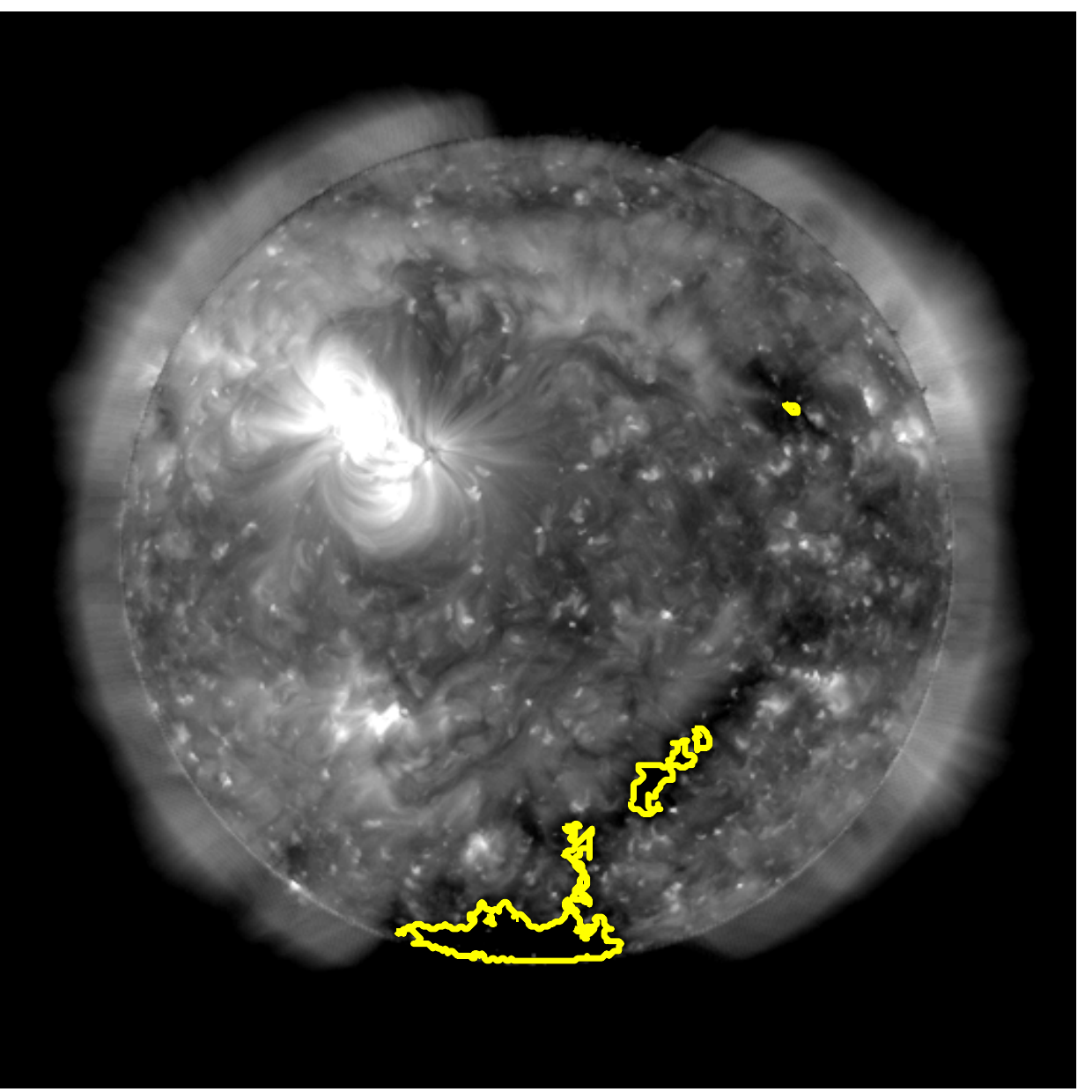}}
  \caption{Converged CHs for different $\lambda_i/\lambda_o$, along with mean $\pm$ standard deviation of intensities inside and outside the CHs as a measure of homogeneity. Initial threshold $t_0=0.3I_{QS}$ for these results.  There is a shift in performance between $\lambda_i/\lambda_o=7$ and $\lambda_i/\lambda_o=8$, with a ratio close to 10 providing a reasonable CH segmentation.}
  \label{fig:foreground}
\end{figure}
Homogeneity parameters $\lambda_i$ and $\lambda_o$ control the direction of movement of the evolving contour based on the homogeneity of intensities inside and outside the contour. Specifically, $\lambda_i$ weights the homogeneity of intensities inside the contour whereas $\lambda_o$ weights the homogeneity outside the contour. From Figure~\ref{fig:foreground}, we note that $\lambda_i/\lambda_o$ should be larger to place preference on keeping the interior of the contour (interior of the CH) more homogeneous. This coincides with the assumption that the interior of CHs will be more homogeneous than the remainder of the Sun taken in aggregate (including QS and active regions). As a measure of the homogeneity of intensities inside and outside the CH boundaries, we specify the standard deviation of intensities in Figure~\ref{fig:foreground}.  We find a smaller standard deviation (larger homogeneity) for intensities within the CHs than outside the CHs and that the homogeneity increases as $\lambda_i/\lambda_o$ increases.  We further note from Figure~\ref{fig:foreground} that a ratio close to 10 appears to provide a reasonable segmentation of the CHs and that the segmentation varies very slowly with respect to $\lambda_i/\lambda_o$ for $\lambda_i/\lambda_o\ge 10$.

\subsubsection{Initialization Threshold Parameter $\alpha$ and Homogeneity Ratio $\lambda_i/\lambda_o$--Quantitative Analysis}
\label{sec:grid_search}
A grid search is conducted to determine a reasonable combination of parameters $\alpha$ and $\lambda_i/\lambda_o$.  The search is conducted over $\alpha\in[0.3,0.7]$ in steps of 0.1 and over $\lambda_i/\lambda_o\in[1,10]$ in steps of 1, $\lambda_i/\lambda_o\in[10,100]$ in steps of 5, and $\lambda_i/\lambda_o\in[100,200]$ in steps of 10.  Four images are used from each of the datasets, i.e., a weekly cadence for CRs 2099, 2106, 2133, and 2150 and the ACWE algorithm is allowed to converge for each combination of parameter values.  We present results for all four CRs, and additionally study the effect of solar cycle on the parameter choices by separating results according to the low activity CRs 2099 and 2106 and the high activity CRs 2133 and 2150.

\begin{figure}
 \centering
 \subfloat[CRs 2099 and 2106.]{\includegraphics[width=0.3\textwidth]{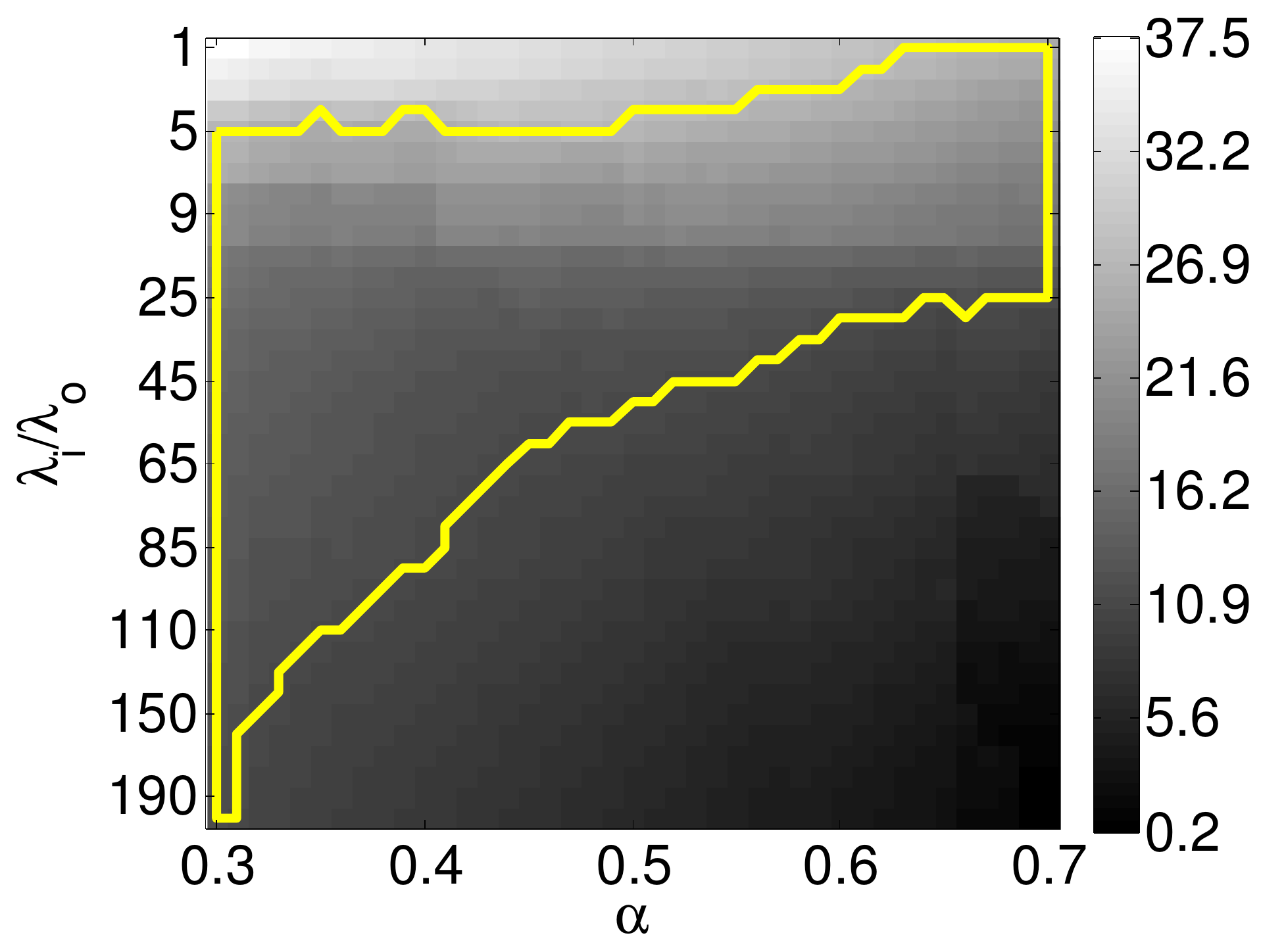}}~~
 \subfloat[CRs 2099 and 2106.]{\includegraphics[width=0.3\textwidth]{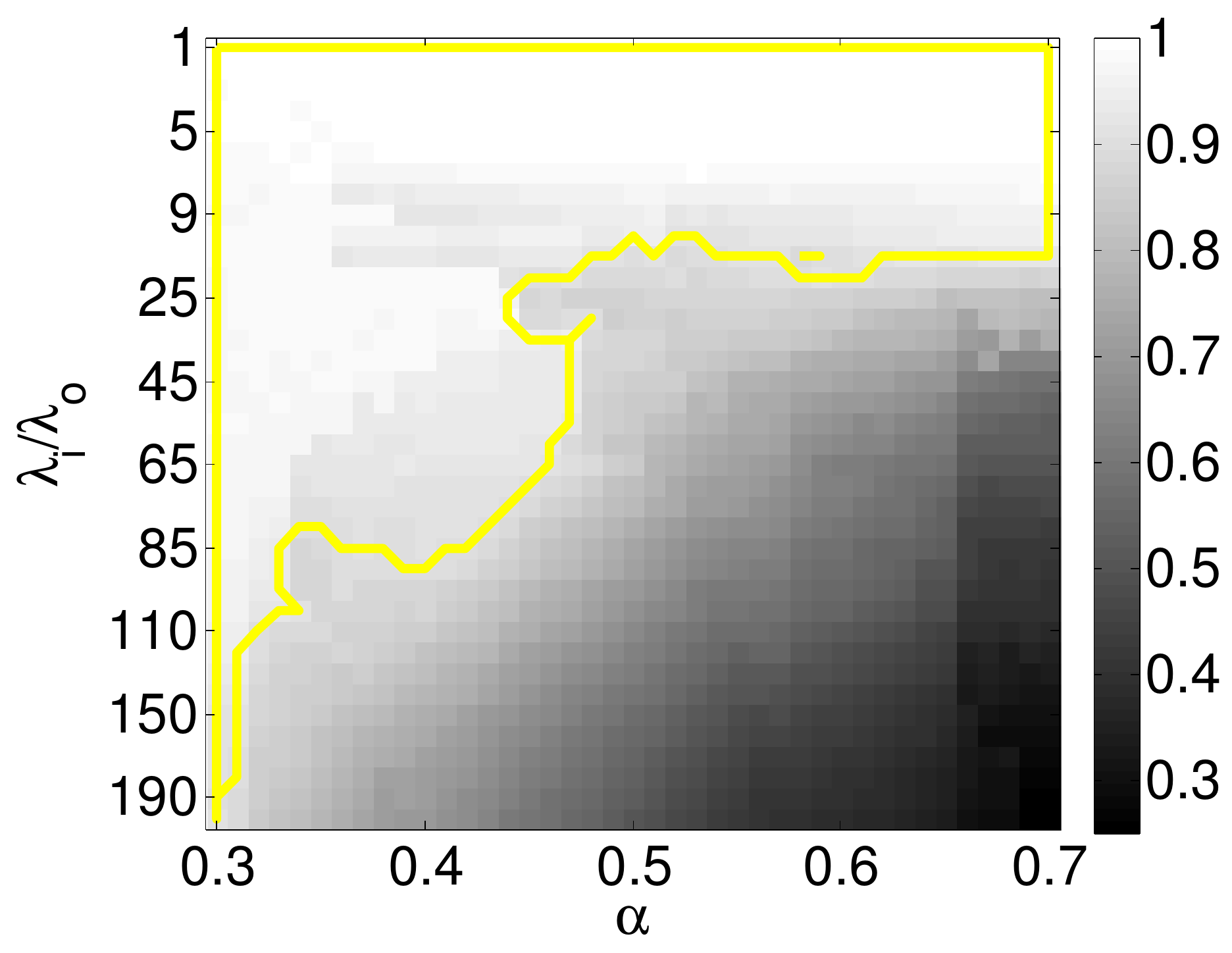}}~~
 \subfloat[CRs 2099 and 2106.]{\includegraphics[width=0.3\textwidth]{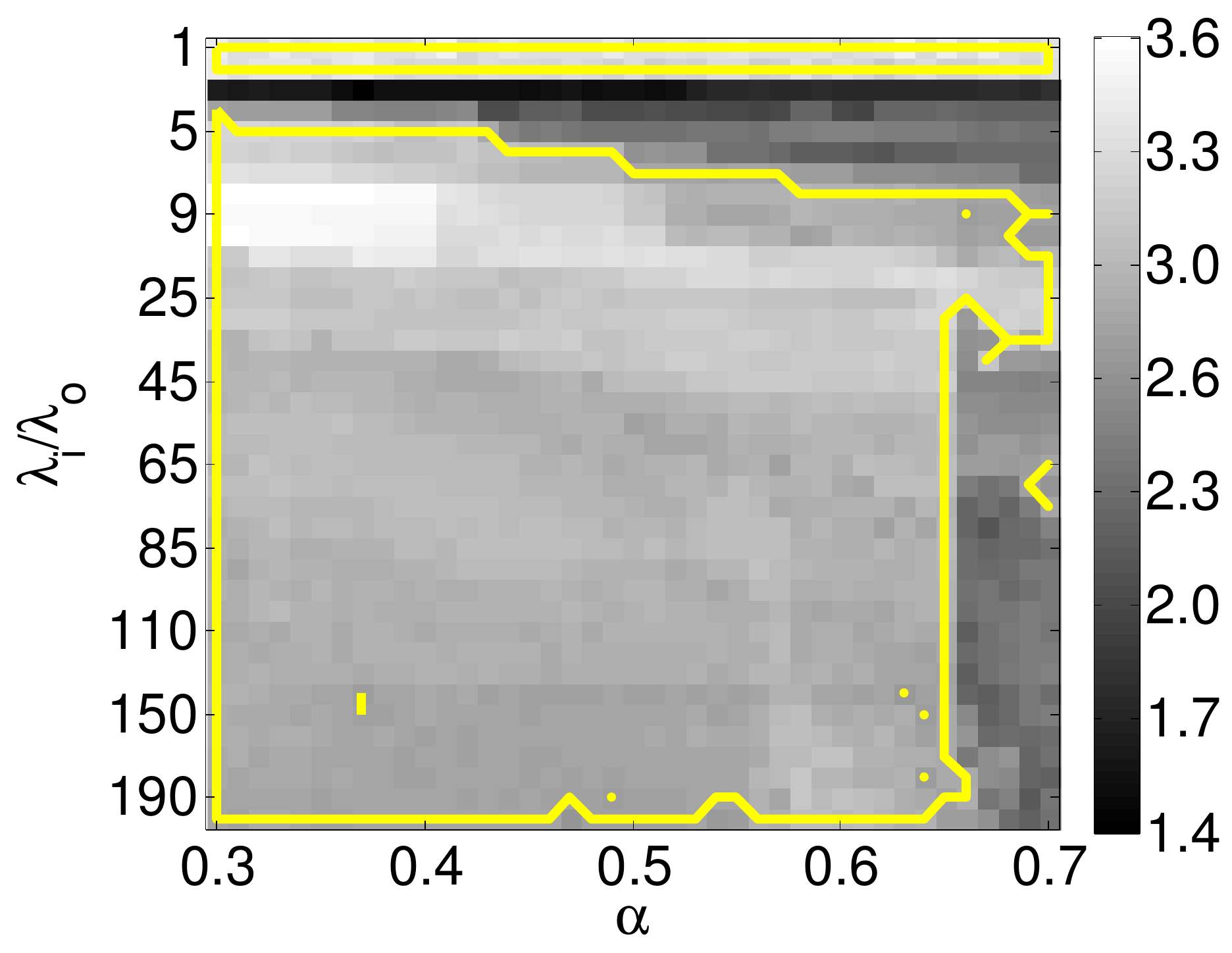}}\\
 \subfloat[CRs 2133 and 2150.]{\includegraphics[width=0.3\textwidth]{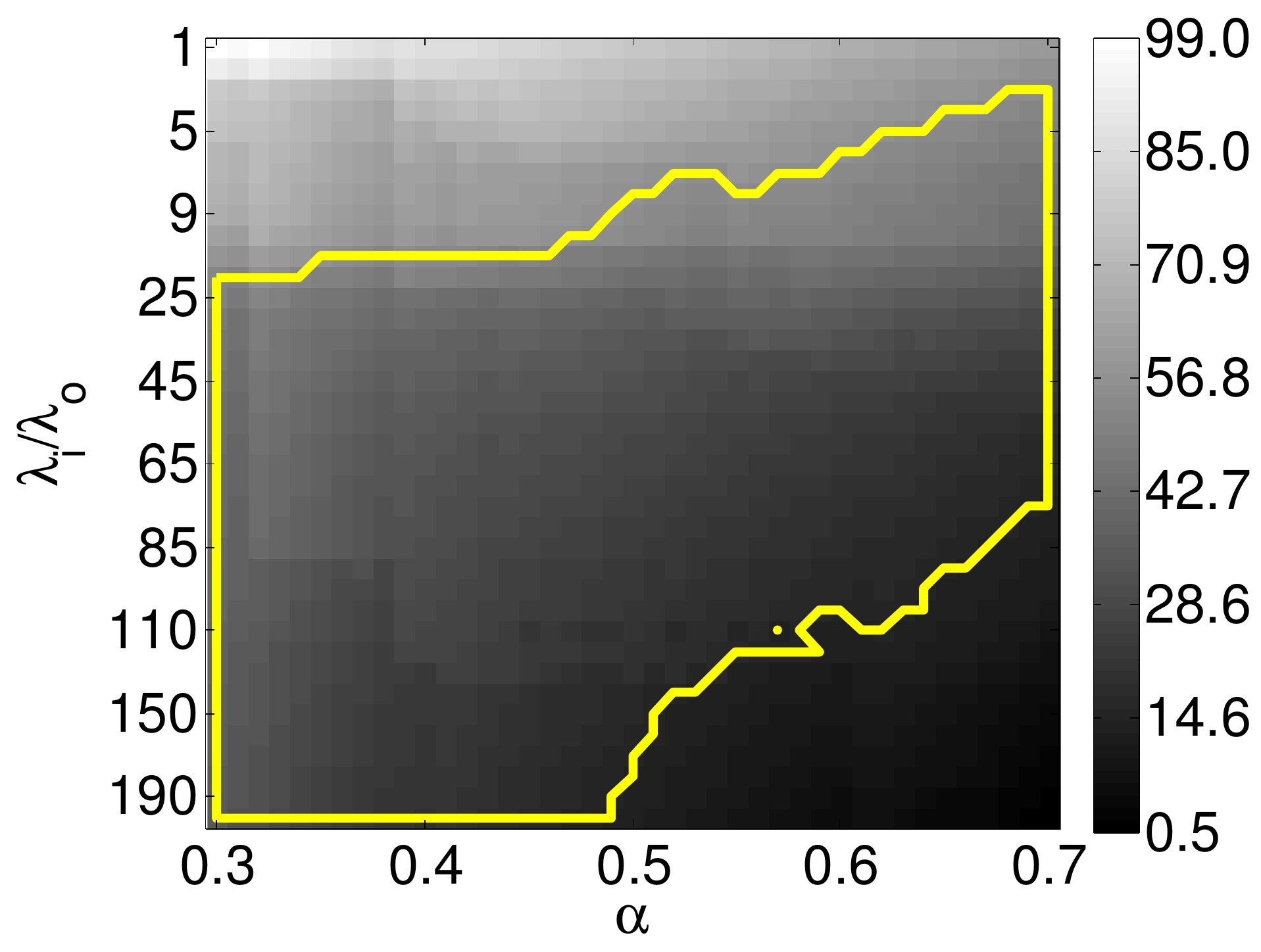}}~~
 \subfloat[CRs 2133 and 2150.]{\includegraphics[width=0.3\textwidth]{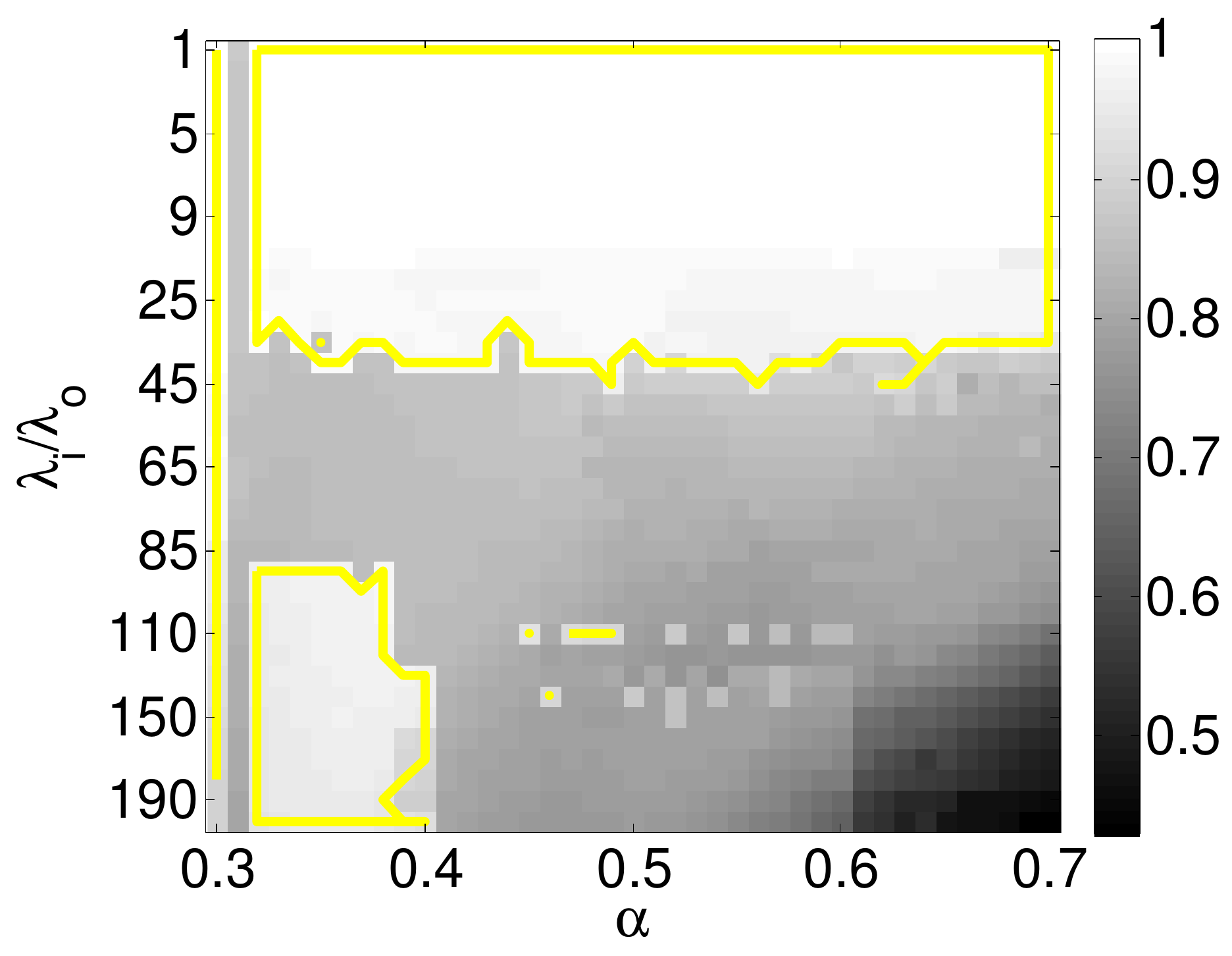}}~~
 \subfloat[CRs 2133 and 2150.]{\includegraphics[width=0.3\textwidth]{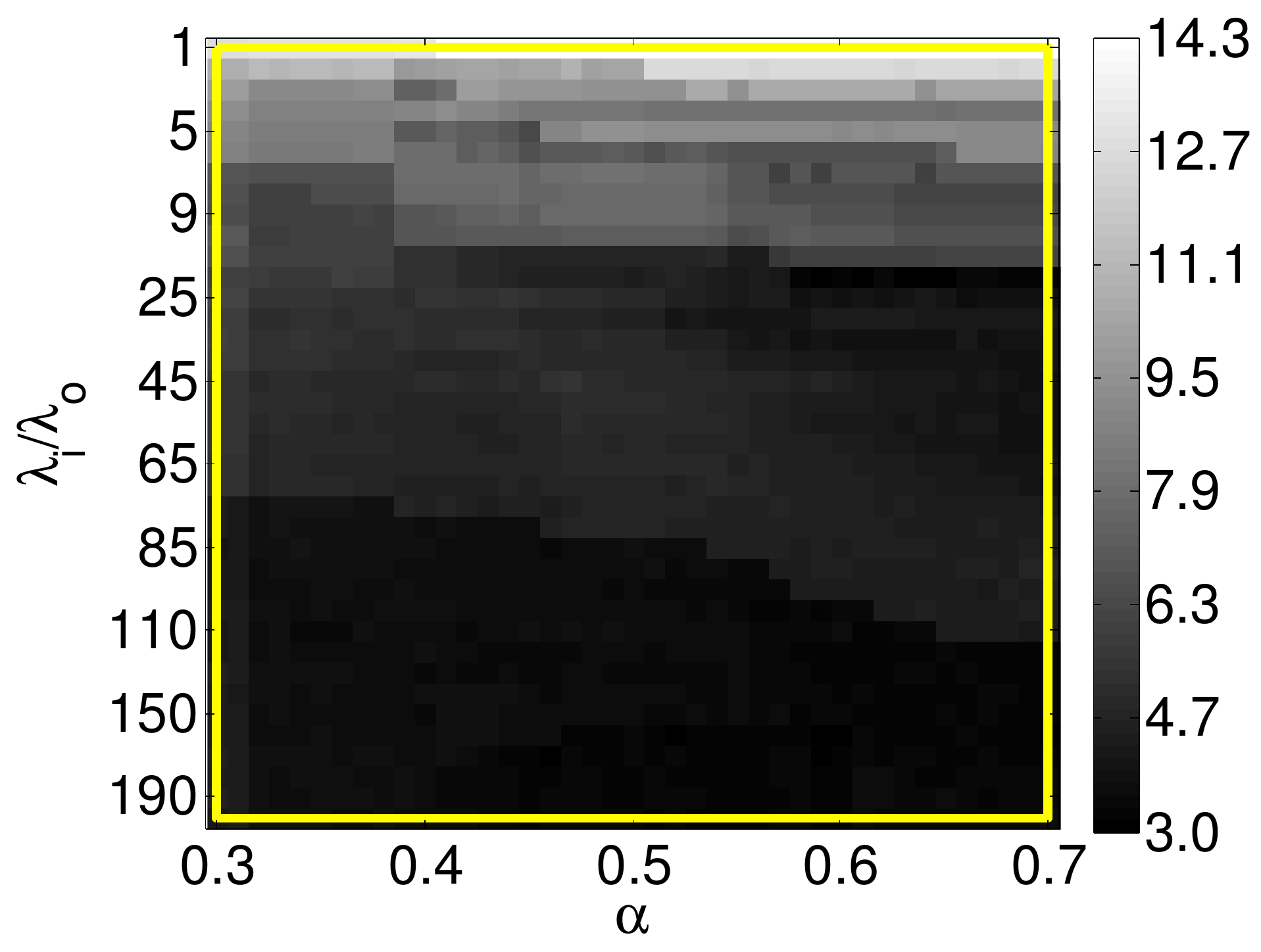}}\\
 \subfloat[All 4 CRs.]{\includegraphics[width=0.3\textwidth]{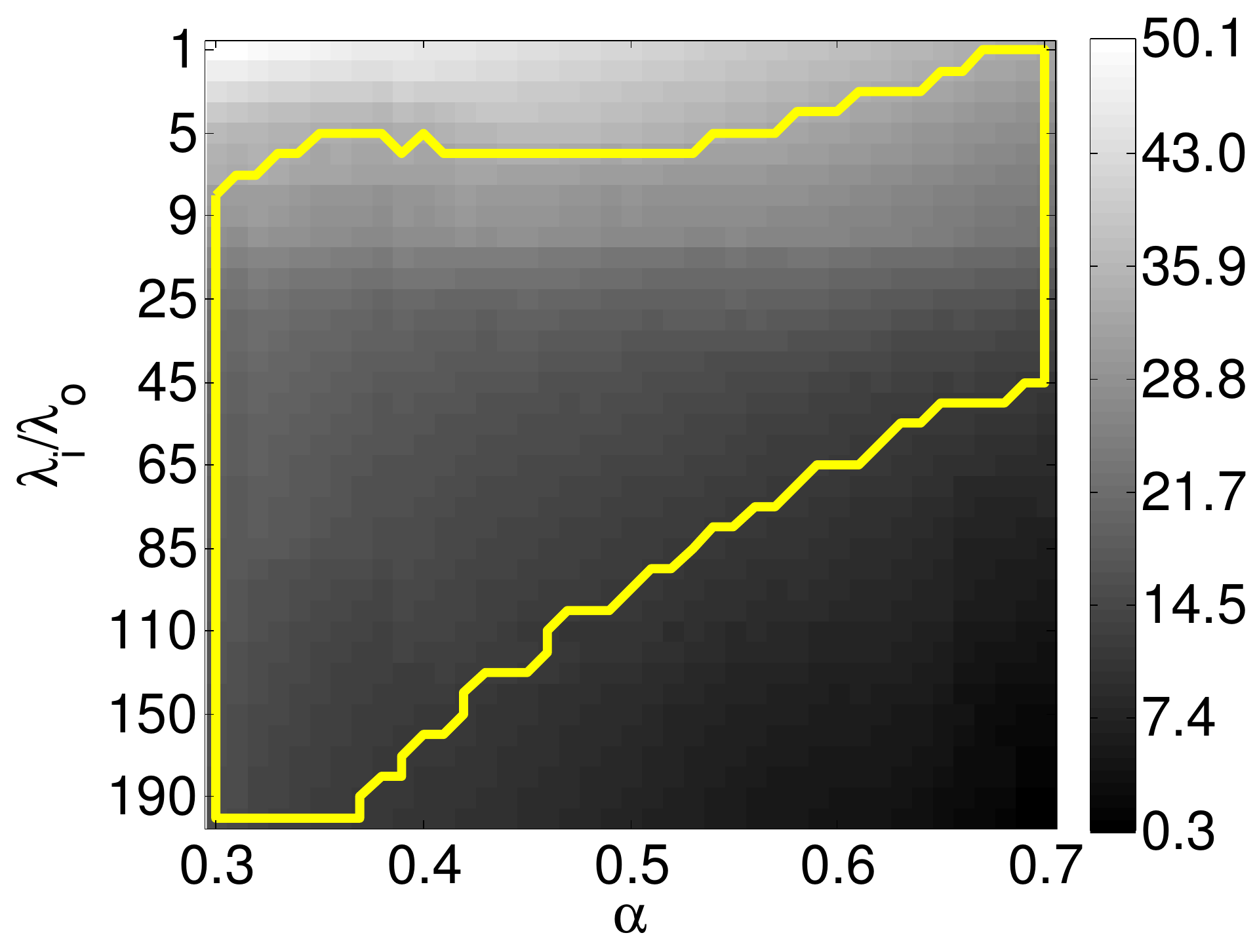}}~~
 \subfloat[All 4 CRs.]{\includegraphics[width=0.3\textwidth]{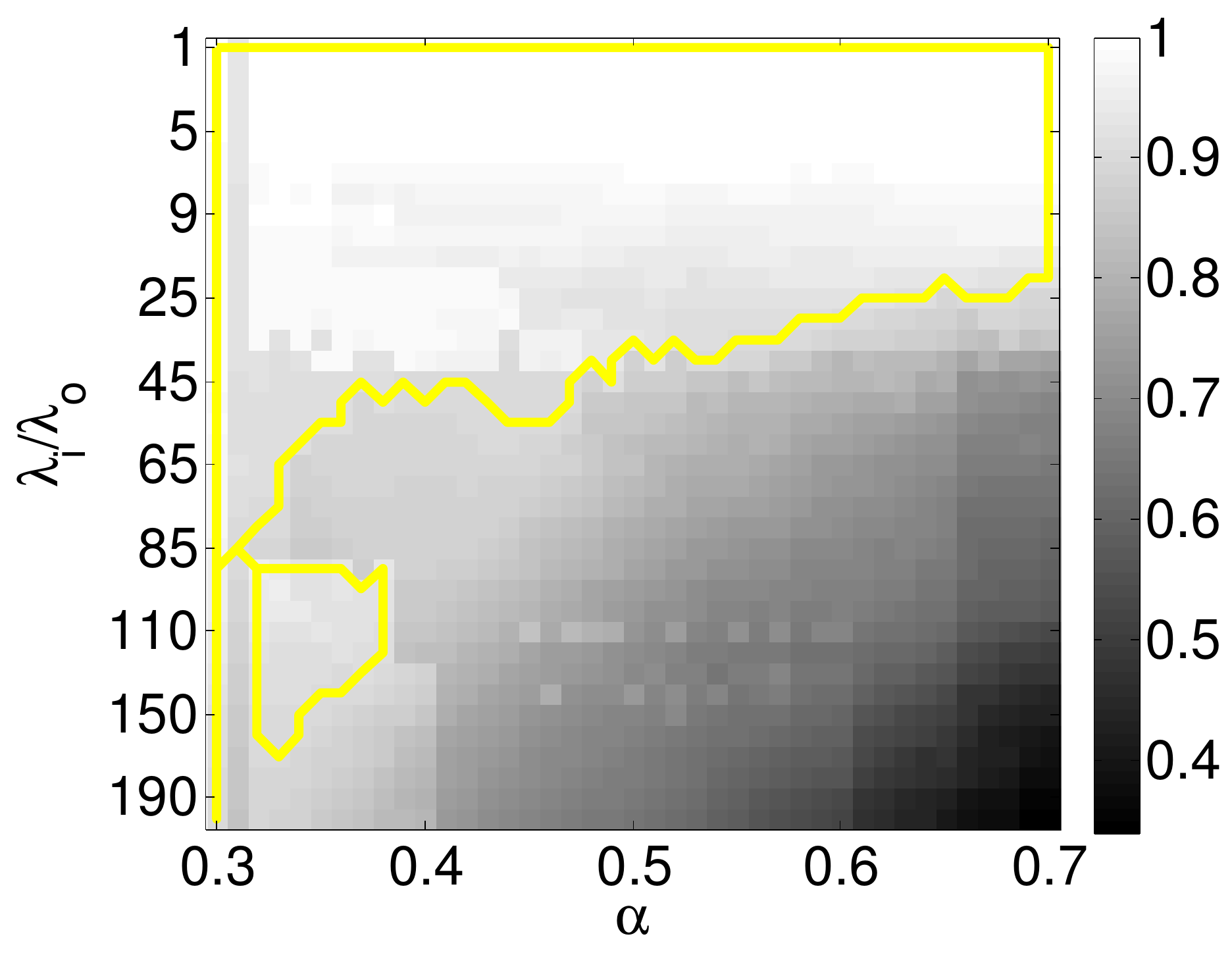}}~~
 \subfloat[All 4 CRs.]{\includegraphics[width=0.3\textwidth]{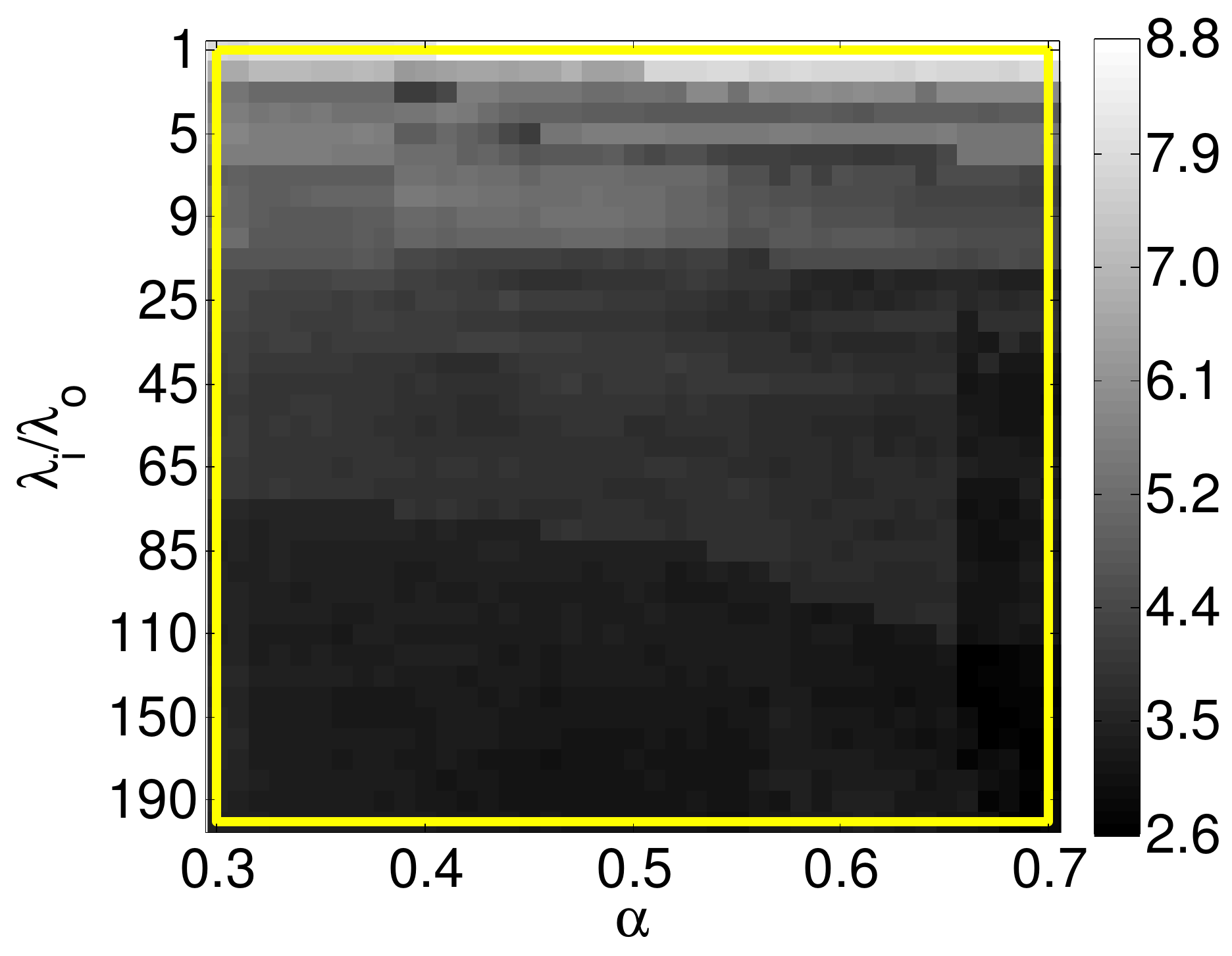}}
 \caption{Results of grid search for reasonable initial threshold parameter $\alpha$ and homogeneity ratio $\lambda_i/\lambda_o$.  Left column: Ratio of converged areas to initial areas $|C_i^{(k)}|/|C_i^{(0)}|$. The yellow line outlines those parameters which result in $1\le|C_i^{(k)}|/|C_i^{(0)}|\le10$. Intensities are logarithmically scaled for better visualization.  Middle column: Proportion of overlap of converged areas to initial areas $|C_i^{(k)}\cap C_i^{(0)}|/|C_i^{(0)}|$.  The yellow line outlines those parameters which result in $|C_i^{(k)}\cap C_i^{(0)}|/|C_i^{(0)}|\ge0.9$.  Right column: Average absolute skewness $\bar{|S|}$.  The yellow line outlines those parameters which result in $\bar{|S|}\ge2.5$.  Intensities are logarithmically scaled for better visualization.}
 \label{fig:grid_search}
\end{figure}

The converged segmentation results are assessed according to three quantities.  First, the ratio of the converged area to the initial area $|C_i^{(k)}|/|C_i^{(0)}|$ is computed where, as in Section~\ref{sec:image_analysis}, $C_i^{(k)}$ is the segmentation at iteration $k$, $|\cdot|$ is the cardinality operator, and the final (converged) value for $k$ is used.  We expect that the ACWE algorithm should extend beyond our conservatively defined CH initialization, and thus that $|C_i^{(k)}|/|C_i^{(0)}|>1$ for a reasonable CH segmentation, but also that $|C_i^{(k)}|/|C_i^{(0)}|$ should not be too large, indicating convergence to areas well outside the CHs (see Figure~\ref{fig:foreground}(a) versus Figure~\ref{fig:foreground}(b)).  Results for the ratio $|C_i^{(k)}|/|C_i^{(0)}|$ are shown in Figures~\ref{fig:grid_search}(a), \ref{fig:grid_search}(d), and \ref{fig:grid_search}(g), with parameter combinations for which $1\le|C_i^{(k)}|/|C_i^{(0)}|\le10$ outlined in yellow.  

Second, the proportion of overlap between the initialization and the converged segmentation $|C_i^{(k)}\cap C_i^{(0)}|/|C_i^{(0)}|$ is computed. Since the algorithm is initialized with conservative estimates of pixels which are highly likely to belong to CHs, we expect that $|C_i^{(k)}\cap C_i^{(0)}|/|C_i^{(0)}|\approx1$.  Results for the proportion $|C_i^{(k)}\cap C_i^{(0)}|/|C_i^{(0)}|$ are shown in Figures~\ref{fig:grid_search}(b), \ref{fig:grid_search}(e), and \ref{fig:grid_search}(h), with parameter combinations for which $|C_i^{(k)}\cap C_i^{(0)}|/|C_i^{(0)}|\ge0.9$ outlined in yellow.

Third, the average absolute skewness $\bar{|S|}$ of the magnetic field underlying the converged segmentation is computed as the third central moment normalized by the cube of the standard deviation, $\text{skewness}(x)=E\left( x-\mu\right)/\sigma^3$, where $\mu$ is the mean, $\sigma$ is the standard deviation, and $E(\cdot)$ is the expectation operator.  To compute this value, concurrent HMI magnetograms are used to determine the underlying magnetic field strength and compute the absolute skewness of each of the $R$ disjoint regions in the segmentation as
\begin{equation}
 |S_r| =  \text{skewness}\left(\text{abs}(H(x,y))|(x,y)\in C_{r}\right),~r=1,\ldots,R,
\end{equation}
where $H$ is the HMI magnetogram and $\cup C_{r}=C_i^{(k)}$.  Then the average absolute skewness $\bar{|S|}$ is the average $|S_r|$ weighted by area:
\begin{equation}
 \bar{|S|}=\frac{\sum\limits_{r=1}^{R}|S_r|}{|C_i^{(k)}|}.
\end{equation}
If the converged segmentation is truly representative of the CHs, we expect that $\bar{|S|}$ is large, indicating unipolar magnetic fields.  Results for the average absolute skewness $\bar{|S|}$ are shown in Figures~\ref{fig:grid_search}(c), \ref{fig:grid_search}(f), and \ref{fig:grid_search}(i), with parameter combinations for which $\bar{|S|}\ge2.5$ outlined in yellow.

We note a few observations regarding the three quantities displayed in Figure~\ref{fig:grid_search}.  First, from Figures~\ref{fig:grid_search}(b), \ref{fig:grid_search}(e), and \ref{fig:grid_search}(h), we note that much of the lower right portion of the $\alpha$-$\lambda_i/\lambda_o$ search space should be discounted since those parameter combinations result in an overlap proportion less than 1.  This conclusion is consistent for both low activity (Figure~\ref{fig:grid_search}(b)) and high activity (Figure~\ref{fig:grid_search}(e)) CRs.  Second, we note that the average absolute skewness (Figures~\ref{fig:grid_search}(c), \ref{fig:grid_search}(f), and \ref{fig:grid_search}(i)) is above 2.5 for the majority of the $\alpha$-$\lambda_i/\lambda_o$ search space.  We note a much larger overall skewness for high activity CRs (note the colorbar in Figure~\ref{fig:grid_search}(f) versus \ref{fig:grid_search}(c)).  We also note larger skewness values for smaller $\lambda_i/\lambda_o$; from Figures~\ref{fig:grid_search}(a), \ref{fig:grid_search}(d), and \ref{fig:grid_search}(g), small values of $\lambda_i/\lambda_o$ correspond to a converged segmentation that covers the majority of the solar surface.  This will include edges of active regions with larger magnetic flux values which results in larger skewness values for the segmented regions, and is consistent for both low and high activity CRs.  Third, considering converged areas, overlap proportion, and skewness simultaneously, we find a large swath of the $\alpha-\lambda_i/\lambda_o$ search space, roughly around the anti-diagonal, appears to yield reasonable segmentations in that small changes in in those parameters will not have a large effect on the final segmentation results.  We choose $\alpha=0.3$ (small seeding threshold) and $\lambda_i/\lambda_o=50$ (allowing reasonable inhomogeneity in the CHs) for results presented in Section~\ref{sec:results}.

\subsection{Computational Considerations}
\label{sec:computation}
Before presenting and discussing the results of ACWE segmentation of CHs, we briefly consider a few computational aspects of the proposed algorithm.  It is difficult to accurately specify the computational cost of the proposed algorithm since it is dependent on many factors including parameter choices (e.g., how close the initialization is to the final segmentation), hardware choices (e.g., CPU architecture), and software choices (e.g., programming language).  The current code, implemented in MATLAB, takes approximately 2 seconds per iteration and 17 iterations per image for an $8\times$ decimated image and is expected to scale linearly with the number of pixels (i.e., $\sim64\times$ slower for \textbf{a full resolution image}); limited empirical study supports the linear scaling of the algorithm with respect to number of pixels.  This implementation cannot, at the moment, be applied in real time on the 12-second cadence AIA images as research has focused on algorithmic proof-of-concept rather than computational efficiency.  We do note, however, that significant improvement in computational efficiency may be possible, for example by implementation in a more efficient language, or by more accurate initialization of the contours; both of these are under current investigation.

\begin{figure}
  \centering
  \subfloat[$8\times$ decimated image.]{\includegraphics[width=0.3\textwidth]{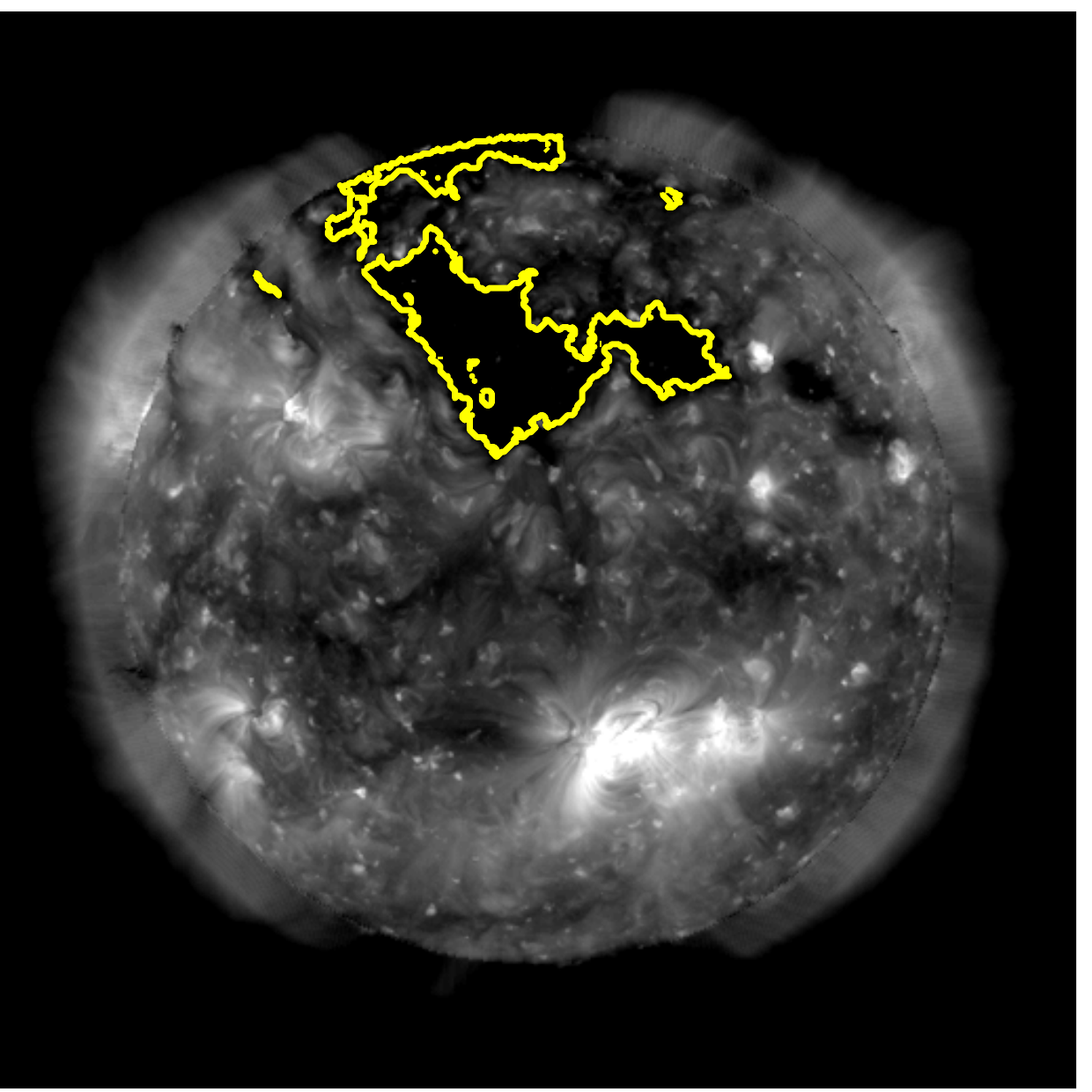}}~~~~~
  \subfloat[Full resolution image.]{\includegraphics[width=0.3\textwidth]{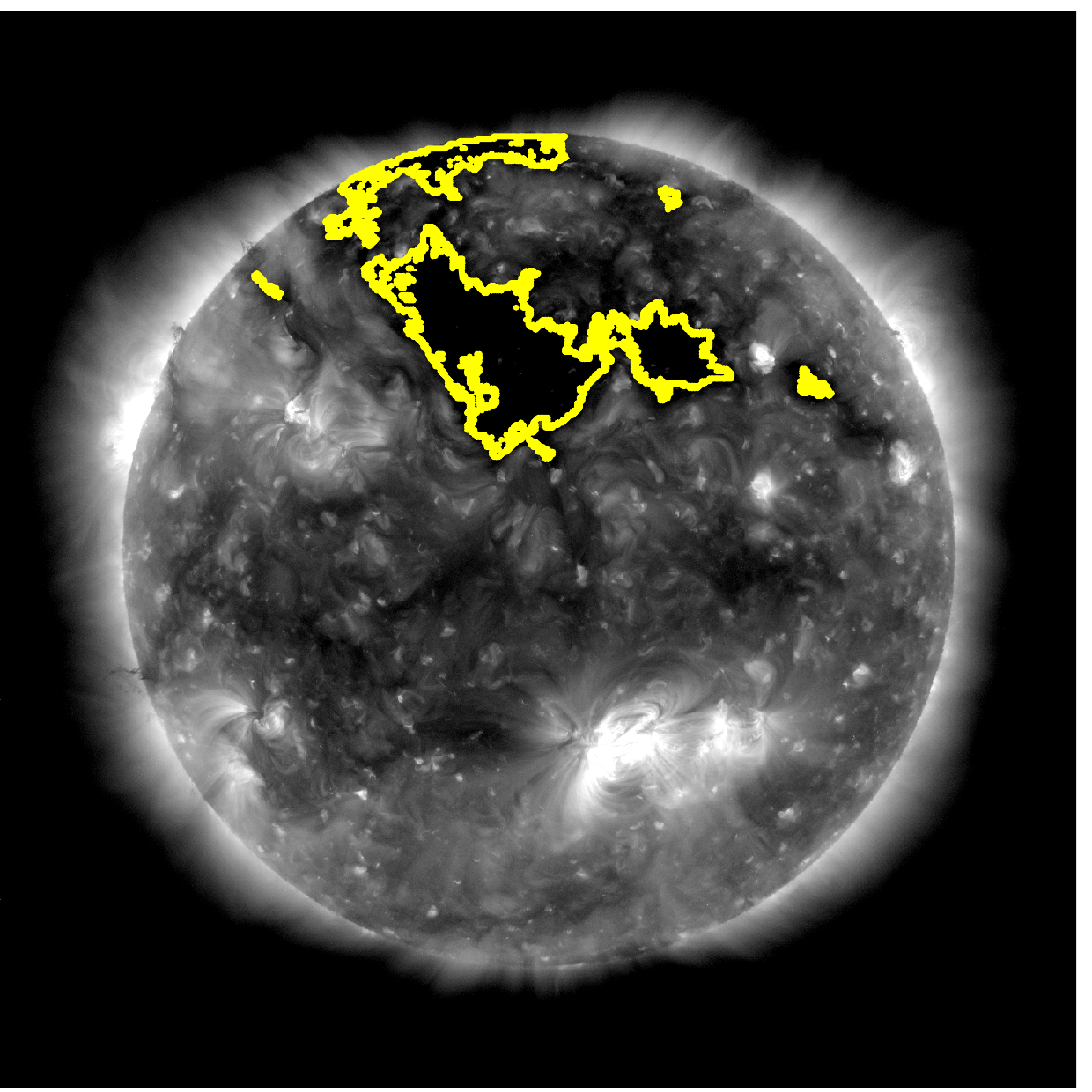}}
  \caption{Comparison of final segmentation for $8\times$ decimated and full resolution image.  The thicker appearance of the CH boundaries in (b) are due to a more tortuous boundary at the higher resolution.  Qualitatively, the segmentations are remarkable similar in appearance with the only difference being the small region to the right.}
  \label{fig:fullres}
\end{figure}

A brief demonstration of the effect of the $8\times$ decimation on the final segmentation is shown in Figure~\ref{fig:fullres}, where we show the converged segmentation for the same image at the two resolutions.  We find the two segmentations to be qualitatively very similar.  The only region which is qualitatively different is the small region to the right in Figure~\ref{fig:fullres}(b) which is missing in Figure~\ref{fig:fullres}(a); a few pixels in the full resolution image do not meet the threshold $0.3I_{QS}$ in the $8\times$ decimated image due to the combination of lowpass filtering and pixel removal of the decimation process.  The overall behavior of the ACWE algorithm, however, appears to be consistent between the $8\times$ decimated and full resolution images.  It should be further noted that the $8\times$ decimated images are used to speed computation in this initial study where we study multiple parameter combinations; future work may study the full resolution segmentations as needed.

\section{Validation of CH Segmentation}
\label{sec:results}
\begin{sidewaysfigure}[p]
  \centering
  \begin{minipage}[l]{0.85\textwidth}
   \subfloat[13 Jul. 2010]{\includegraphics[width=0.13\textwidth]{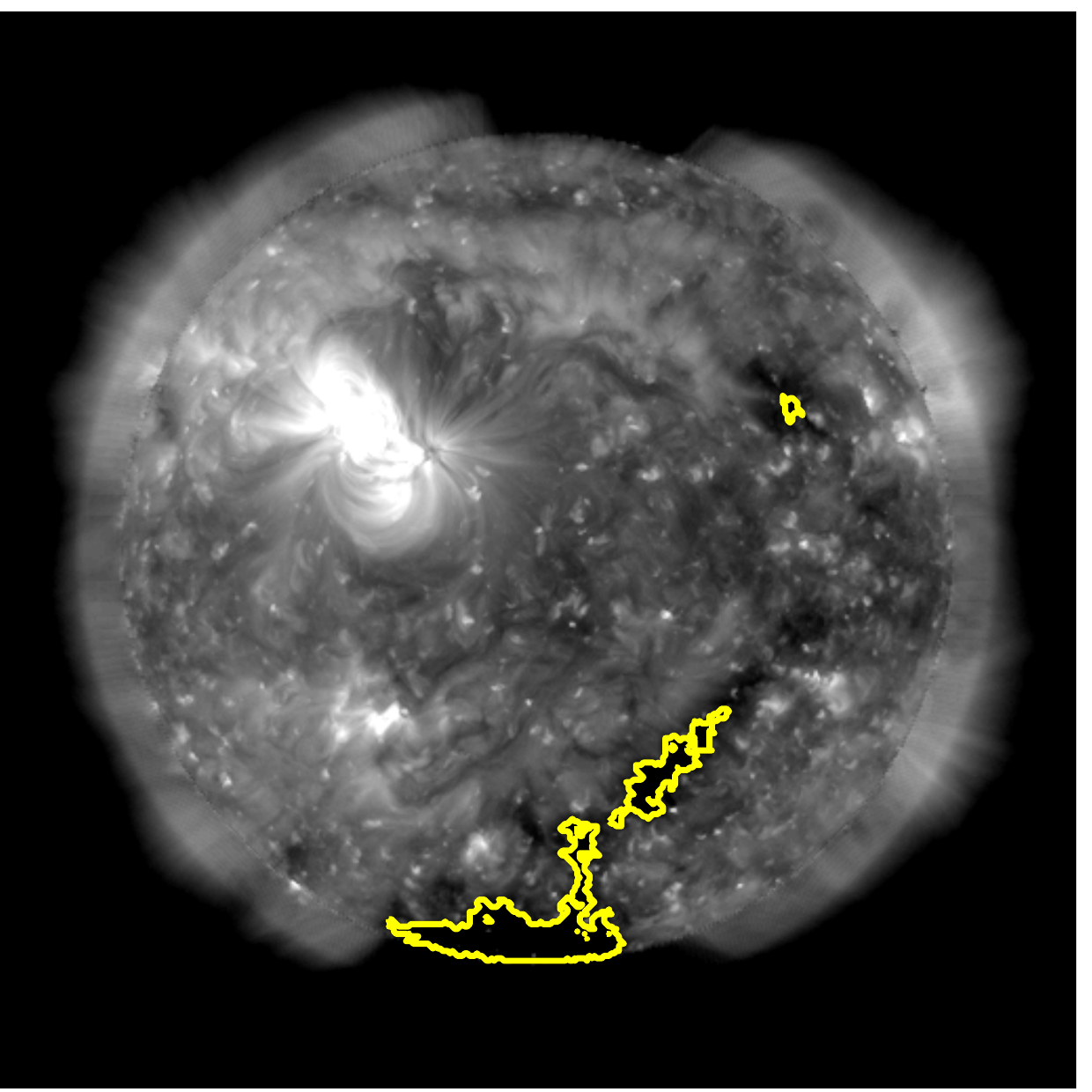}}~
   \subfloat[14 Jul. 2010]{\includegraphics[width=0.13\textwidth]{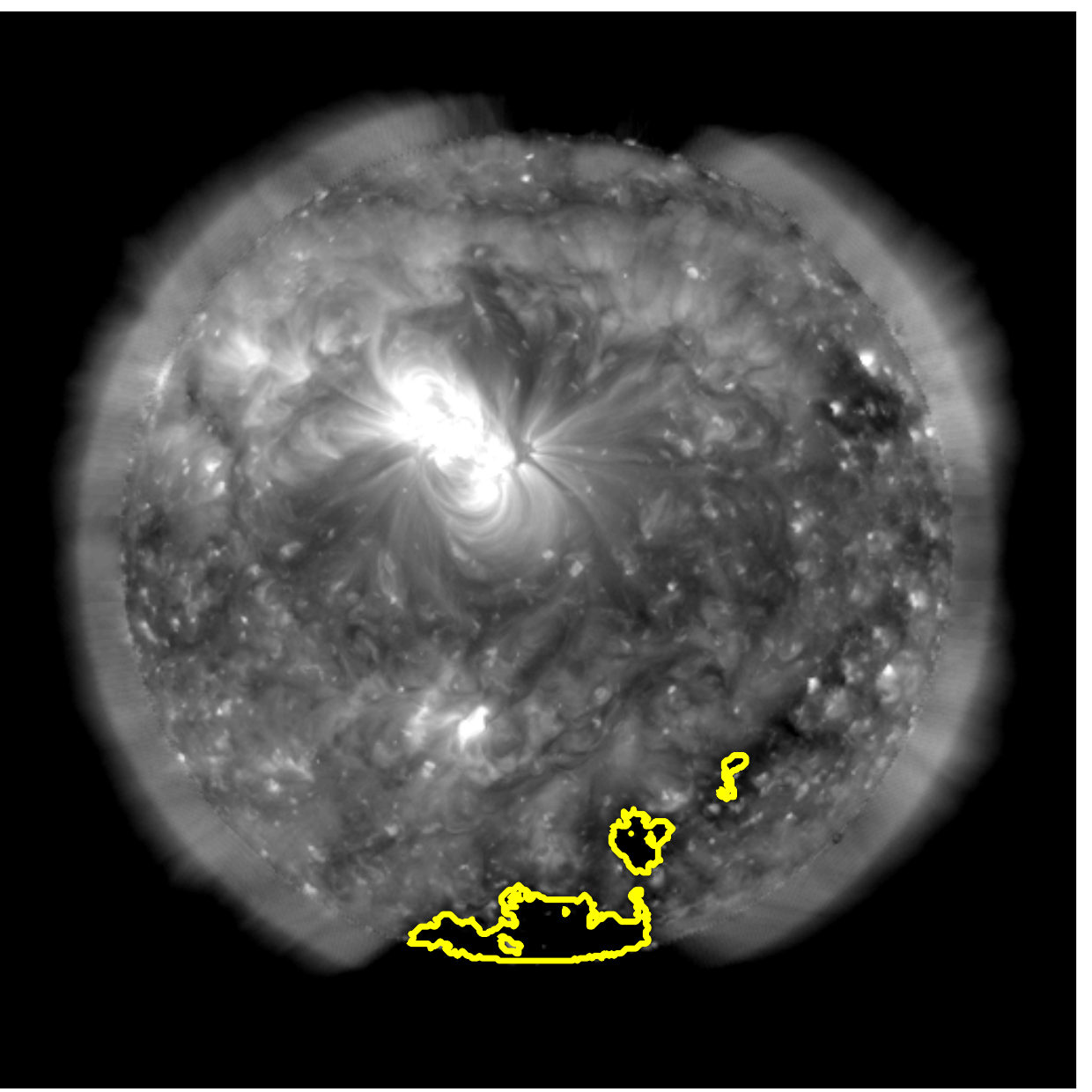}}~
   \subfloat[15 Jul. 2010]{\includegraphics[width=0.13\textwidth]{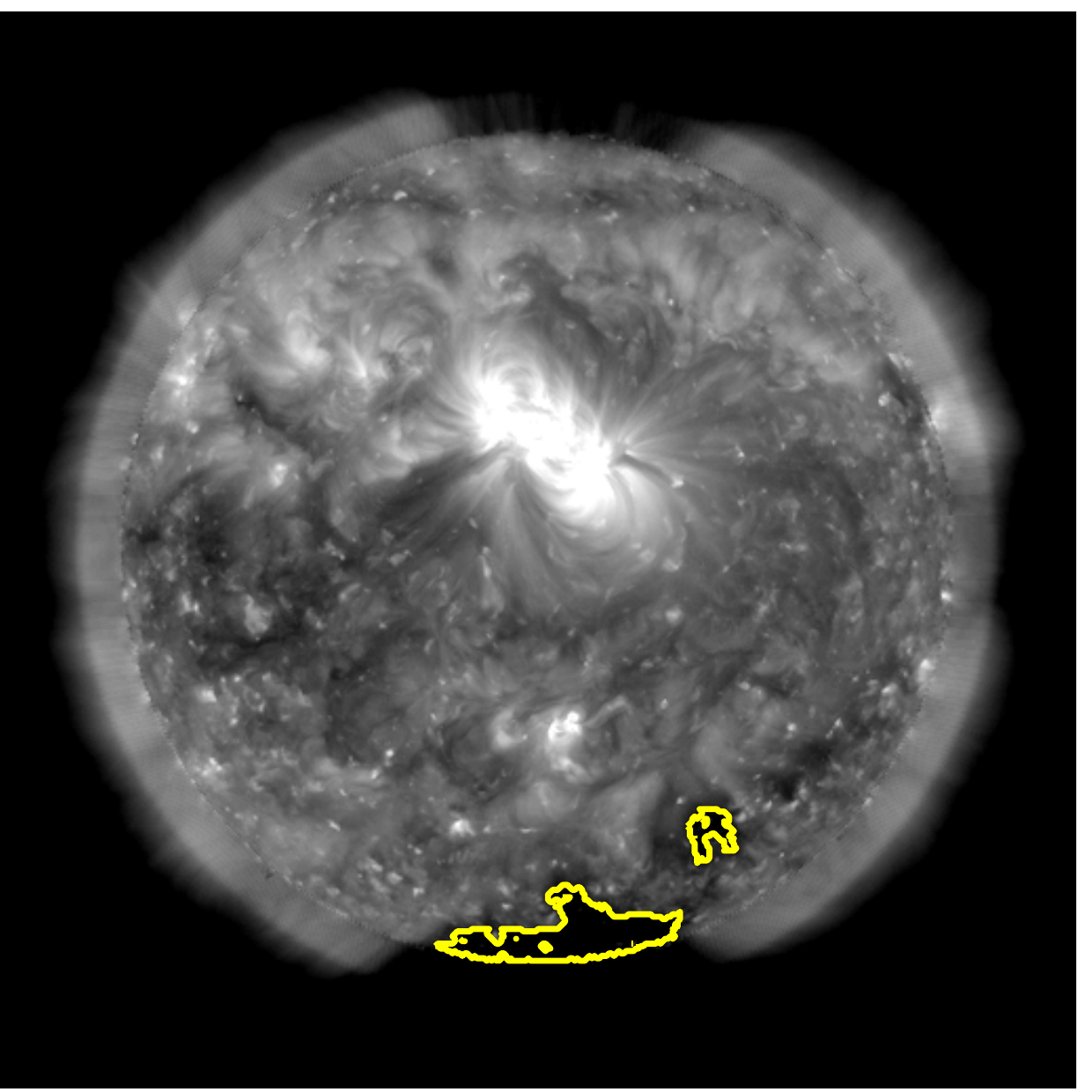}}~
   \subfloat[16 Jul. 2010]{\includegraphics[width=0.13\textwidth]{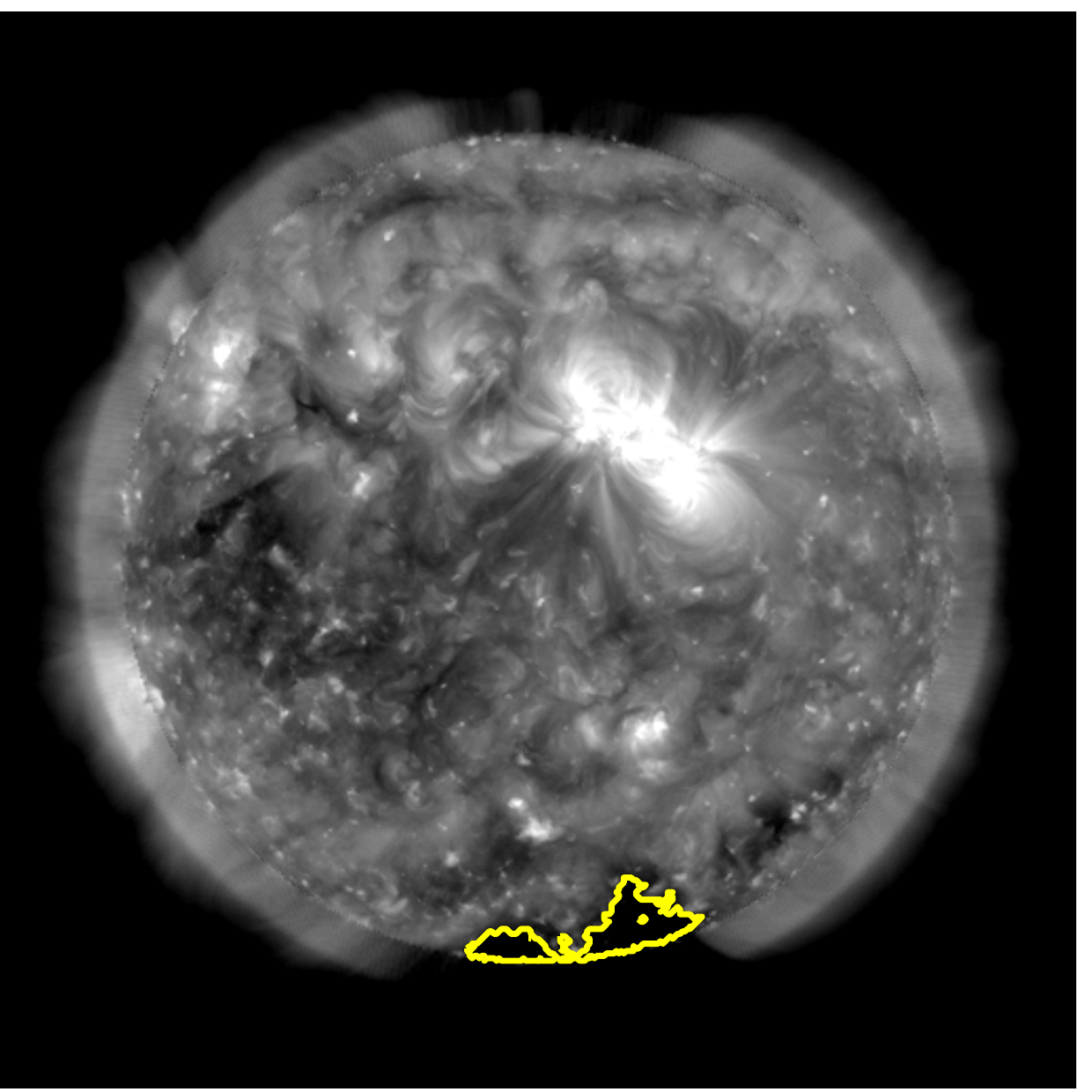}}~
   \subfloat[17 Jul. 2010]{\includegraphics[width=0.13\textwidth]{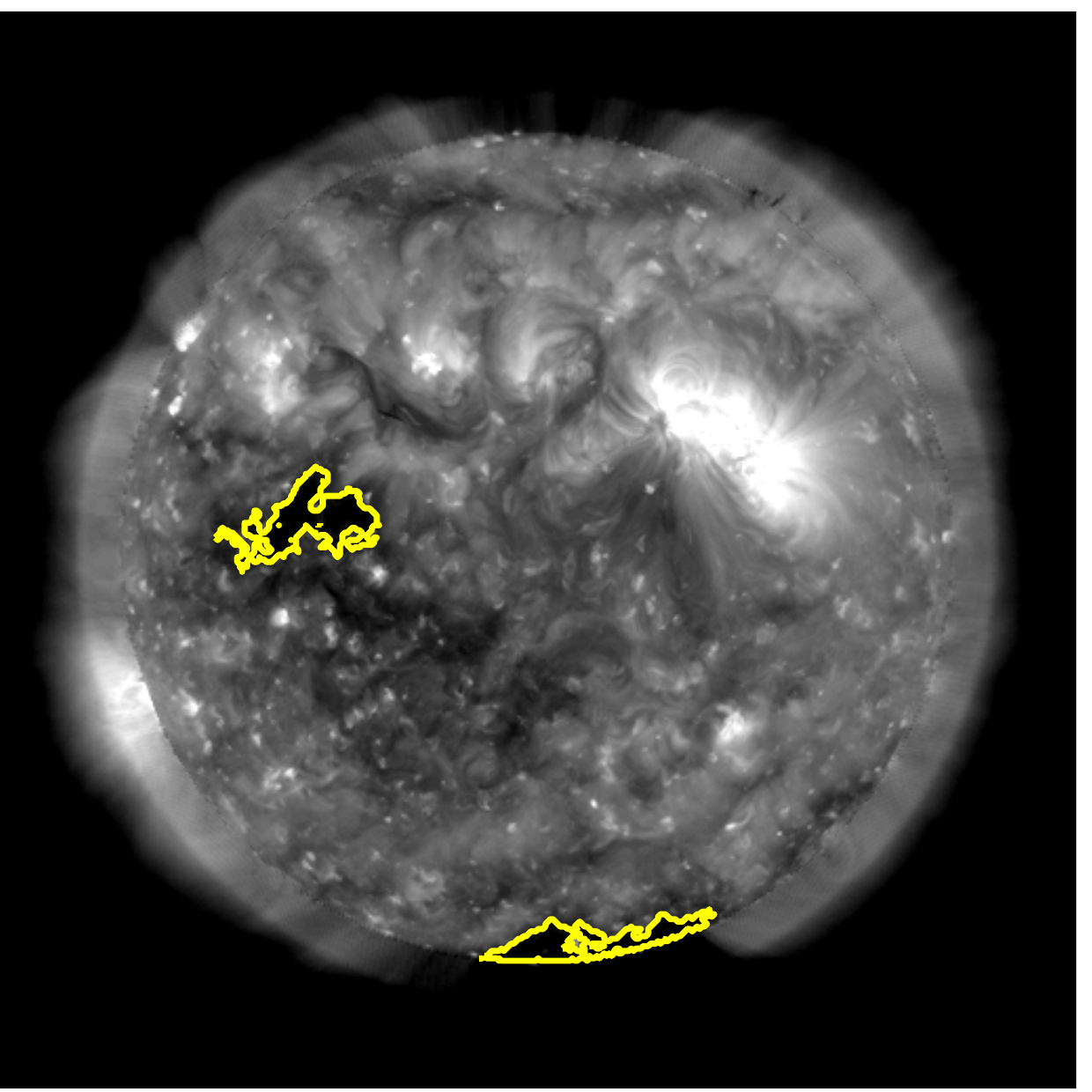}}~
   \subfloat[18 Jul. 2010]{\includegraphics[width=0.13\textwidth]{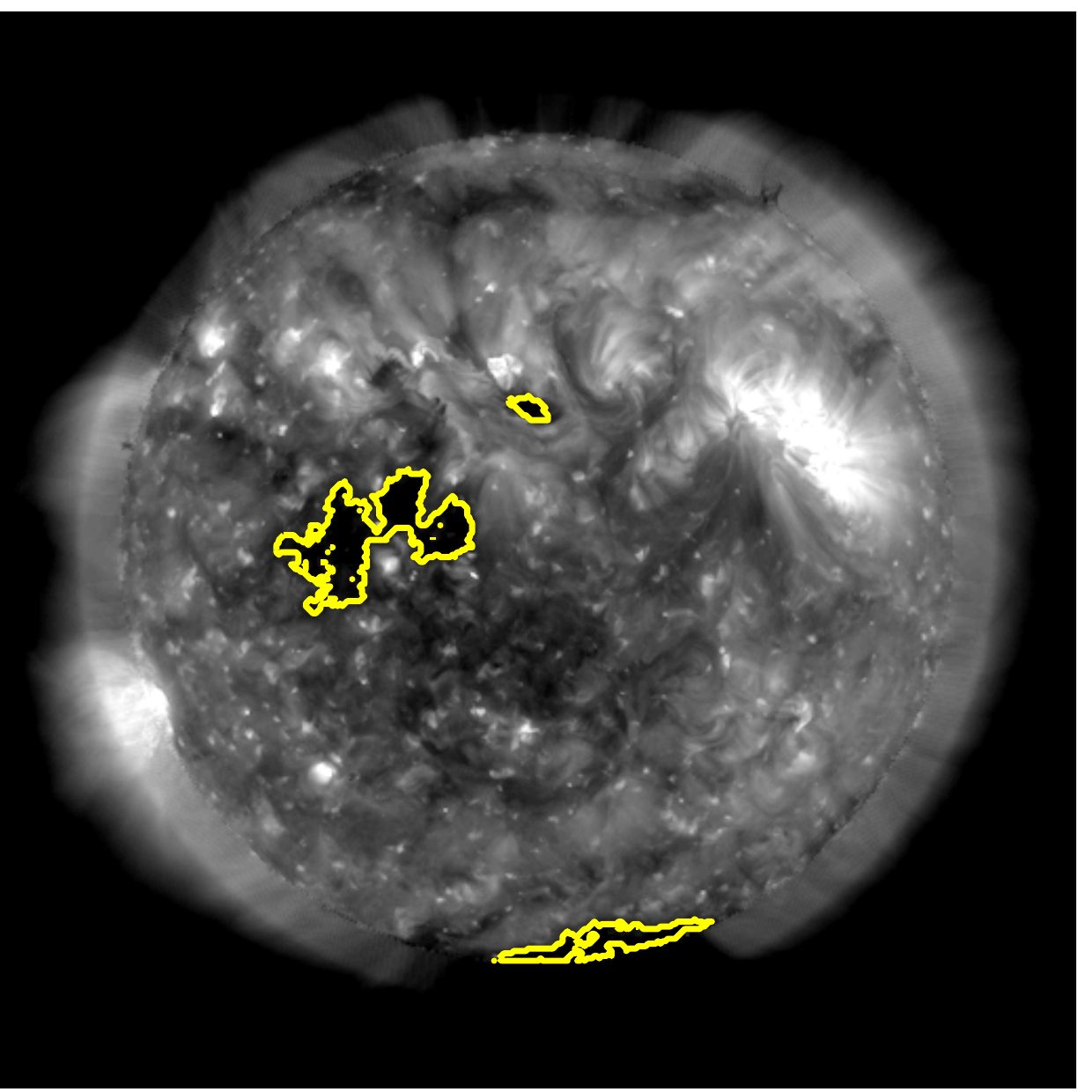}}~
   \subfloat[19 Jul. 2010]{\includegraphics[width=0.13\textwidth]{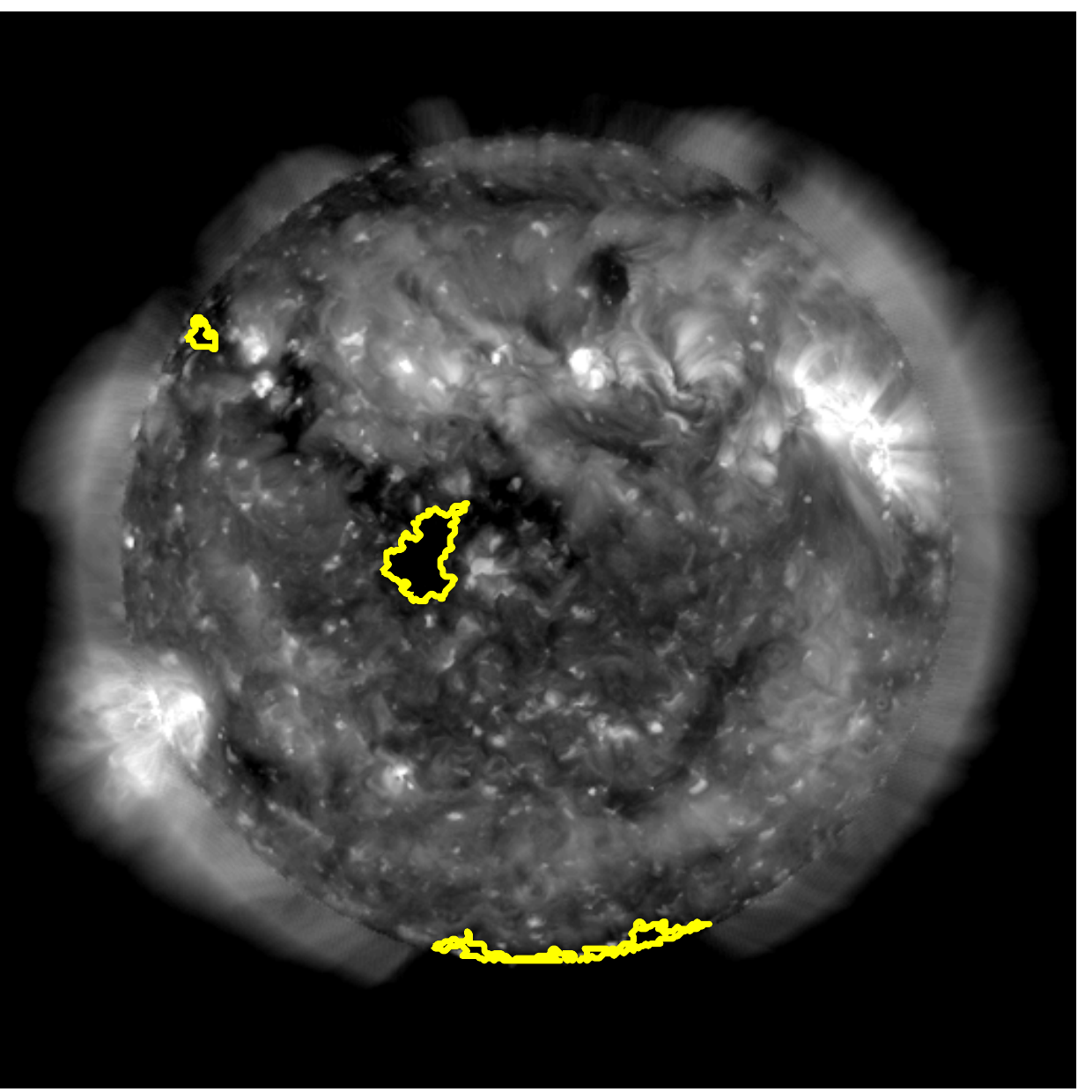}}\\[-2ex]\\
   \subfloat[20 Jul. 2010]{\includegraphics[width=0.13\textwidth]{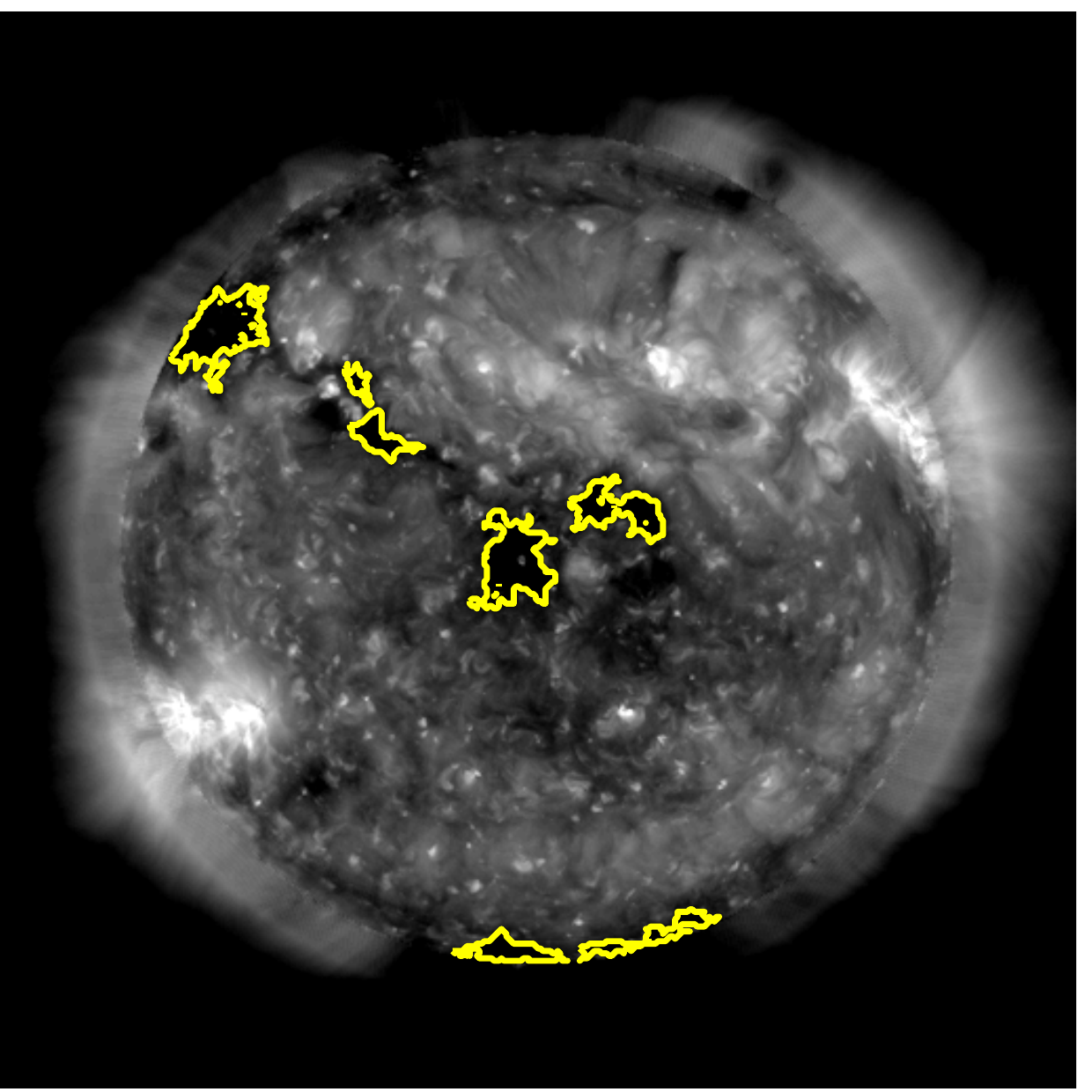}}~
   \subfloat[21 Jul. 2010]{\includegraphics[width=0.13\textwidth]{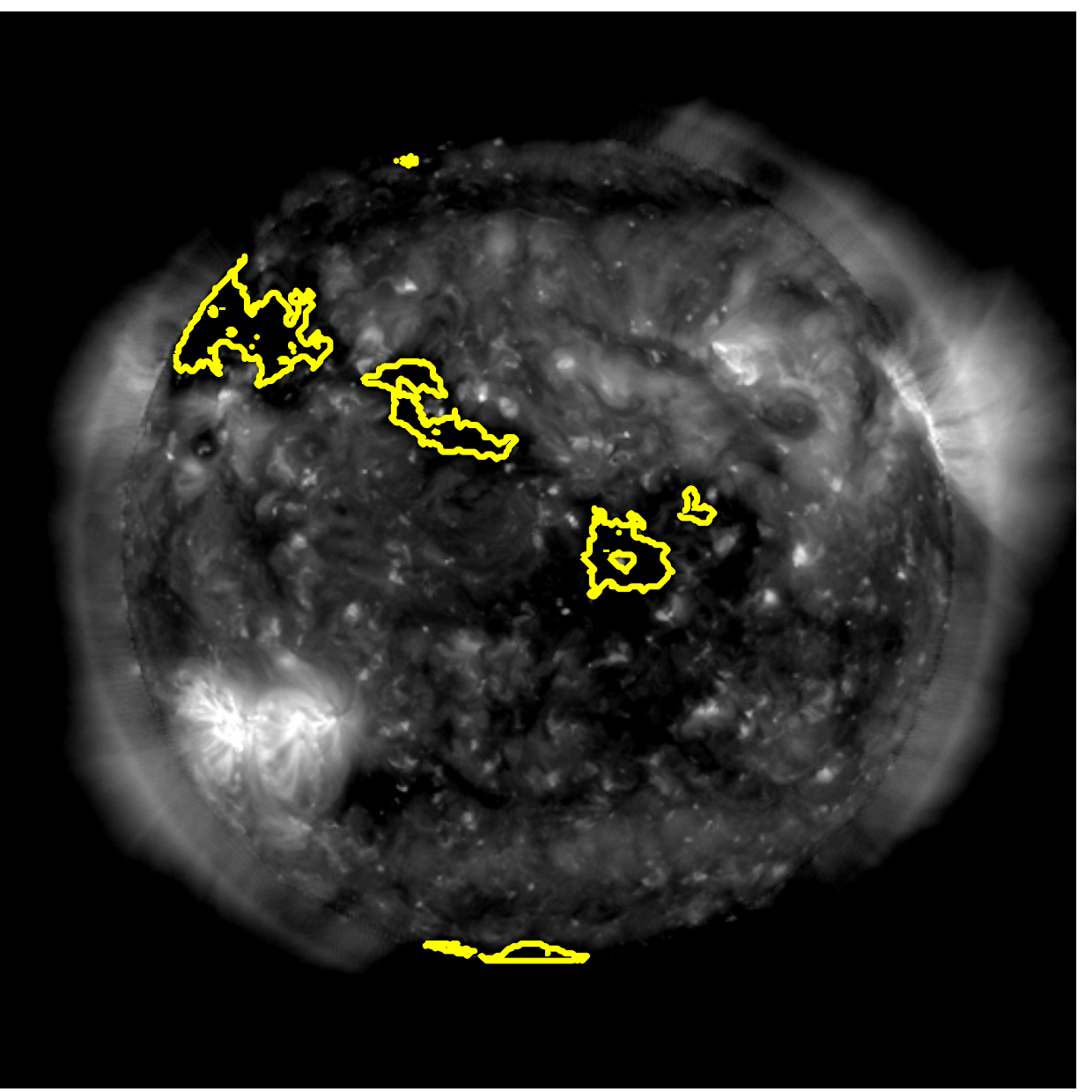}}~
   \subfloat[22 Jul. 2010]{\includegraphics[width=0.13\textwidth]{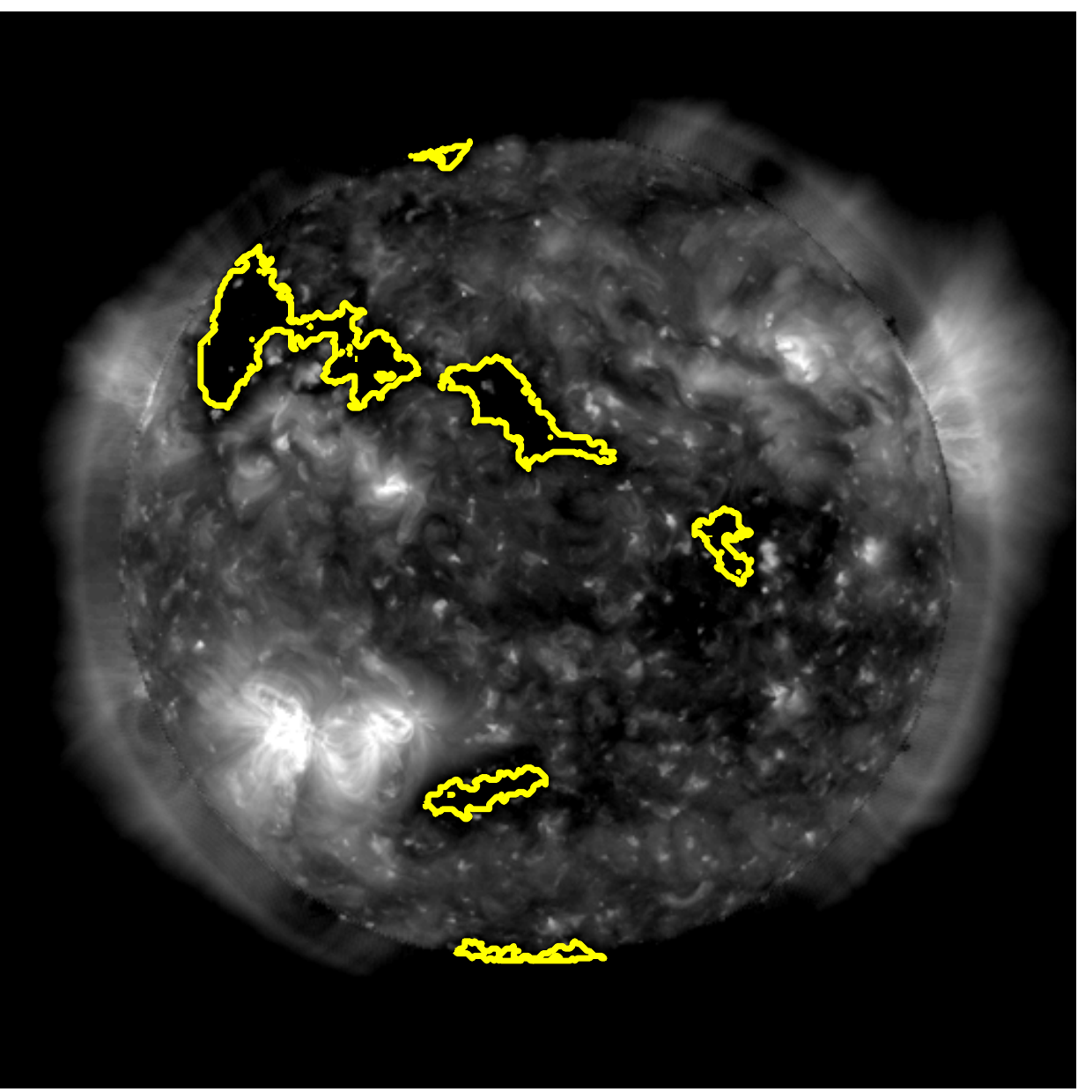}}~
   \subfloat[23 Jul. 2010]{\includegraphics[width=0.13\textwidth]{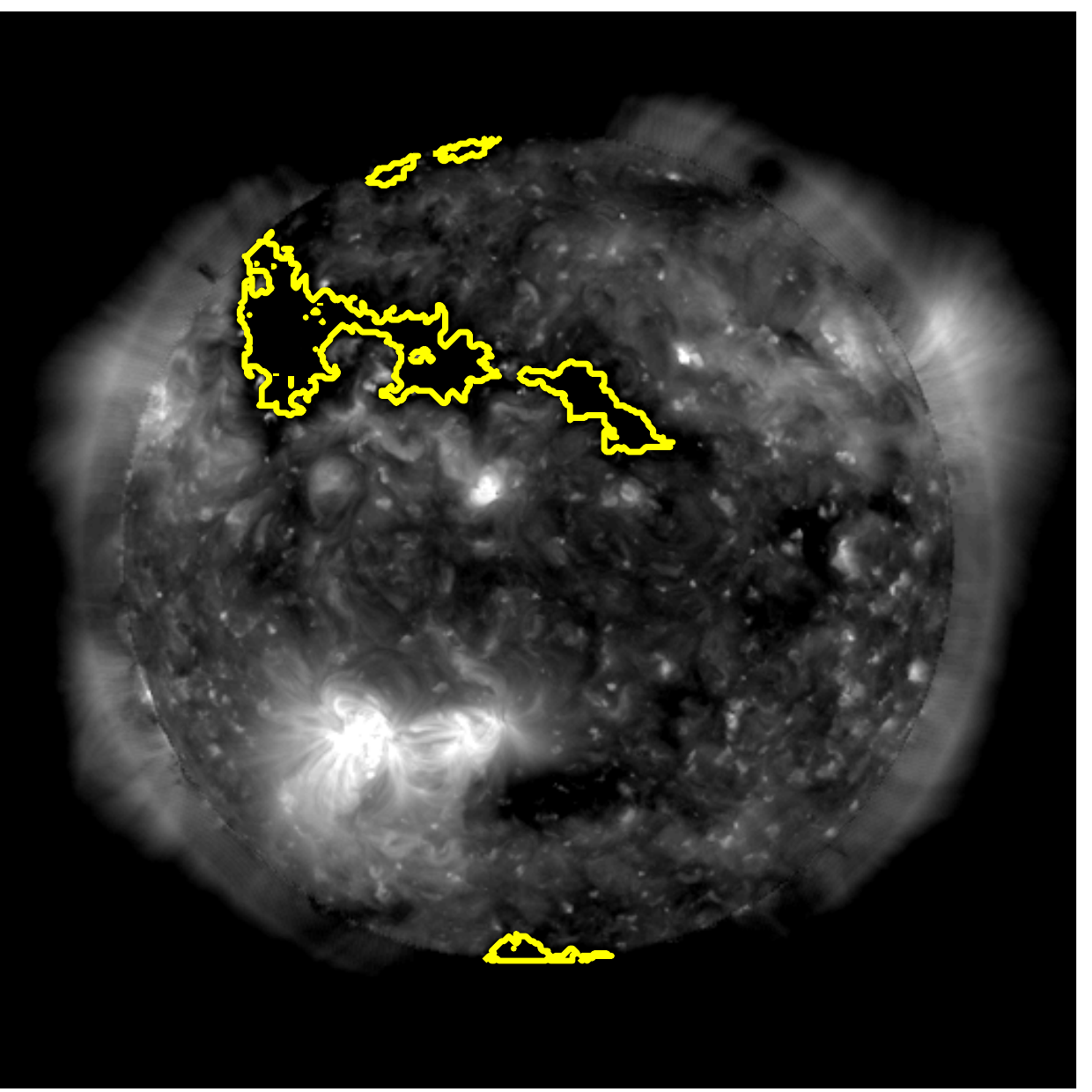}}~
   \subfloat[24 Jul. 2010]{\includegraphics[width=0.13\textwidth]{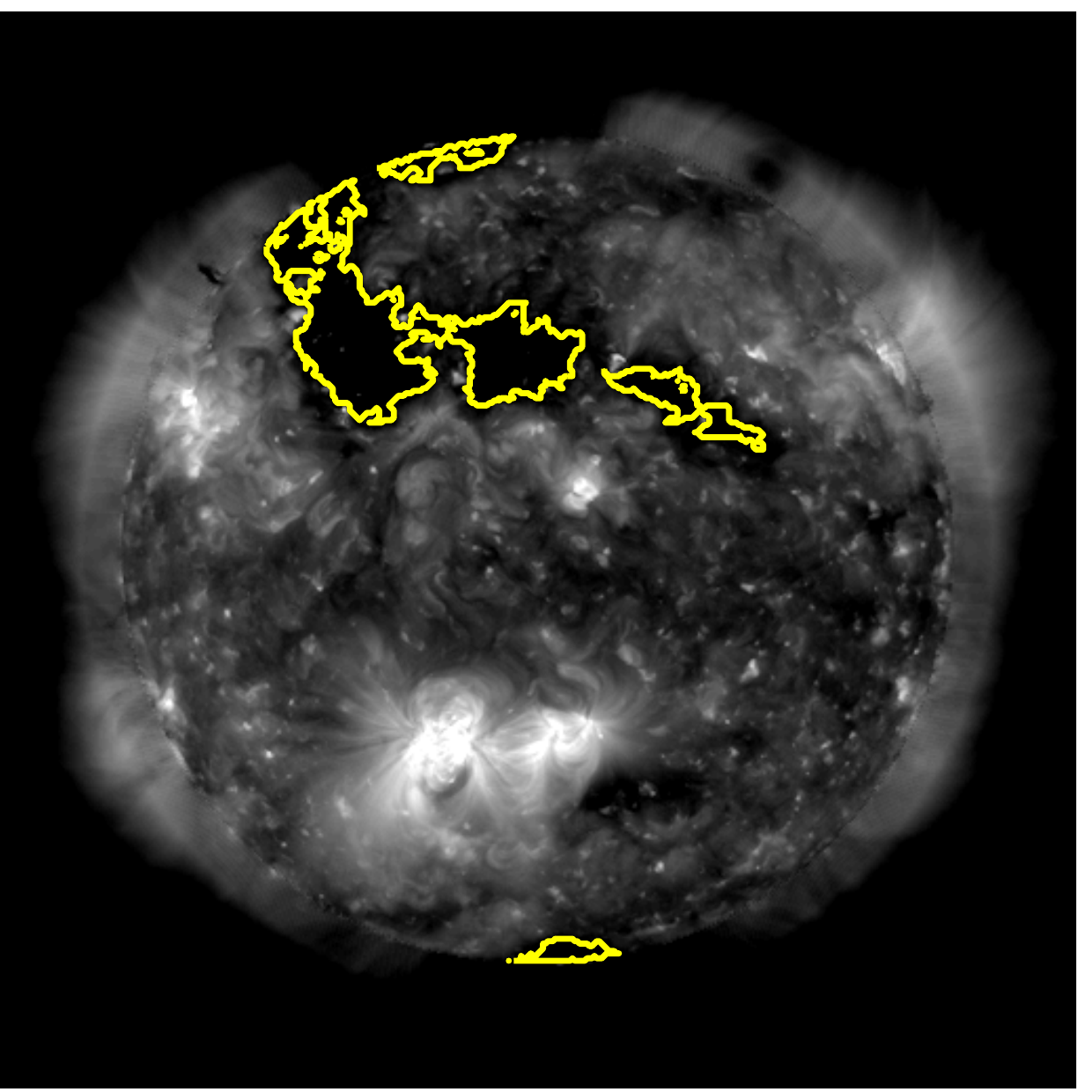}}~
   \subfloat[25 Jul. 2010]{\includegraphics[width=0.13\textwidth]{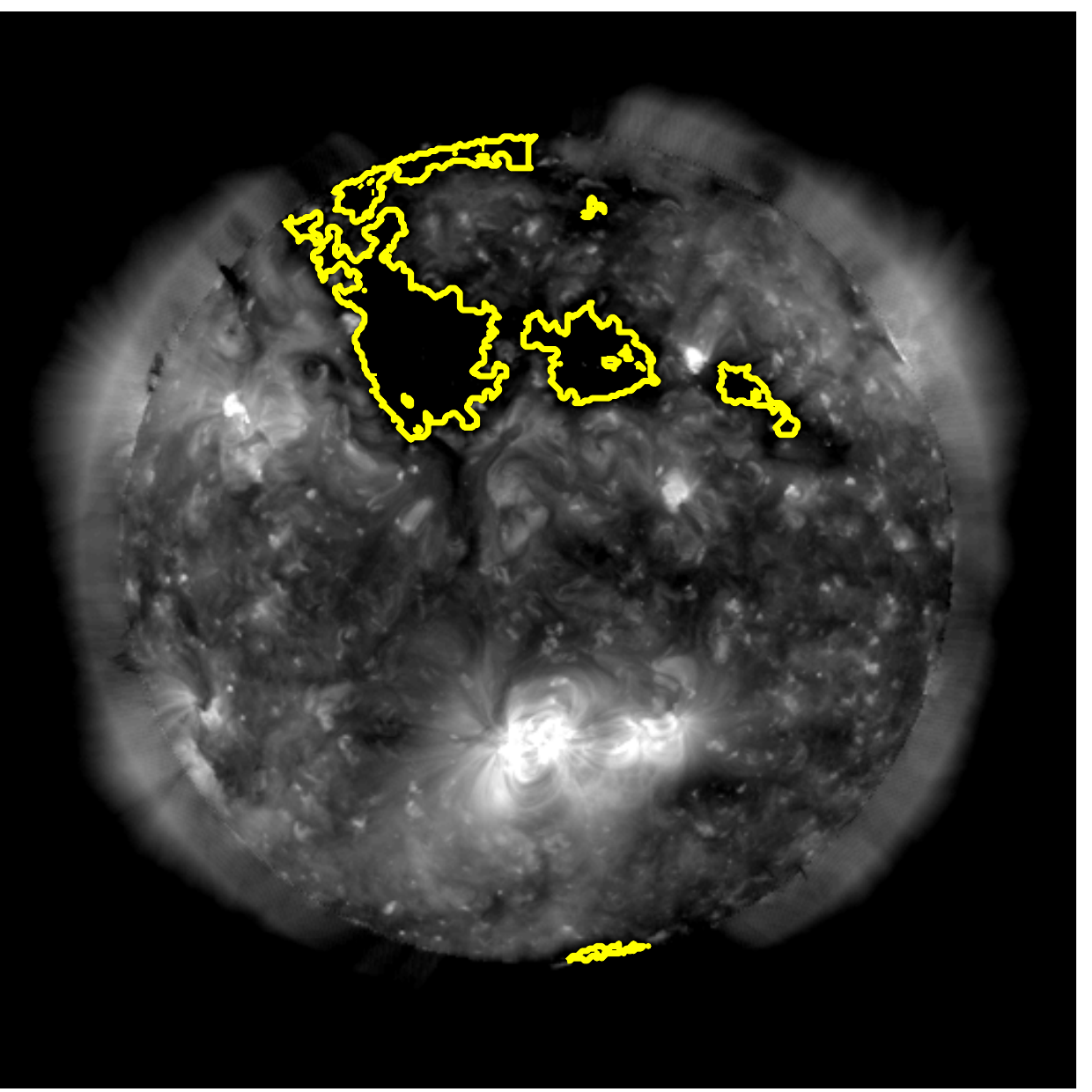}}~
   \subfloat[26 Jul. 2010]{\includegraphics[width=0.13\textwidth]{figures/CR2099_file14_seg}}\\[-2ex]\\
   \subfloat[27 Jul. 2010]{\includegraphics[width=0.13\textwidth]{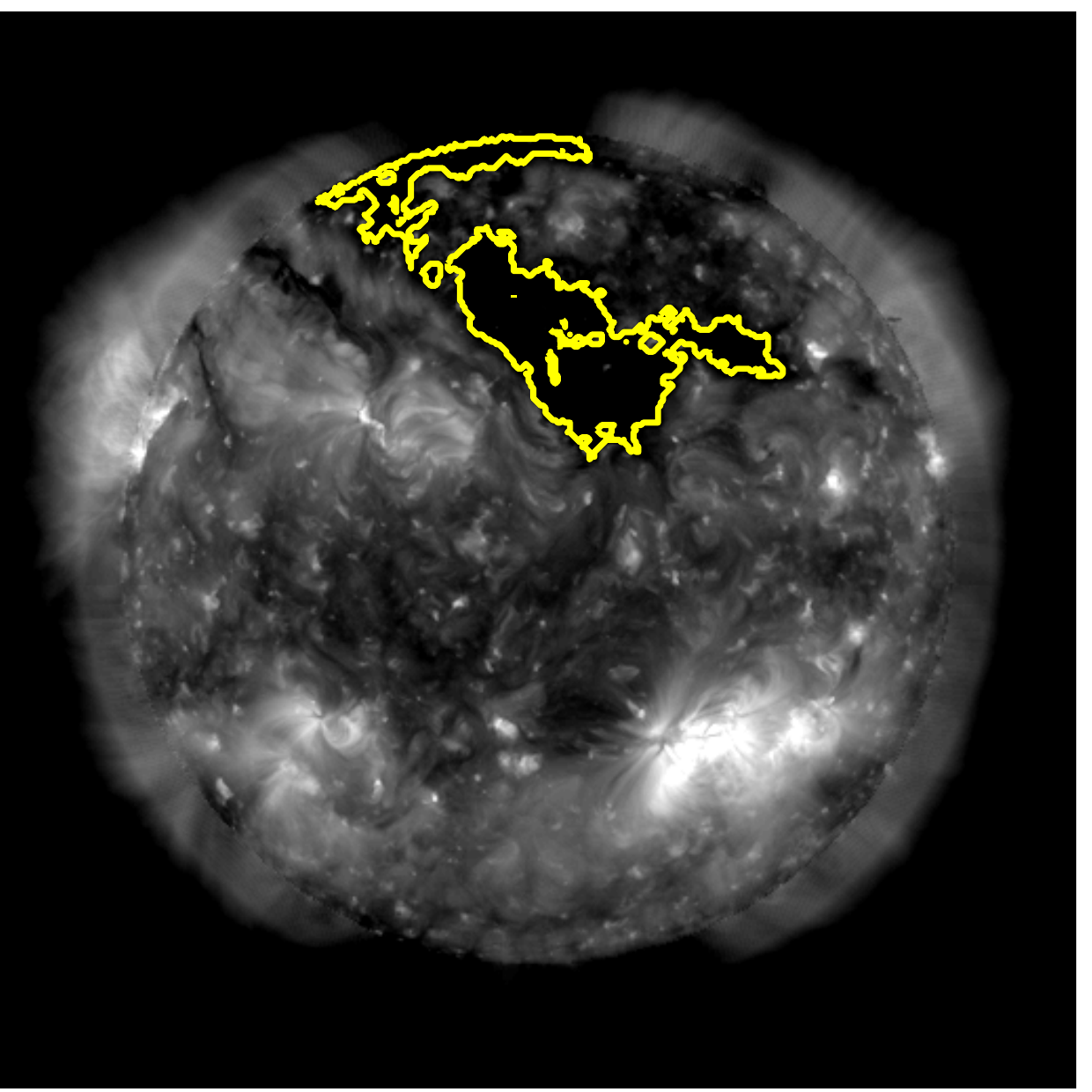}}~
   \subfloat[28 Jul. 2010]{\includegraphics[width=0.13\textwidth]{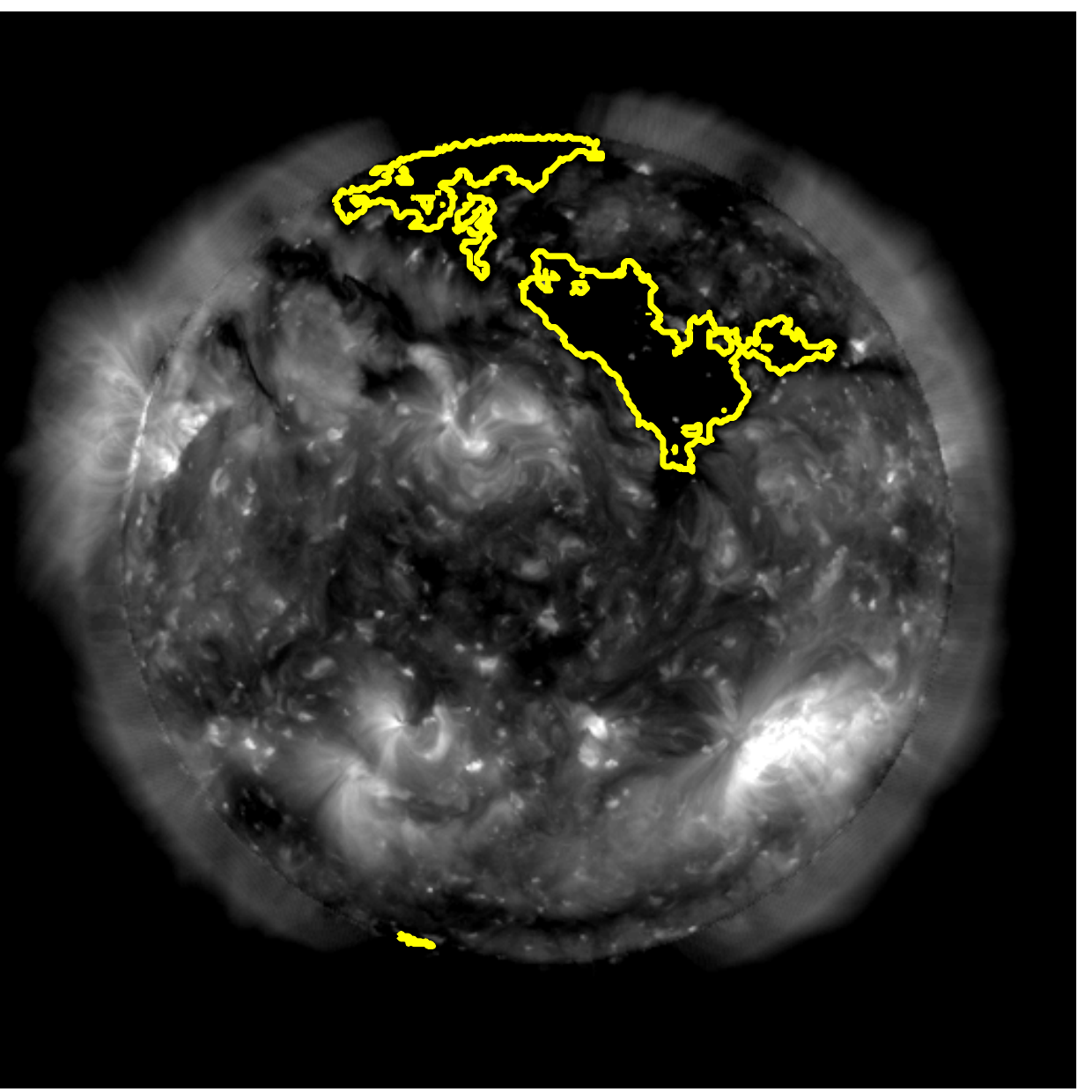}}~
   \subfloat[29 Jul. 2010]{\includegraphics[width=0.13\textwidth]{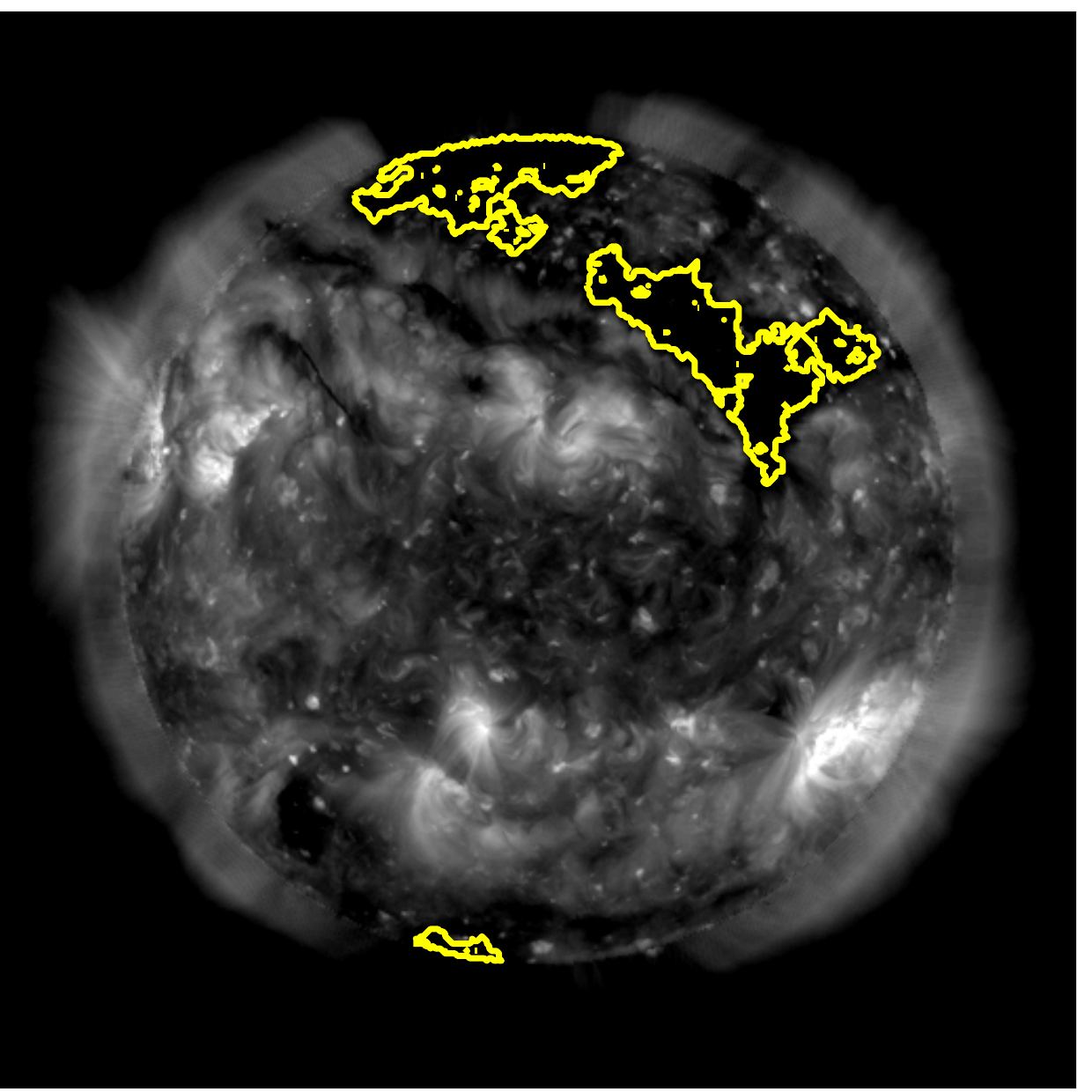}}~
   \subfloat[30 Jul. 2010]{\includegraphics[width=0.13\textwidth]{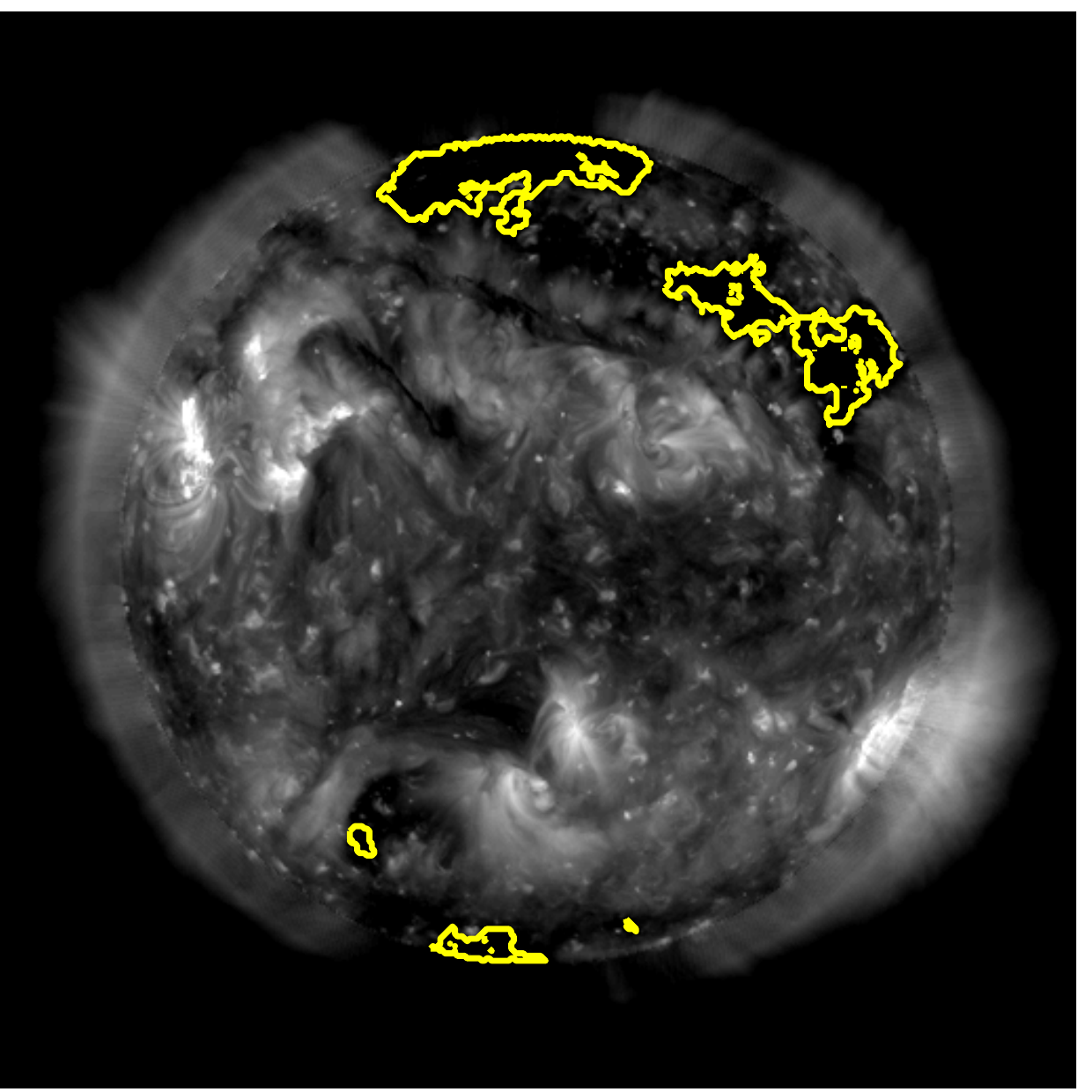}}~
   \subfloat[31 Jul. 2010]{\includegraphics[width=0.13\textwidth]{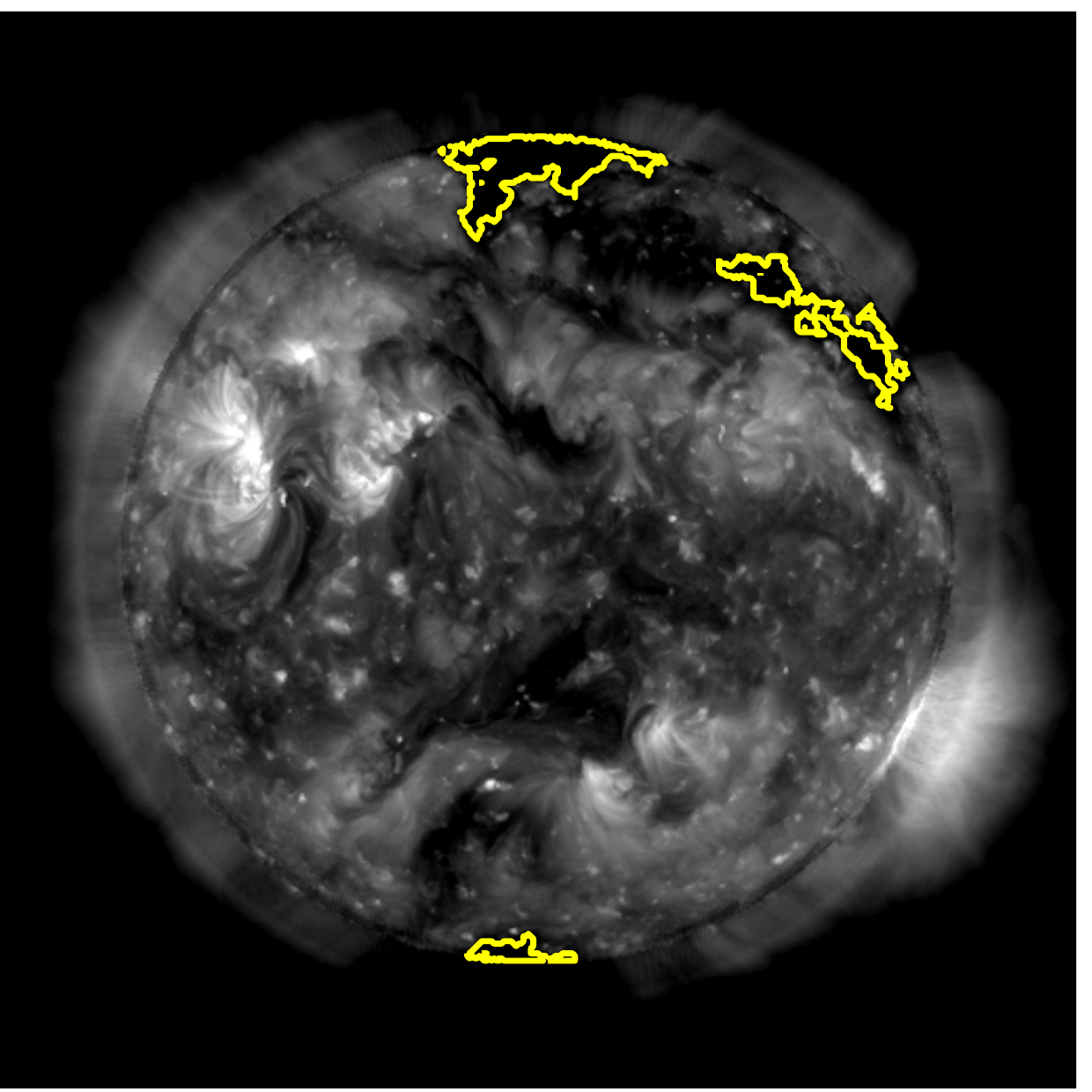}}~
   \subfloat[01 Aug. 2010]{\includegraphics[width=0.13\textwidth]{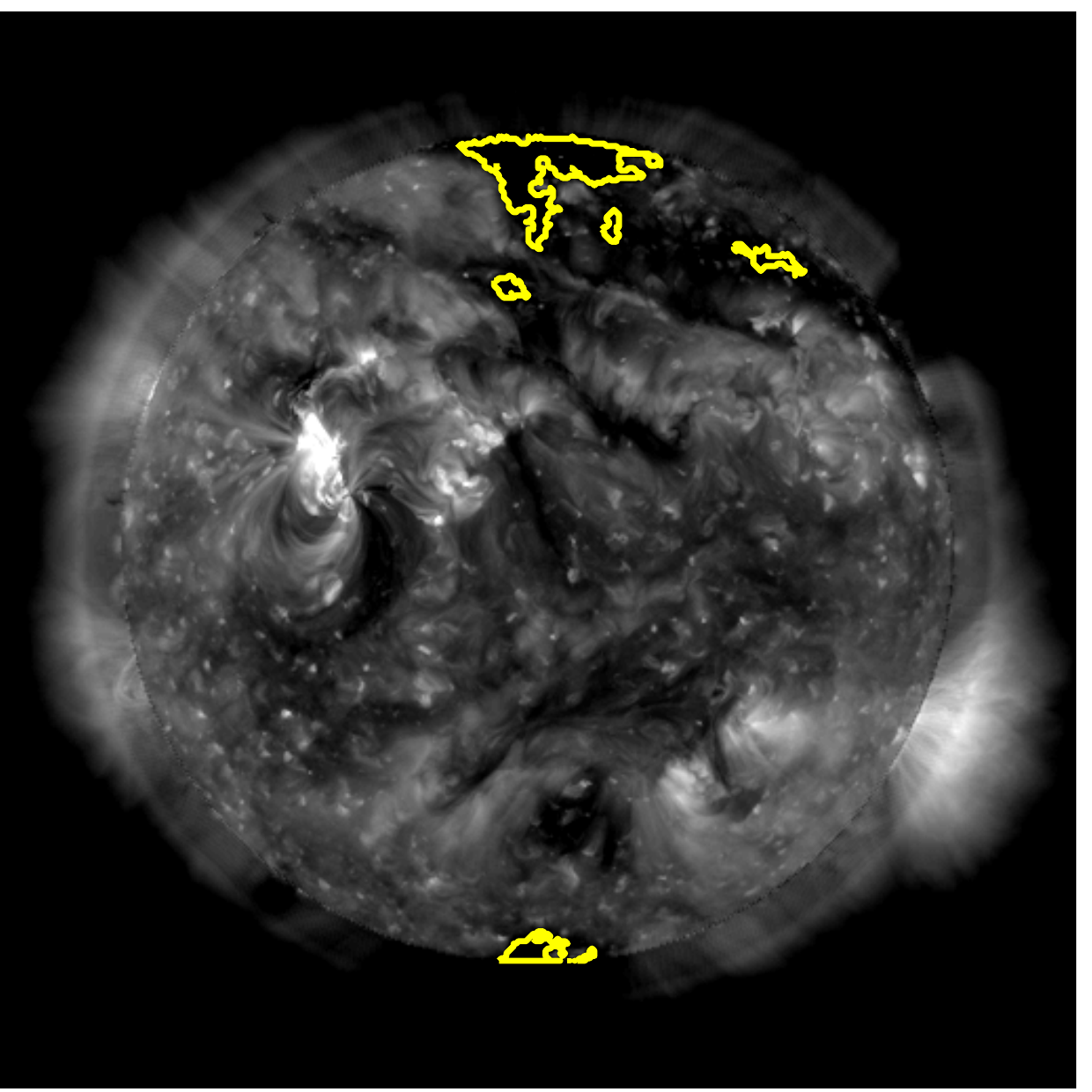}}~
   \subfloat[02 Aug. 2010]{\includegraphics[width=0.13\textwidth]{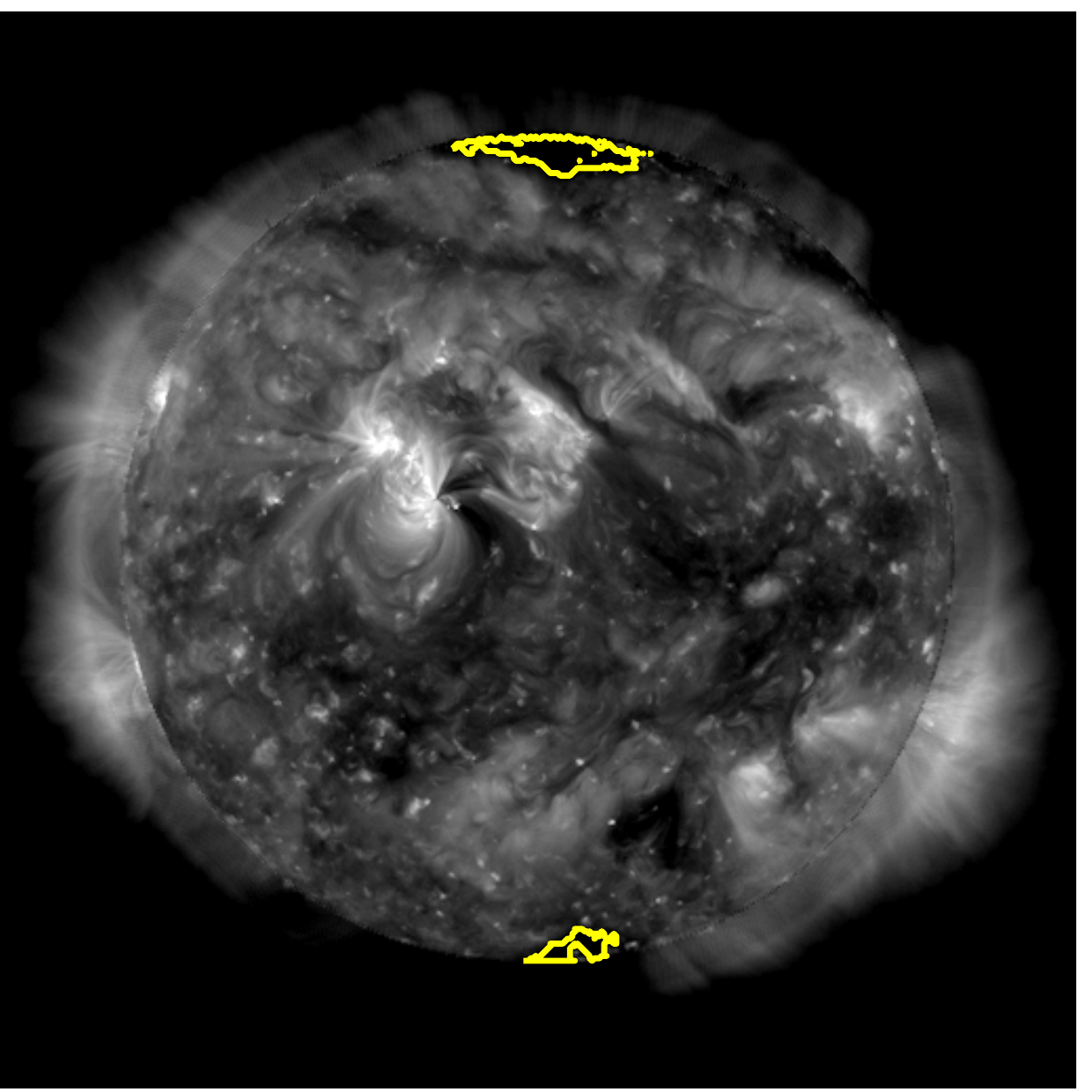}}\\[-2ex]\\
   \subfloat[03 Aug. 2010]{\includegraphics[width=0.13\textwidth]{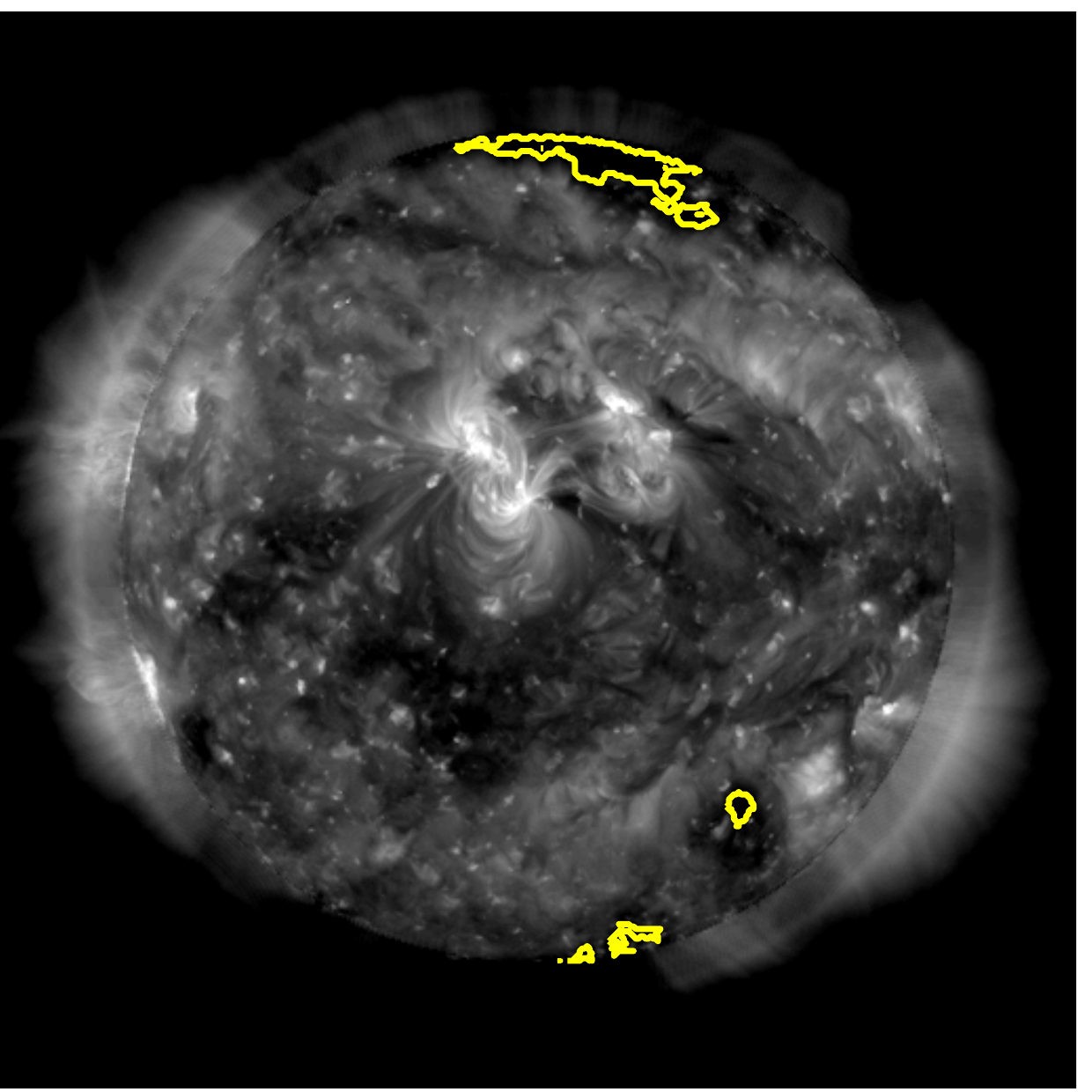}}~
   \subfloat[04 Aug. 2010]{\includegraphics[width=0.13\textwidth]{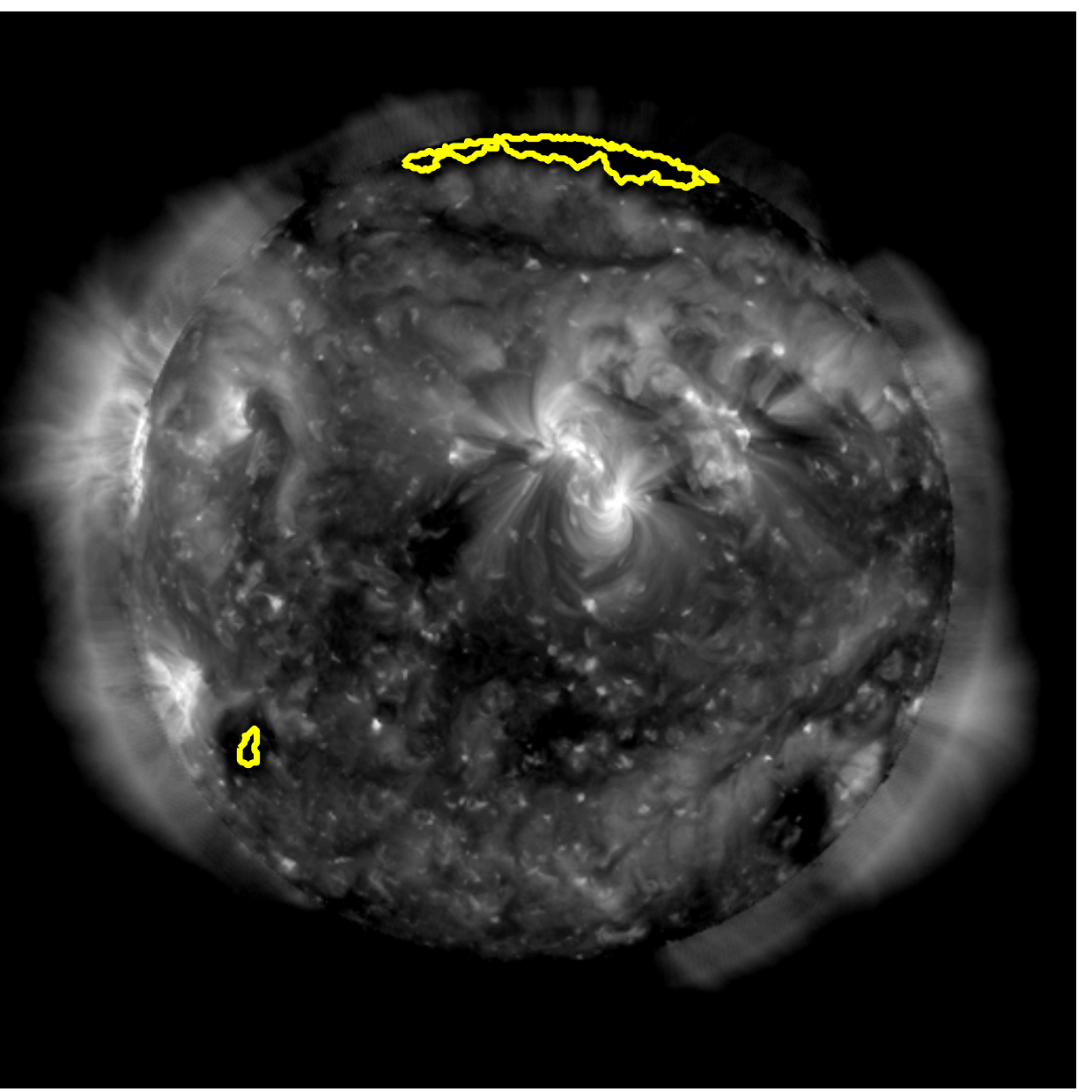}}~
   \subfloat[05 Aug. 2010]{\includegraphics[width=0.13\textwidth]{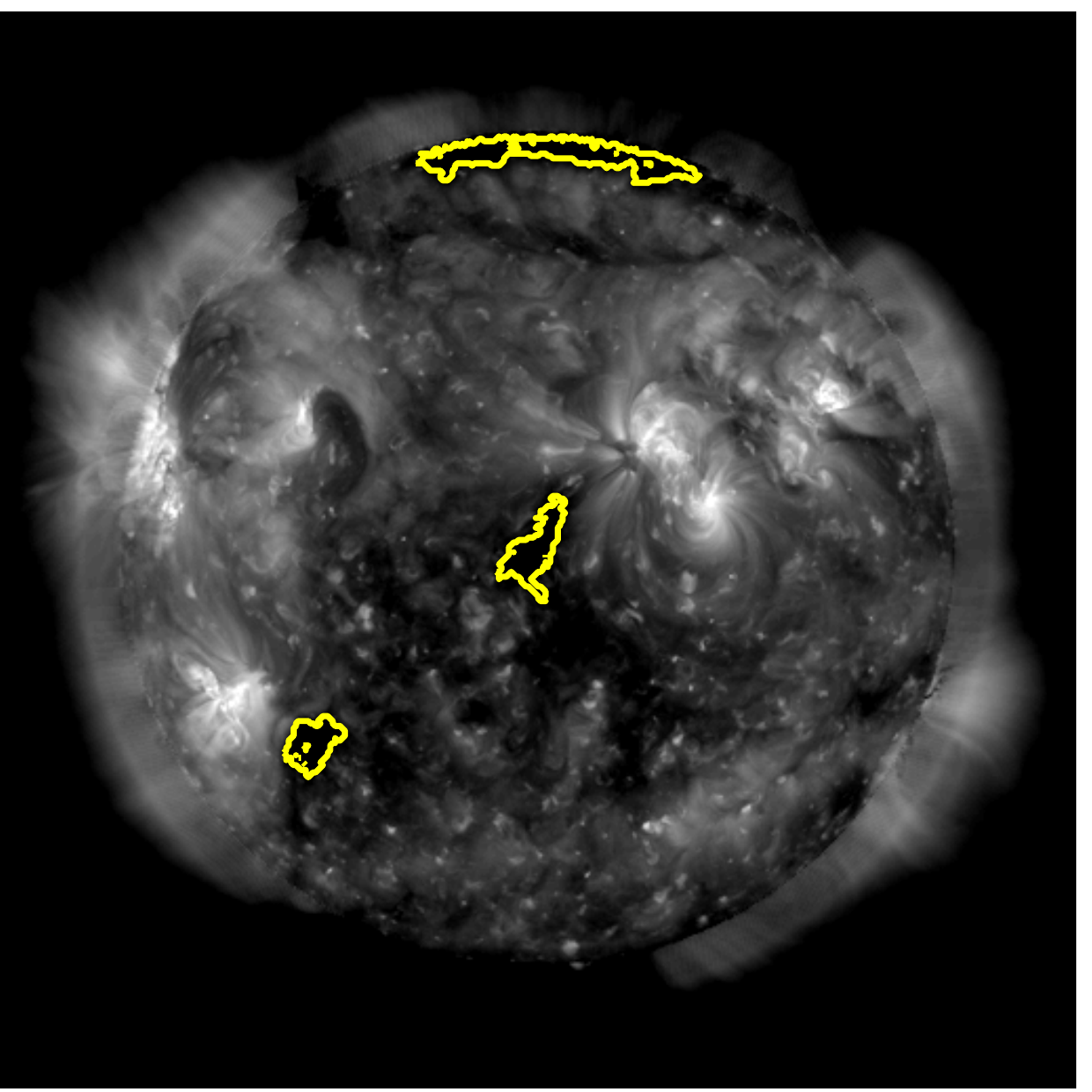}}~
   \subfloat[06 Aug. 2010]{\includegraphics[width=0.13\textwidth]{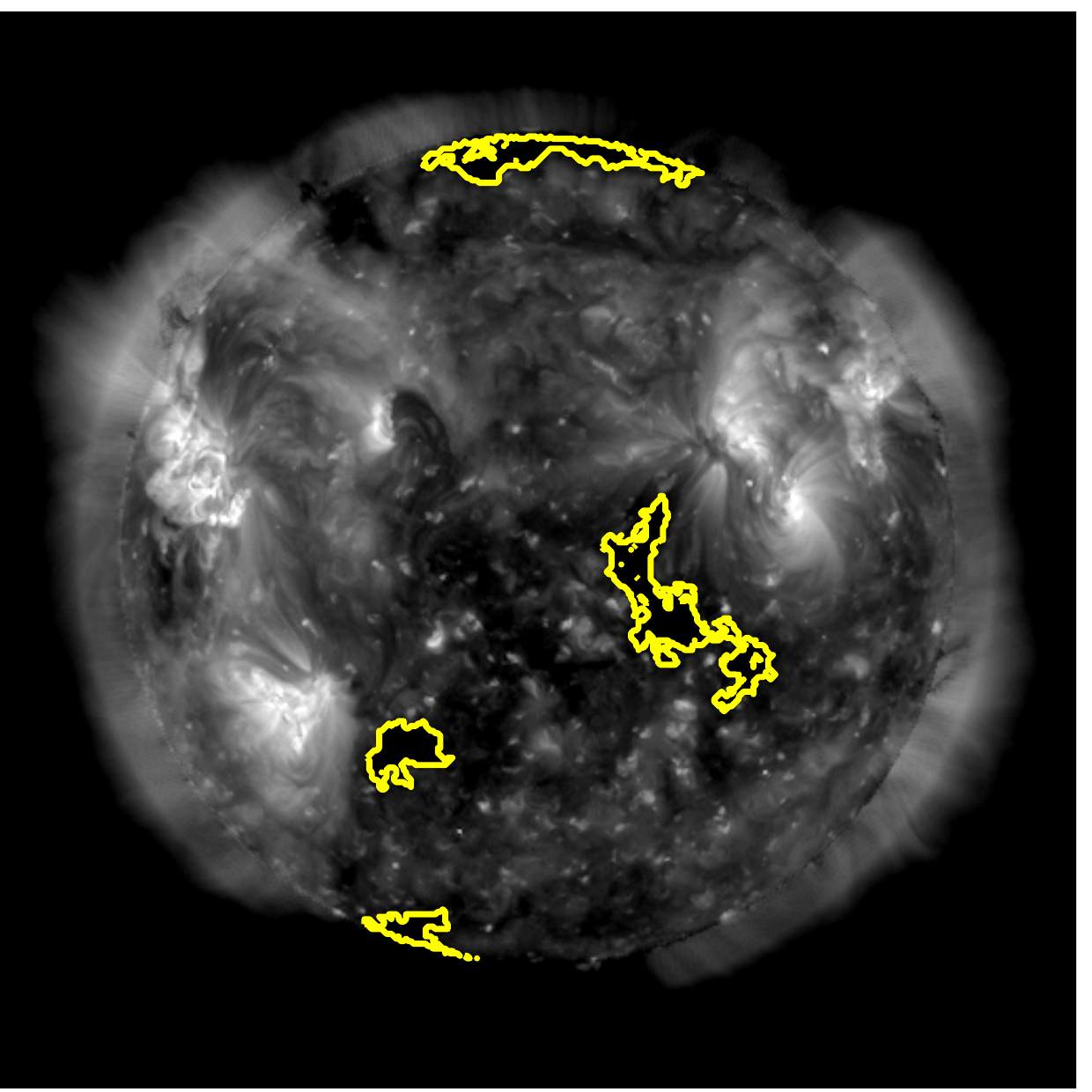}}~
   \subfloat[07 Aug. 2010]{\includegraphics[width=0.13\textwidth]{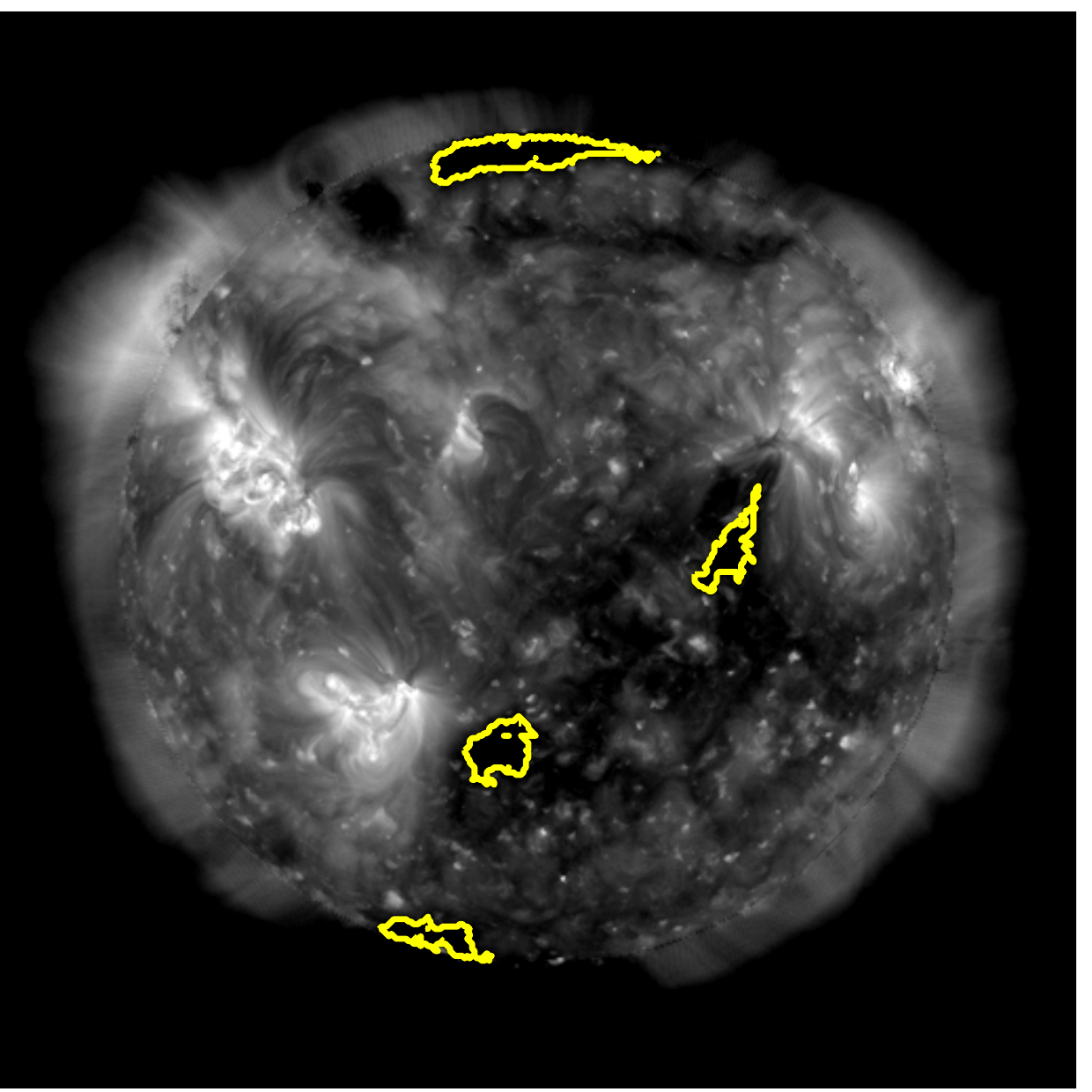}}~
   \subfloat[08 Aug. 2010]{\includegraphics[width=0.13\textwidth]{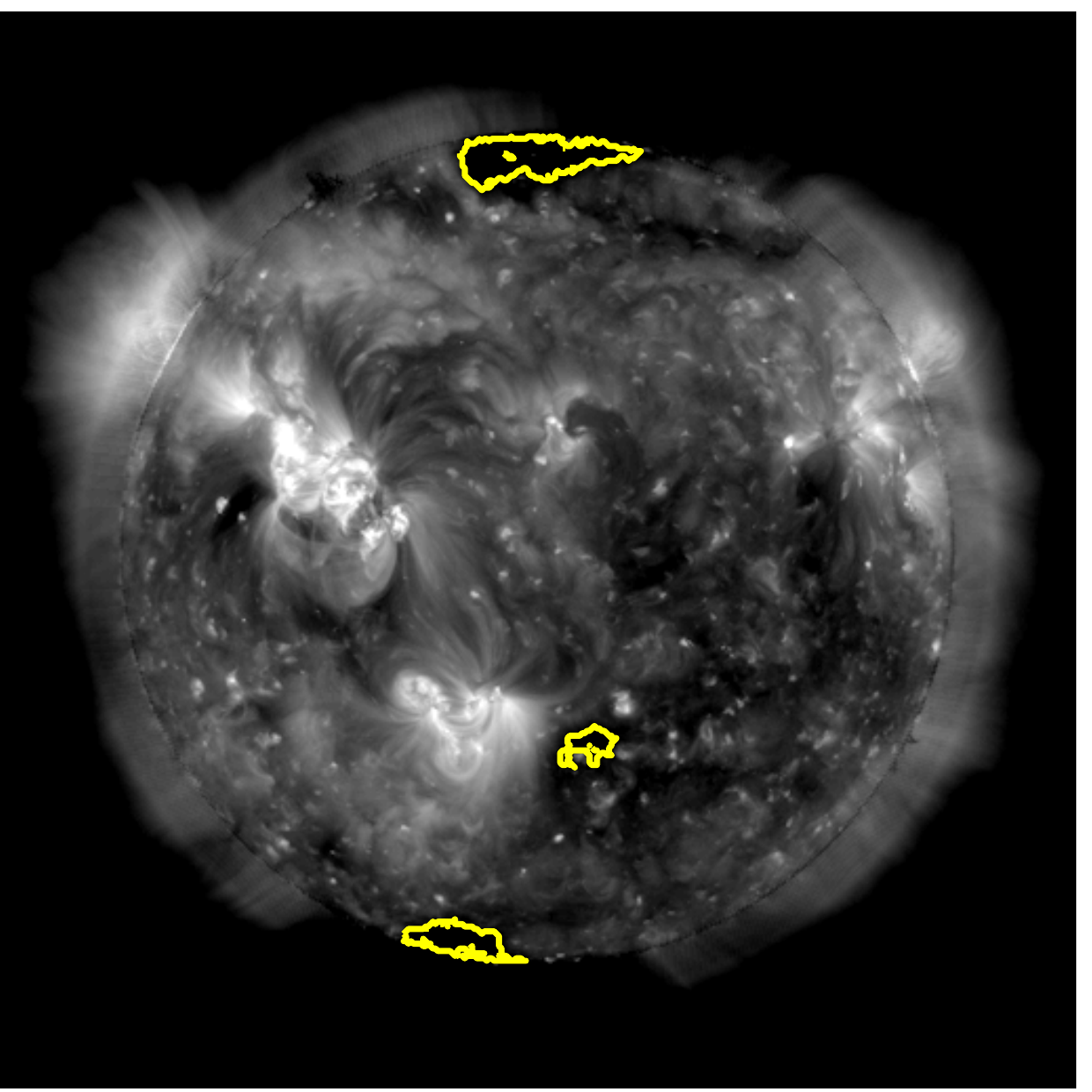}}~
   \subfloat[09 Aug. 2010]{\includegraphics[width=0.13\textwidth]{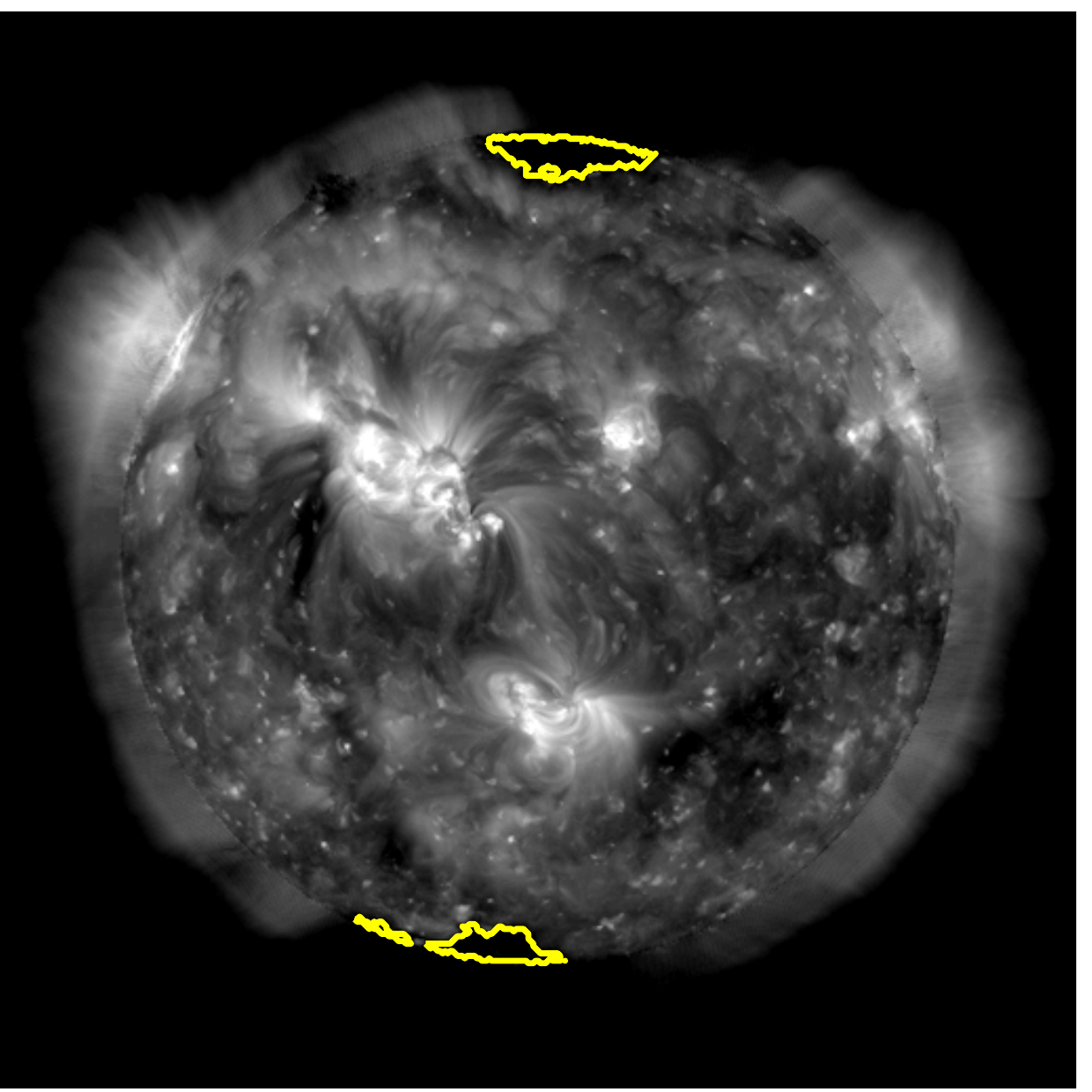}}\\[-2ex]
   \caption{Carrington rotation 2099 193 \AA~images.  Images are clipped to $[100,2500]$ and log scaled. Coronal holes are denoted by the yellow outline for $\lambda_i/\lambda_o=50$ and $\alpha=0.3$.}
   \label{fig:CR2099_segs}
  \end{minipage}
\end{sidewaysfigure}

\begin{sidewaysfigure}[p]
  \centering
  \begin{minipage}[l]{0.85\textwidth}
   \subfloat[25 Jan. 2013]{\includegraphics[width=0.13\textwidth]{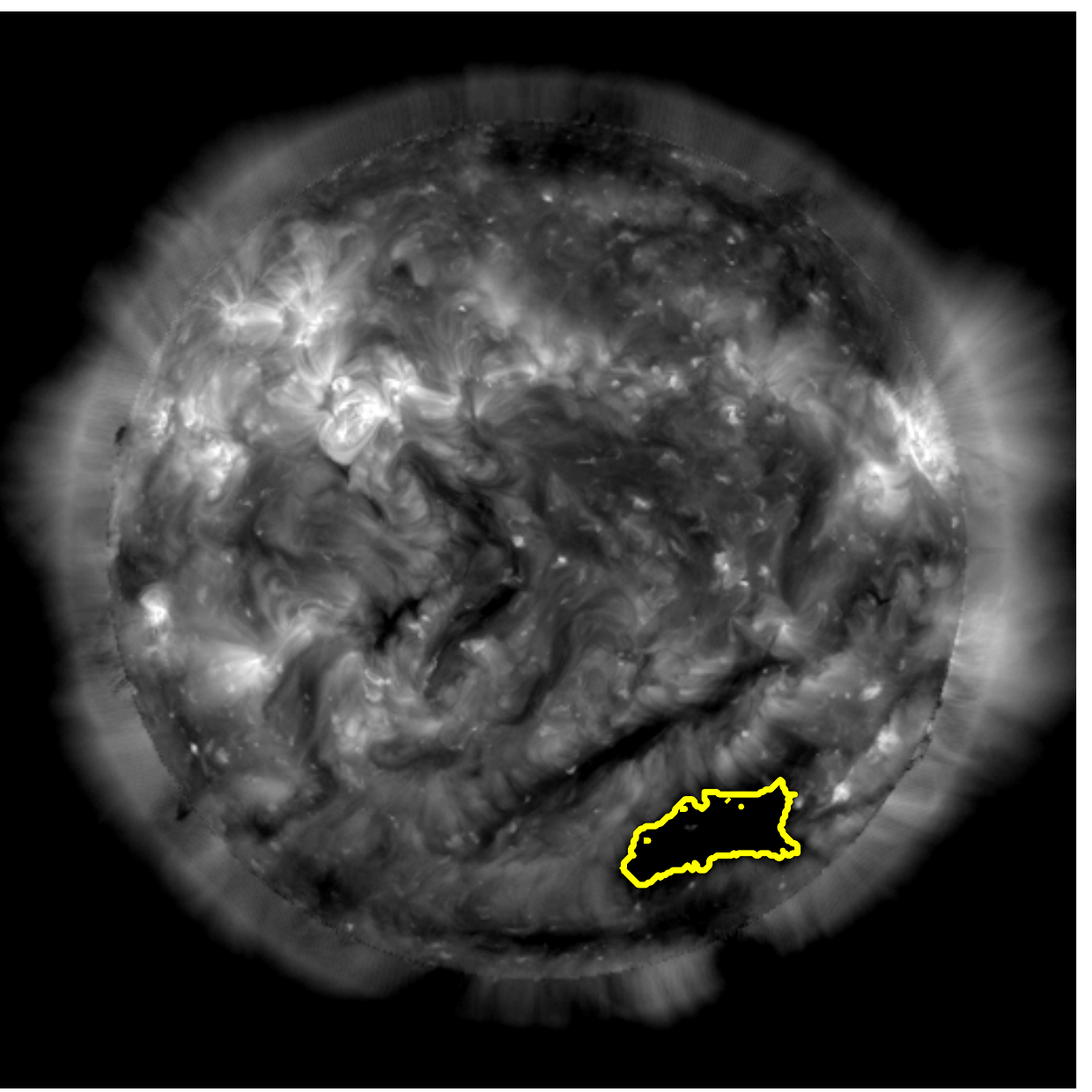}}~
   \subfloat[26 Jan. 2013]{\includegraphics[width=0.13\textwidth]{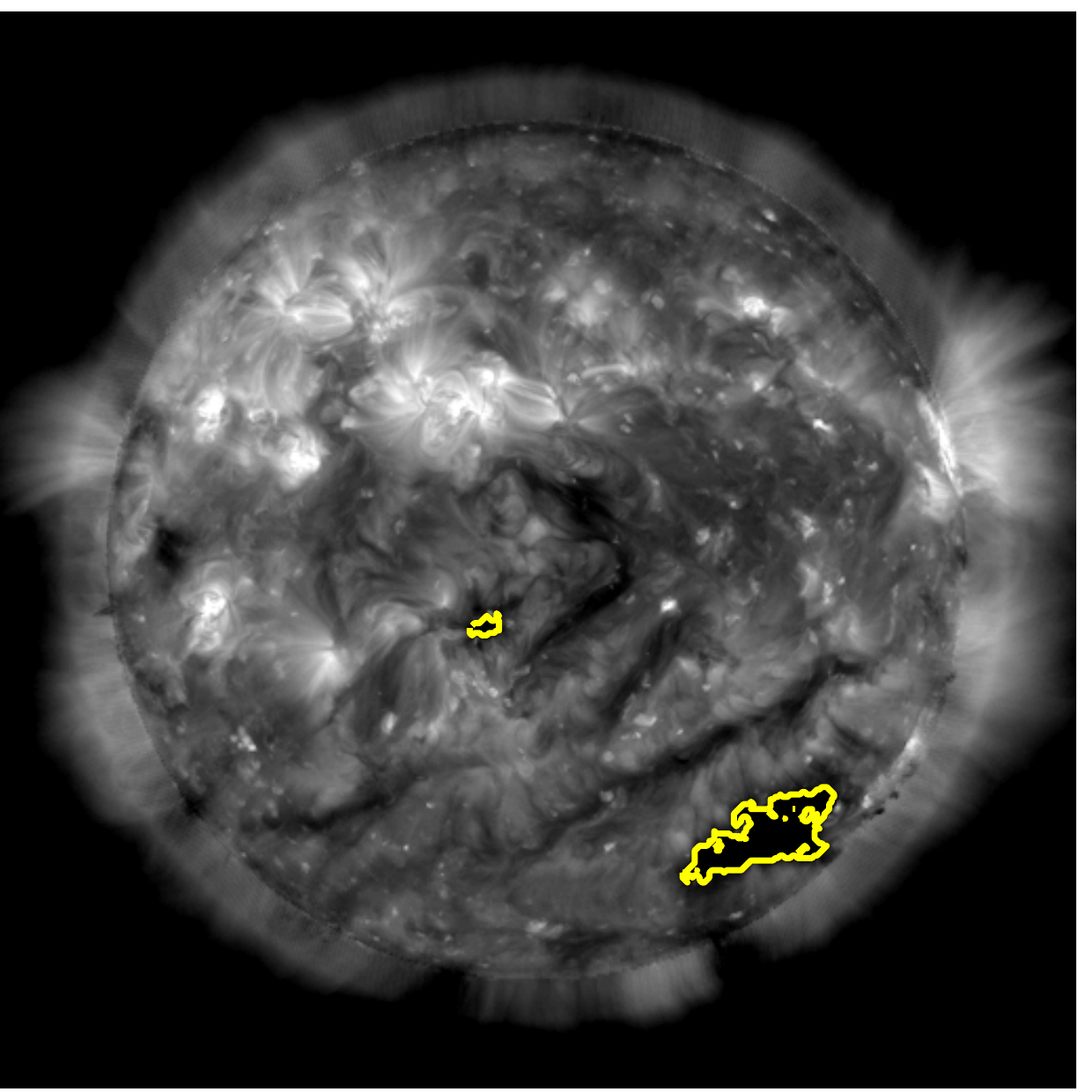}}~
   \subfloat[27 Jan. 2013]{\includegraphics[width=0.13\textwidth]{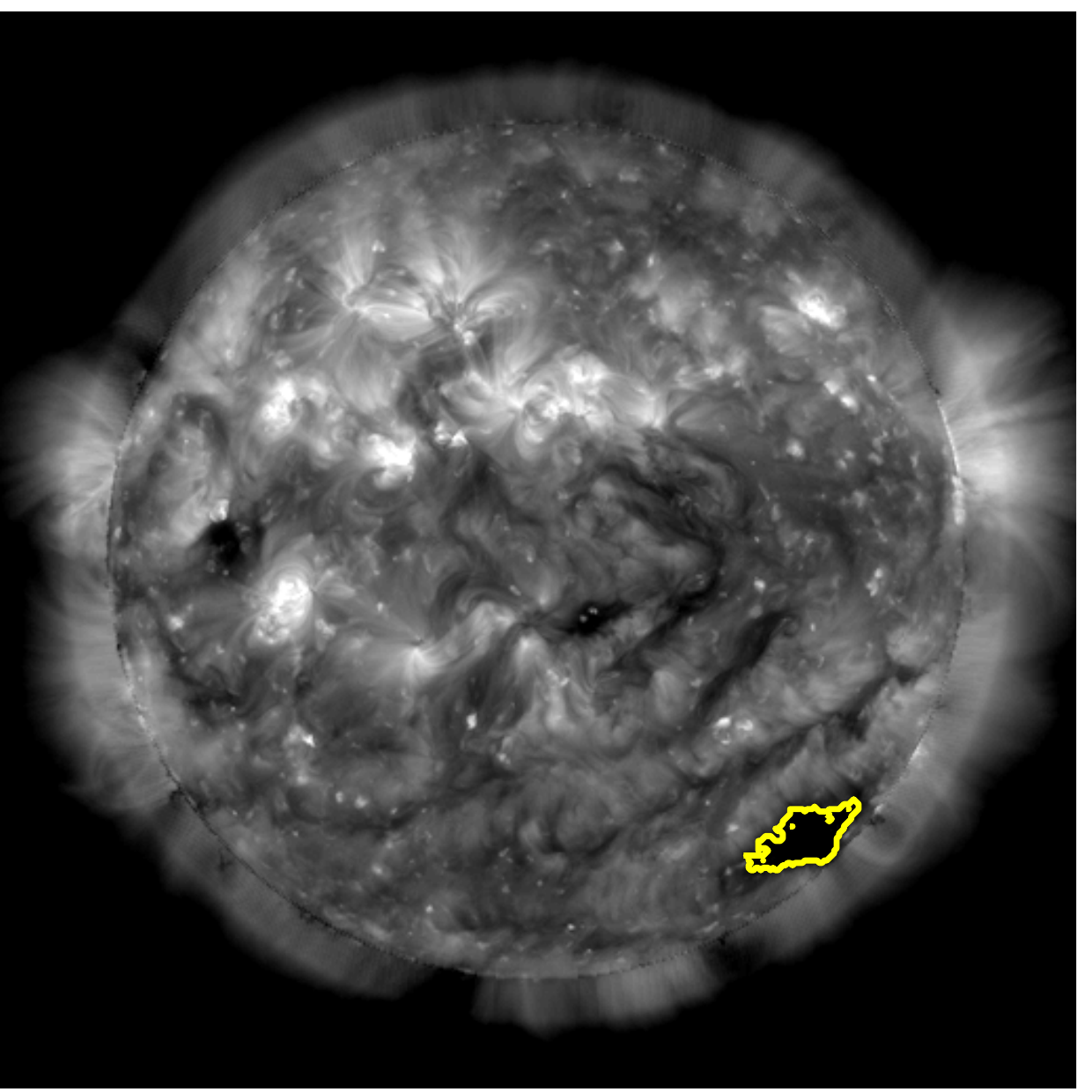}}~
   \subfloat[28 Jan. 2013]{\includegraphics[width=0.13\textwidth]{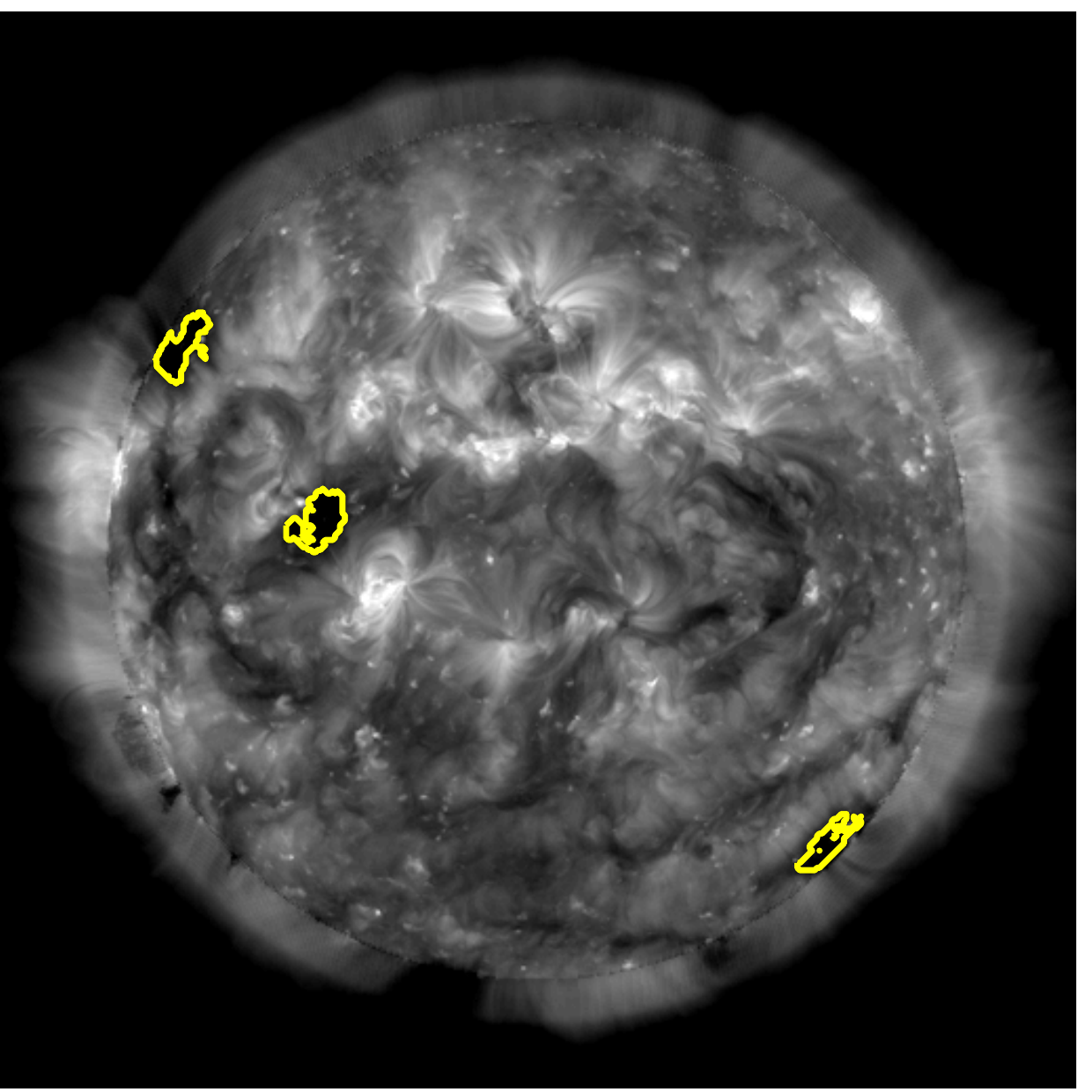}}~
   \subfloat[29 Jan. 2013]{\includegraphics[width=0.13\textwidth]{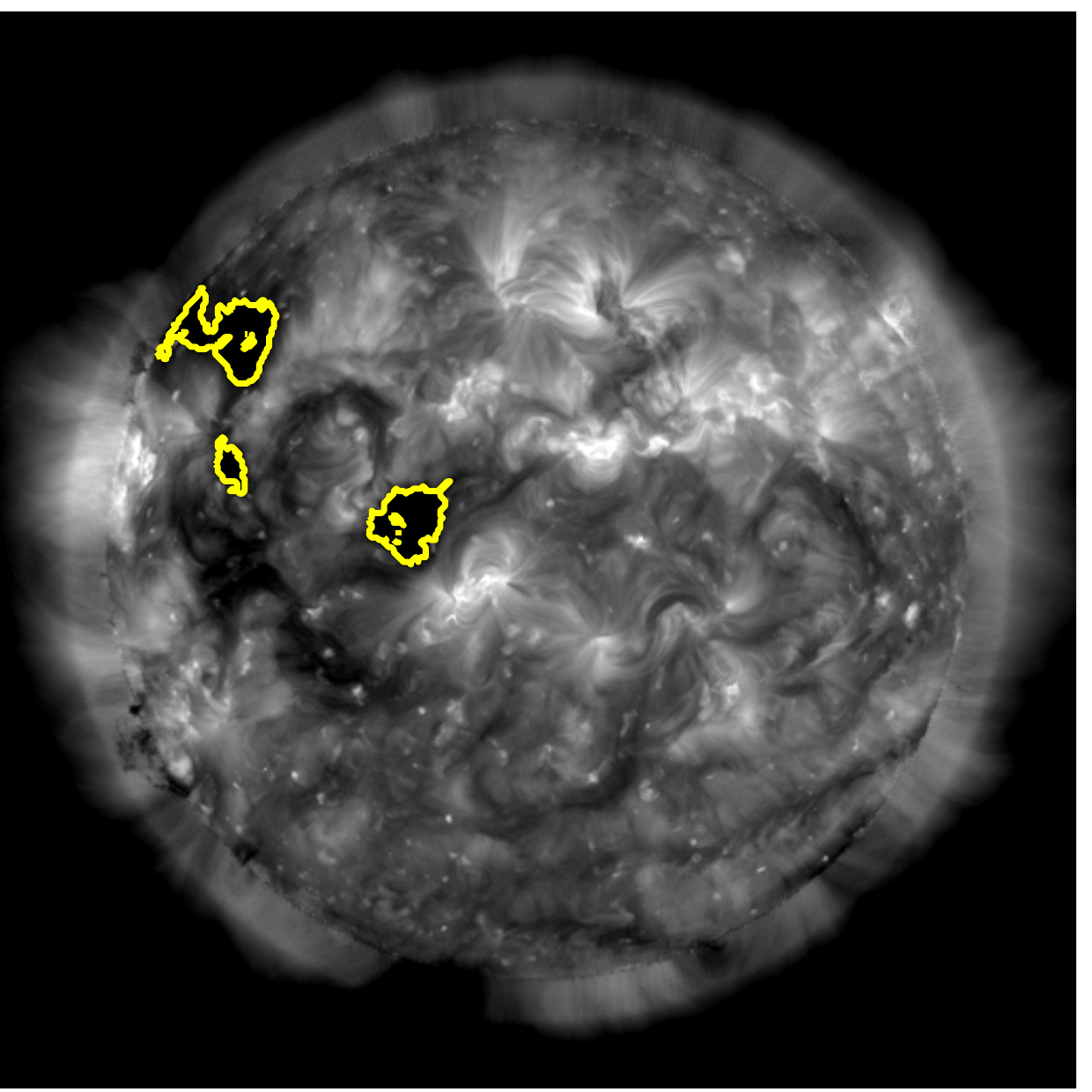}}~
   \subfloat[30 Jan. 2013]{\includegraphics[width=0.13\textwidth]{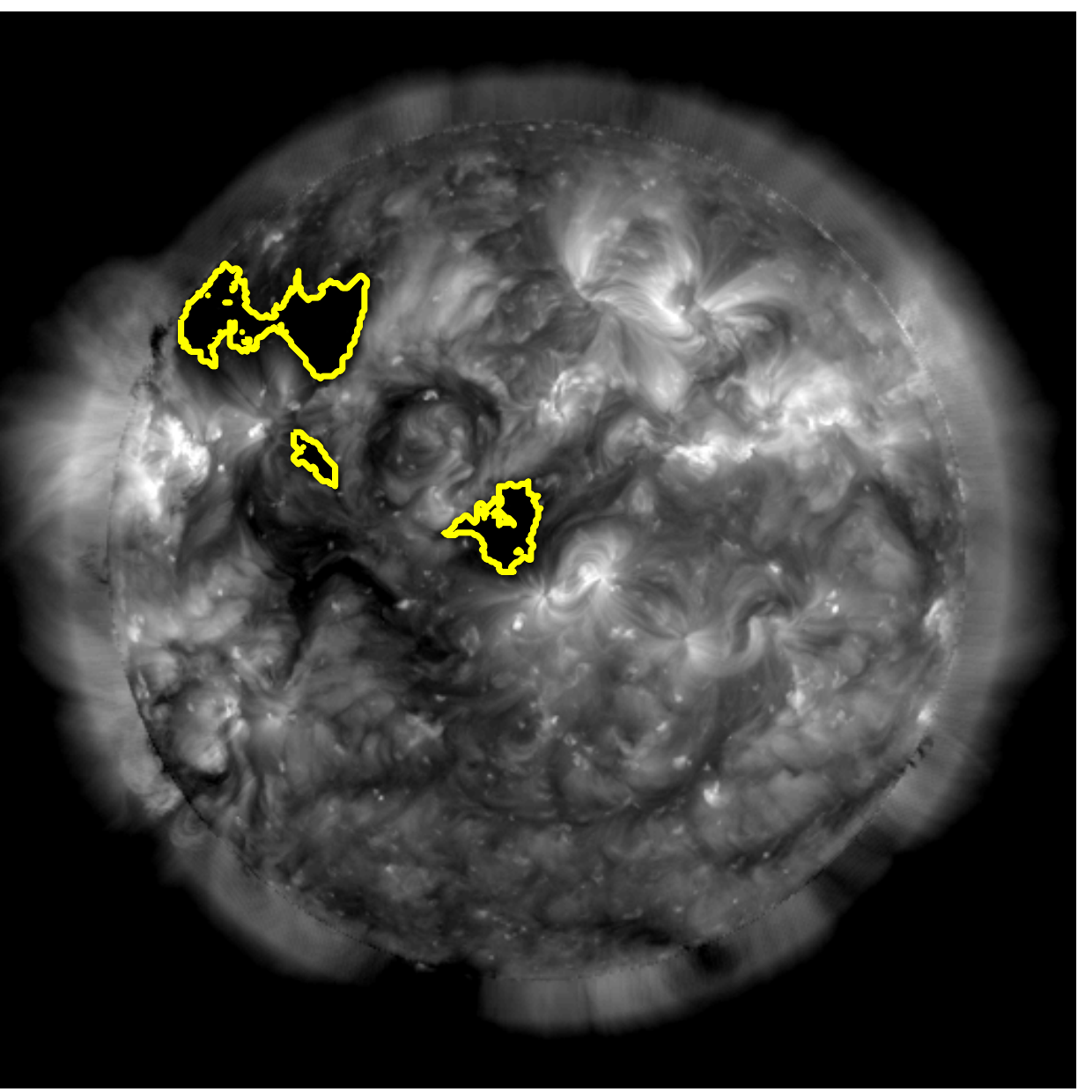}}~
   \subfloat[31 Jan. 2013]{\includegraphics[width=0.13\textwidth]{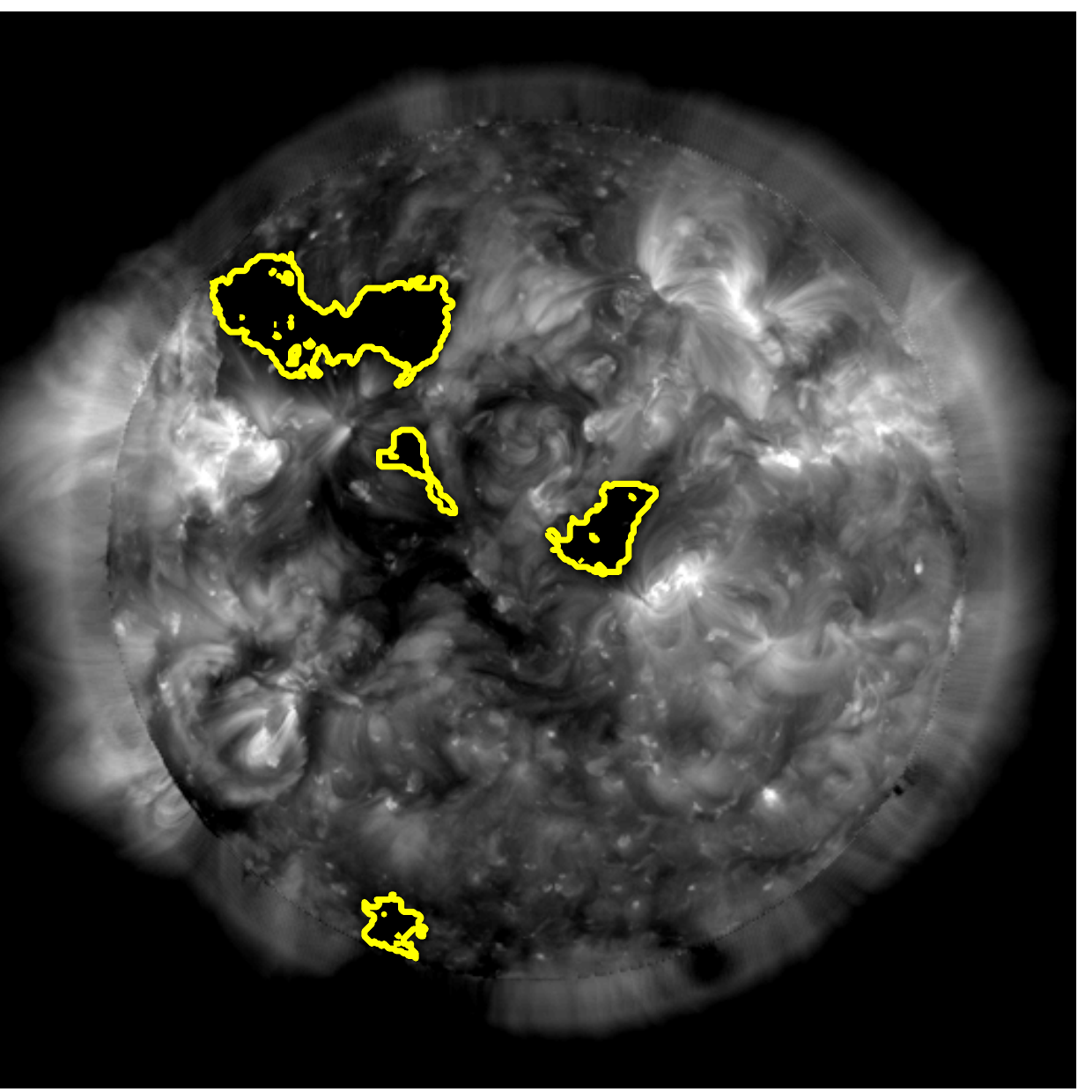}}\\[-2ex]\\
   \subfloat[01 Feb. 2013]{\includegraphics[width=0.13\textwidth]{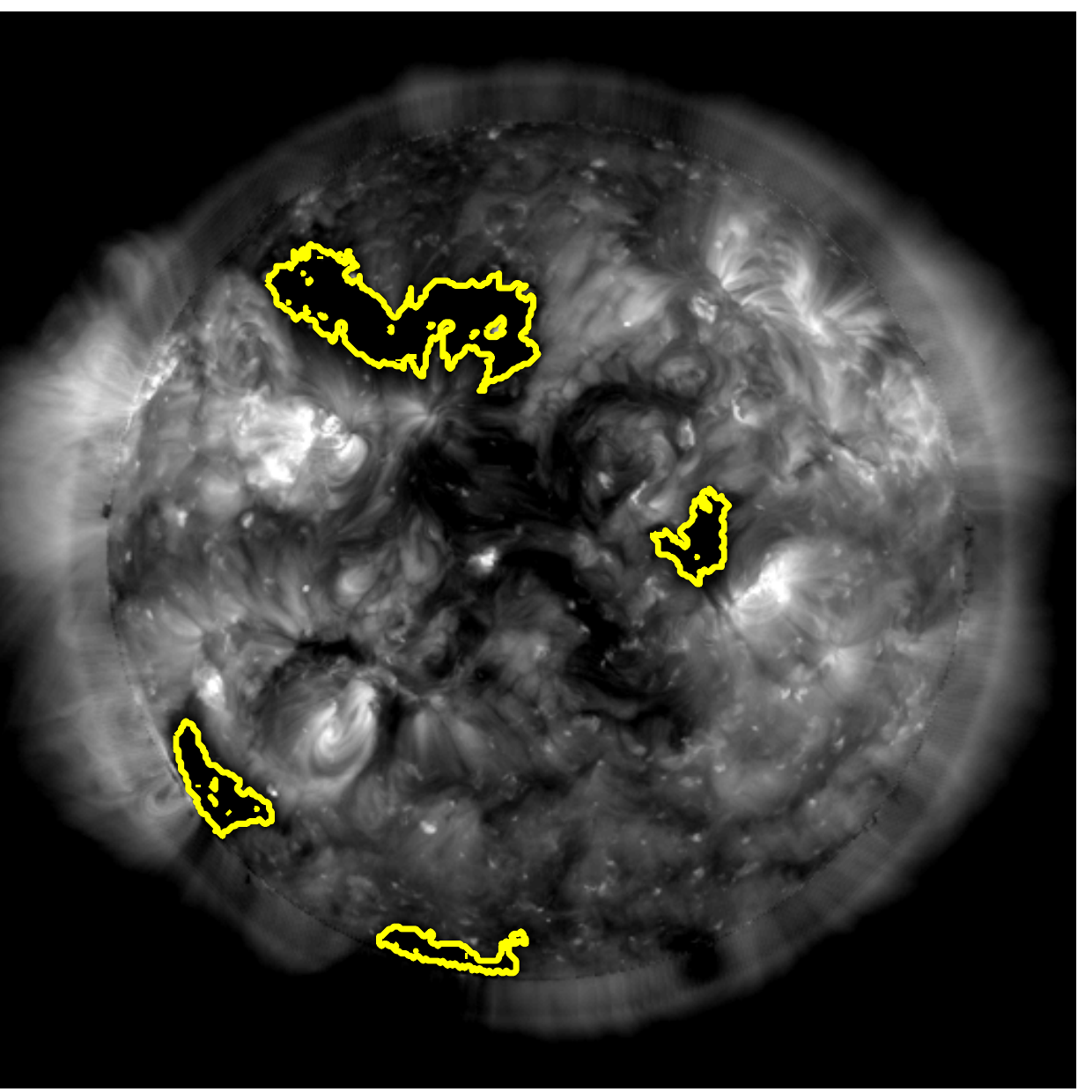}}~
   \subfloat[02 Feb. 2013]{\includegraphics[width=0.13\textwidth]{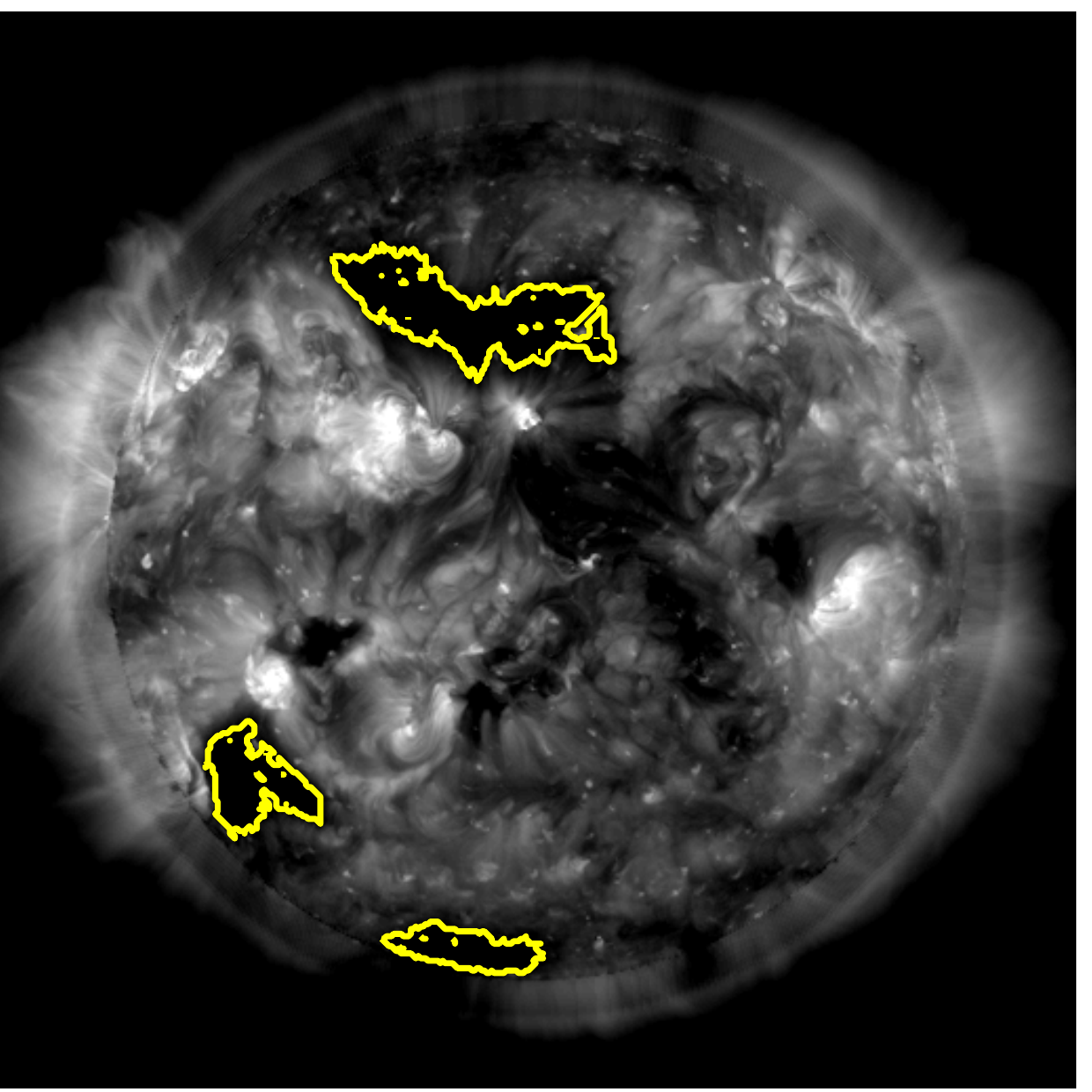}}~
   \subfloat[03 Feb. 2013]{\includegraphics[width=0.13\textwidth]{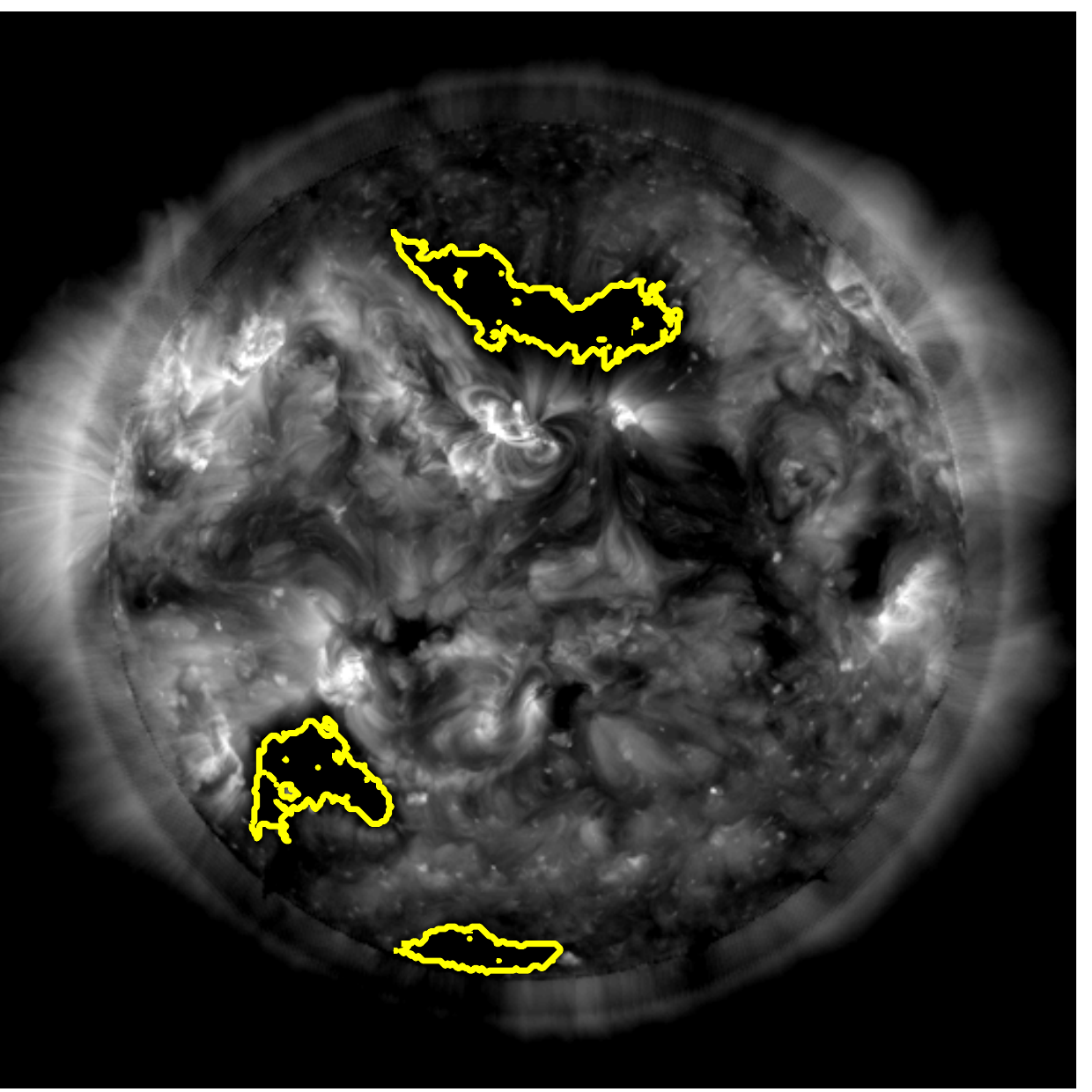}}~
   \subfloat[04 Feb. 2013]{\includegraphics[width=0.13\textwidth]{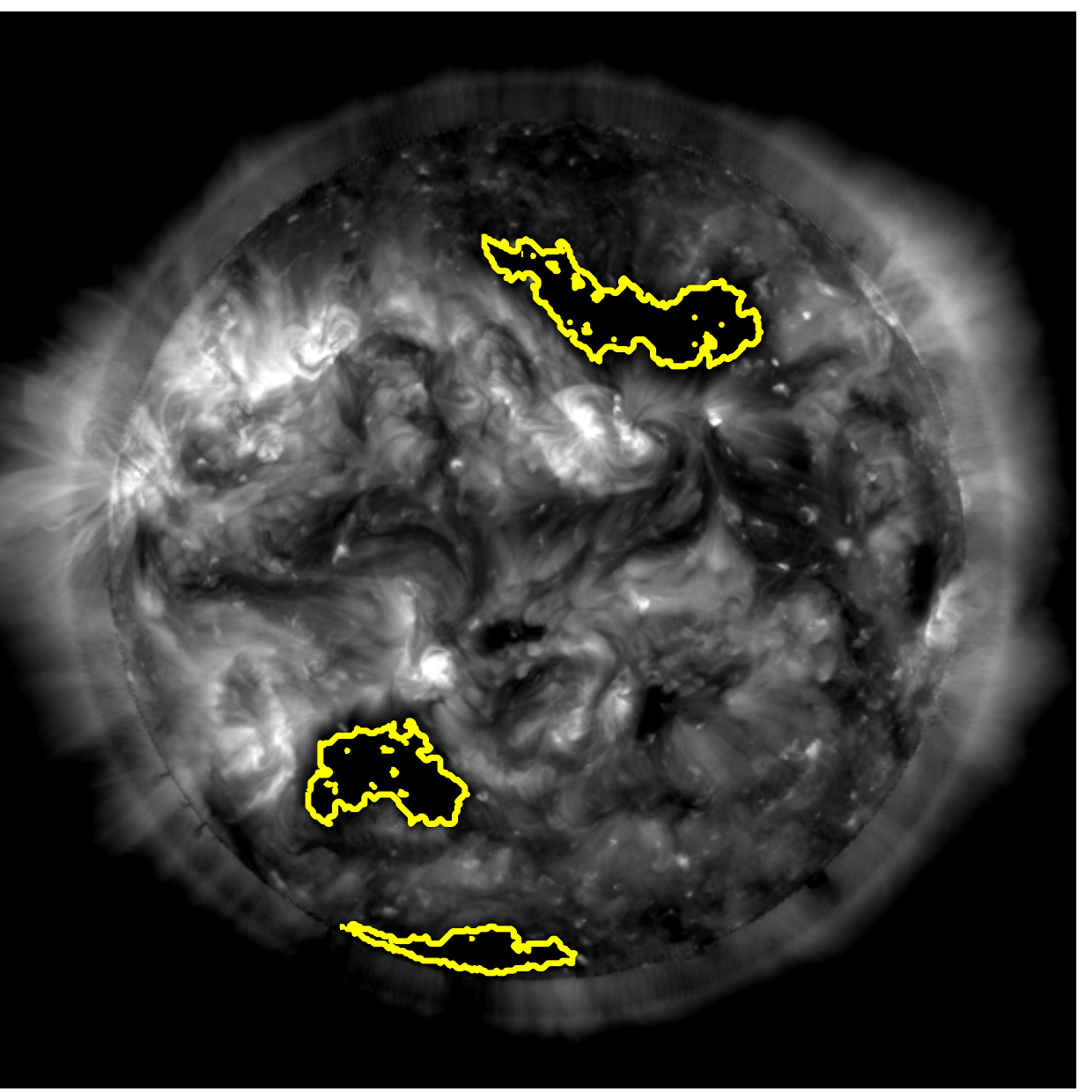}}~
   \subfloat[05 Feb. 2013]{\includegraphics[width=0.13\textwidth]{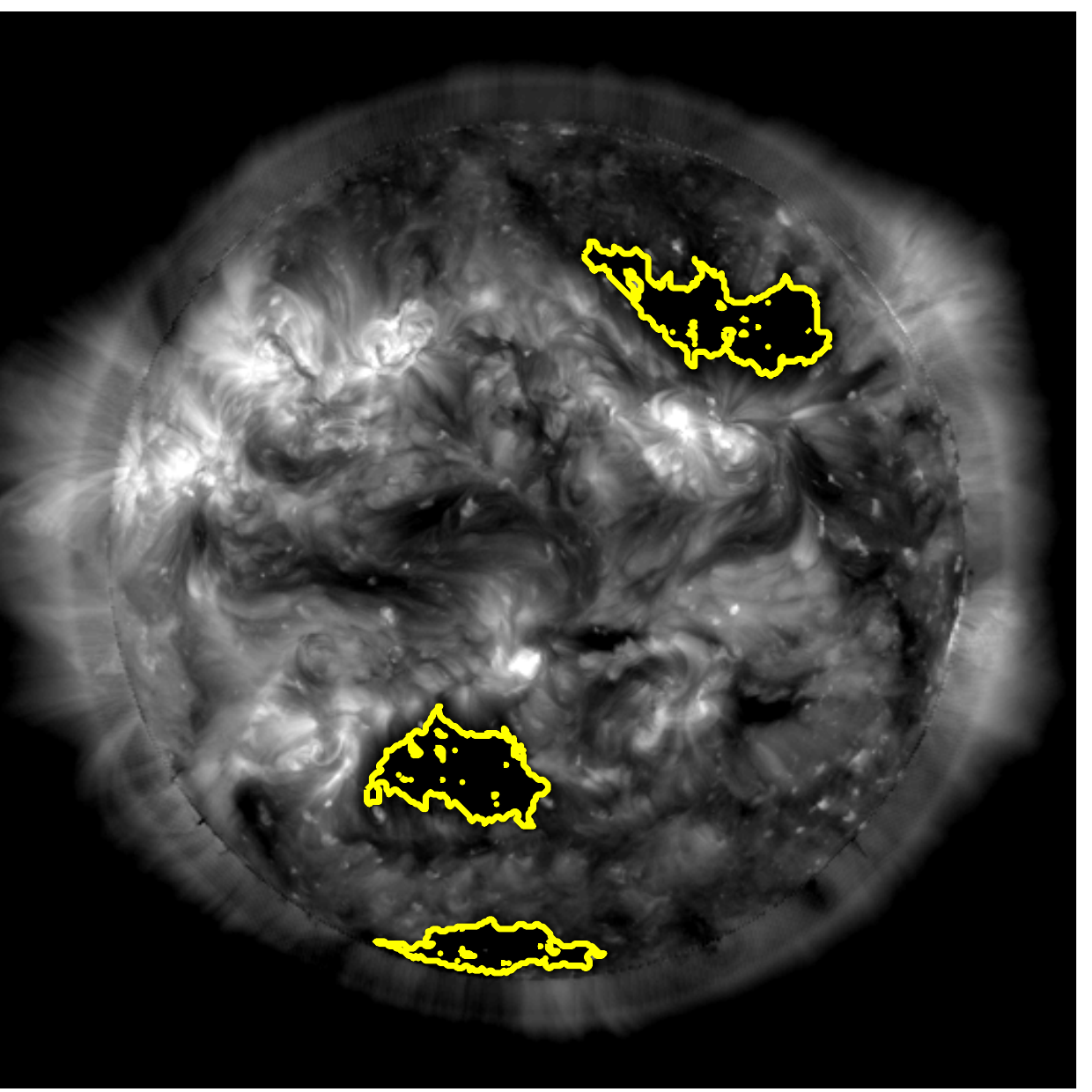}}~
   \subfloat[06 Feb. 2013]{\includegraphics[width=0.13\textwidth]{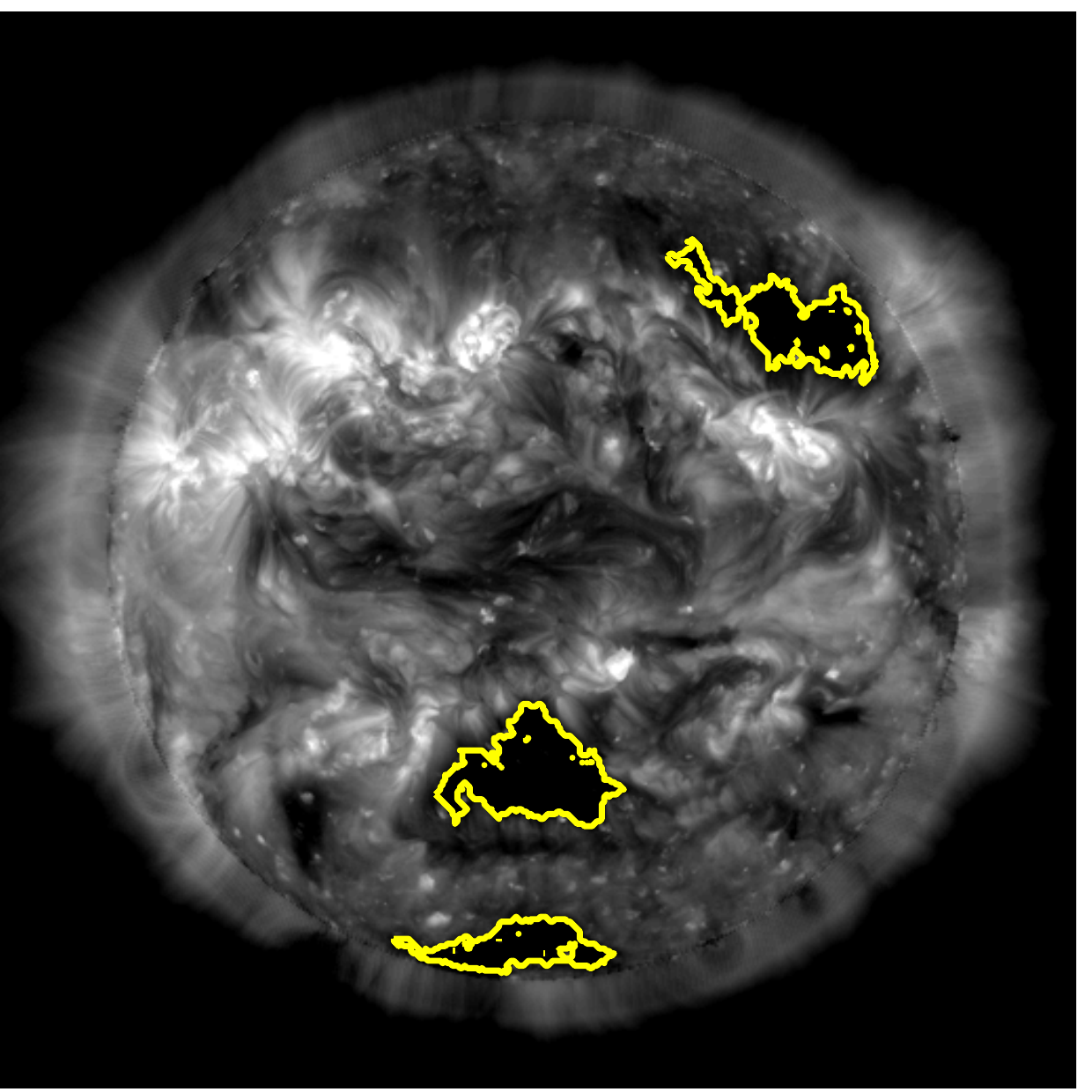}}~
   \subfloat[07 Feb. 2013]{\includegraphics[width=0.13\textwidth]{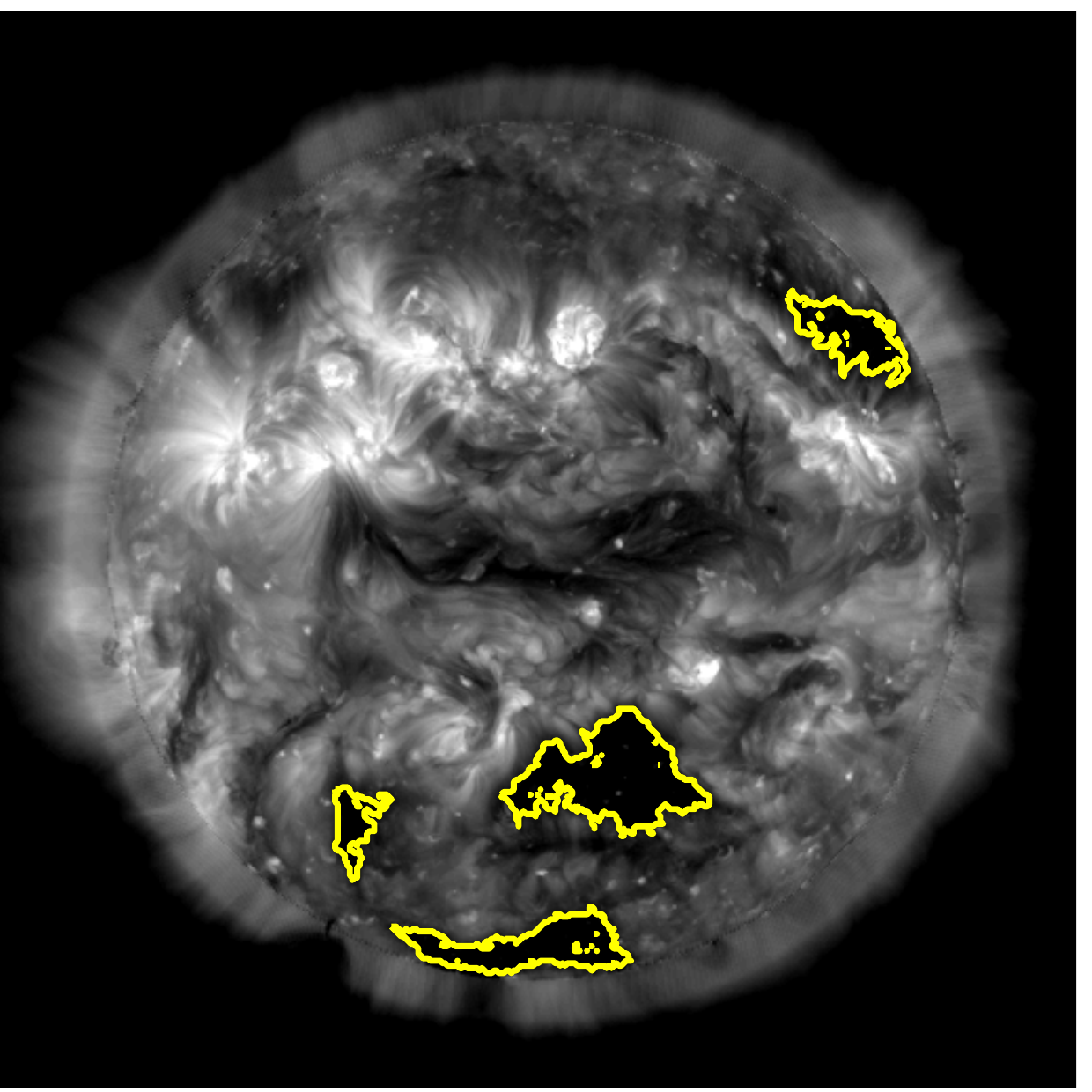}}\\[-2ex]\\
   \subfloat[08 Feb. 2013]{\includegraphics[width=0.13\textwidth]{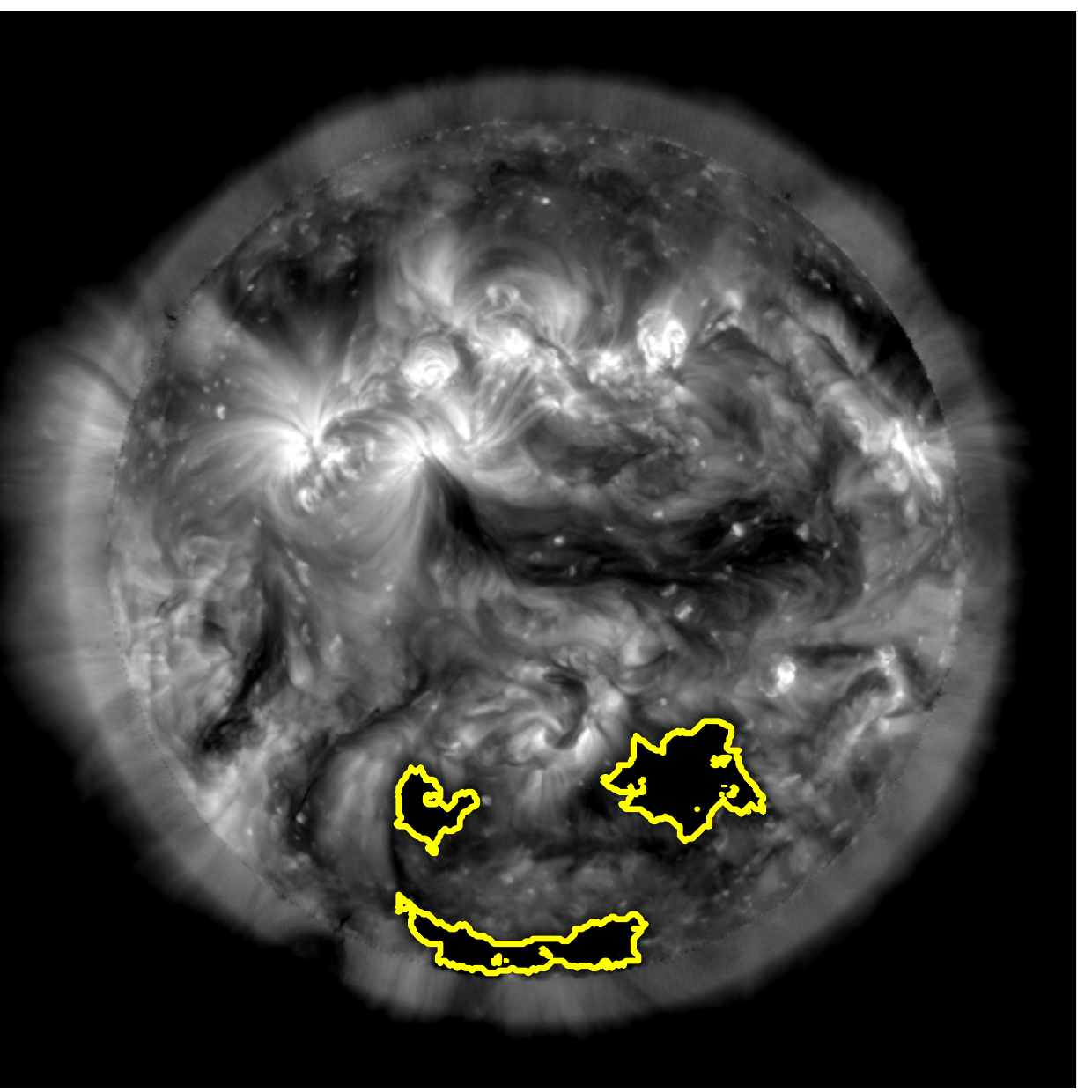}}~
   \subfloat[09 Feb. 2013]{\includegraphics[width=0.13\textwidth]{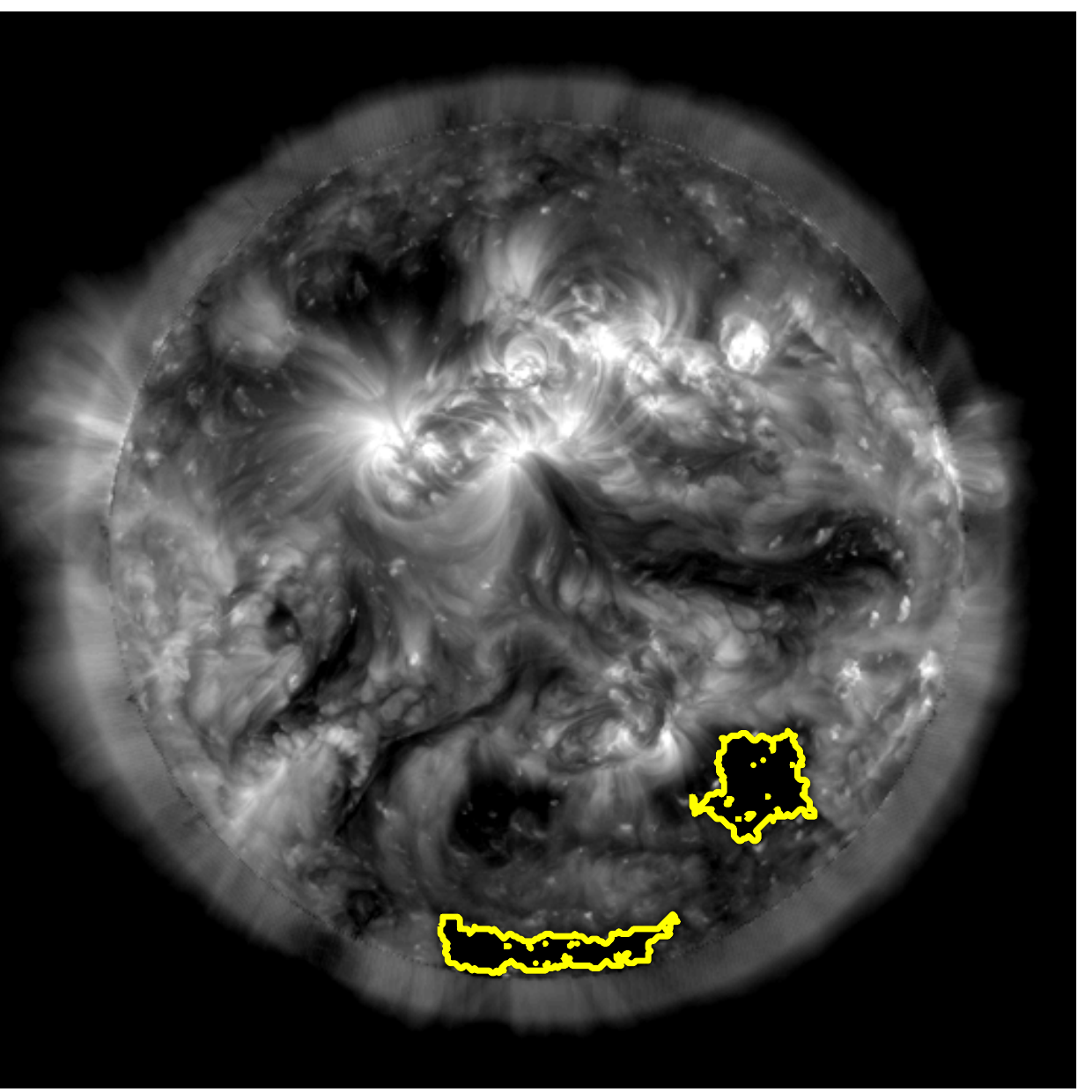}}~
   \subfloat[10 Feb. 2013]{\includegraphics[width=0.13\textwidth]{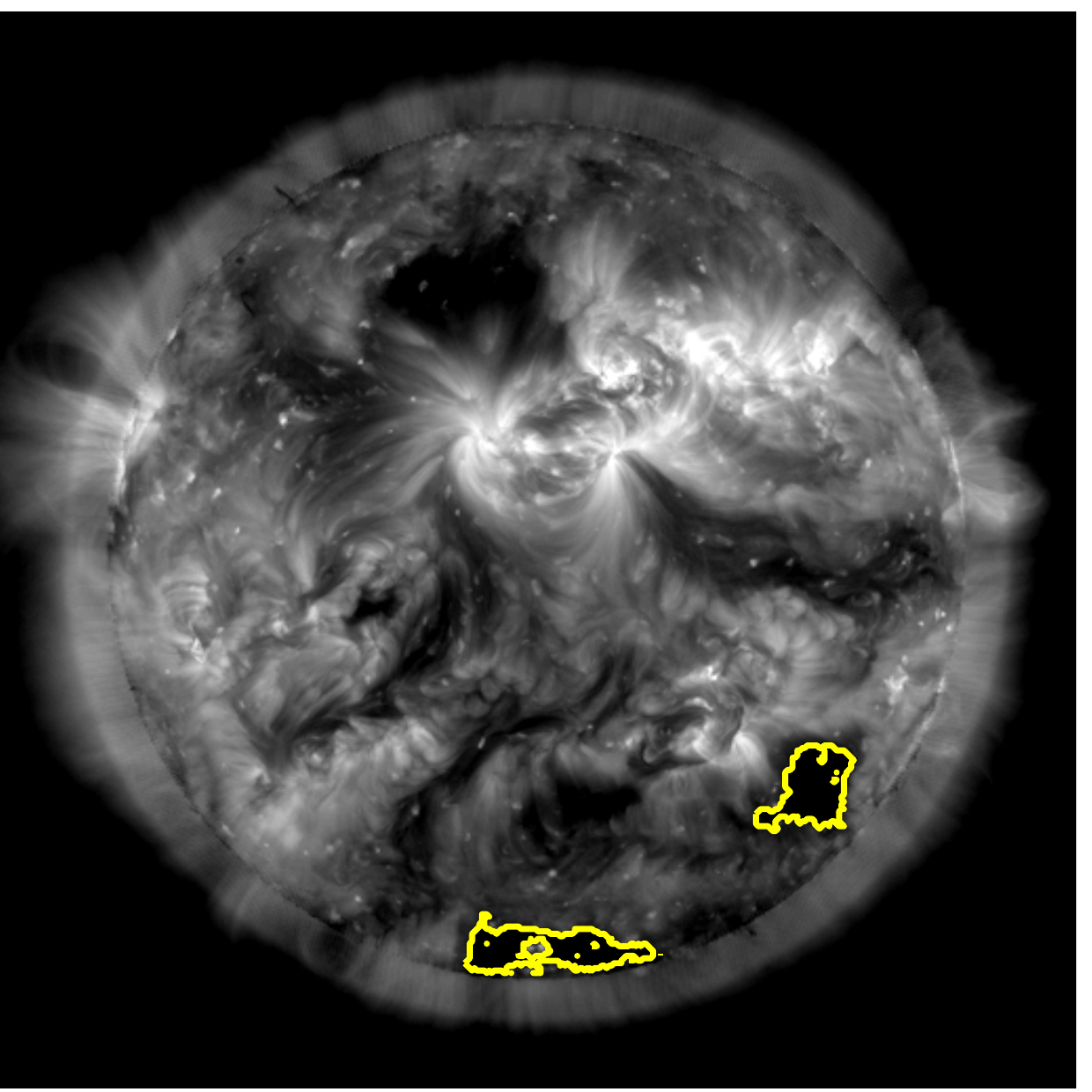}}~
   \subfloat[11 Feb. 2013]{\includegraphics[width=0.13\textwidth]{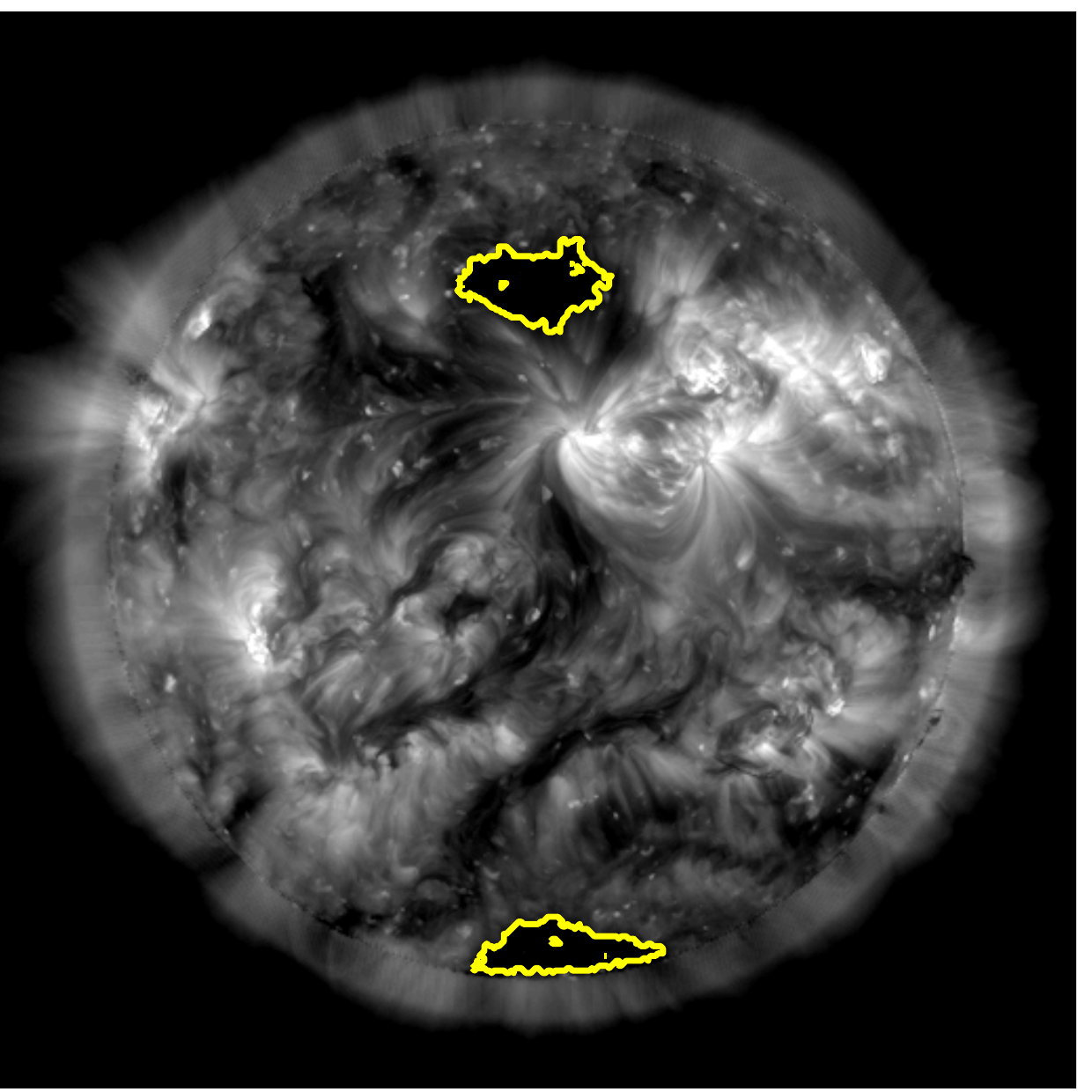}}~
   \subfloat[12 Feb. 2013]{\includegraphics[width=0.13\textwidth]{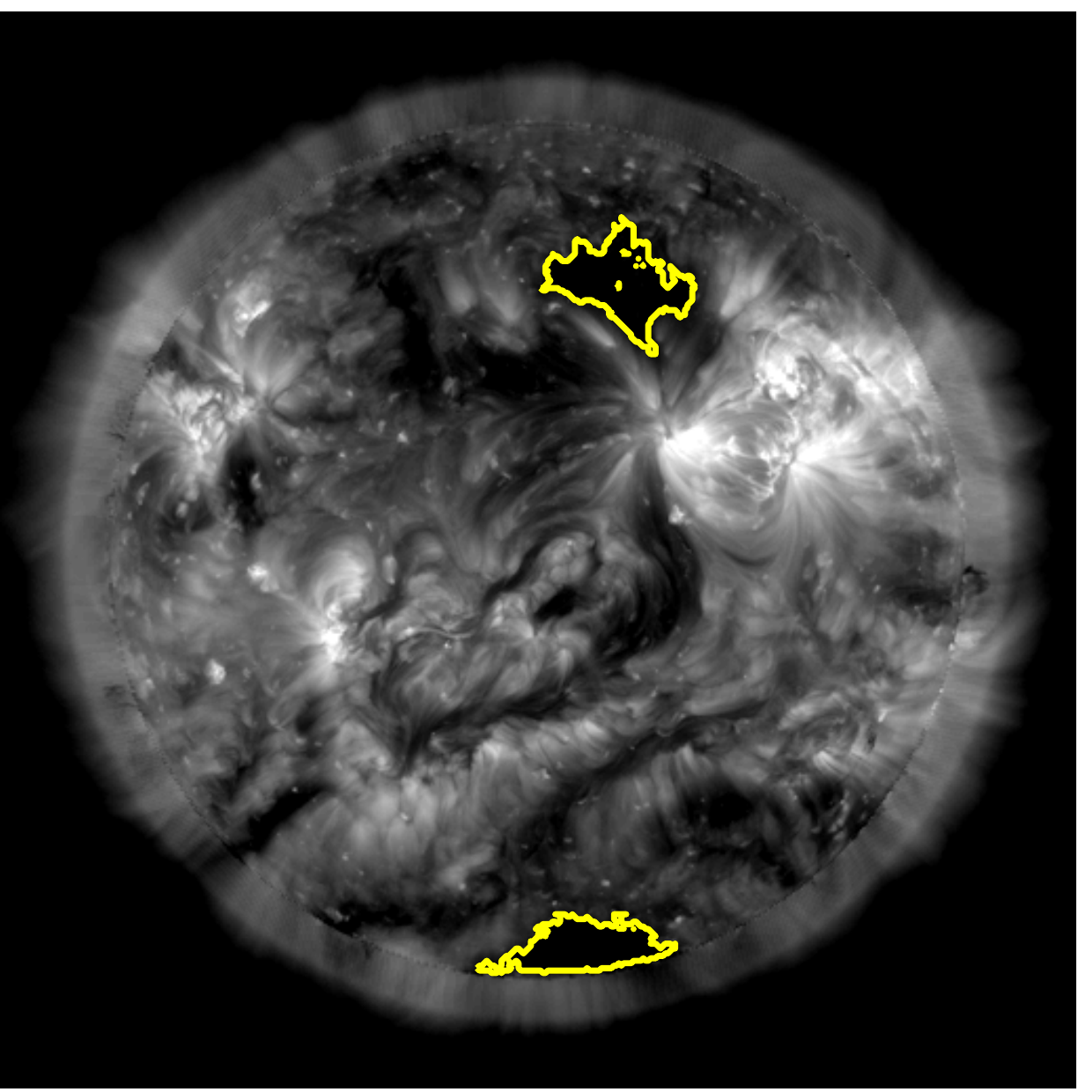}}~
   \subfloat[13 Feb. 2013]{\includegraphics[width=0.13\textwidth]{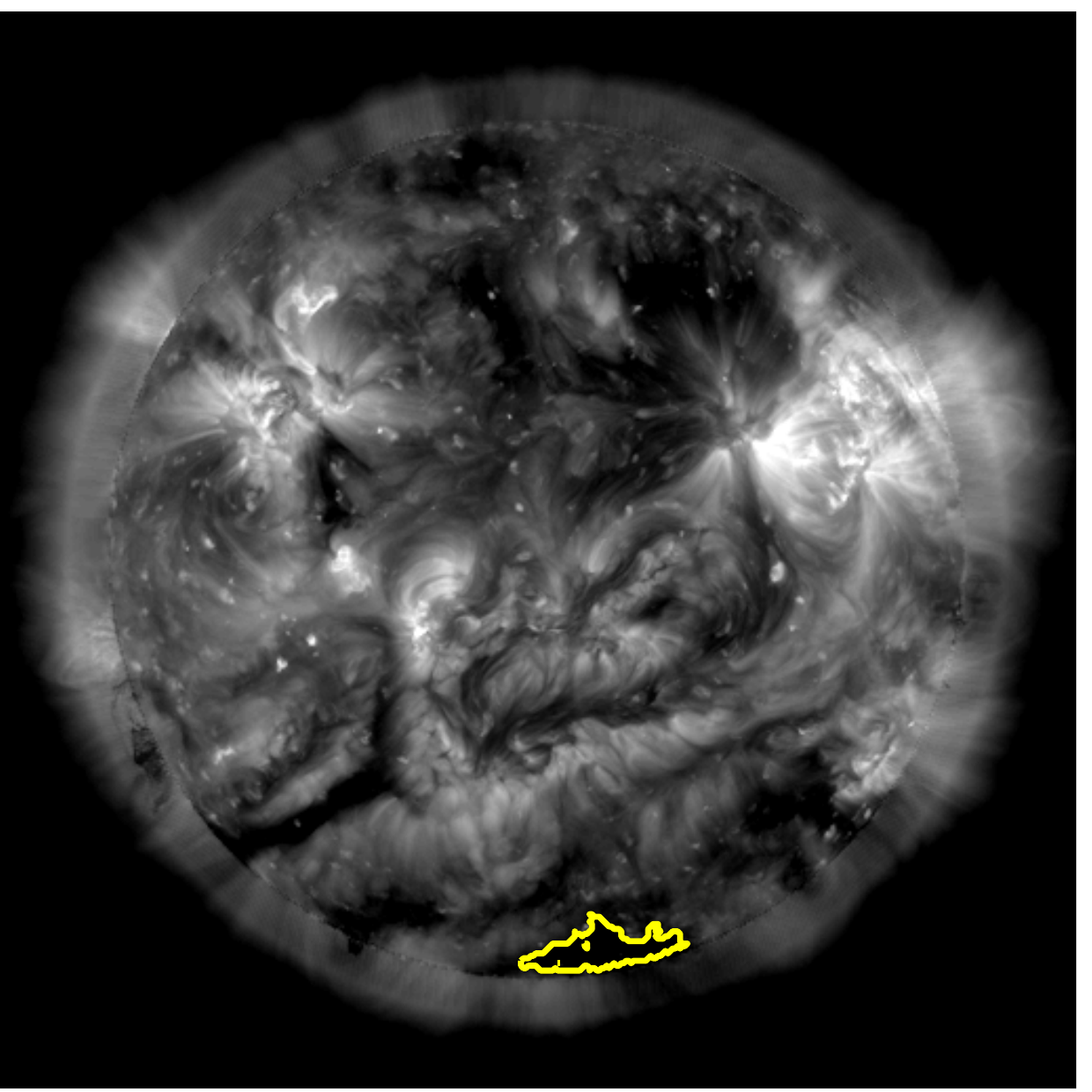}}~
   \subfloat[14 Feb. 2013]{\includegraphics[width=0.13\textwidth]{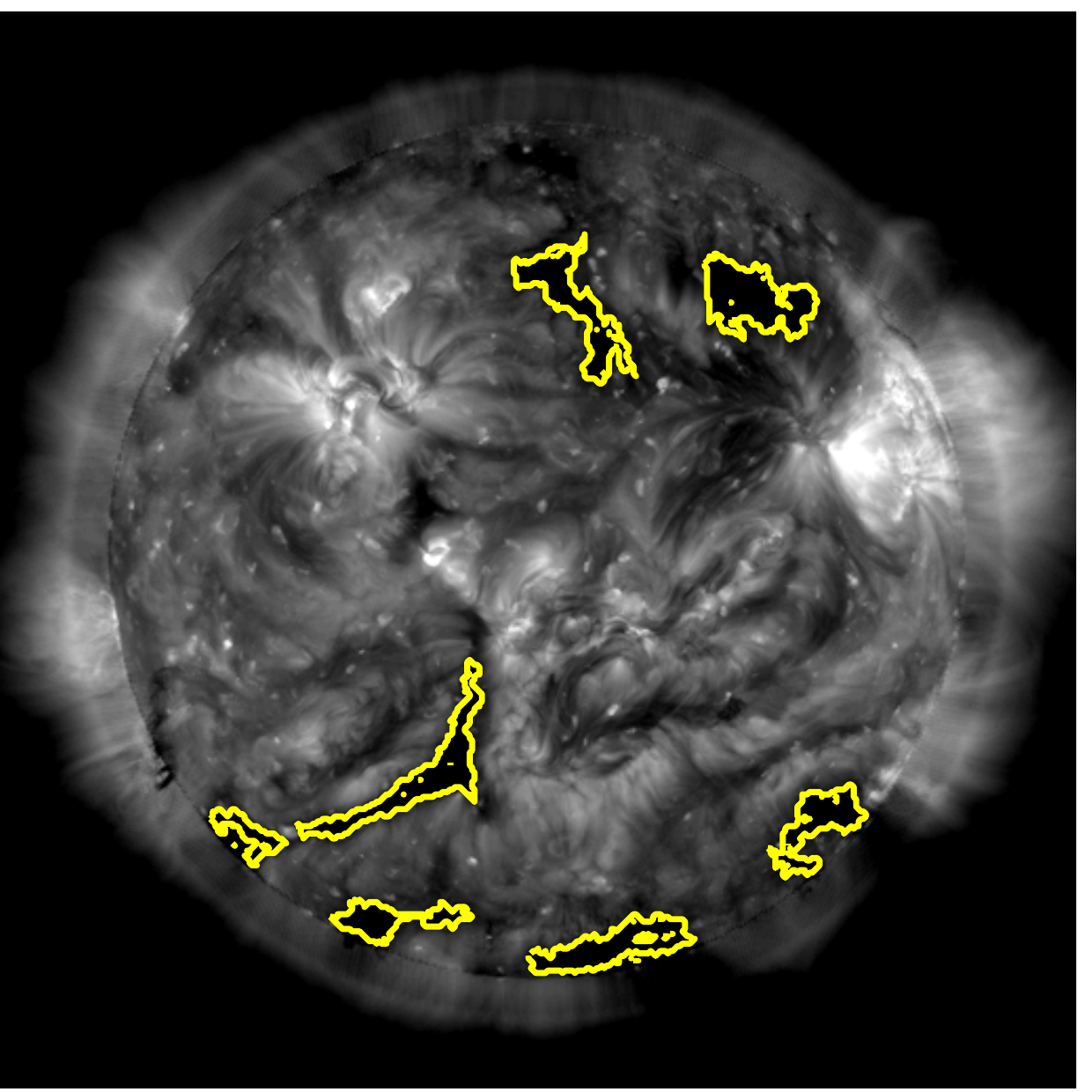}}\\[-2ex]\\
   \subfloat[15 Feb. 2013]{\includegraphics[width=0.13\textwidth]{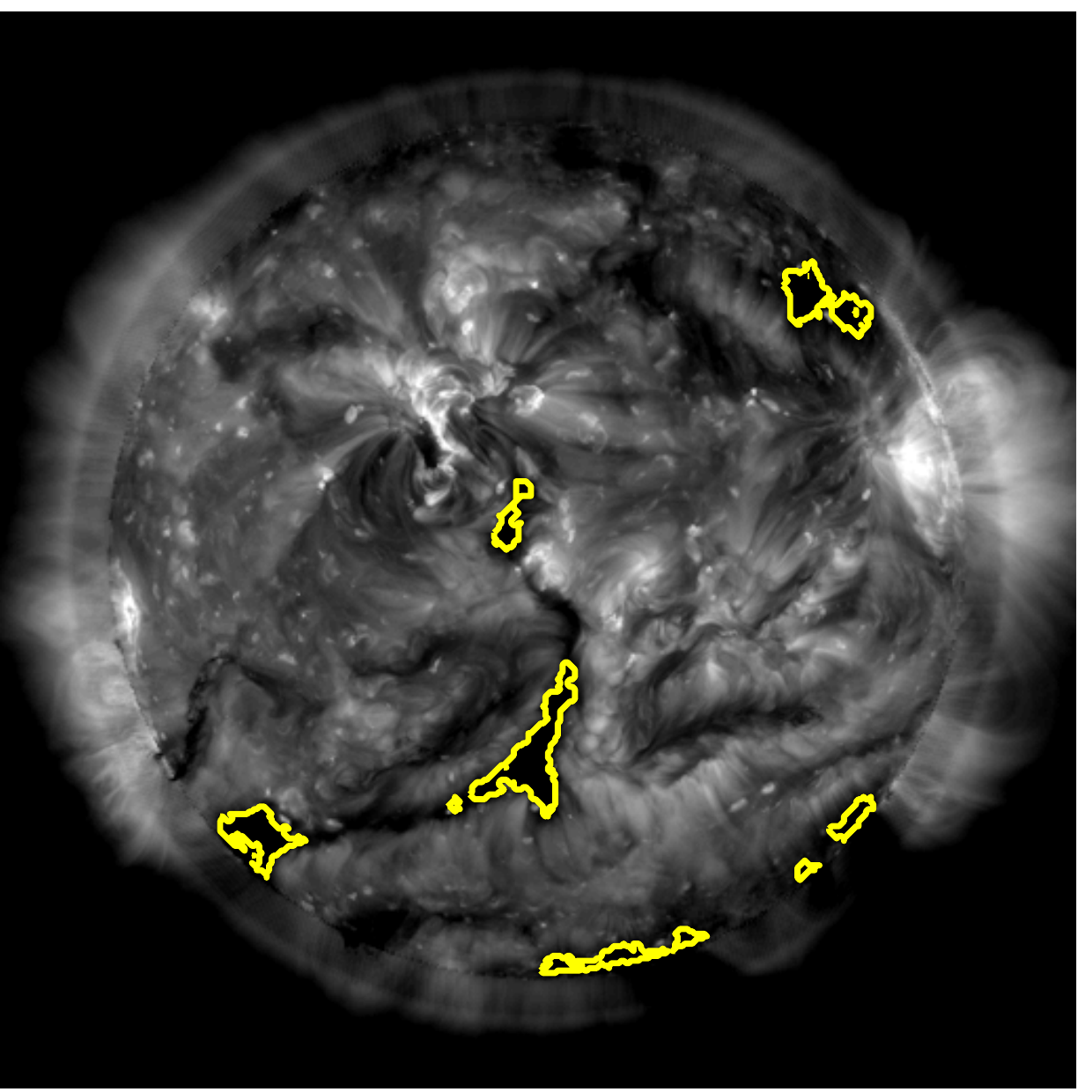}}~
   \subfloat[16 Feb. 2013]{\includegraphics[width=0.13\textwidth]{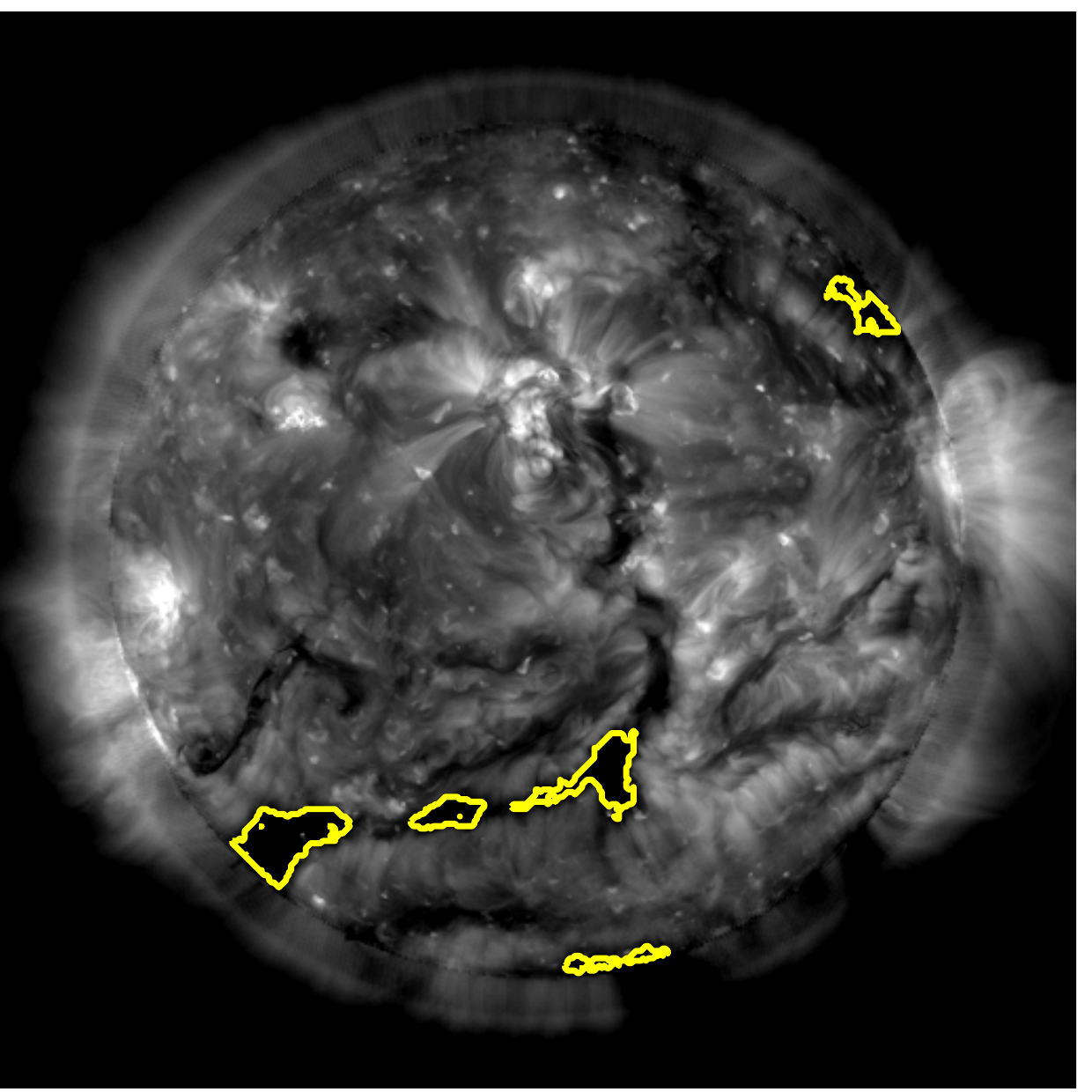}}~
   \subfloat[17 Feb. 2013]{\includegraphics[width=0.13\textwidth]{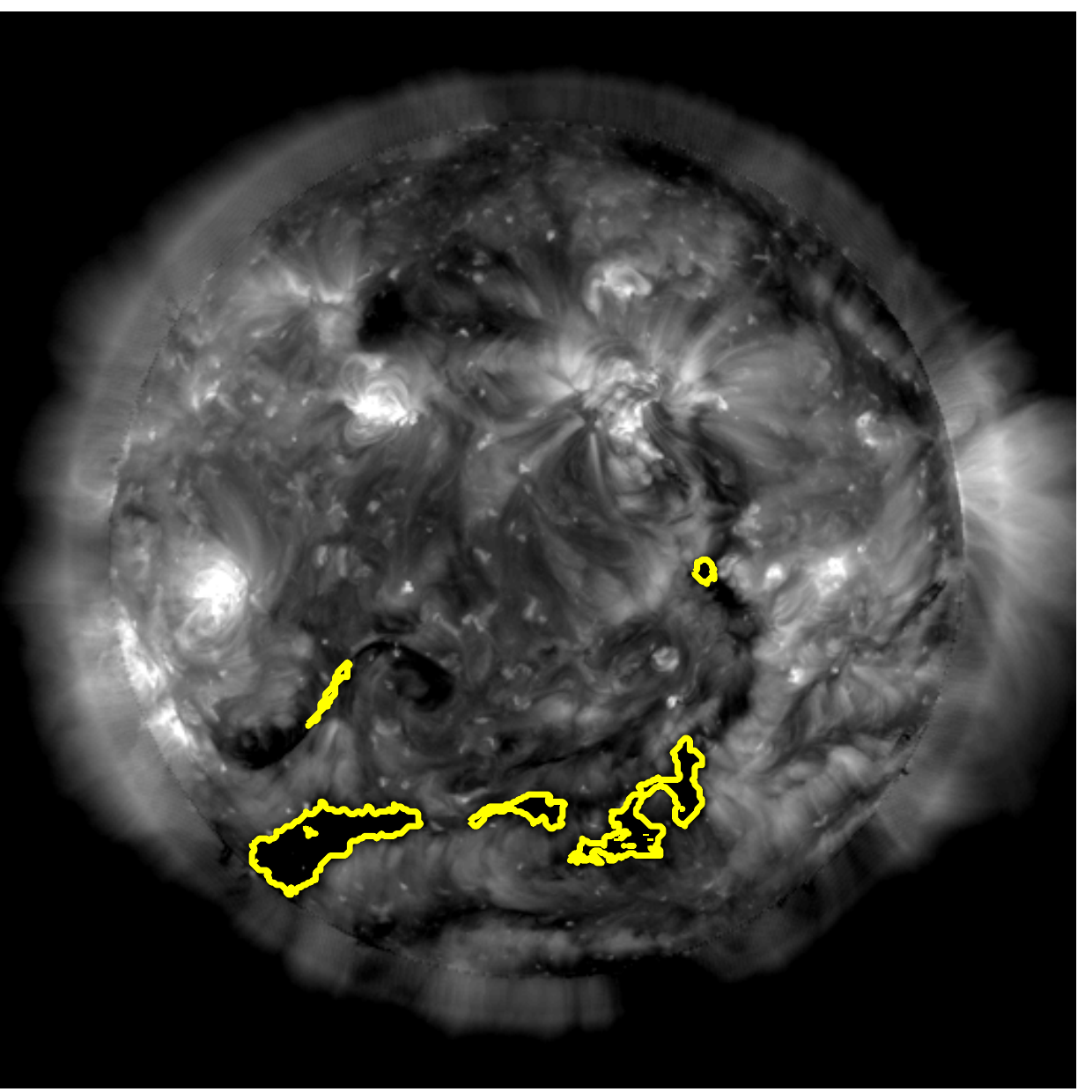}}~
   \subfloat[18 Feb. 2013]{\includegraphics[width=0.13\textwidth]{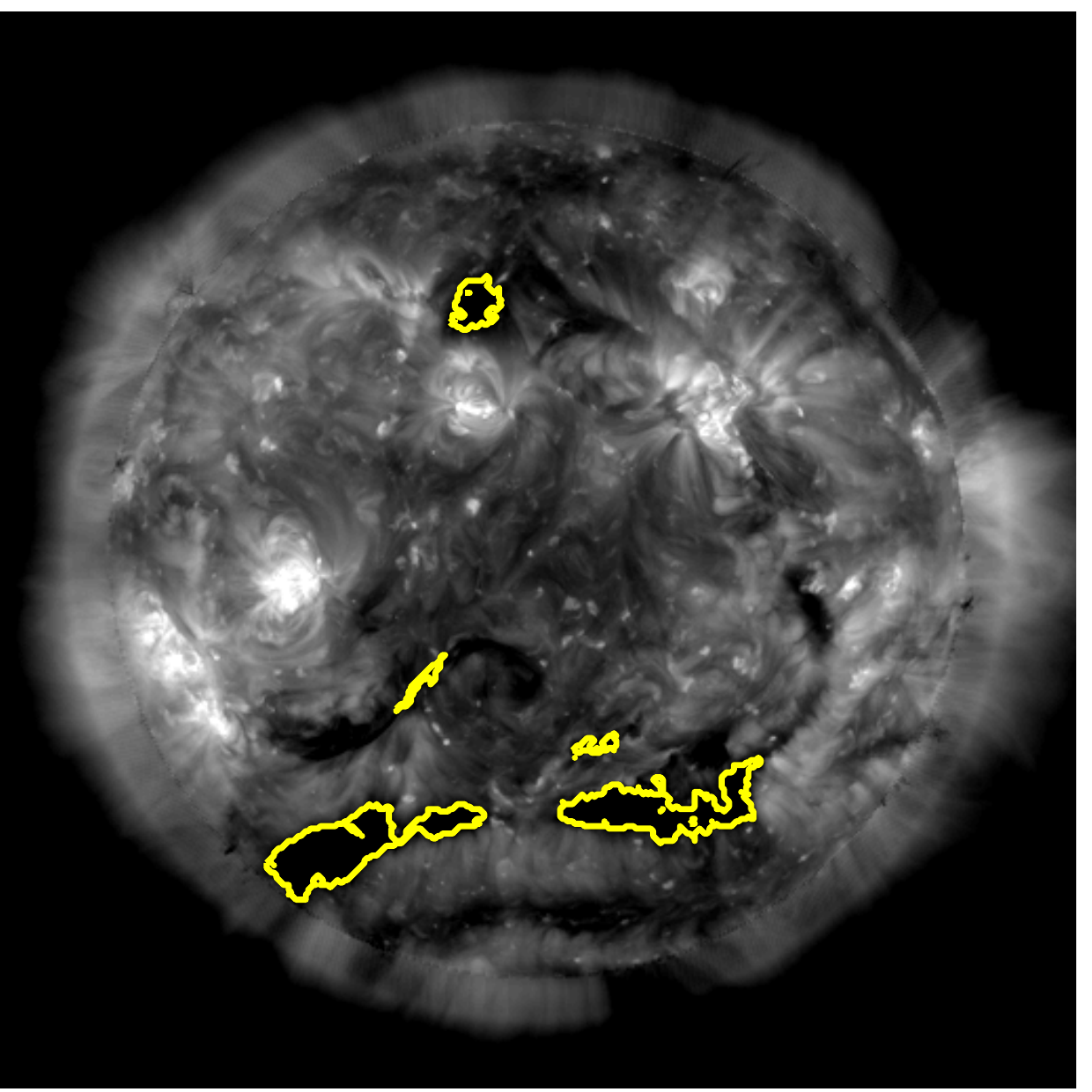}}~
   \subfloat[19 Feb. 2013]{\includegraphics[width=0.13\textwidth]{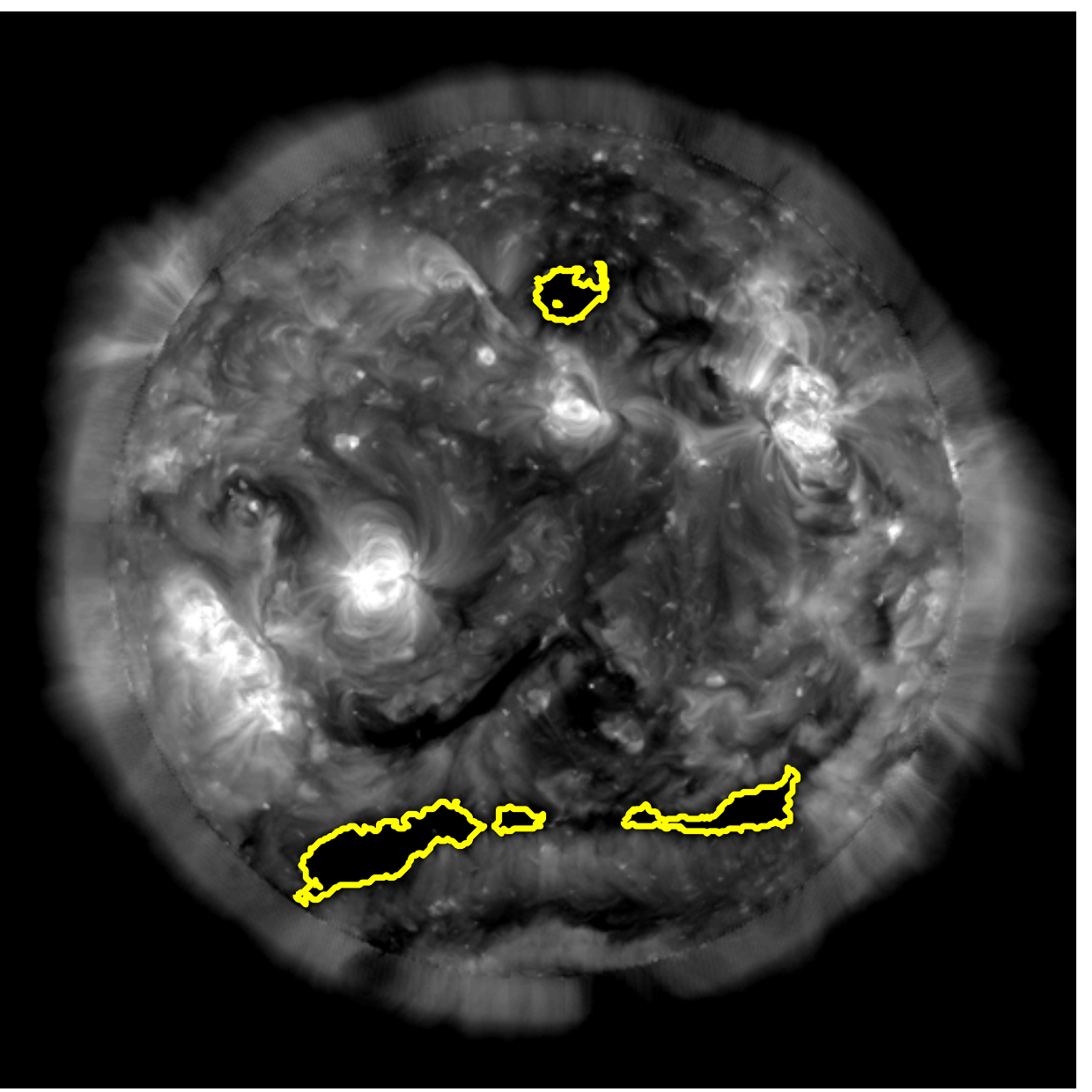}}~
   \subfloat[20 Feb. 2013]{\includegraphics[width=0.13\textwidth]{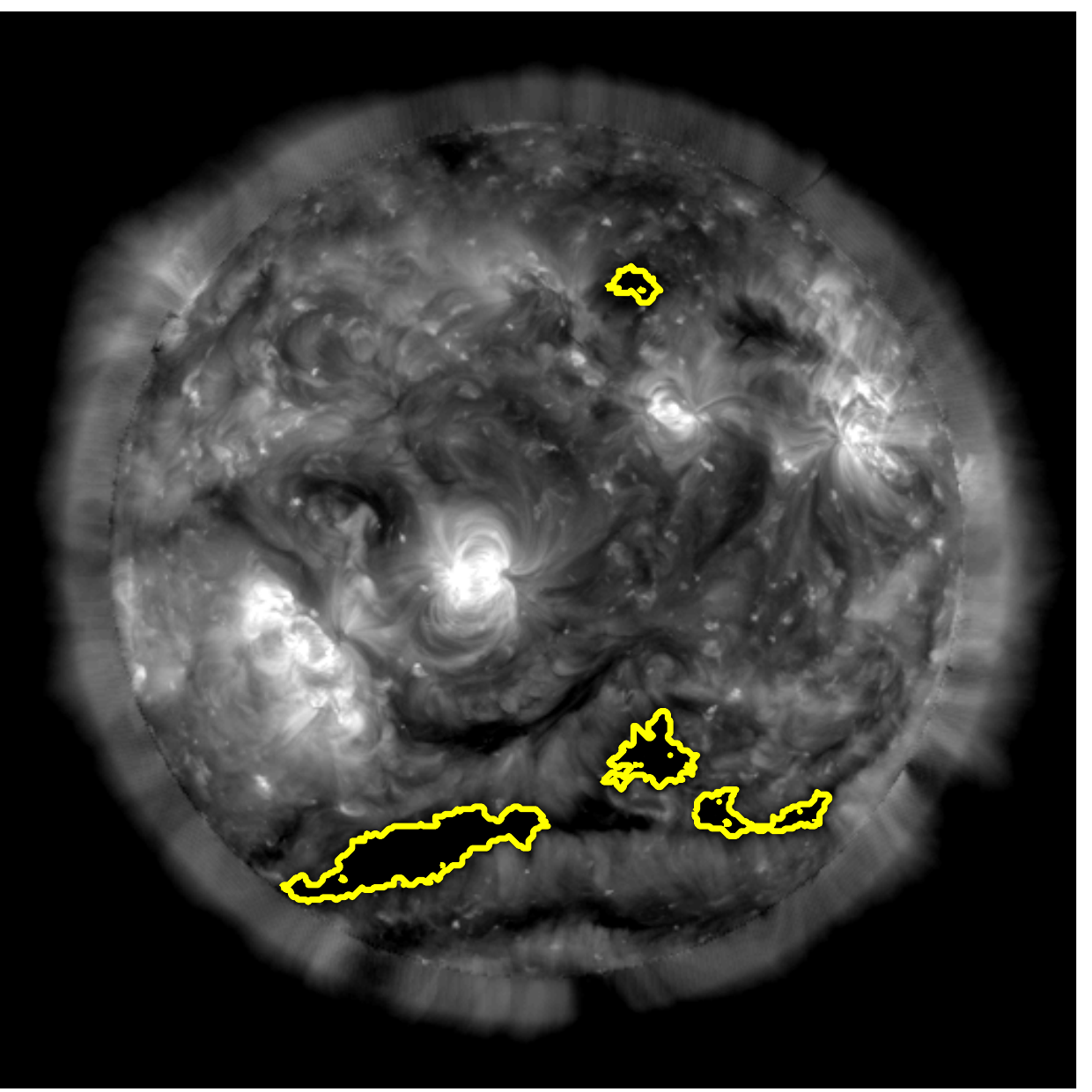}}~
   \subfloat[21 Feb. 2013]{\includegraphics[width=0.13\textwidth]{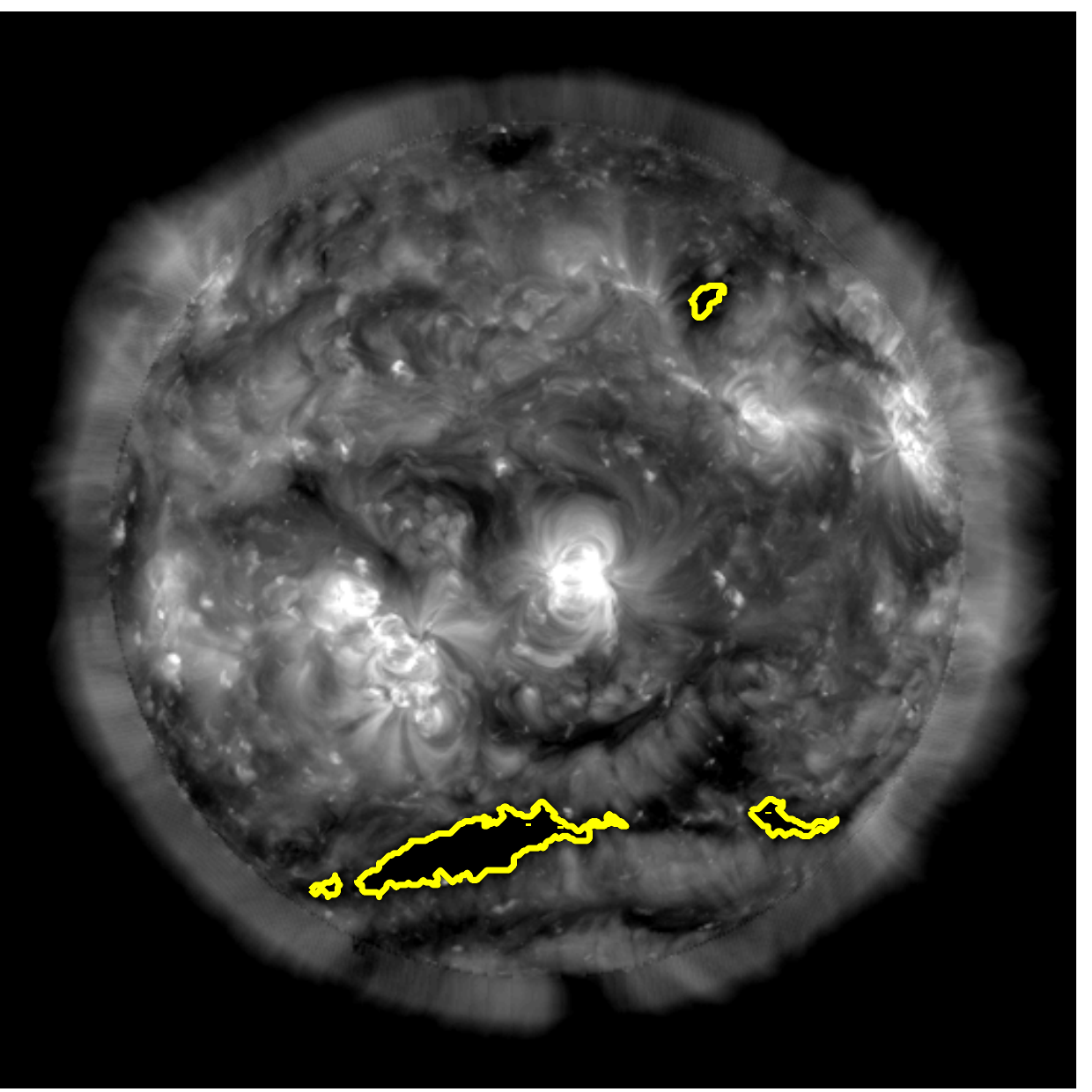}}\\[-2ex]
   \caption{Carrington rotation 2133 193 \AA~images.  Images are clipped to $[100,2500]$ and log scaled. Coronal holes are denoted by the yellow outline for $\lambda_i/\lambda_o=50$ and $\alpha=0.3$.}
   \label{fig:CR2133_segs}
  \end{minipage}
\end{sidewaysfigure}

Segmentation results are shown in Figure~\ref{fig:CR2099_segs} for CR 2099 images and Figure~\ref{fig:CR2133_segs} for CR 2133 images using $\alpha=0.3$ and $\lambda_i/\lambda_o=50$; segmentation results for CRs 2106 and 2150 are included as supplemental figures.  In this section, we use several quantitative and qualitative methods to study the accuracy of the CH segmentations. We consider the unipolarity of the underlying photospheric magnetic field, compare CH location with high-speed solar wind data, and briefly qualitatively compare to SPoCA~\citep{verbeeck2014}. 

\captionsetup[subfigure]{labelformat=empty}

\subsection{Magnetic Unipolarity}
\label{sec:mdi}
\begin{sidewaysfigure}[p]
  \centering
  \begin{minipage}[l]{0.85\textwidth}
   \subfloat[13 Jul. 2010]{\includegraphics[width=0.13\textwidth]{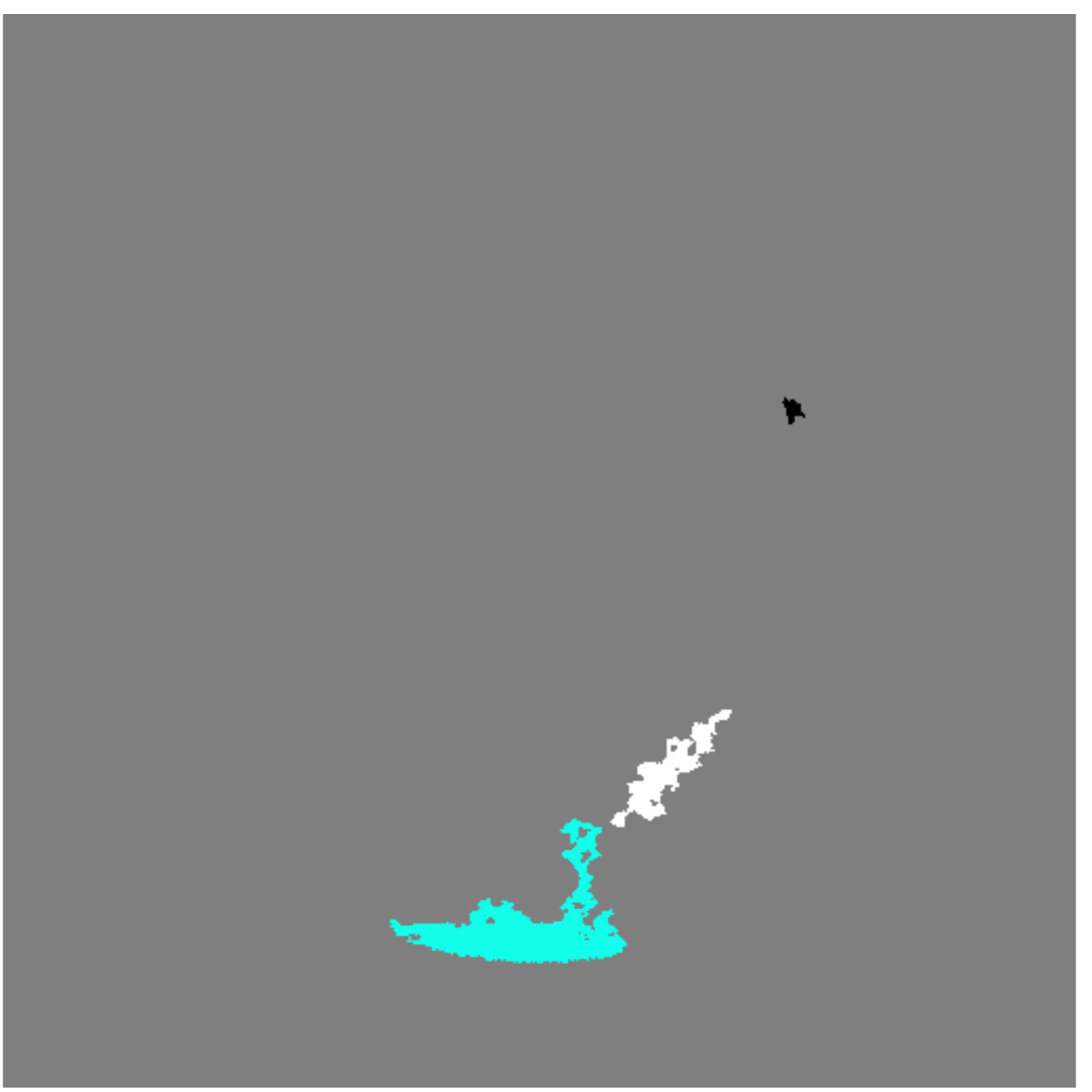}}~
   \subfloat[14 Jul. 2010]{\includegraphics[width=0.13\textwidth]{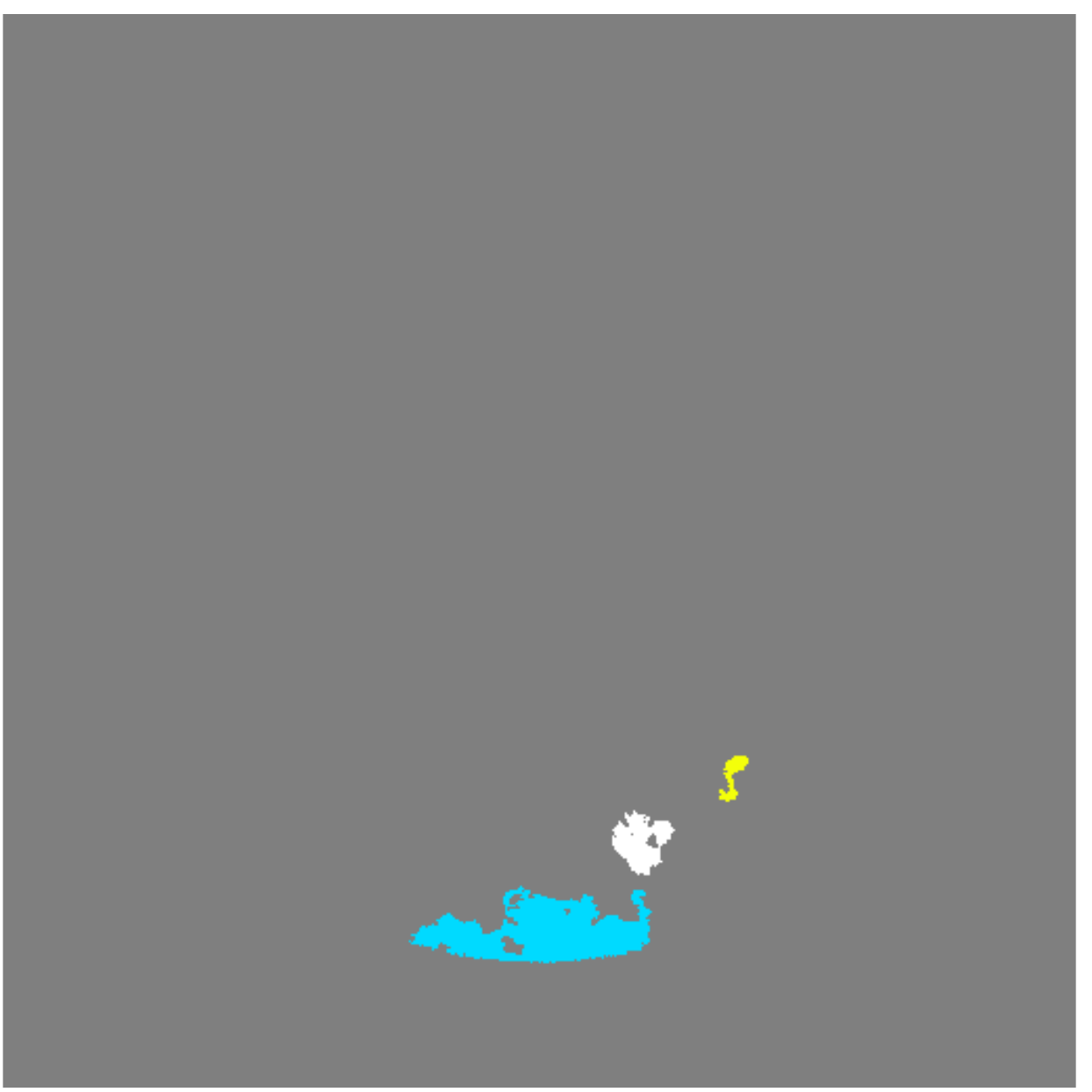}}~
   \subfloat[15 Jul. 2010]{\includegraphics[width=0.13\textwidth]{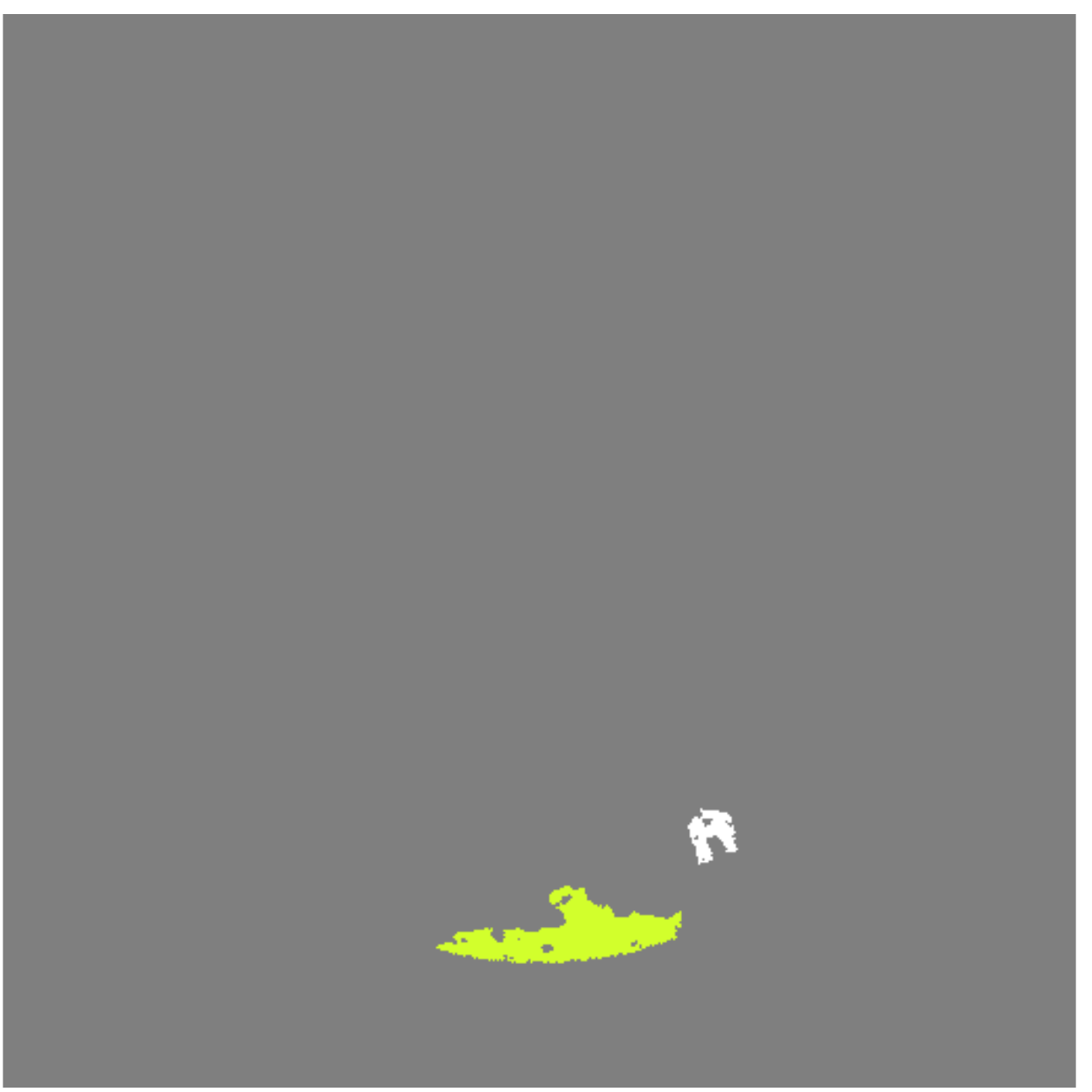}}~
   \subfloat[16 Jul. 2010]{\includegraphics[width=0.13\textwidth]{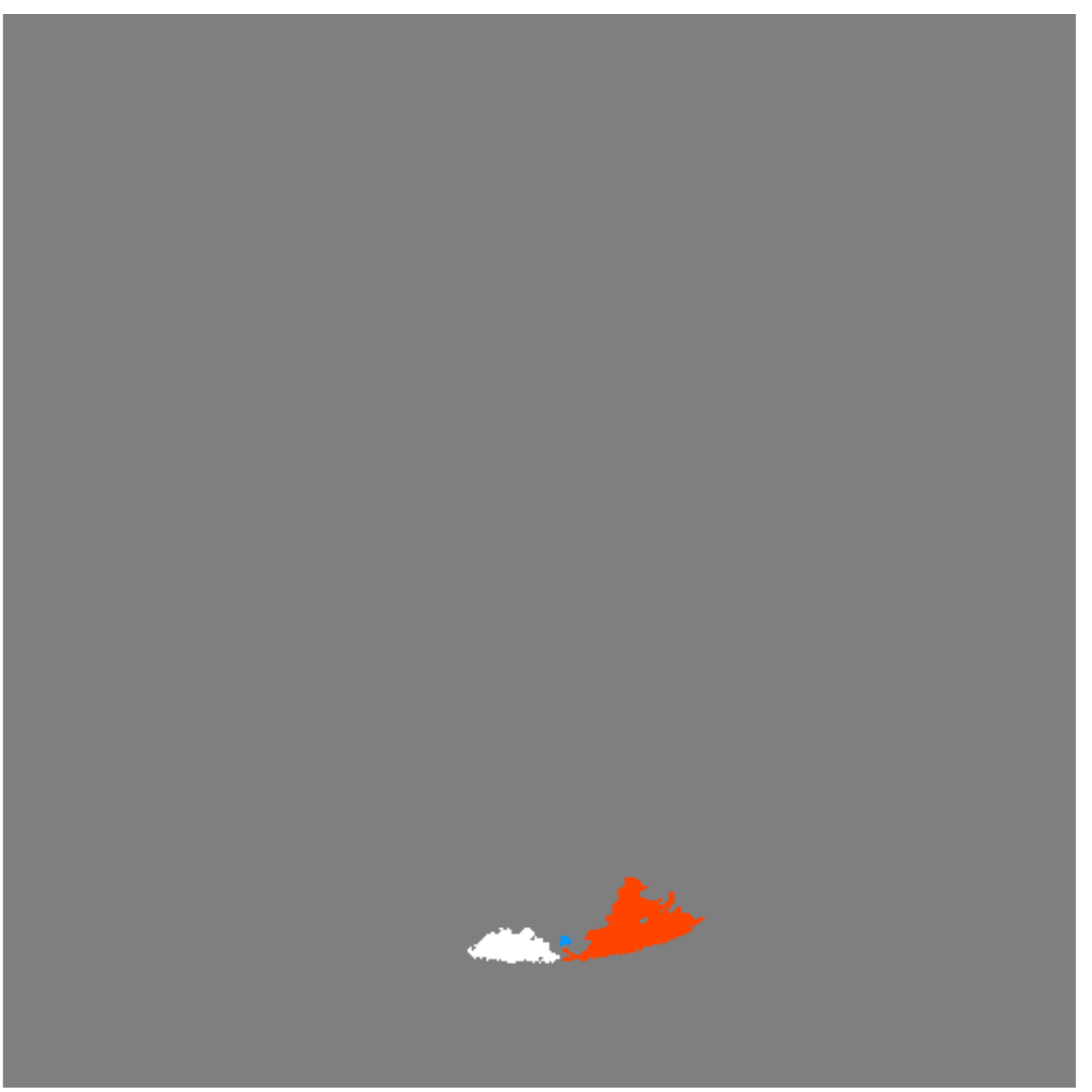}}~
   \subfloat[17 Jul. 2010]{\includegraphics[width=0.13\textwidth]{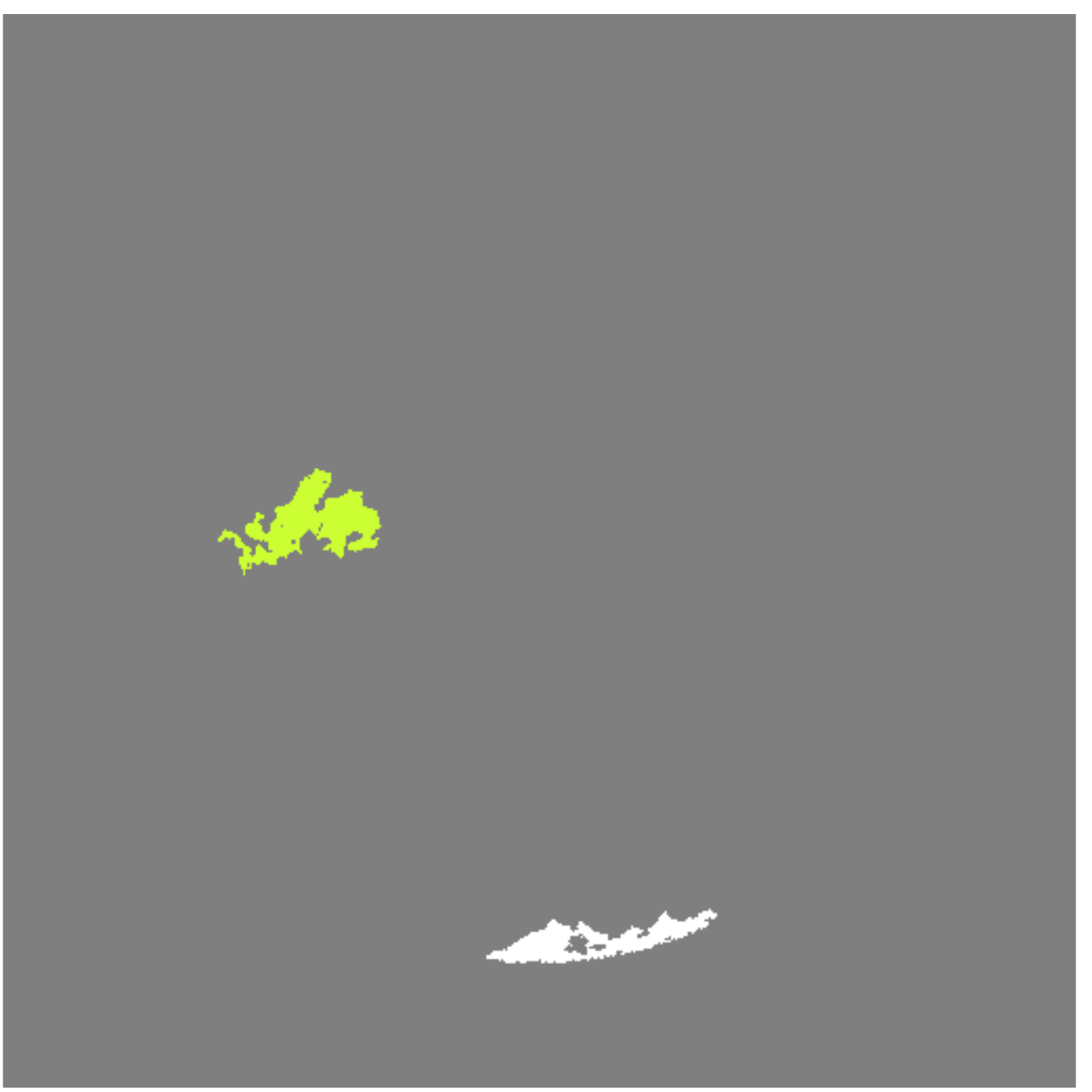}}~
   \subfloat[18 Jul. 2010]{\includegraphics[width=0.13\textwidth]{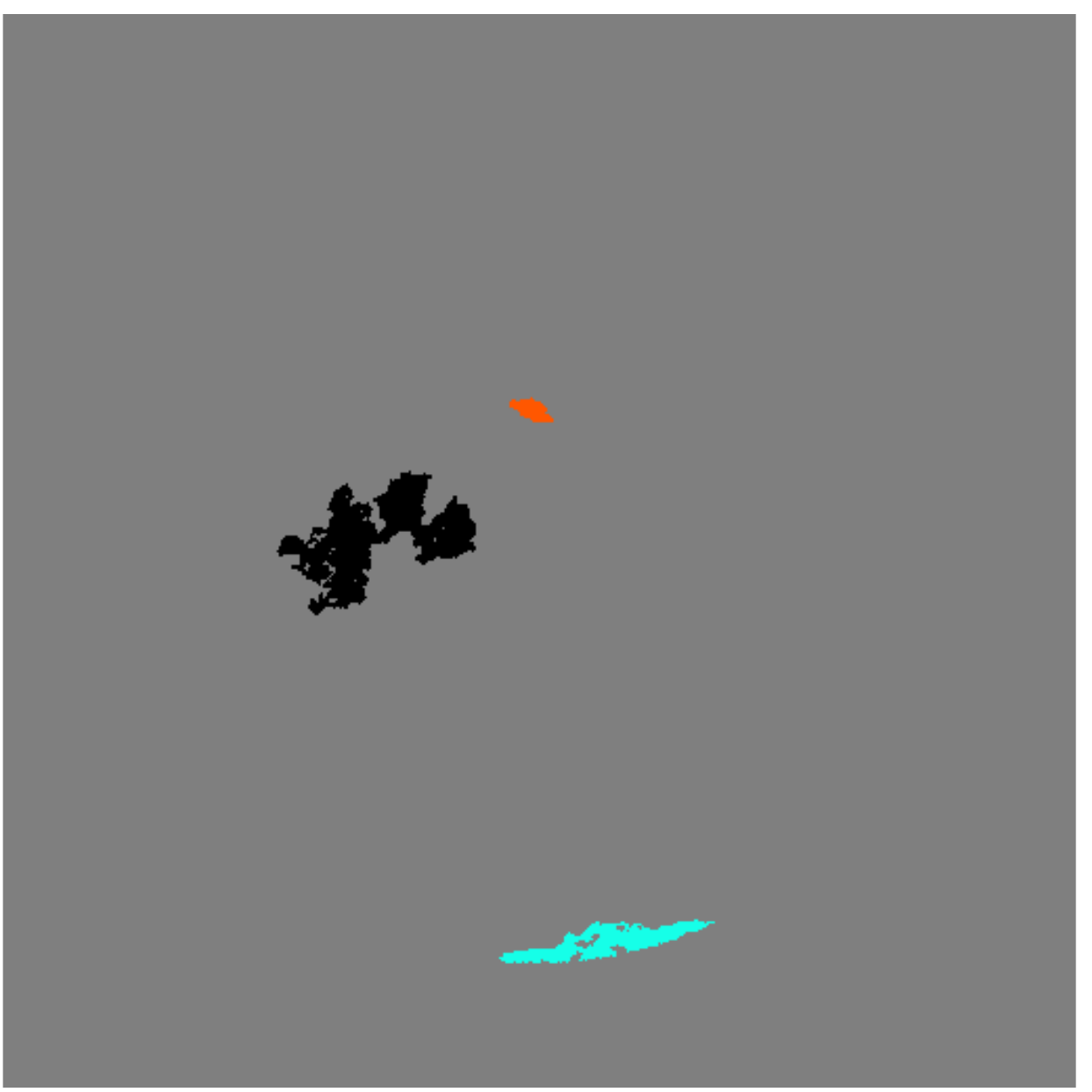}}~
   \subfloat[19 Jul. 2010]{\includegraphics[width=0.13\textwidth]{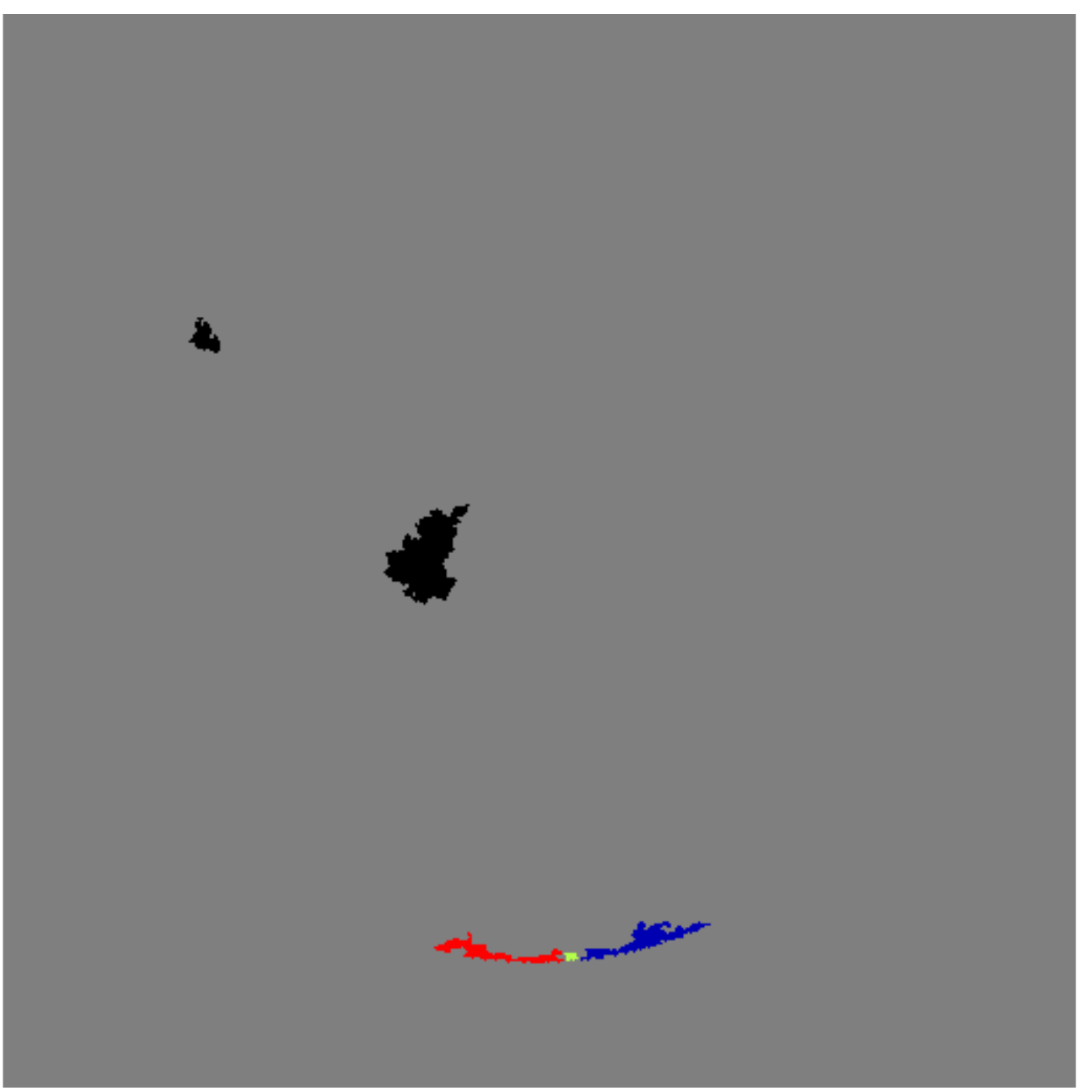}}\\[-2ex]\\
   \subfloat[20 Jul. 2010]{\includegraphics[width=0.13\textwidth]{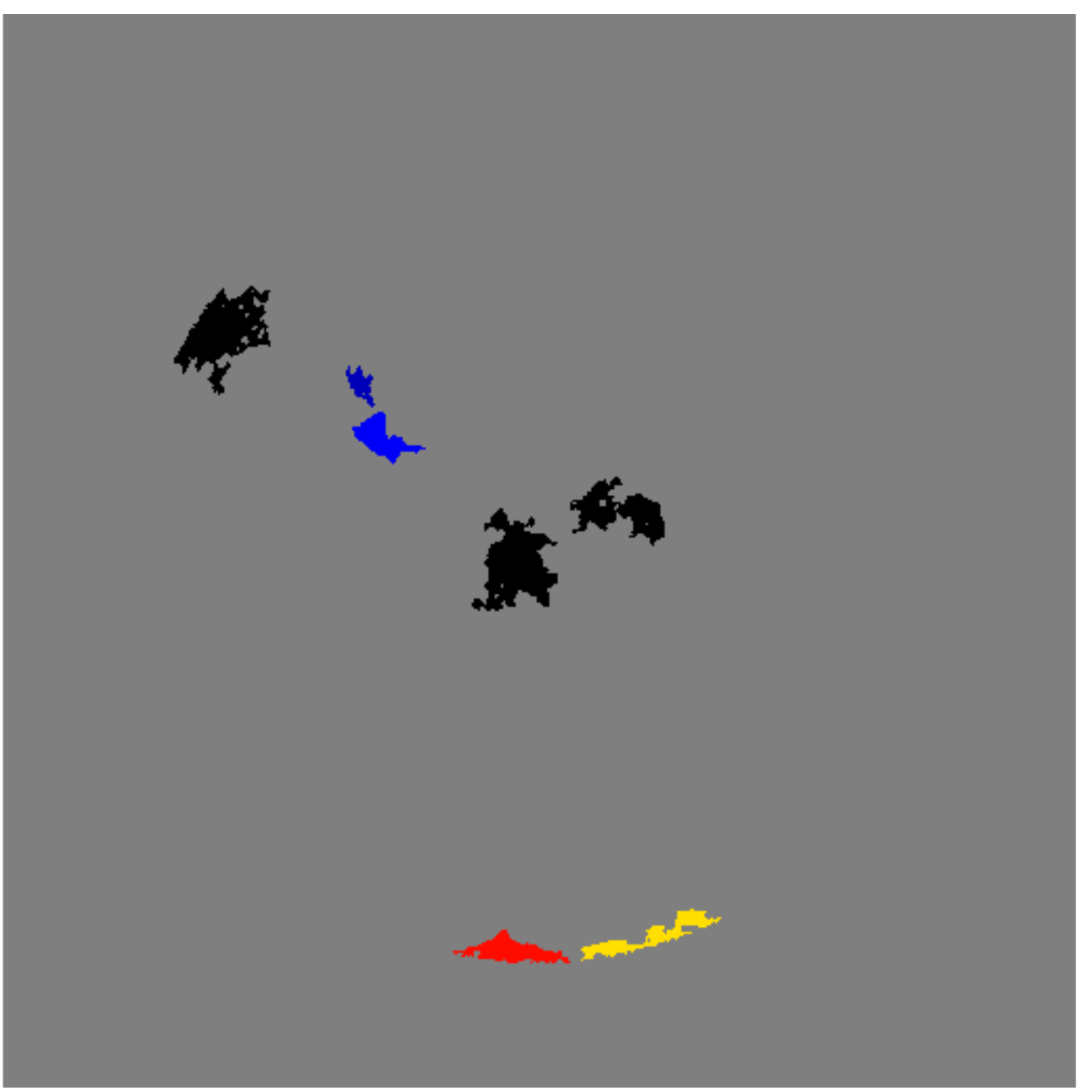}}~
   \subfloat[21 Jul. 2010]{\includegraphics[width=0.13\textwidth]{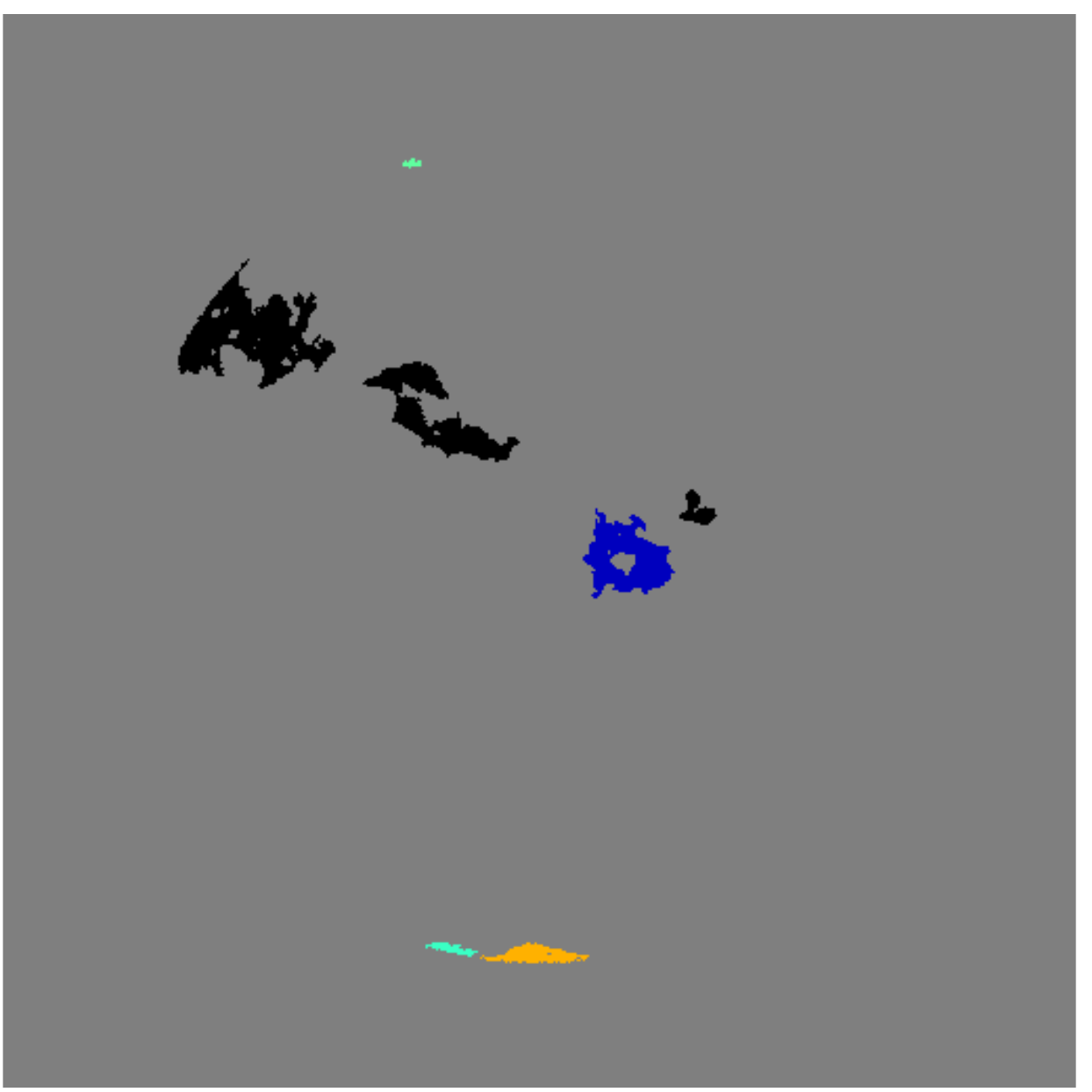}}~
   \subfloat[22 Jul. 2010]{\includegraphics[width=0.13\textwidth]{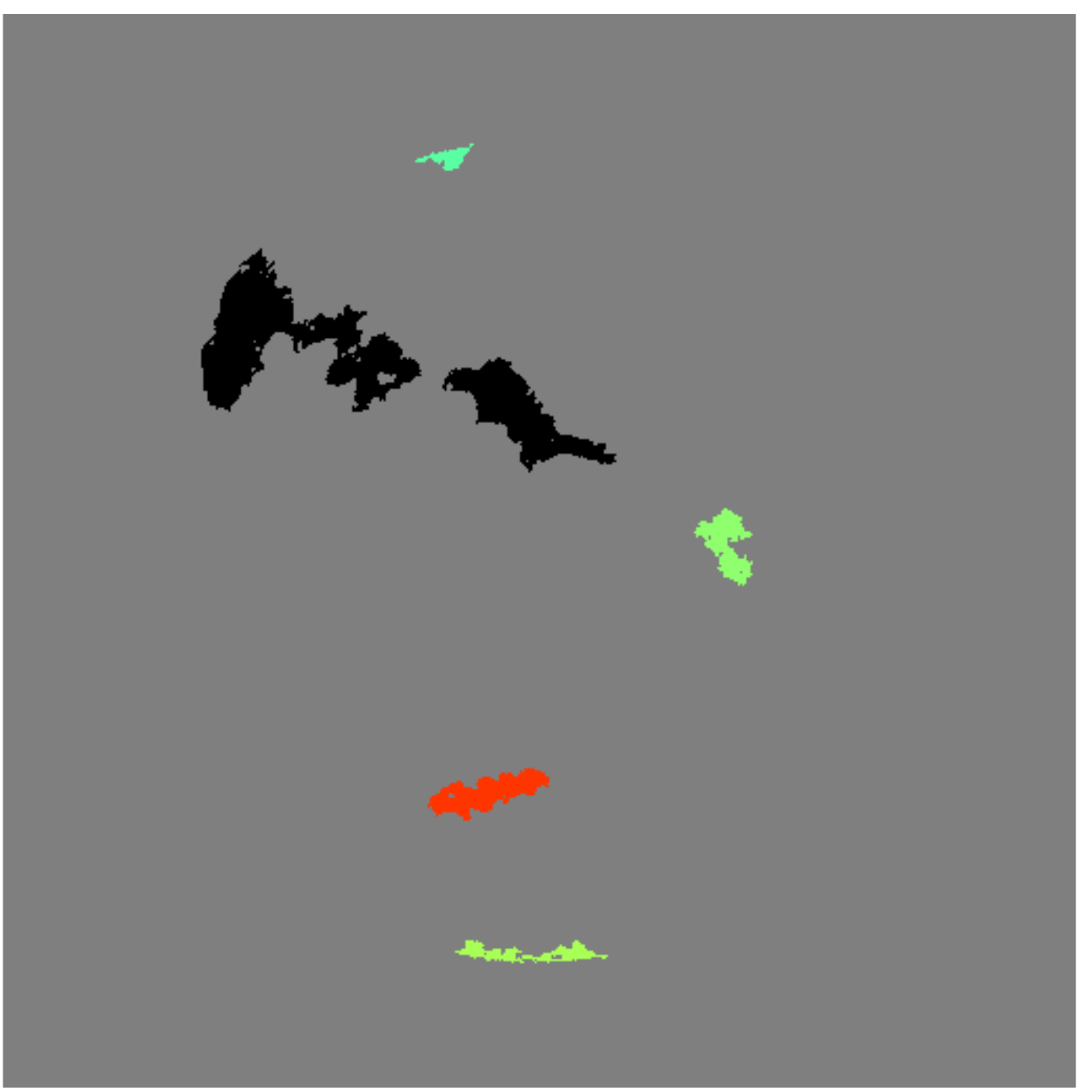}}~
   \subfloat[23 Jul. 2010]{\includegraphics[width=0.13\textwidth]{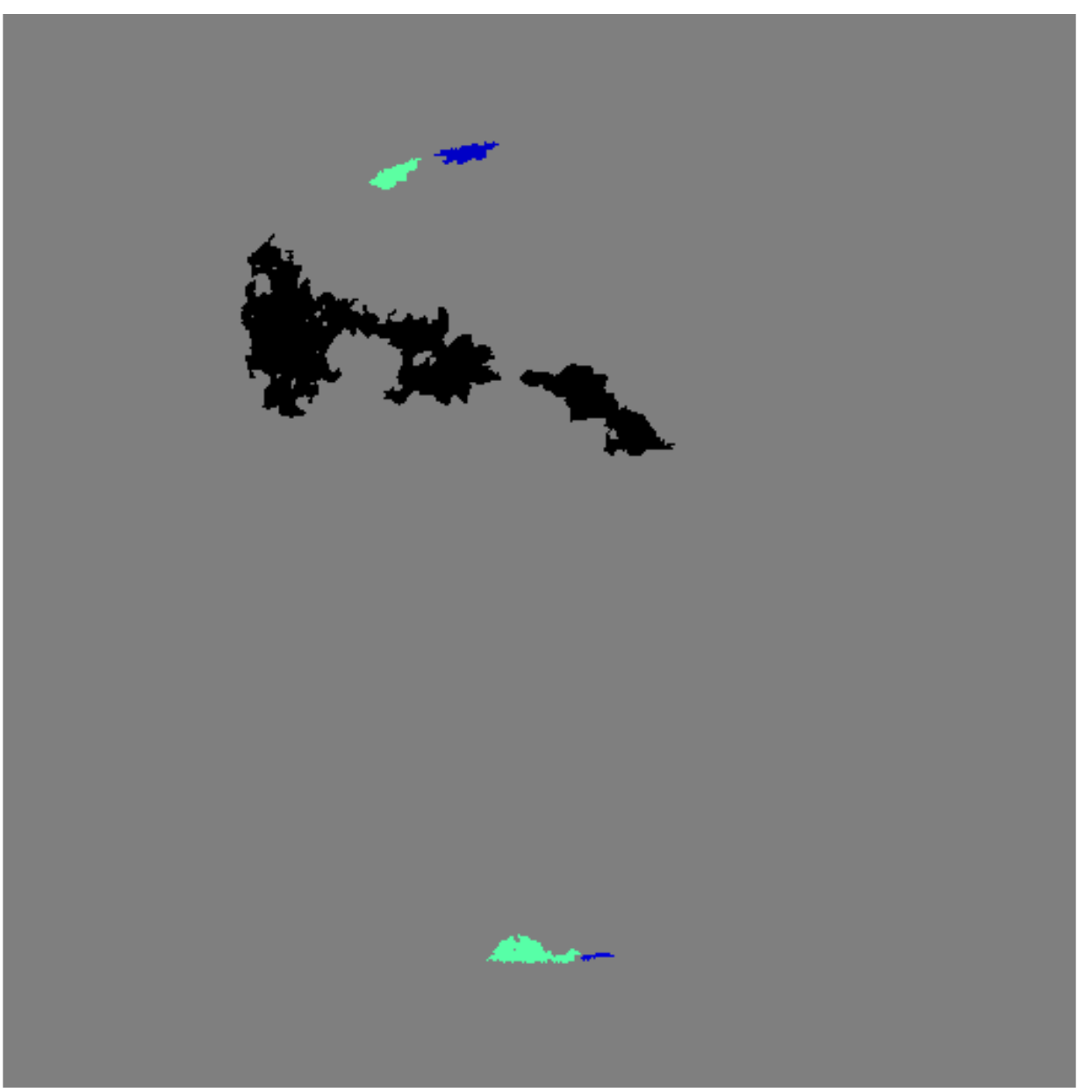}}~
   \subfloat[24 Jul. 2010]{\includegraphics[width=0.13\textwidth]{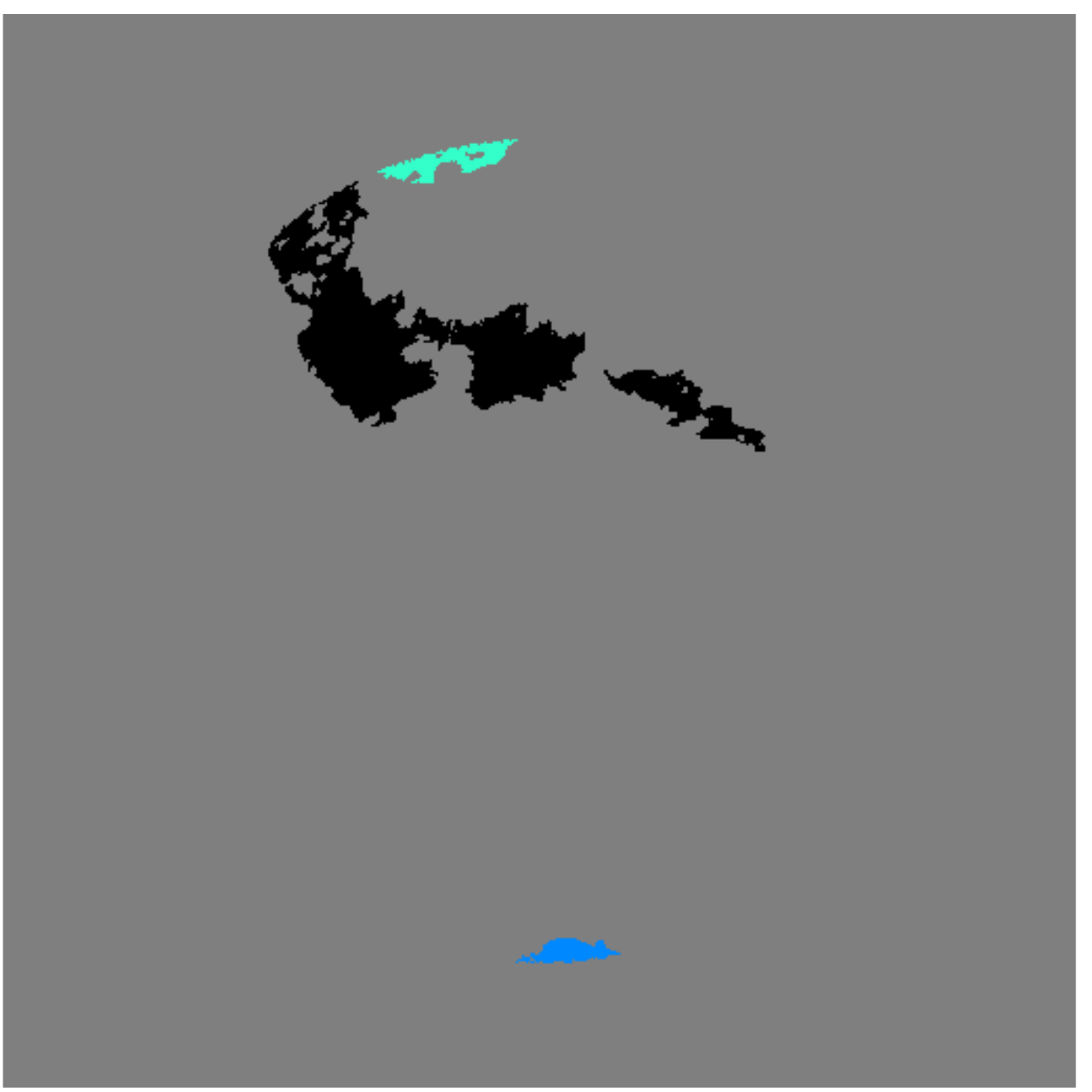}}~
   \subfloat[25 Jul. 2010]{\includegraphics[width=0.13\textwidth]{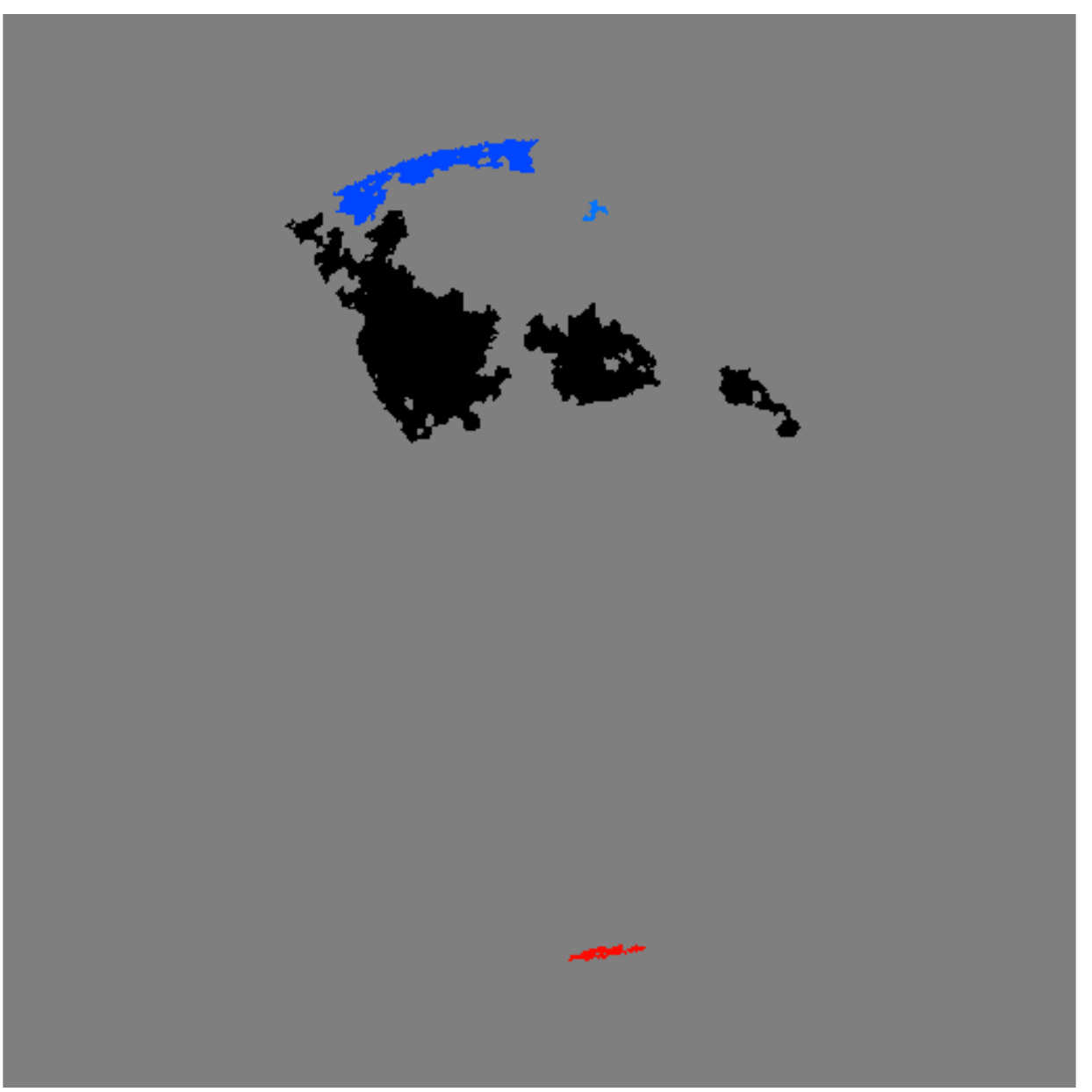}}~
   \subfloat[26 Jul. 2010]{\includegraphics[width=0.13\textwidth]{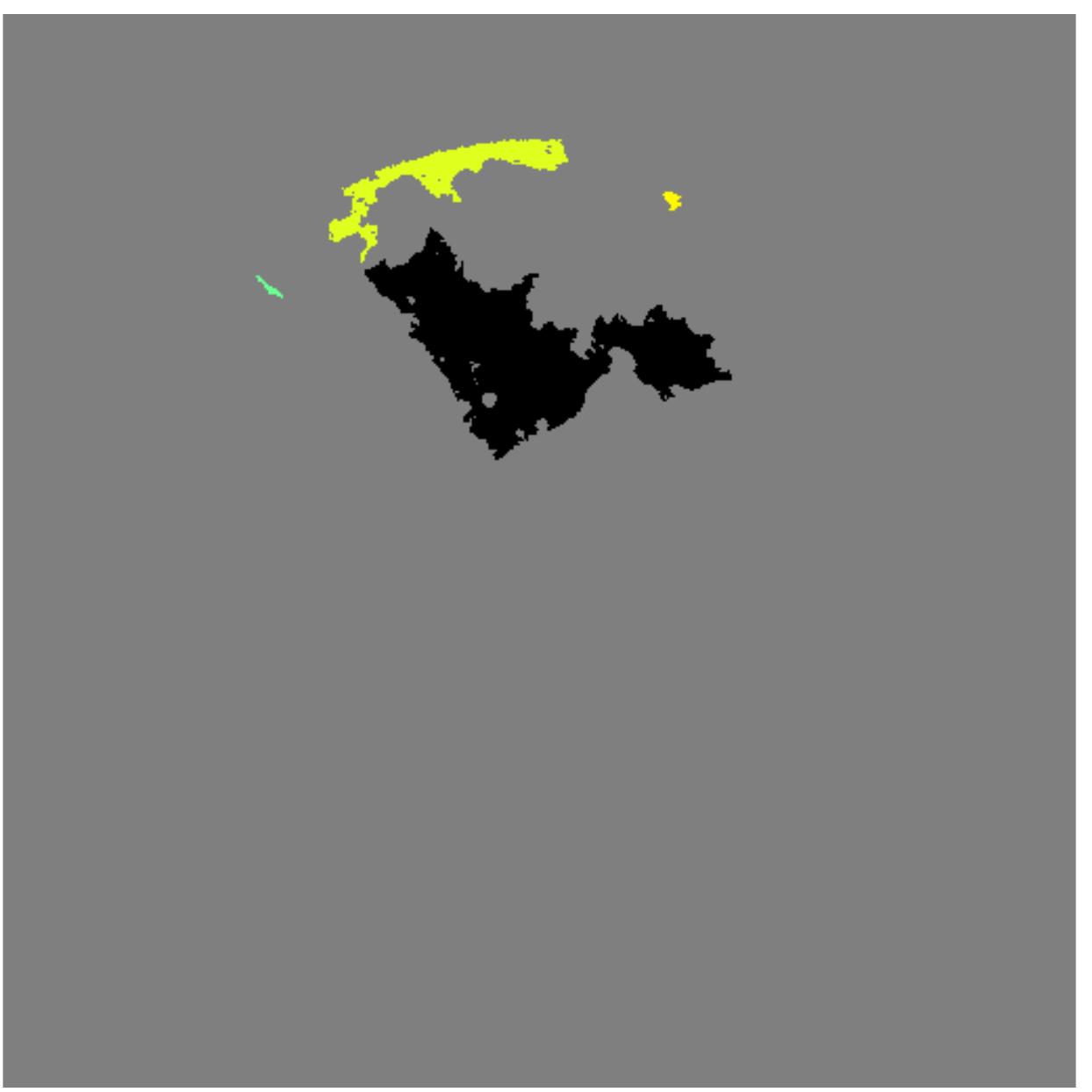}}\\[-2ex]\\
   \subfloat[27 Jul. 2010]{\includegraphics[width=0.13\textwidth]{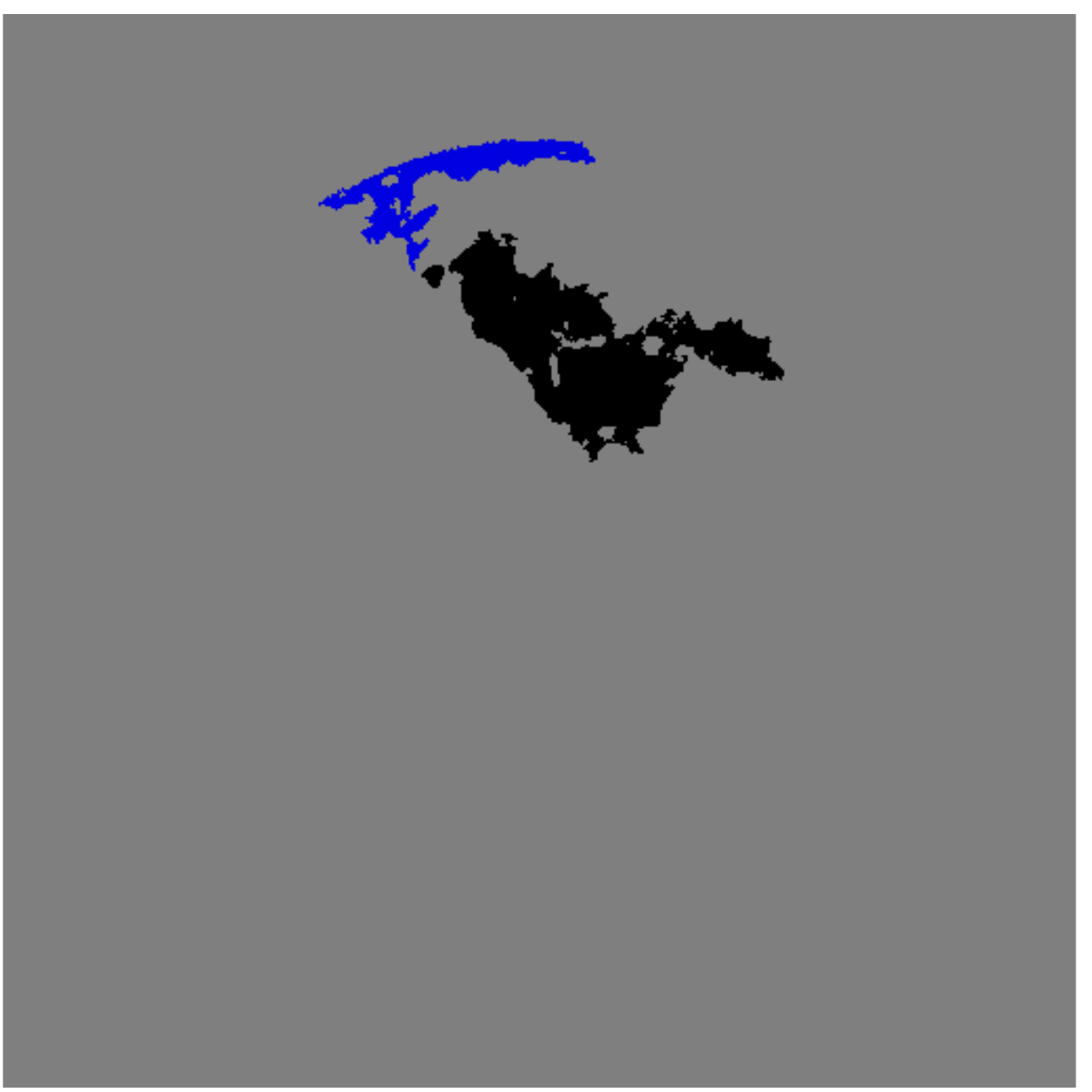}}~
   \subfloat[28 Jul. 2010]{\includegraphics[width=0.13\textwidth]{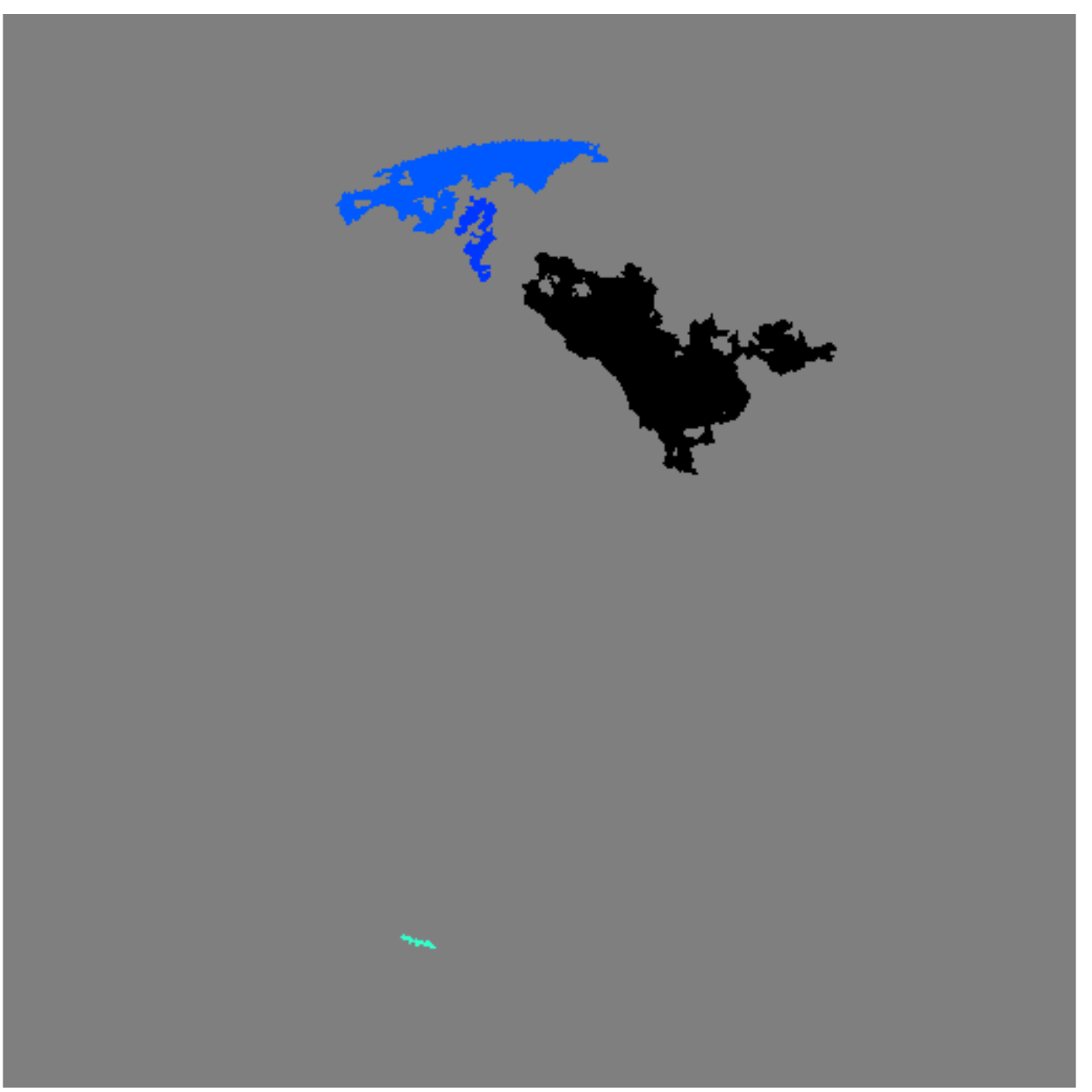}}~
   \subfloat[29 Jul. 2010]{\includegraphics[width=0.13\textwidth]{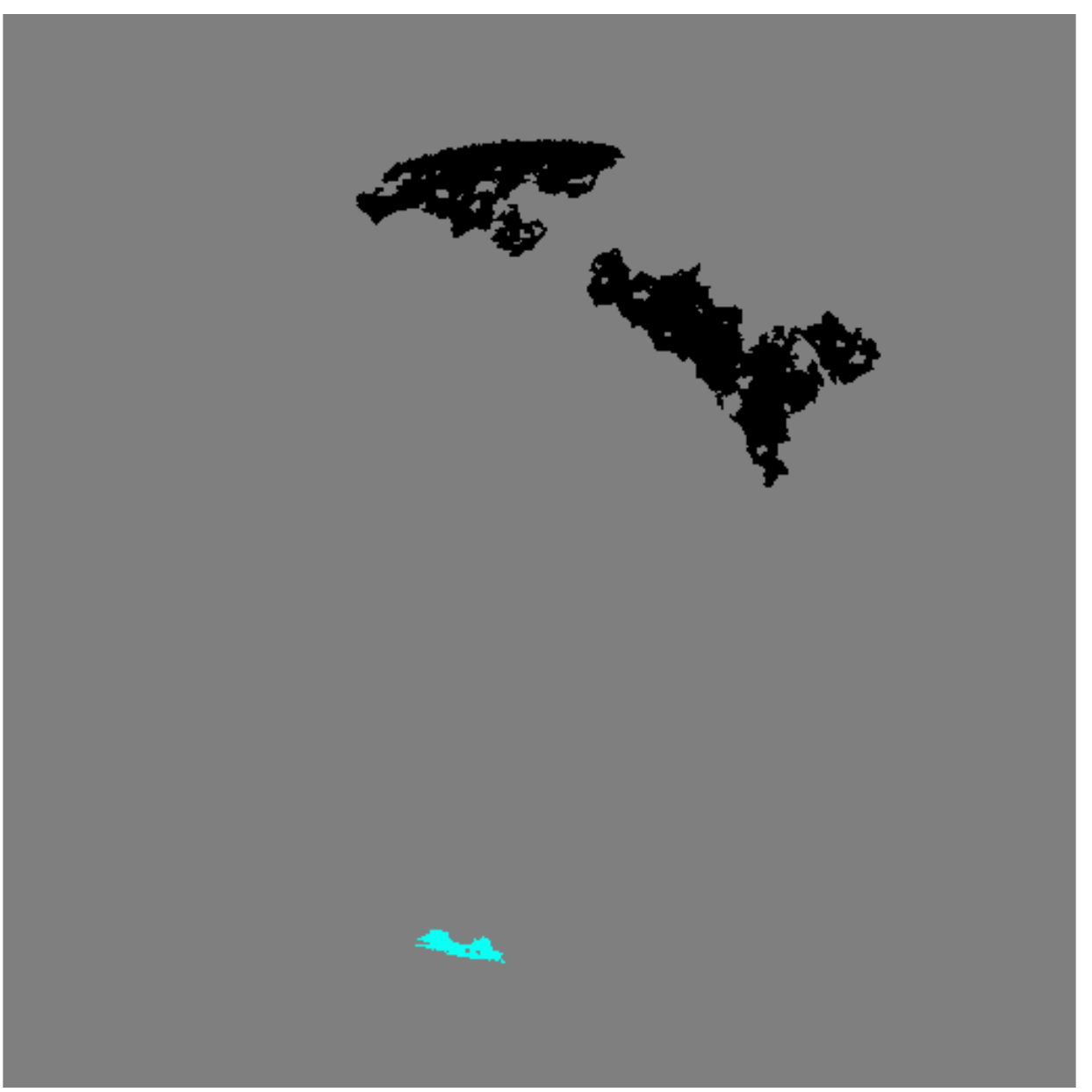}}~
   \subfloat[30 Jul. 2010]{\includegraphics[width=0.13\textwidth]{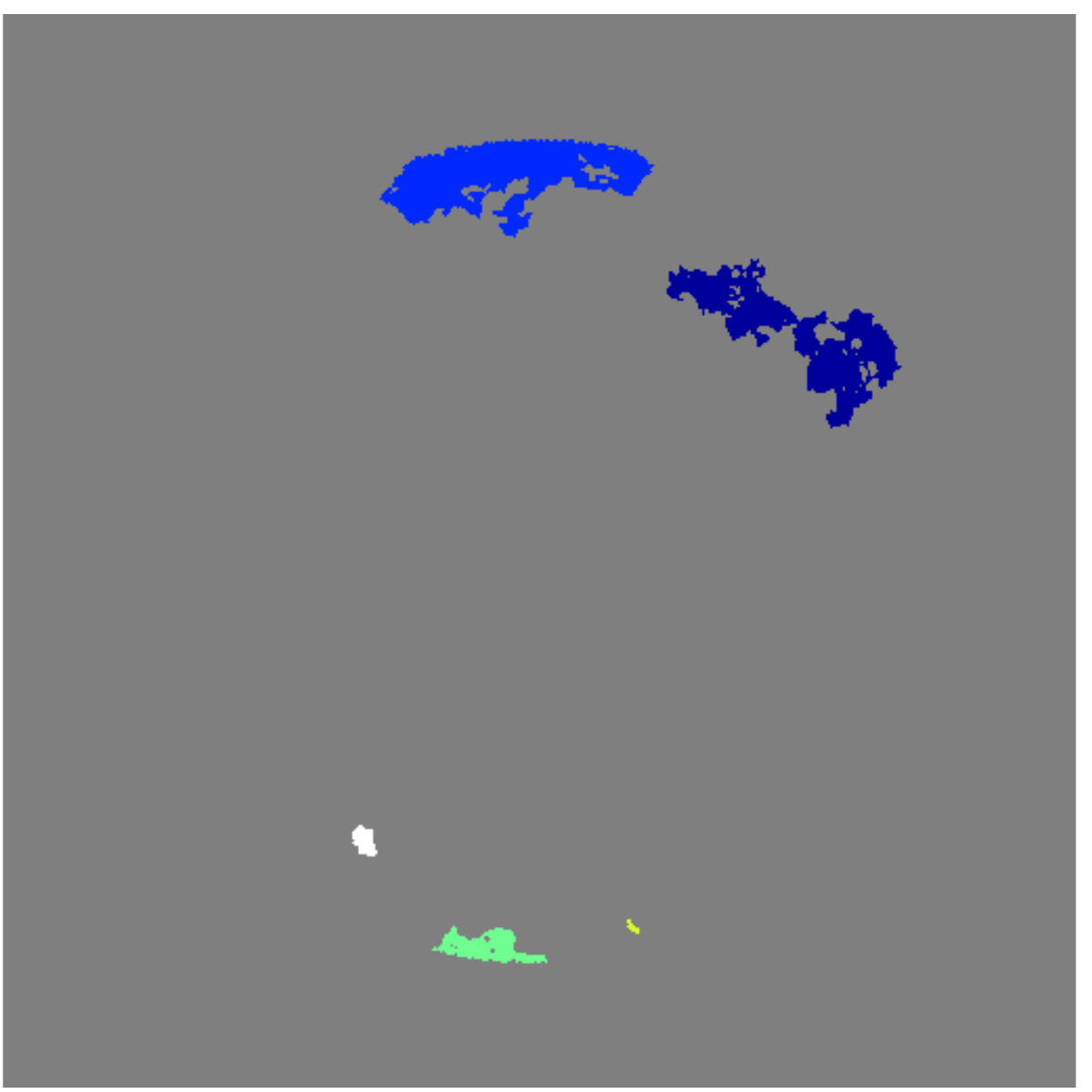}}~
   \subfloat[31 Jul. 2010]{\includegraphics[width=0.13\textwidth]{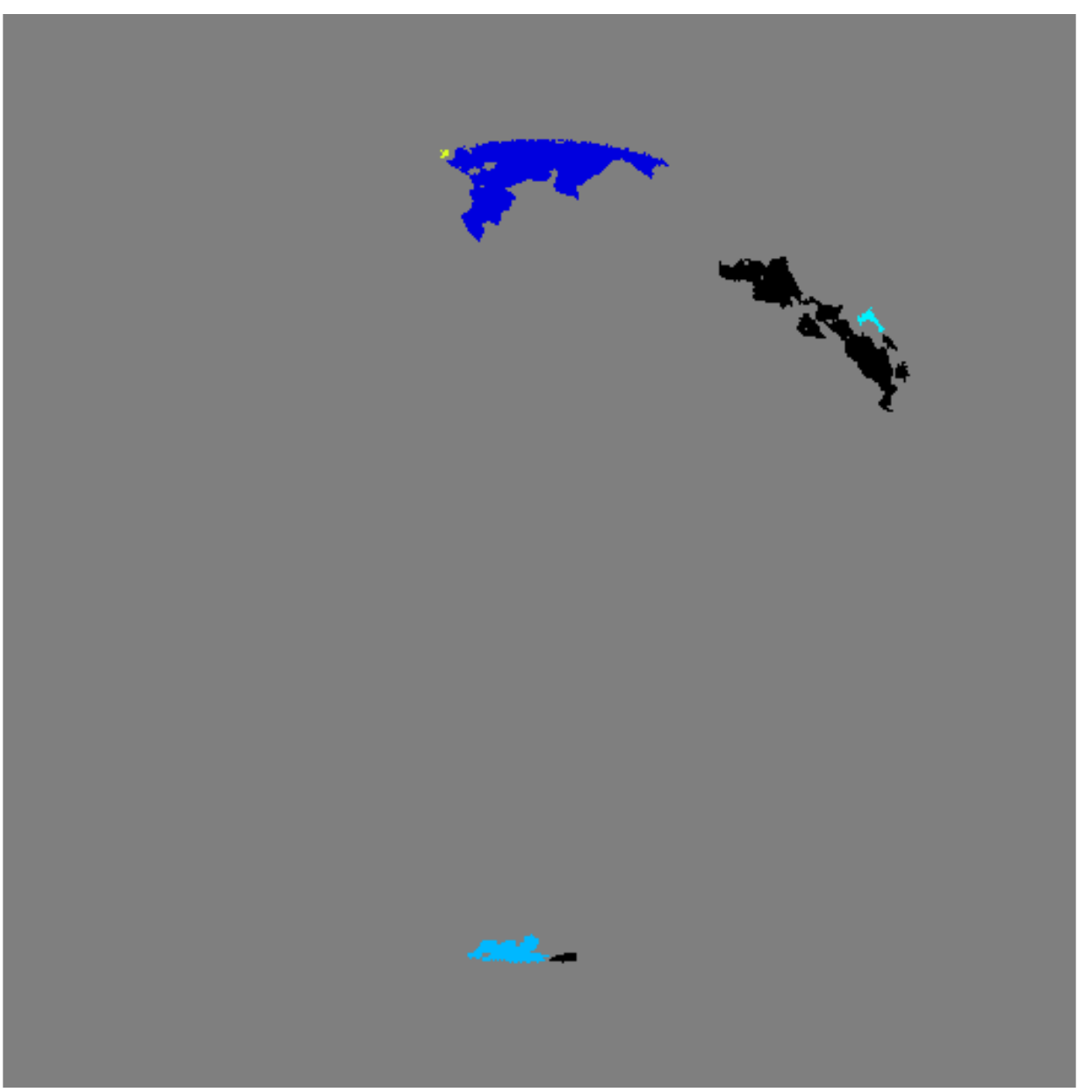}}~
   \subfloat[01 Aug. 2010]{\includegraphics[width=0.13\textwidth]{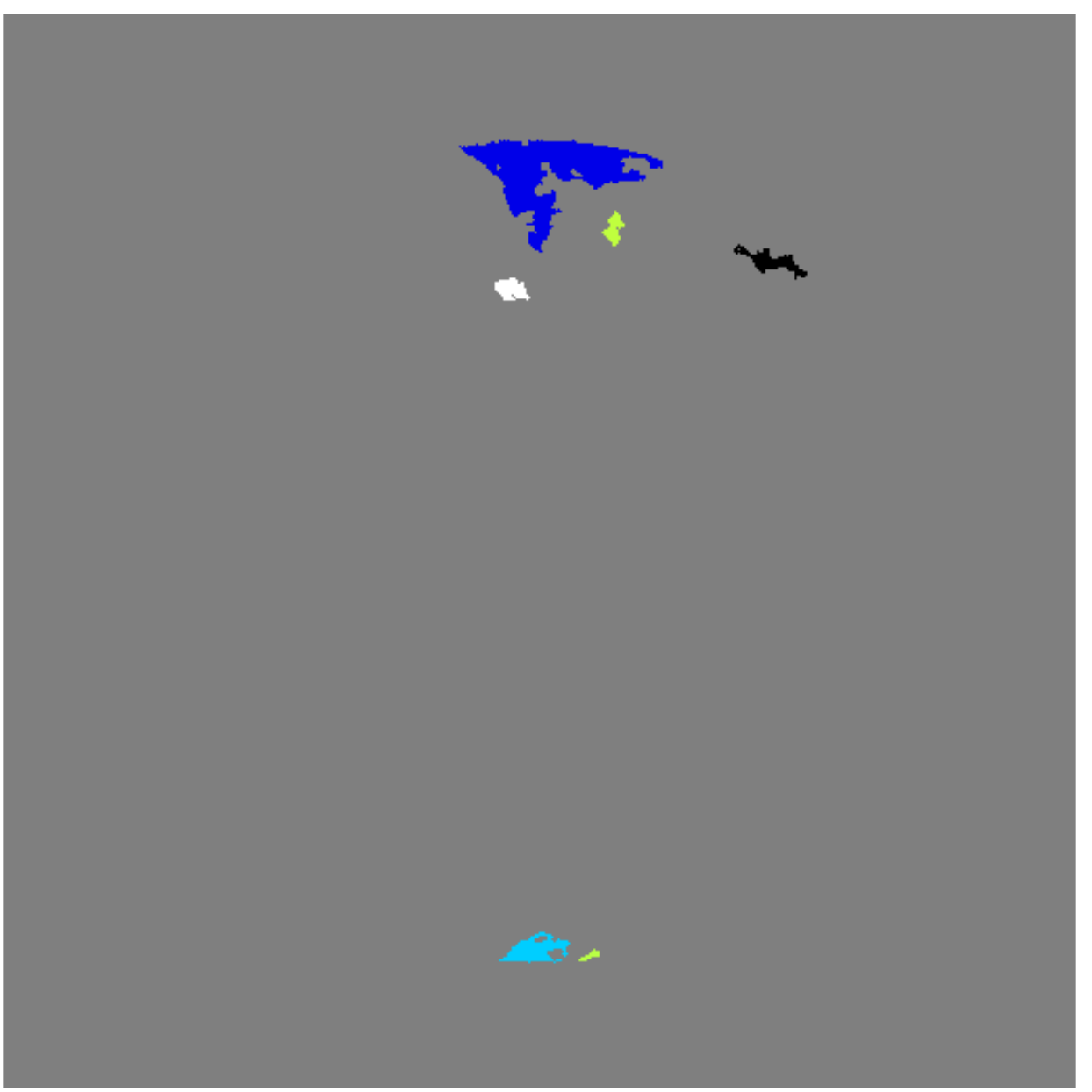}}~
   \subfloat[02 Aug. 2010]{\includegraphics[width=0.13\textwidth]{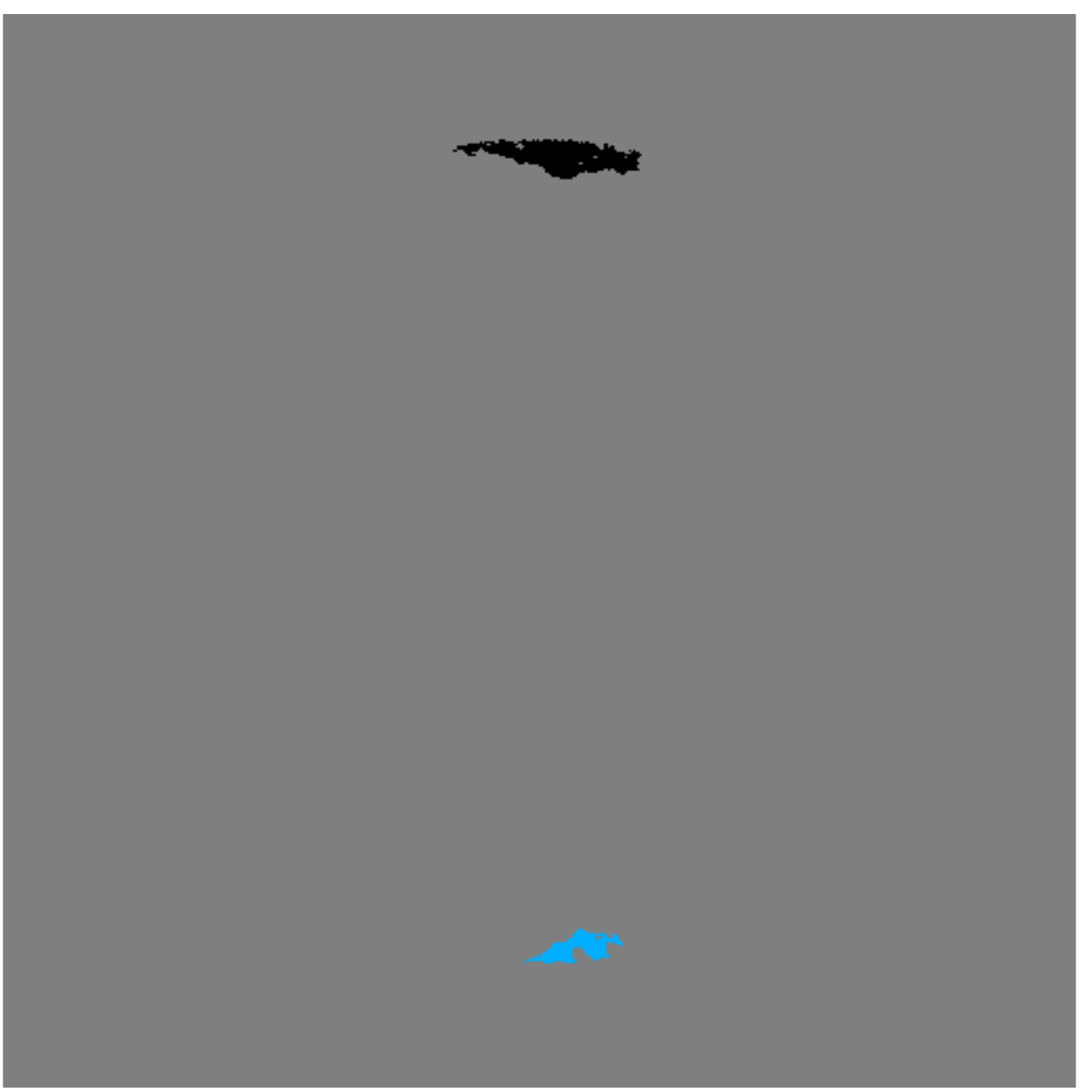}}\\[-2ex]\\
   \subfloat[03 Aug. 2010]{\includegraphics[width=0.13\textwidth]{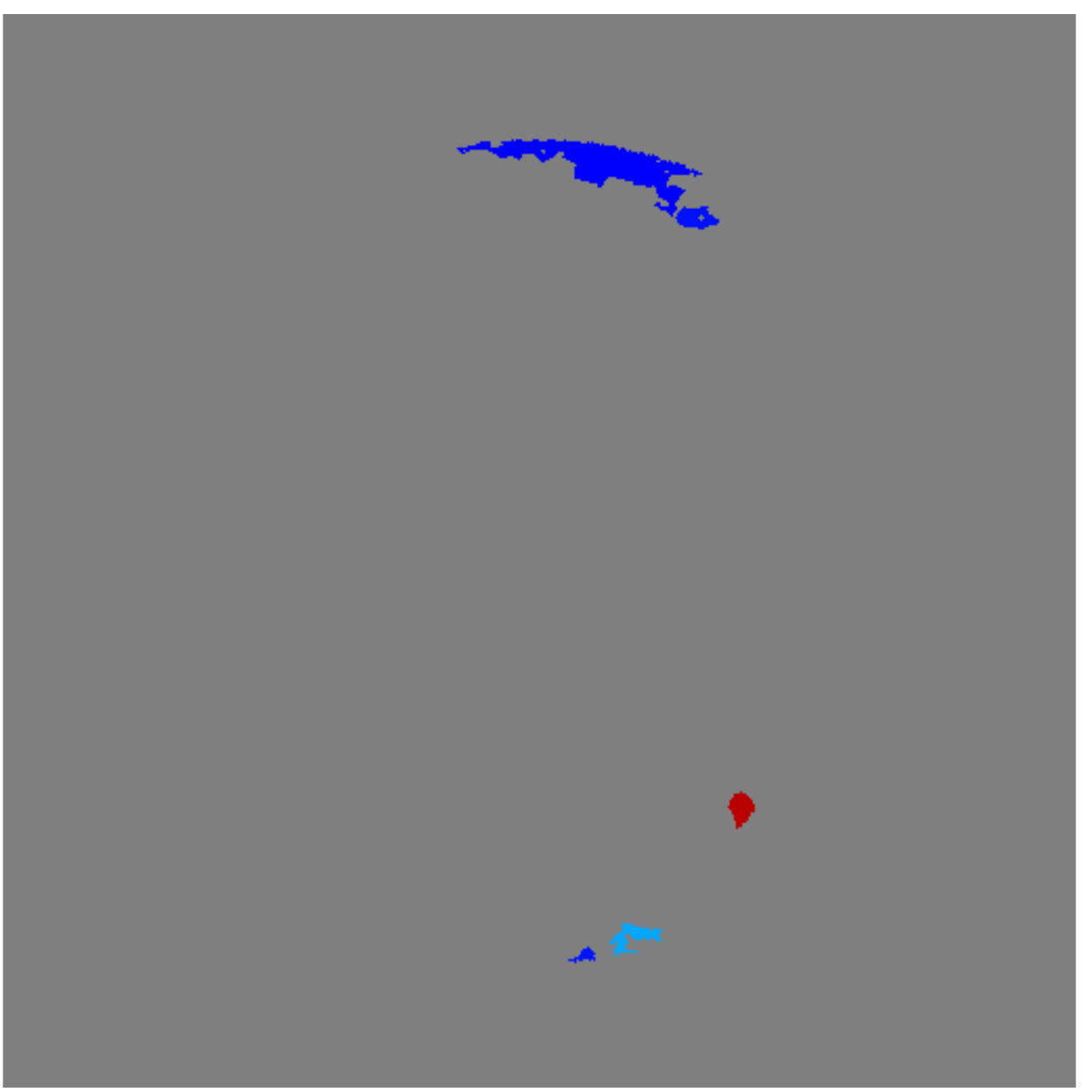}}~
   \subfloat[04 Aug. 2010]{\includegraphics[width=0.13\textwidth]{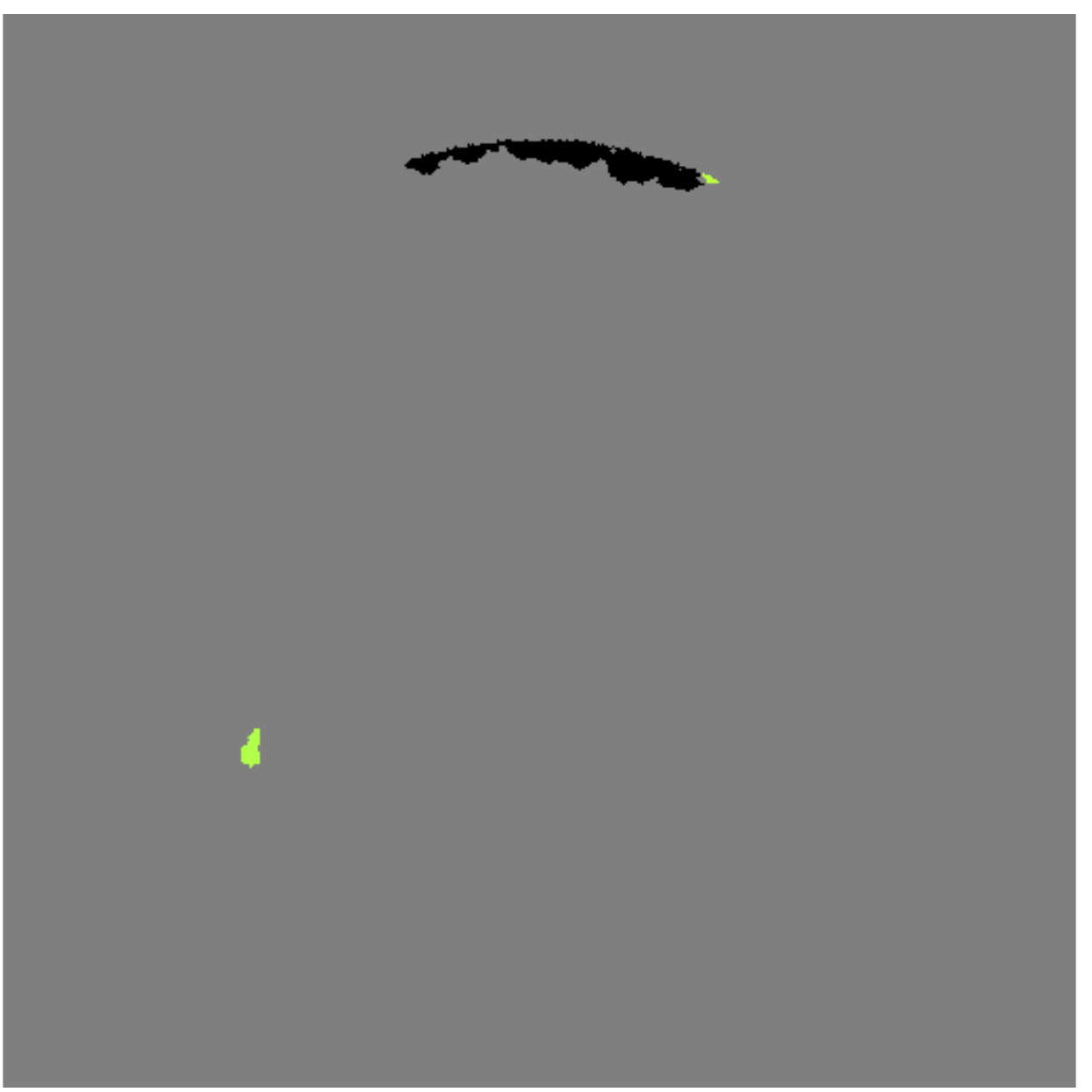}}~
   \subfloat[05 Aug. 2010]{\includegraphics[width=0.13\textwidth]{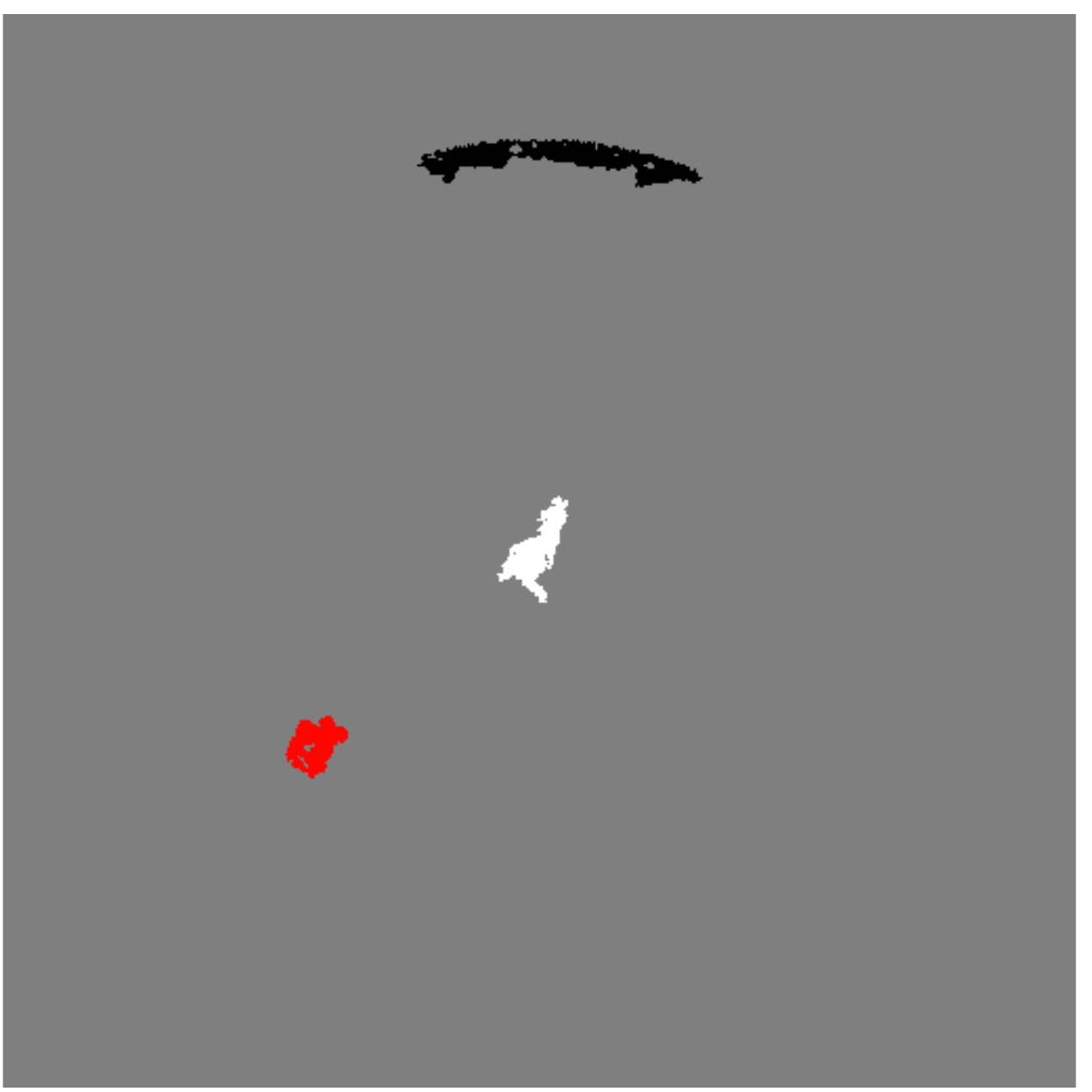}}~
   \subfloat[06 Aug. 2010]{\includegraphics[width=0.13\textwidth]{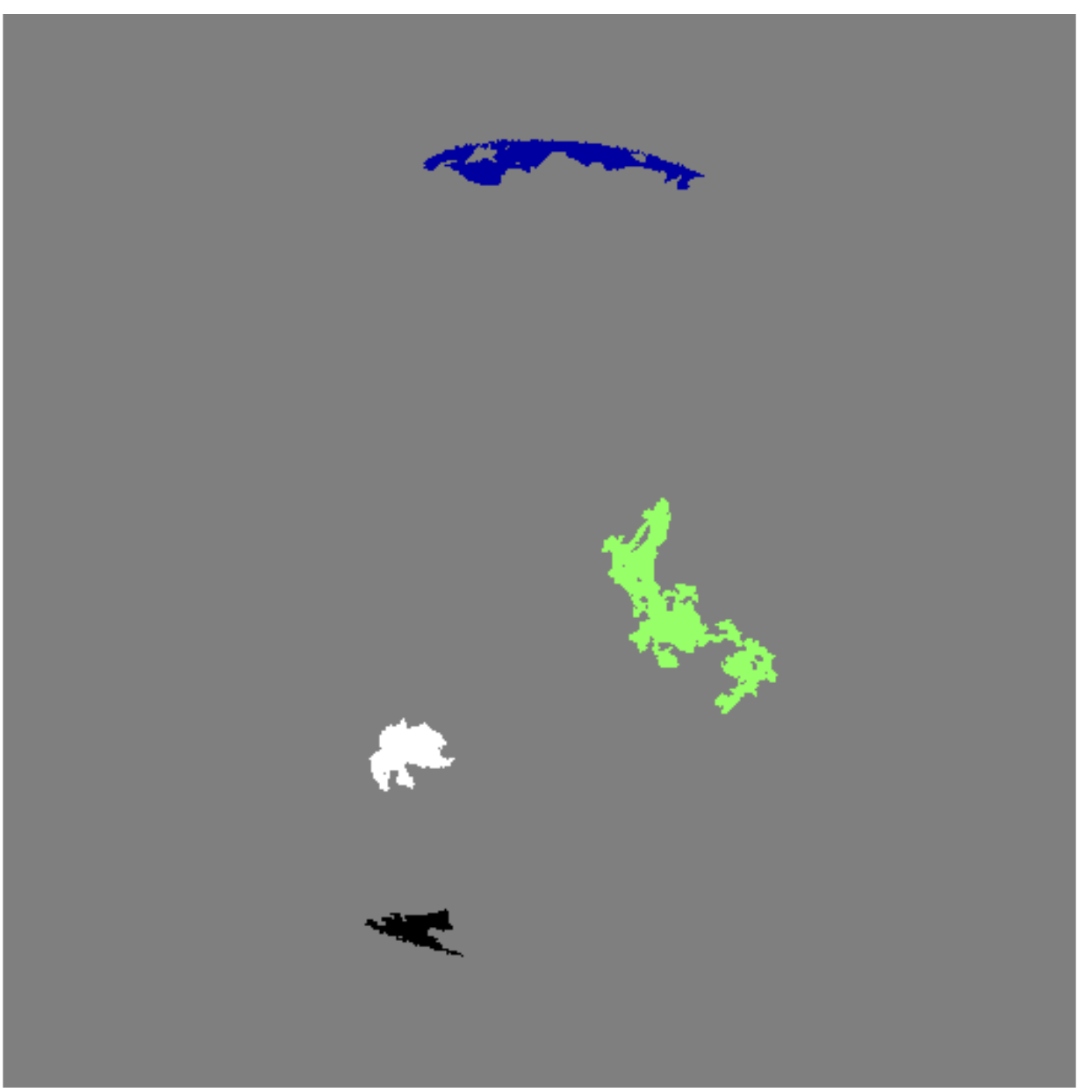}}~
   \subfloat[07 Aug. 2010]{\includegraphics[width=0.13\textwidth]{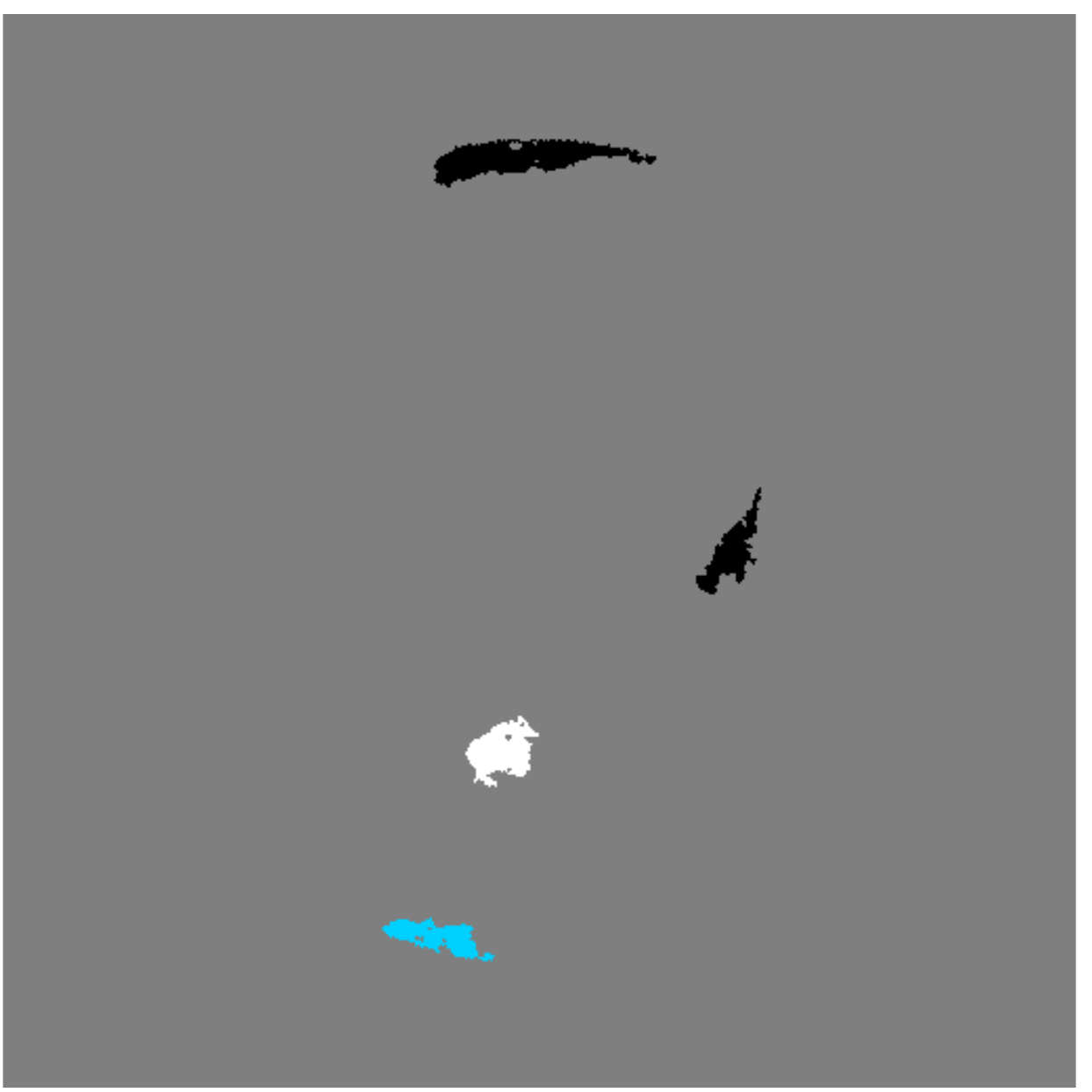}}~
   \subfloat[08 Aug. 2010]{\includegraphics[width=0.13\textwidth]{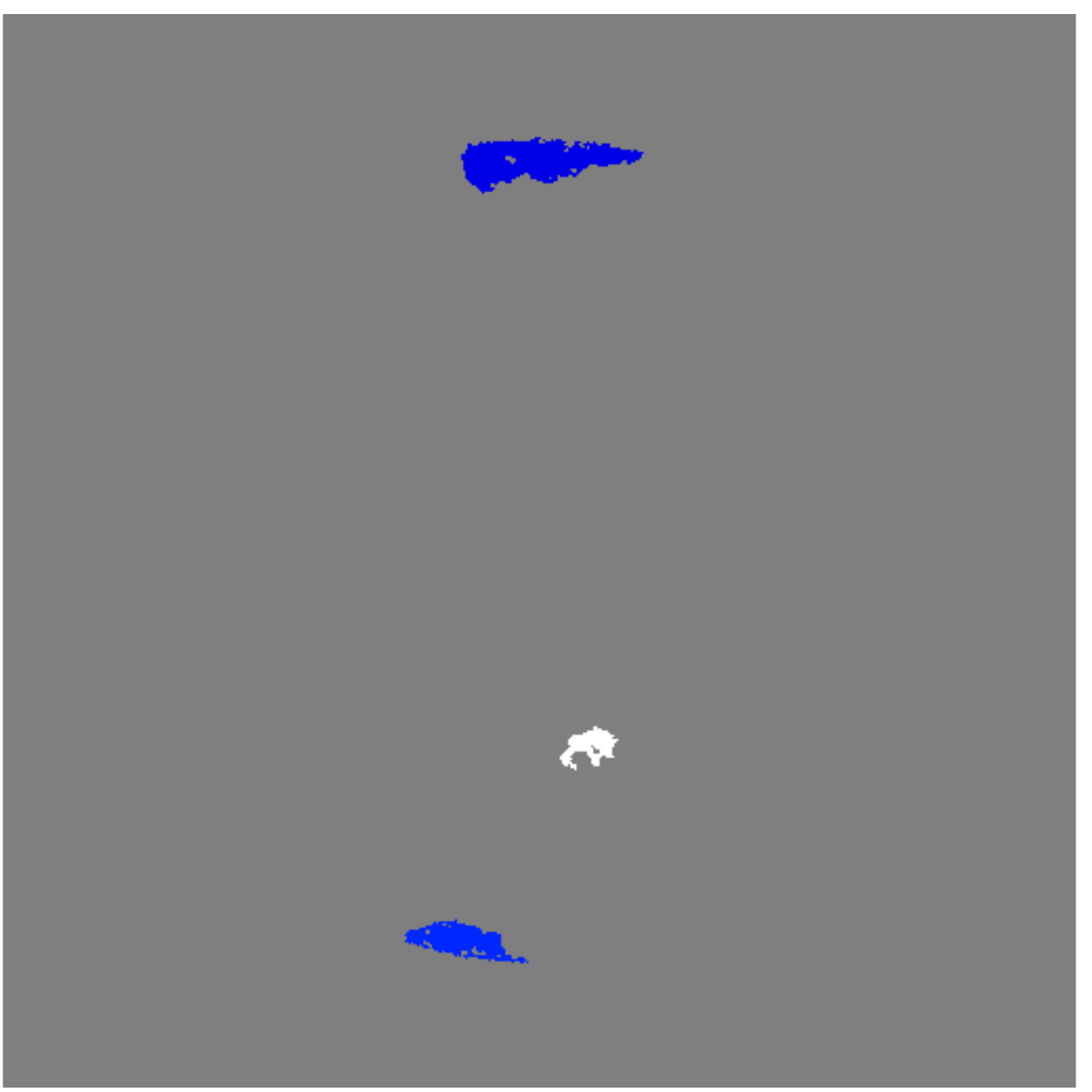}}~
   \subfloat[09 Aug. 2010]{\includegraphics[width=0.13\textwidth]{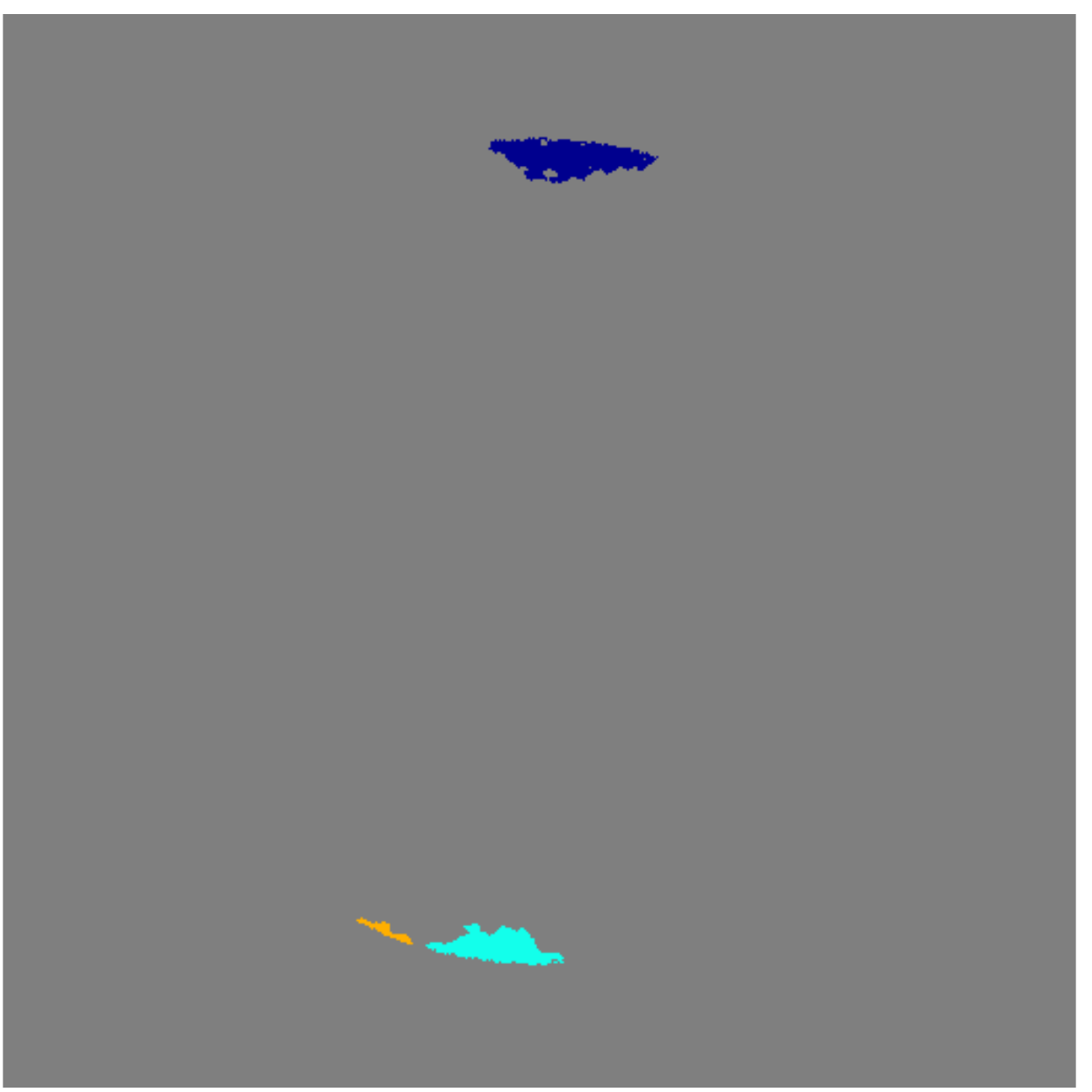}}\\[-2ex]
  \end{minipage}
  \begin{minipage}[r]{0.075\textwidth}
   \subfloat{\includegraphics[width=0.5in]{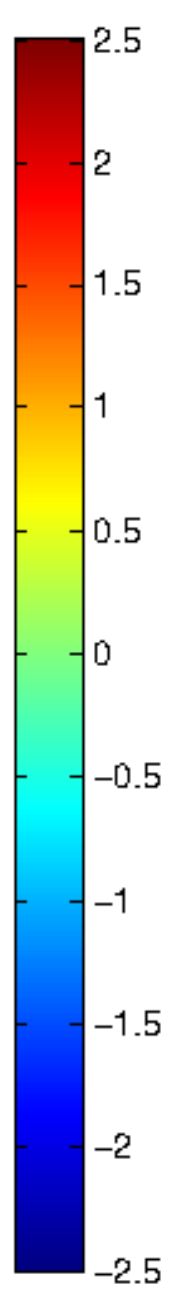}}
  \end{minipage}
  \caption{Carrington rotation 2099 skewness images for $\lambda_i/\lambda_o=50$ and $\alpha=0.3$. The colorbar is shown to the right; skewness values $\ge2.5$ are set to white and skewness values $\le-2.5$ are set to black.}
  \label{fig:CR2099_skewness}
\end{sidewaysfigure}

\begin{sidewaysfigure}[p]
  \centering
  \begin{minipage}[l]{0.85\textwidth}
   \subfloat[25 Jan. 2013]{\includegraphics[width=0.13\textwidth]{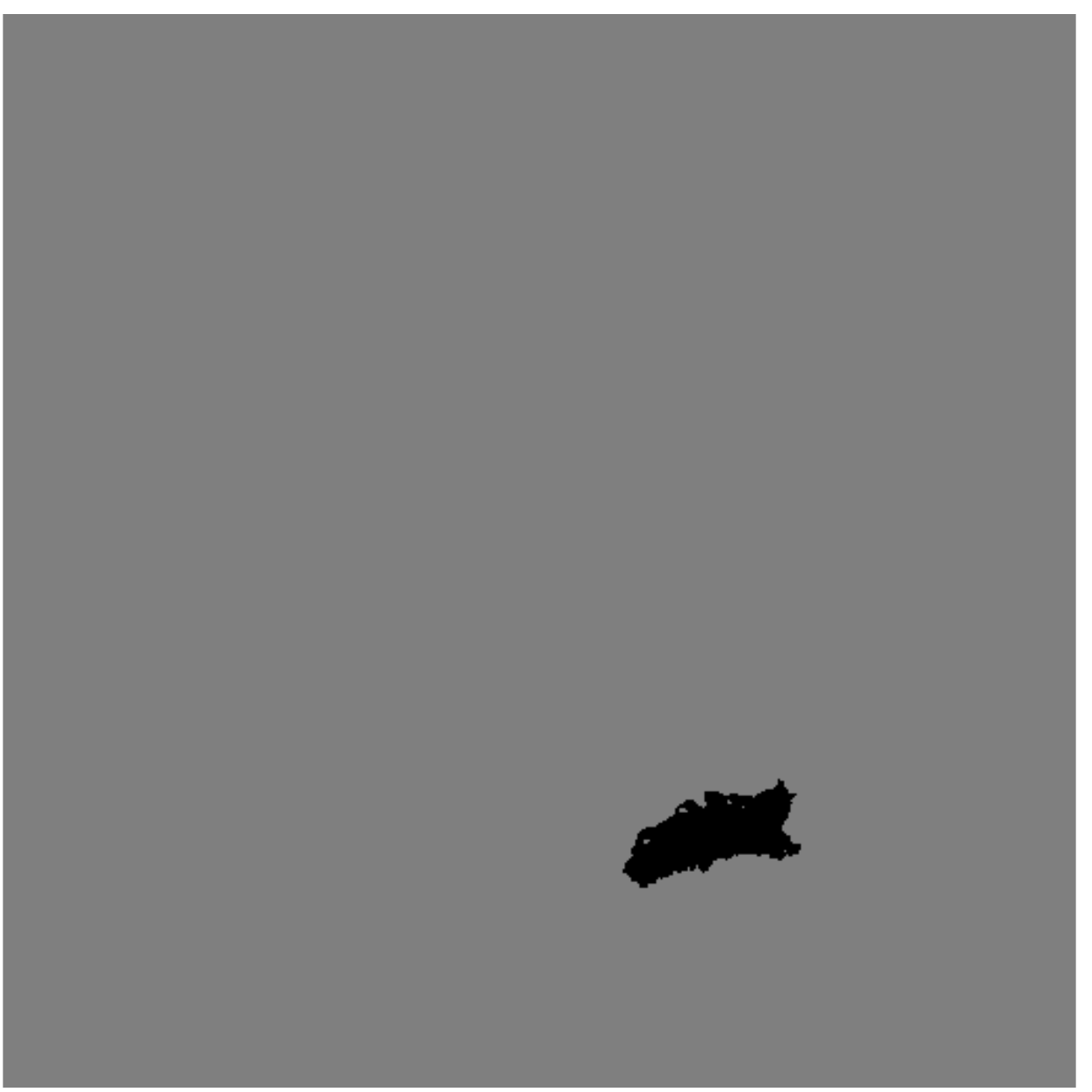}}~
   \subfloat[26 Jan. 2013]{\includegraphics[width=0.13\textwidth]{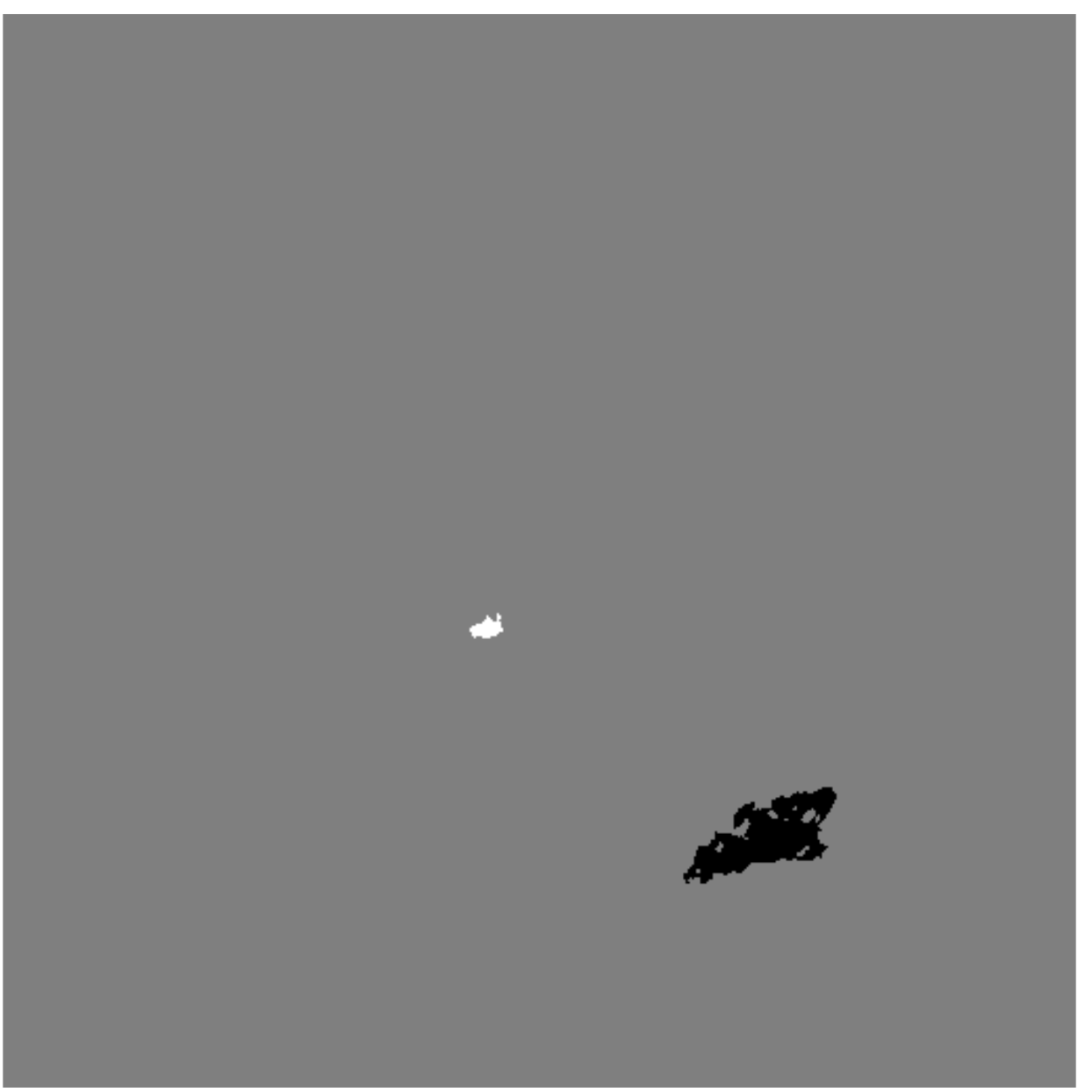}}~
   \subfloat[27 Jan. 2013]{\includegraphics[width=0.13\textwidth]{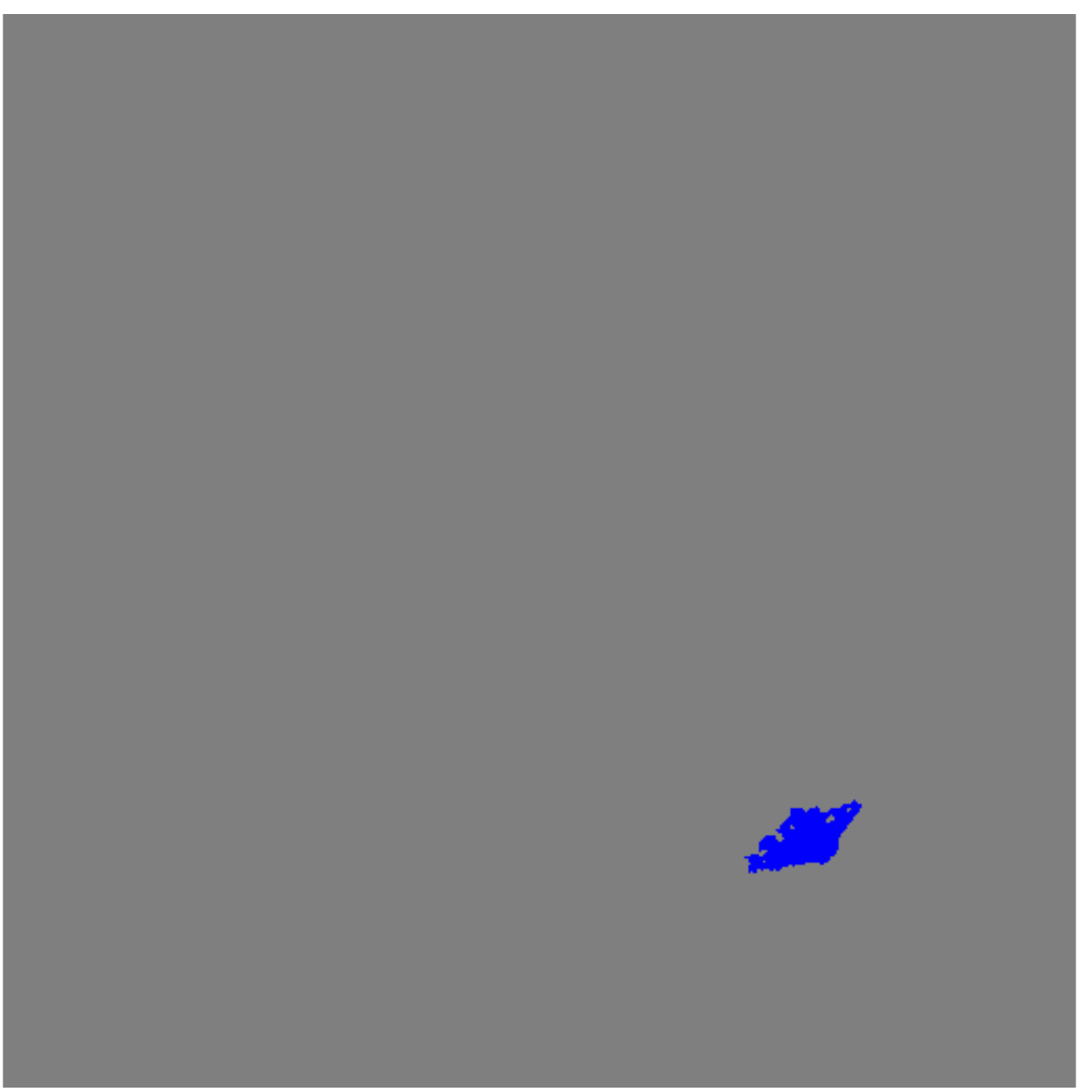}}~
   \subfloat[28 Jan. 2013]{\includegraphics[width=0.13\textwidth]{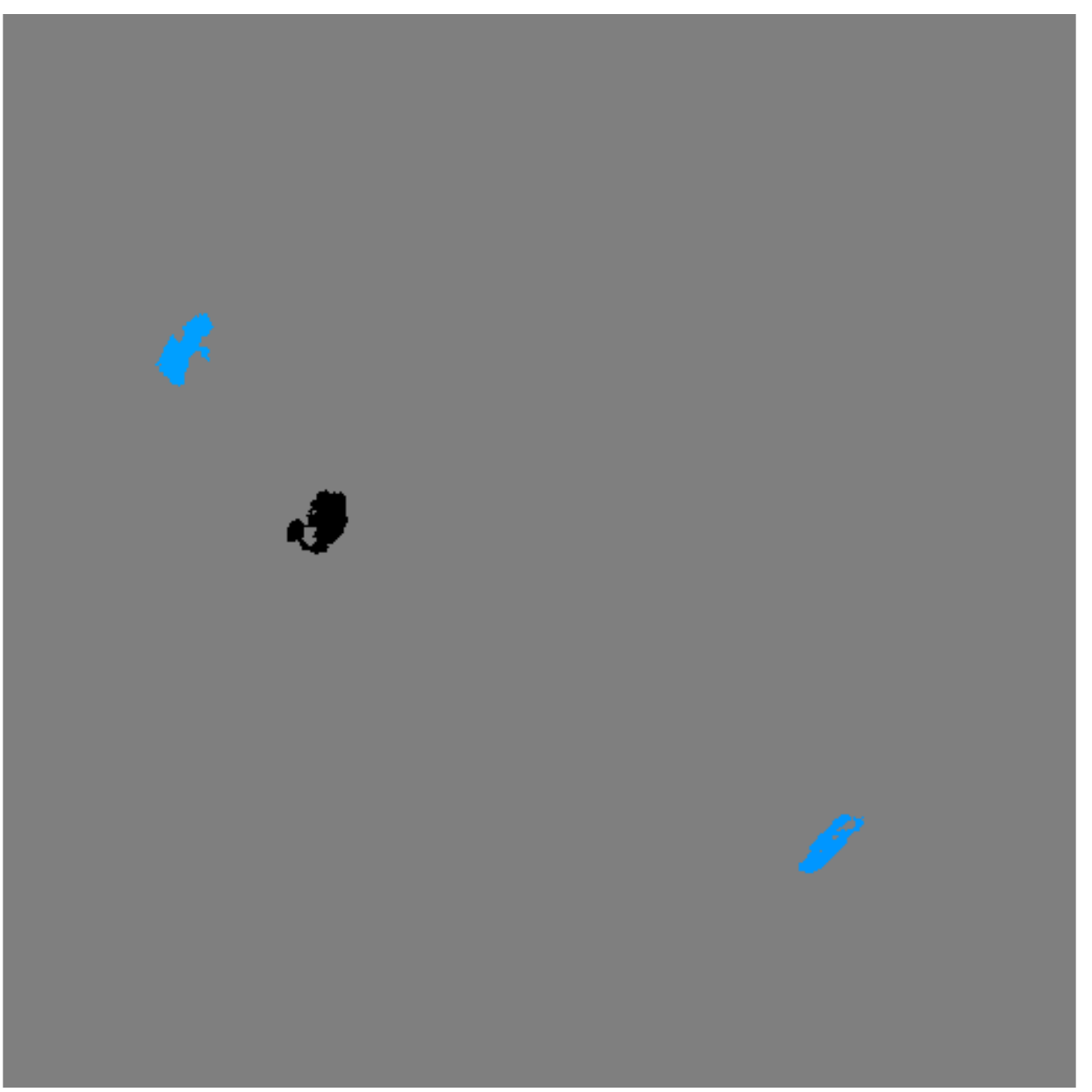}}~
   \subfloat[29 Jan. 2013]{\includegraphics[width=0.13\textwidth]{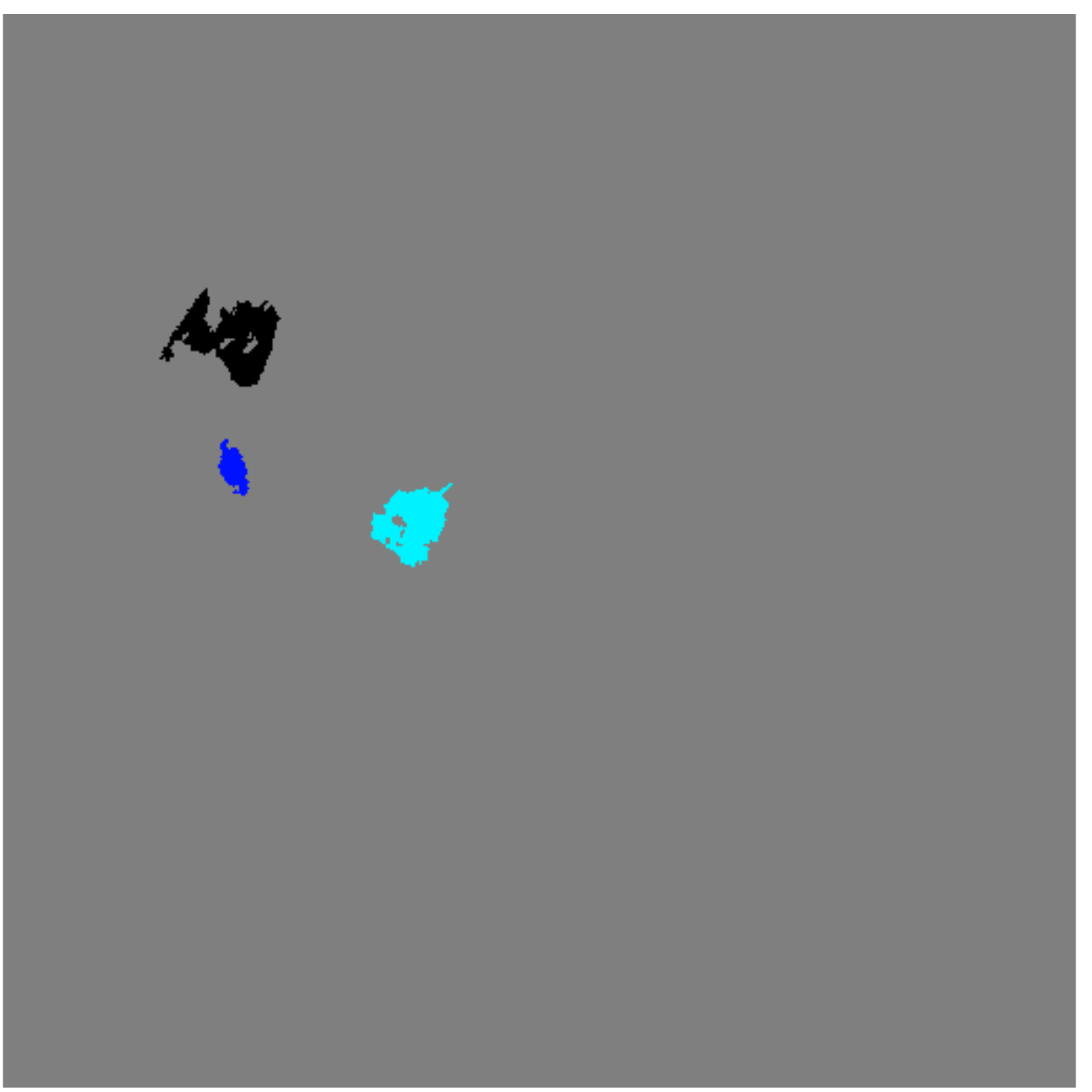}}~
   \subfloat[30 Jan. 2013]{\includegraphics[width=0.13\textwidth]{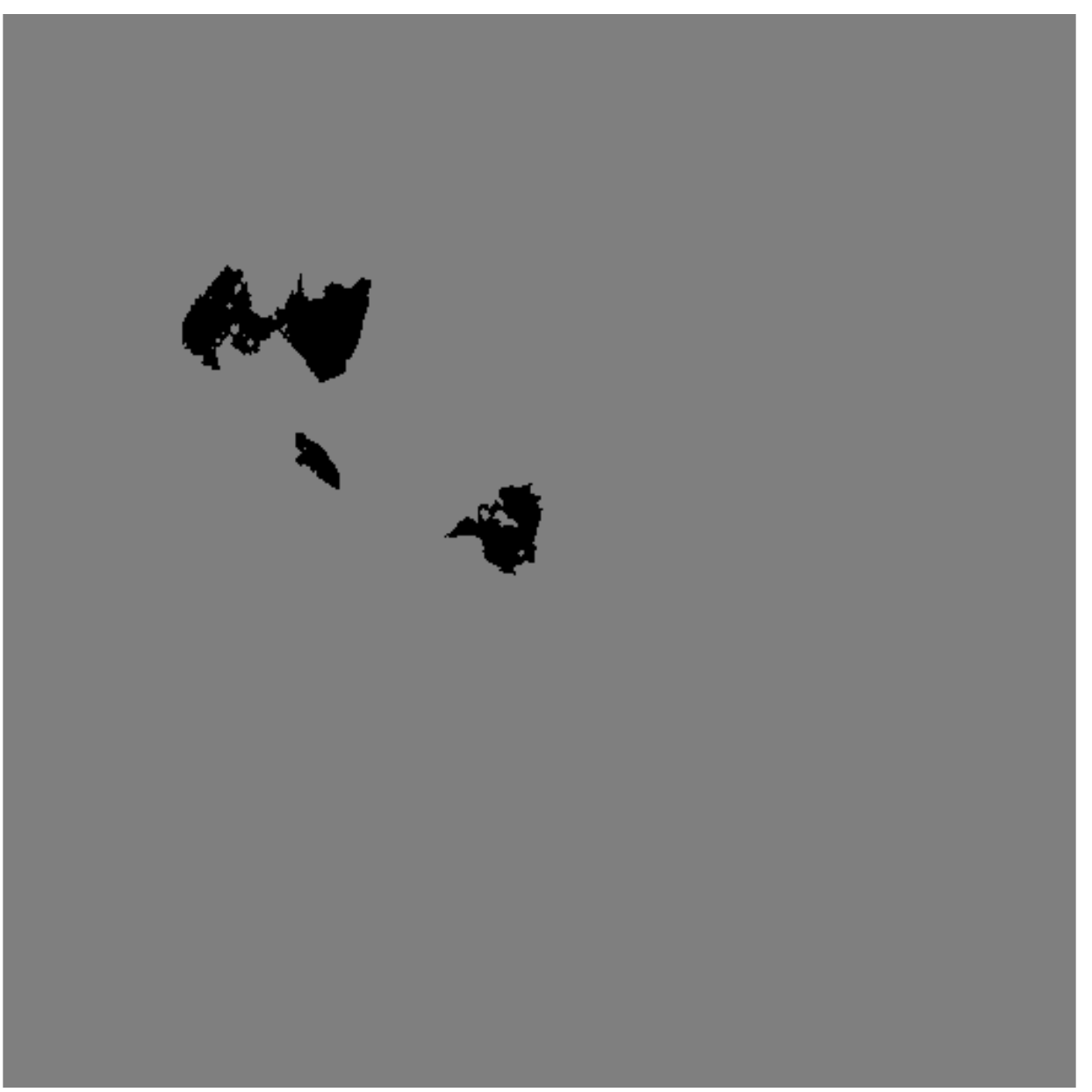}}~
   \subfloat[31 Jan. 2013]{\includegraphics[width=0.13\textwidth]{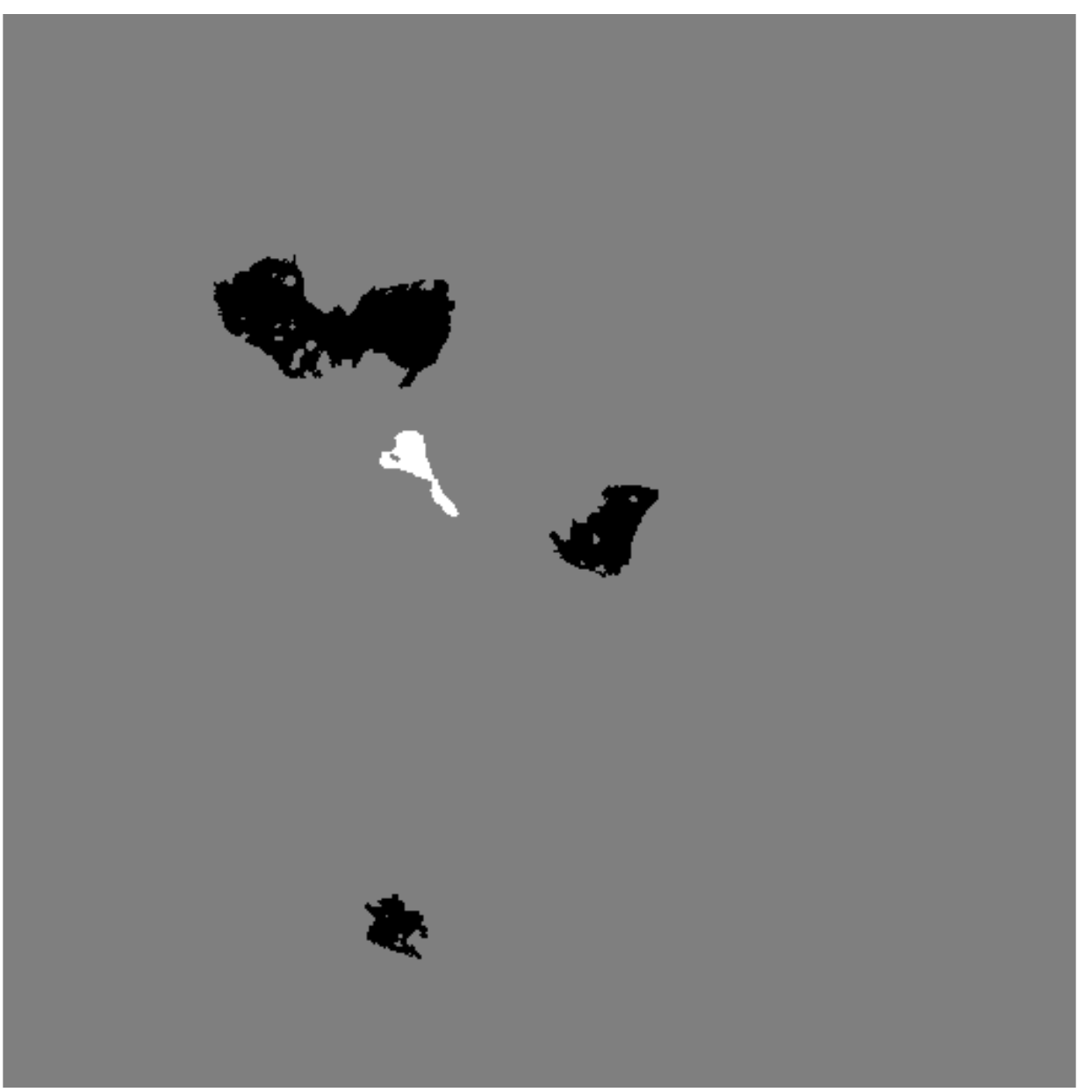}}\\[-2ex]\\
   \subfloat[01 Feb. 2013]{\includegraphics[width=0.13\textwidth]{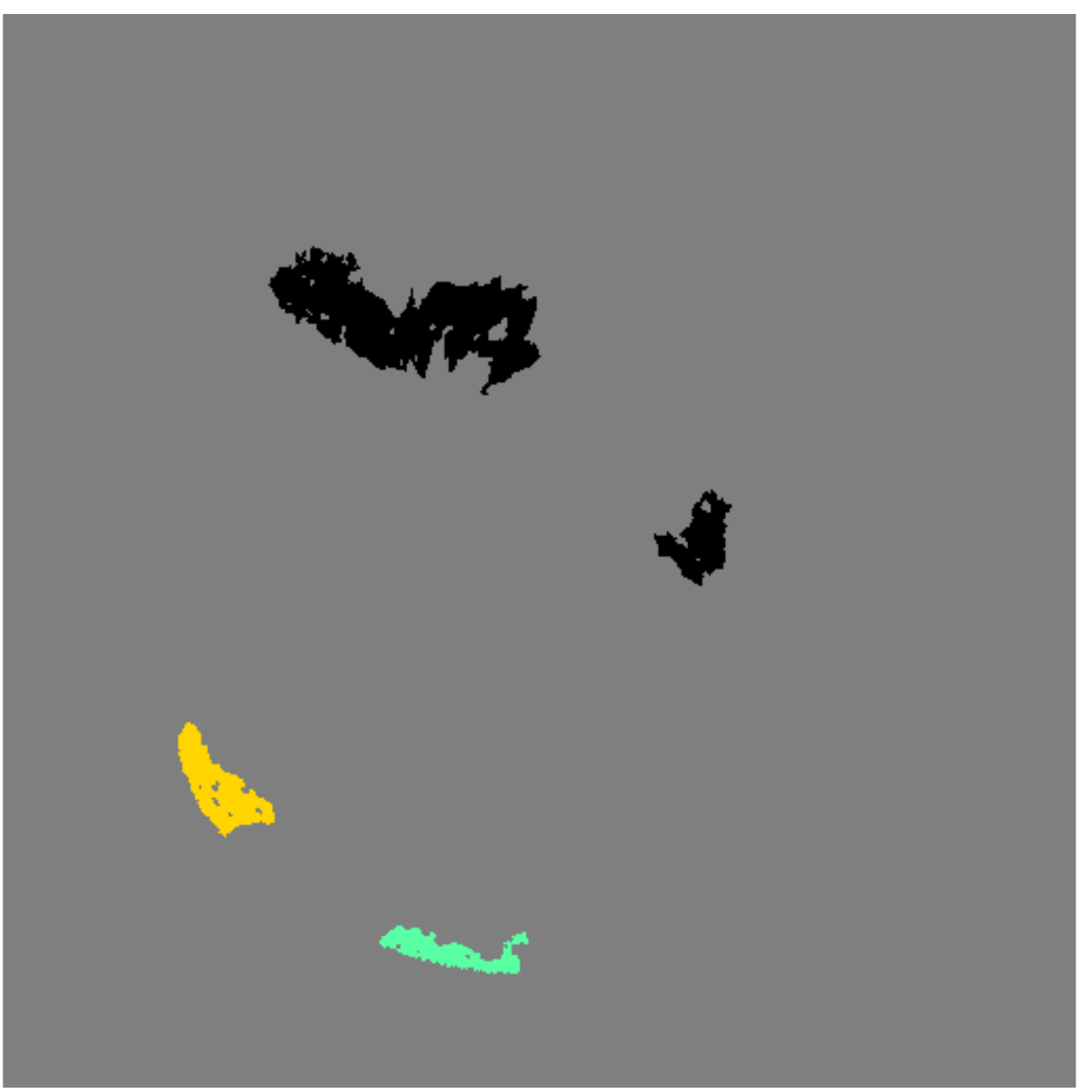}}~
   \subfloat[02 Feb. 2013]{\includegraphics[width=0.13\textwidth]{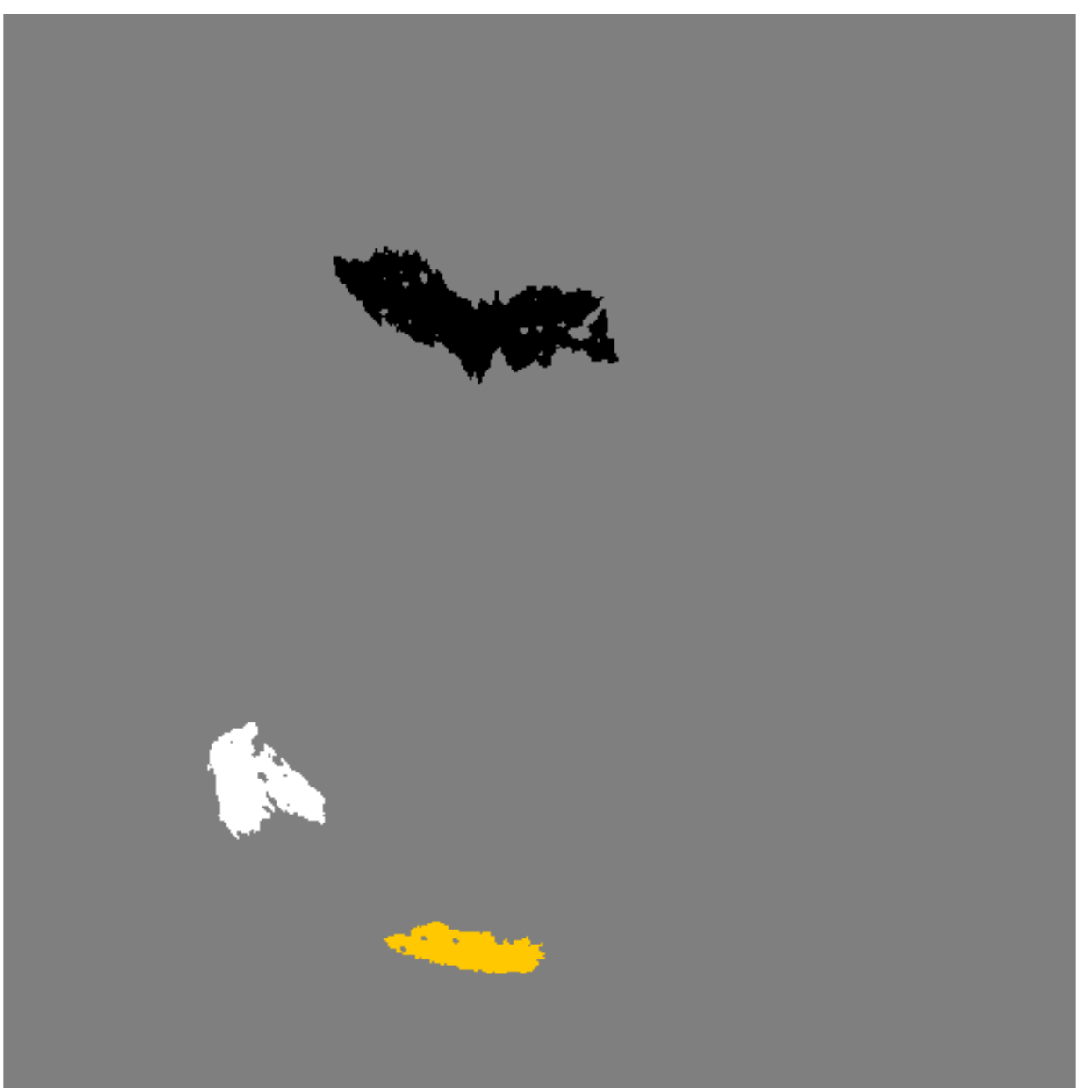}}~
   \subfloat[03 Feb. 2013]{\includegraphics[width=0.13\textwidth]{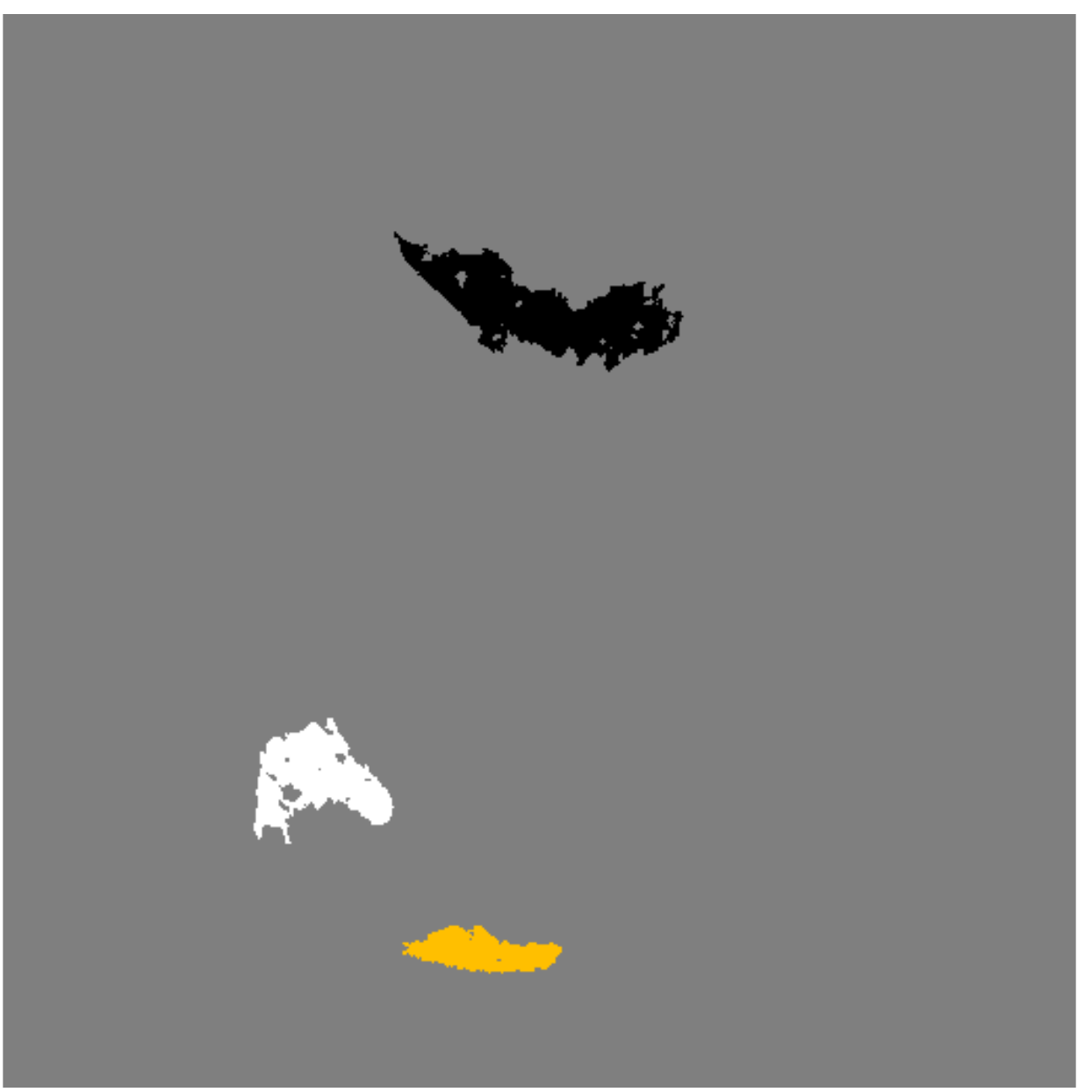}}~
   \subfloat[04 Feb. 2013]{\includegraphics[width=0.13\textwidth]{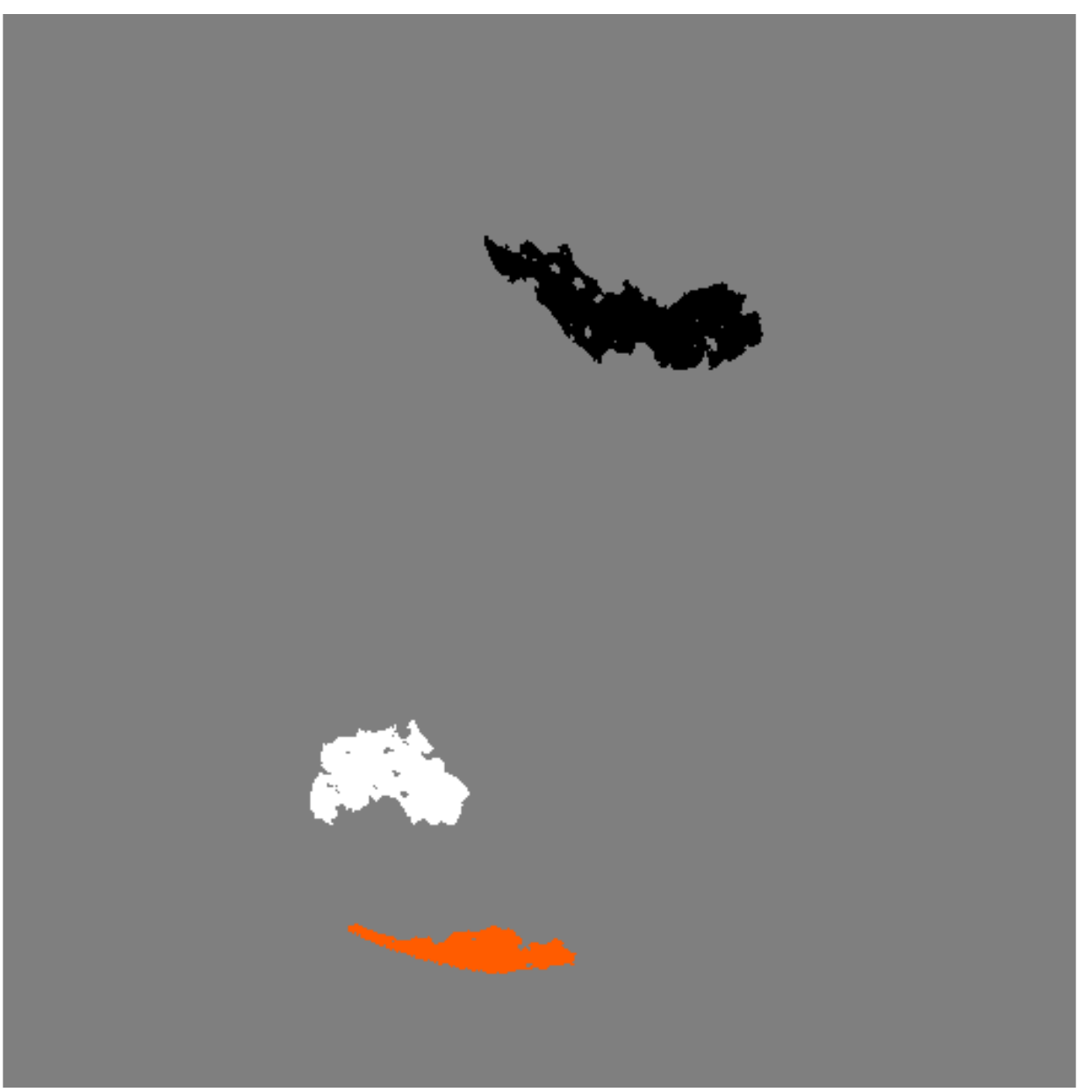}}~
   \subfloat[05 Feb. 2013]{\includegraphics[width=0.13\textwidth]{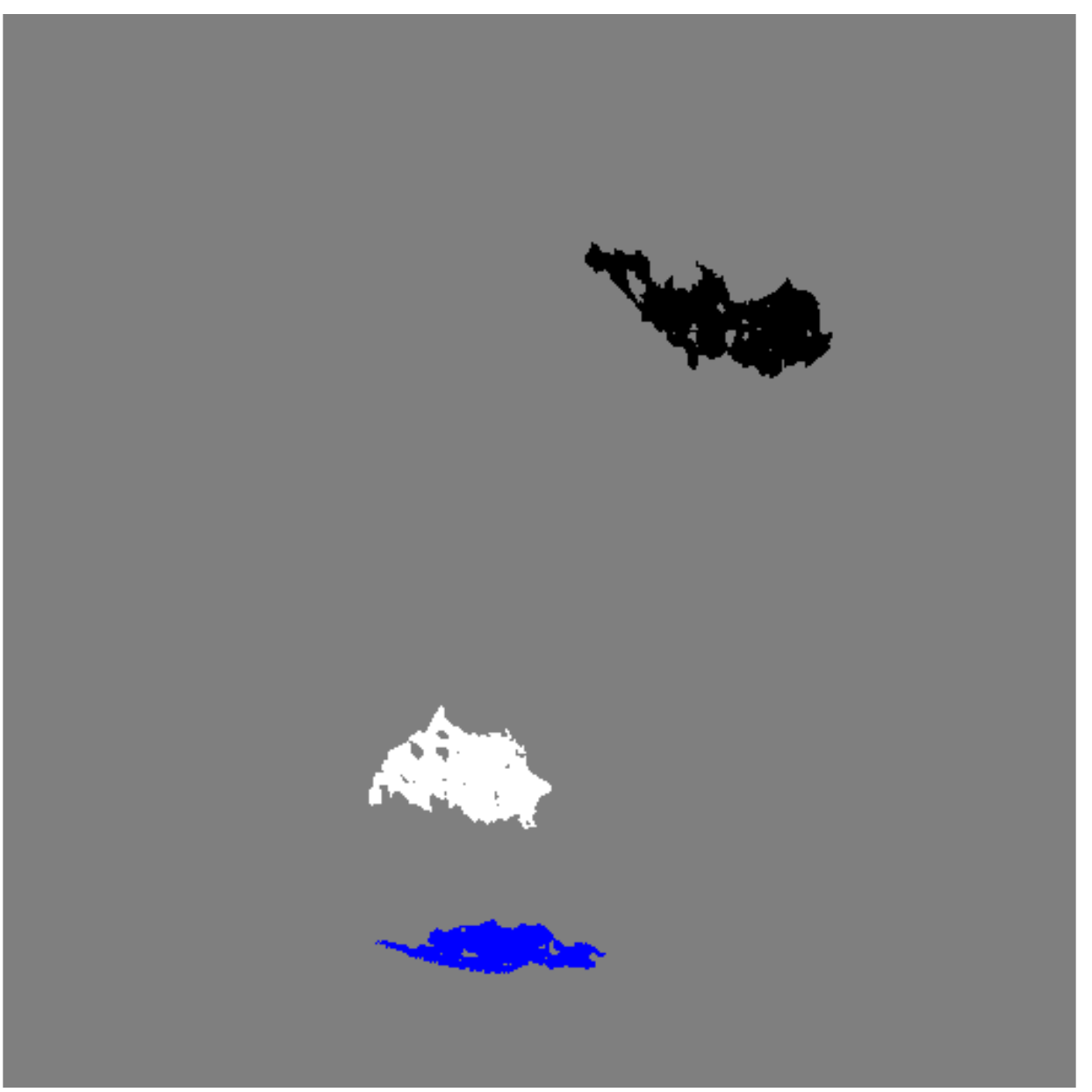}}~
   \subfloat[06 Feb. 2013]{\includegraphics[width=0.13\textwidth]{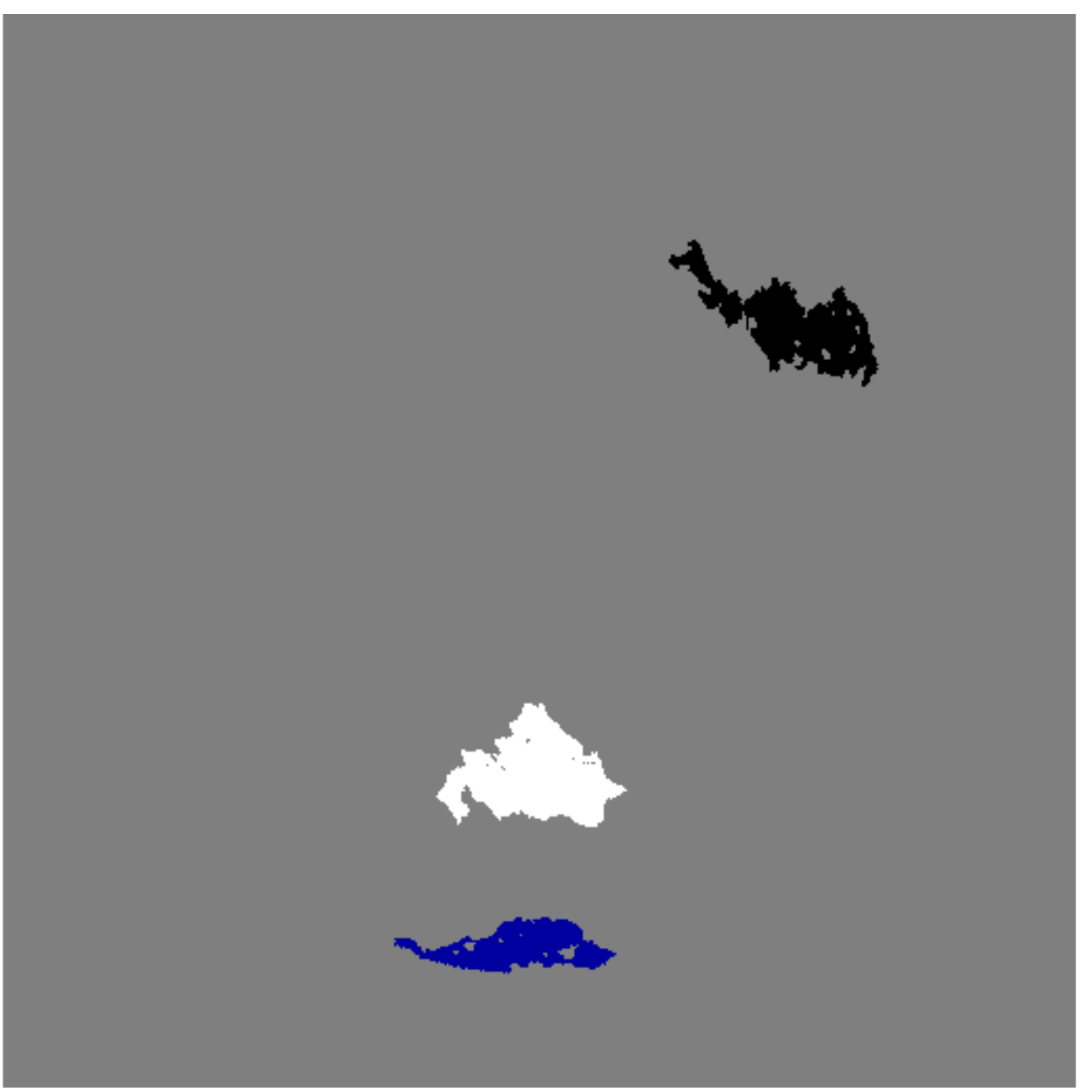}}~
   \subfloat[07 Feb. 2013]{\includegraphics[width=0.13\textwidth]{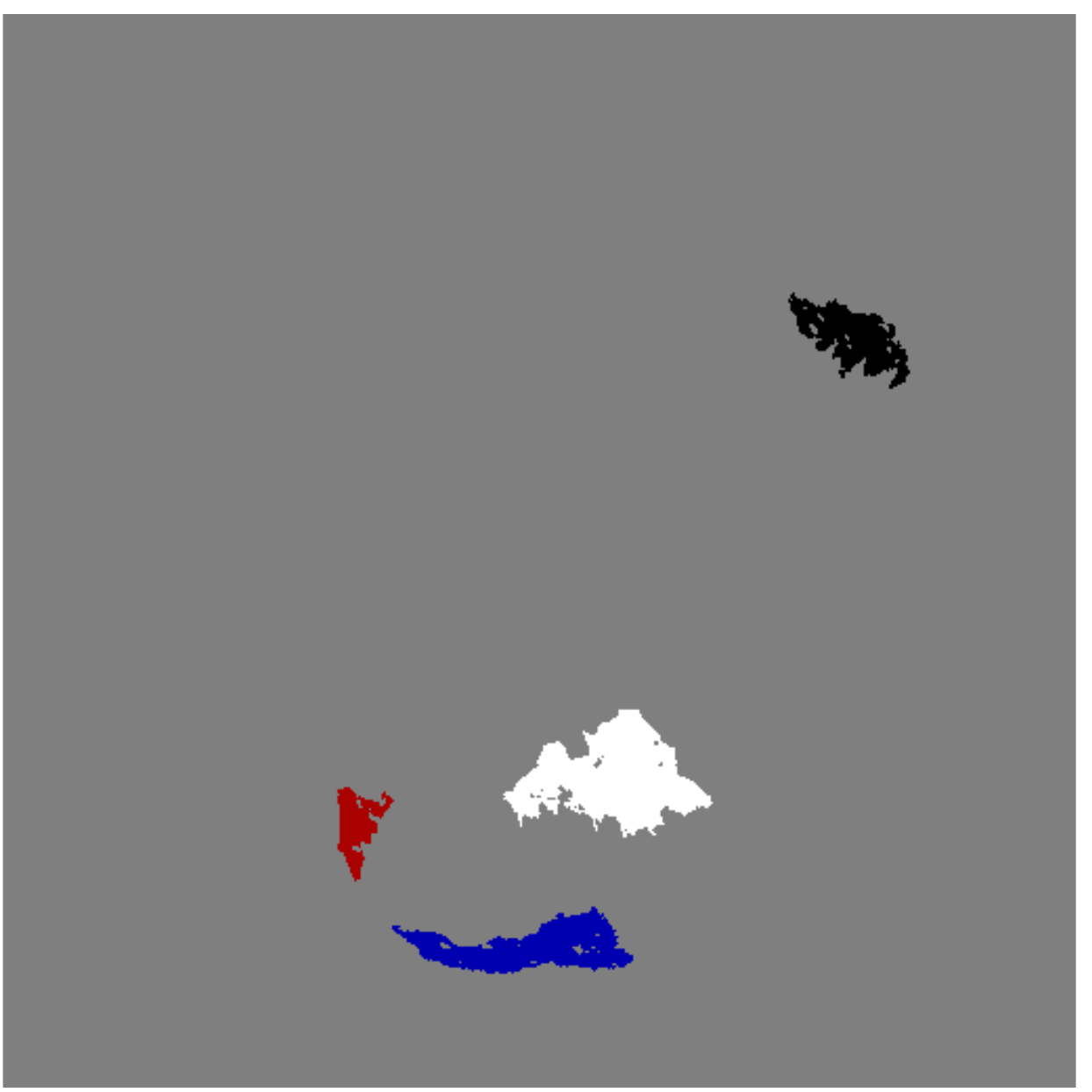}}\\[-2ex]\\
   \subfloat[08 Feb. 2013]{\includegraphics[width=0.13\textwidth]{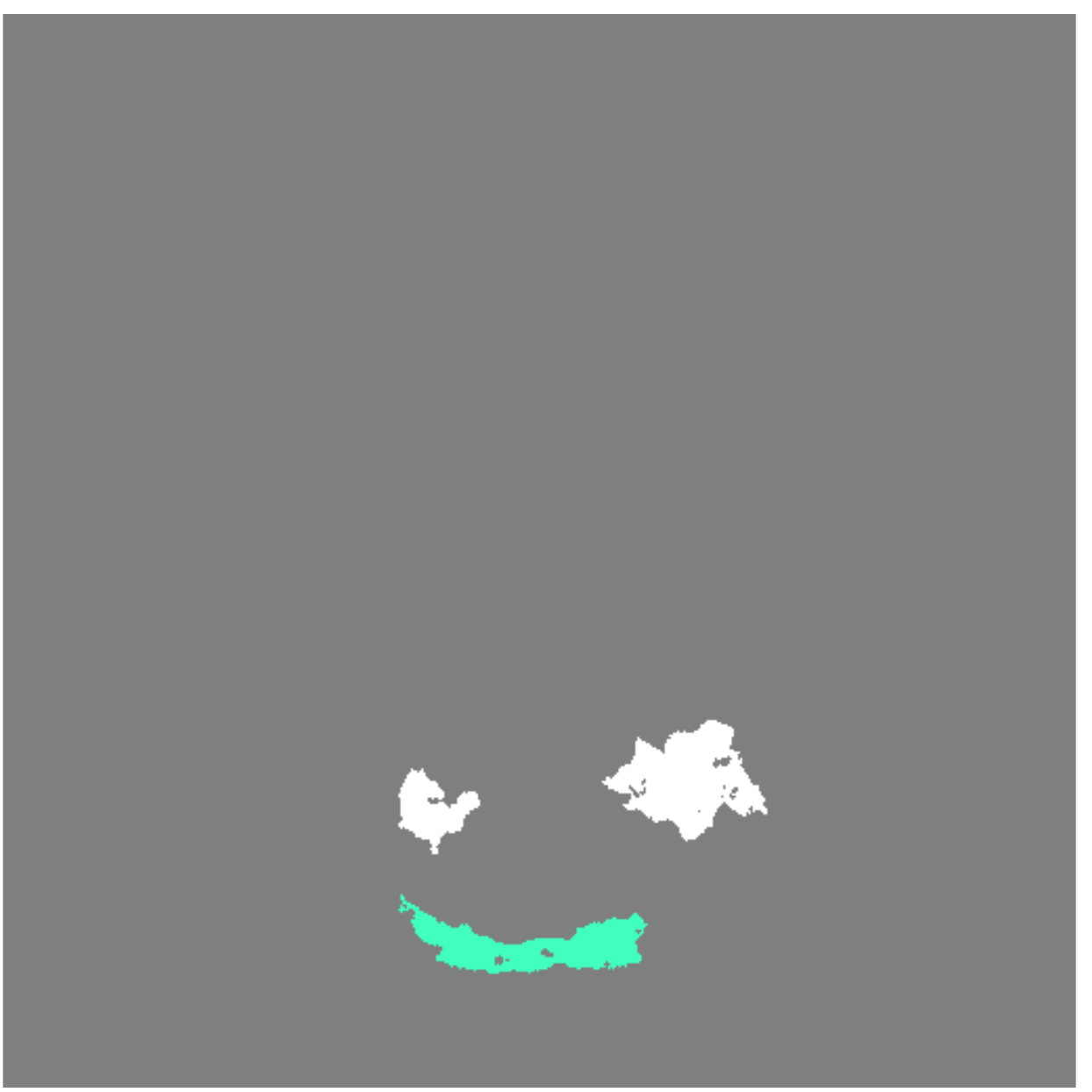}}~
   \subfloat[09 Feb. 2013]{\includegraphics[width=0.13\textwidth]{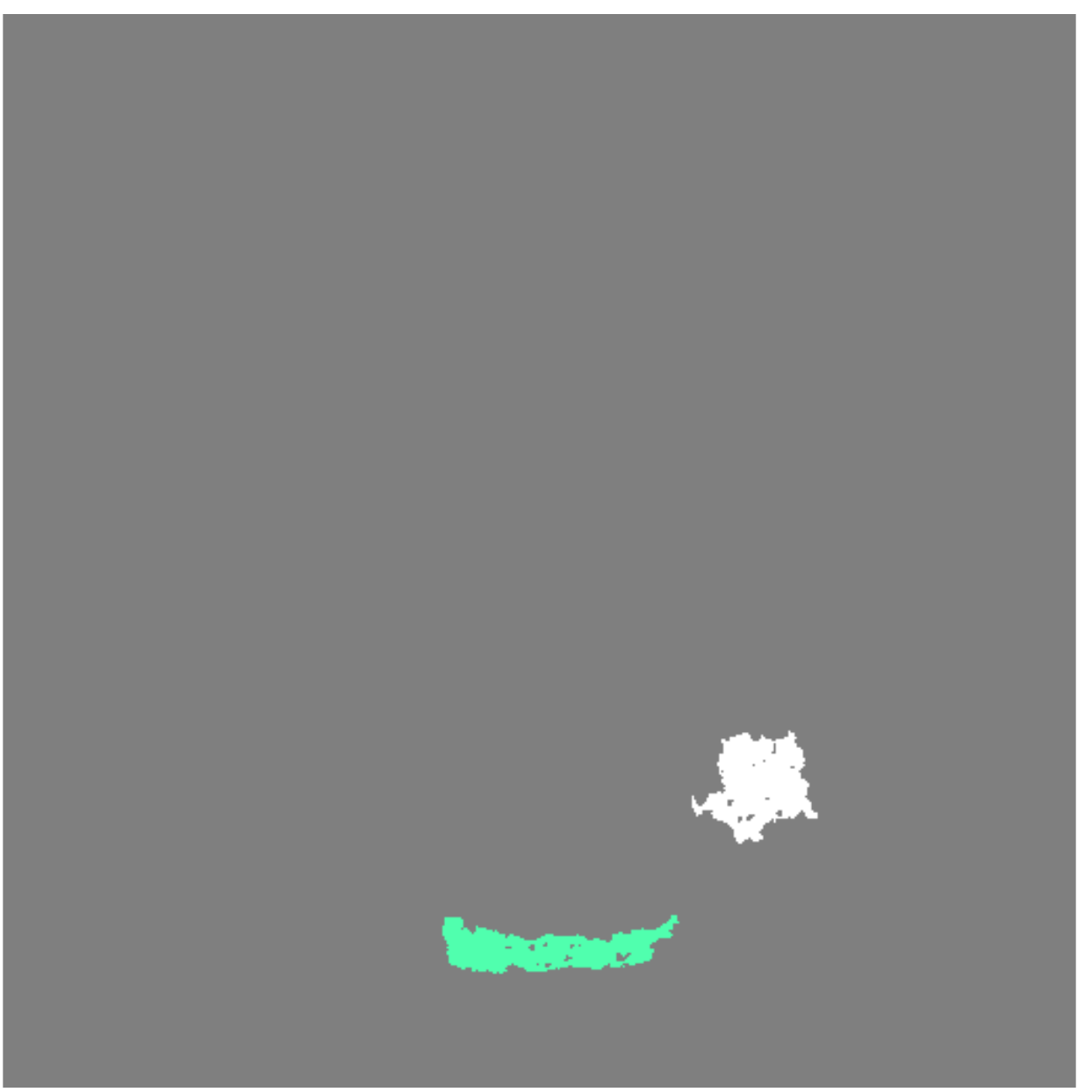}}~
   \subfloat[10 Feb. 2013]{\includegraphics[width=0.13\textwidth]{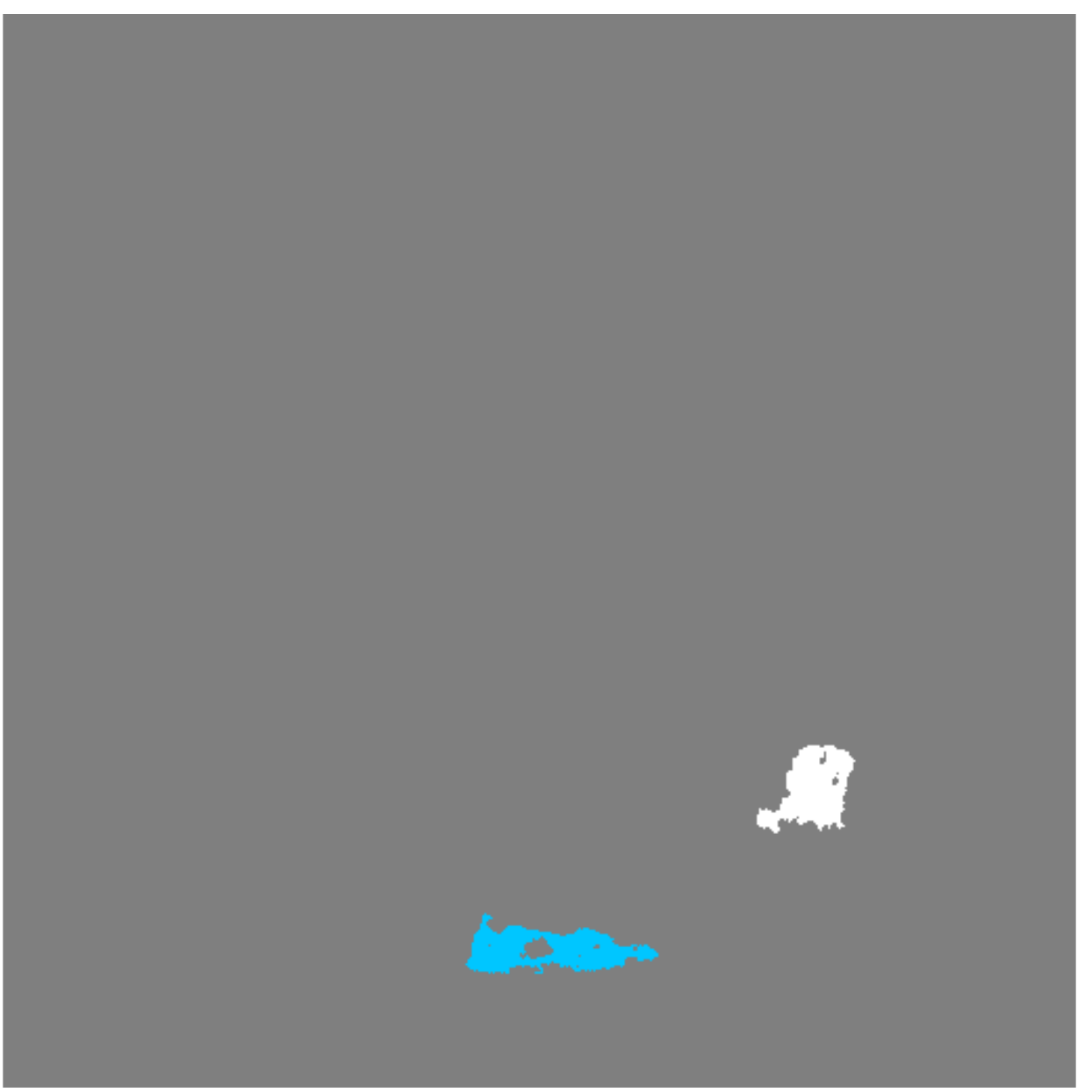}}~
   \subfloat[11 Feb. 2013]{\includegraphics[width=0.13\textwidth]{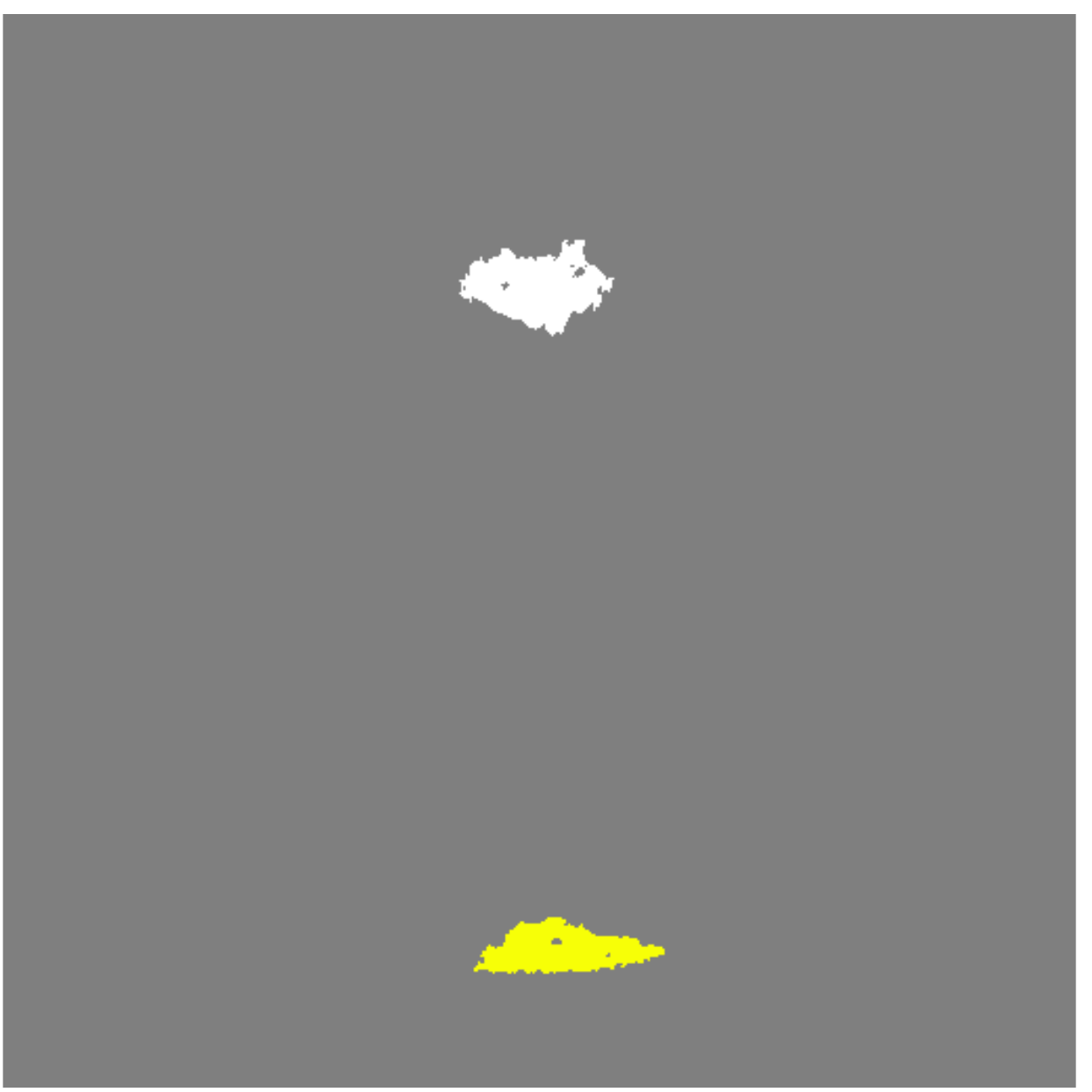}}~
   \subfloat[12 Feb. 2013]{\includegraphics[width=0.13\textwidth]{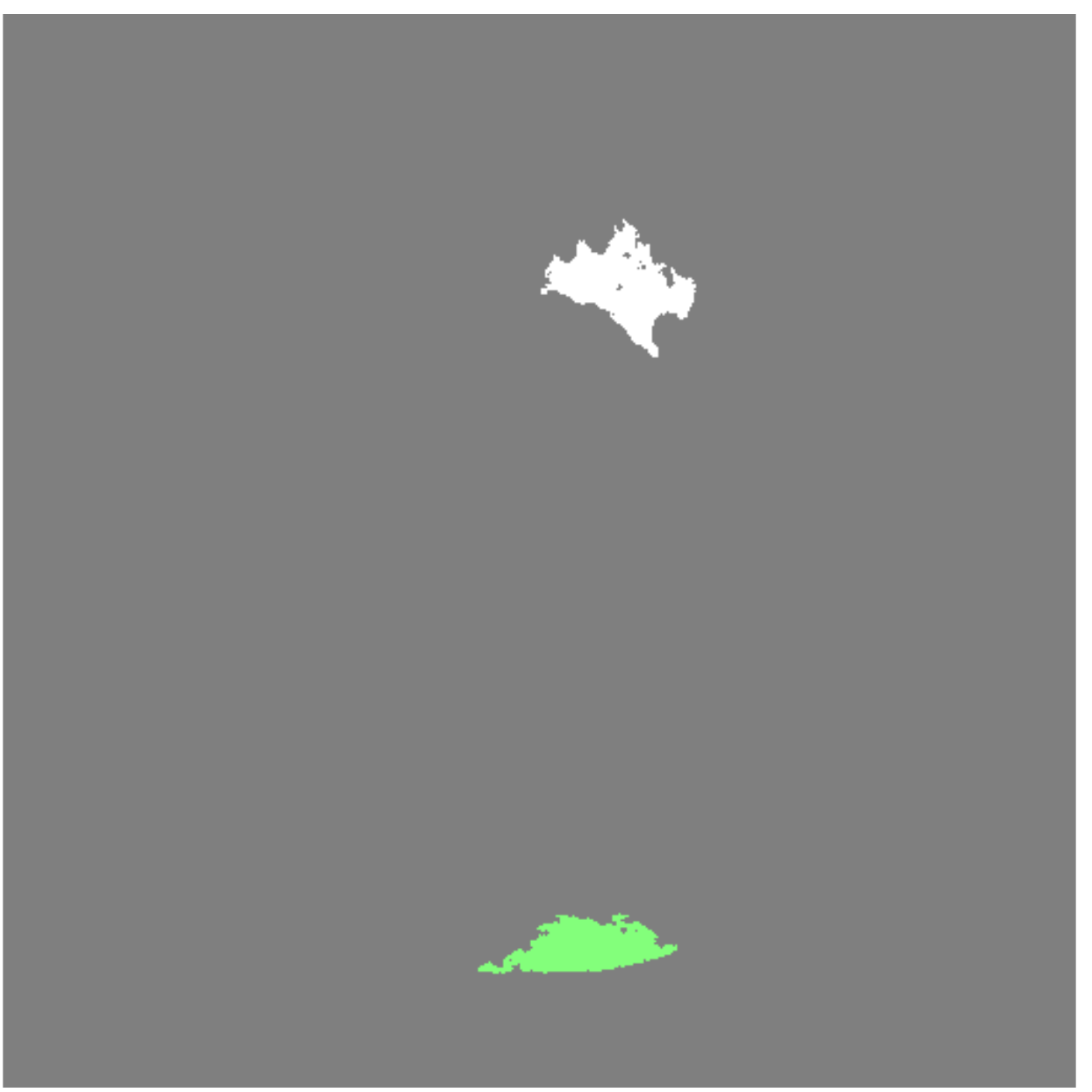}}~
   \subfloat[13 Feb. 2013]{\includegraphics[width=0.13\textwidth]{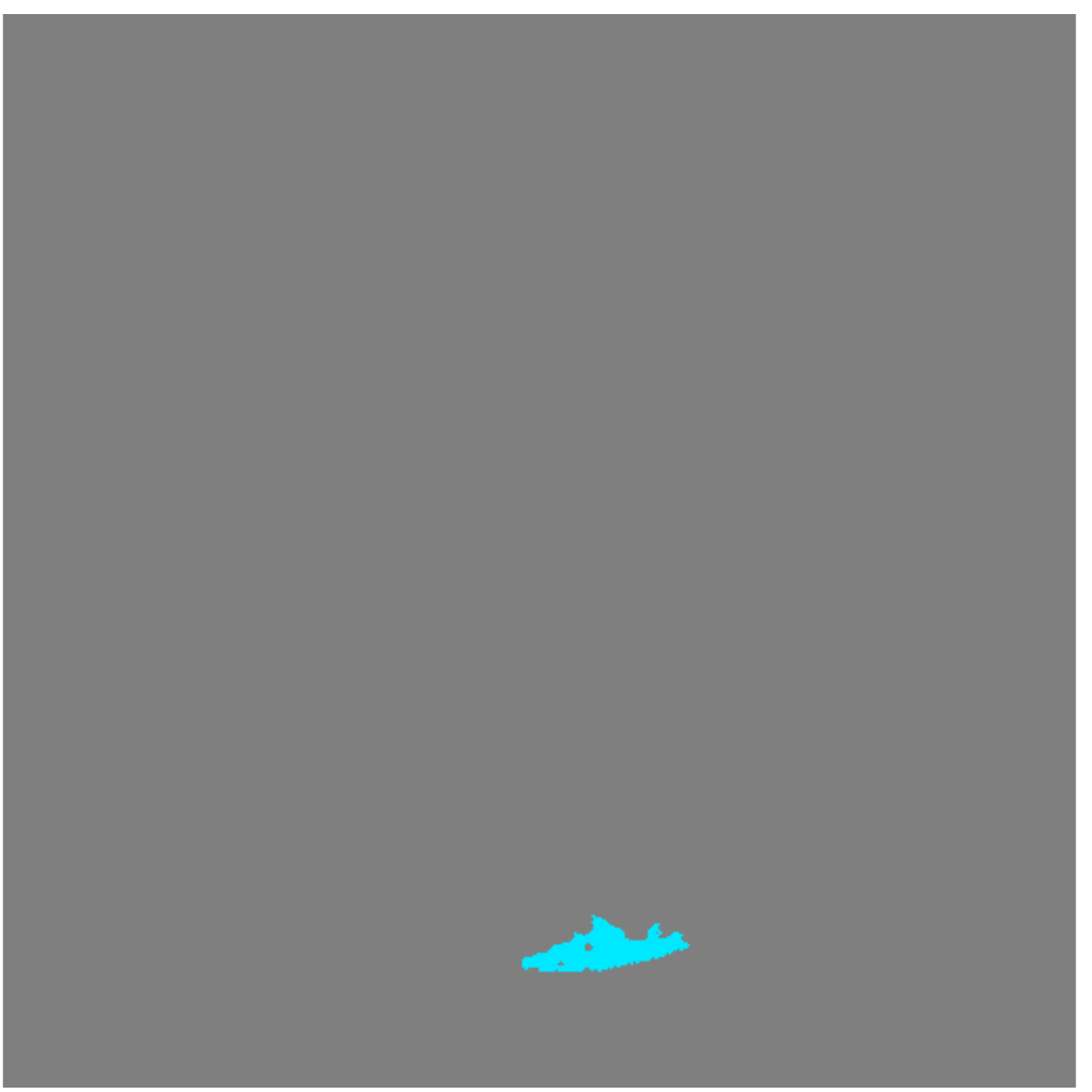}}~
   \subfloat[14 Feb. 2013]{\includegraphics[width=0.13\textwidth]{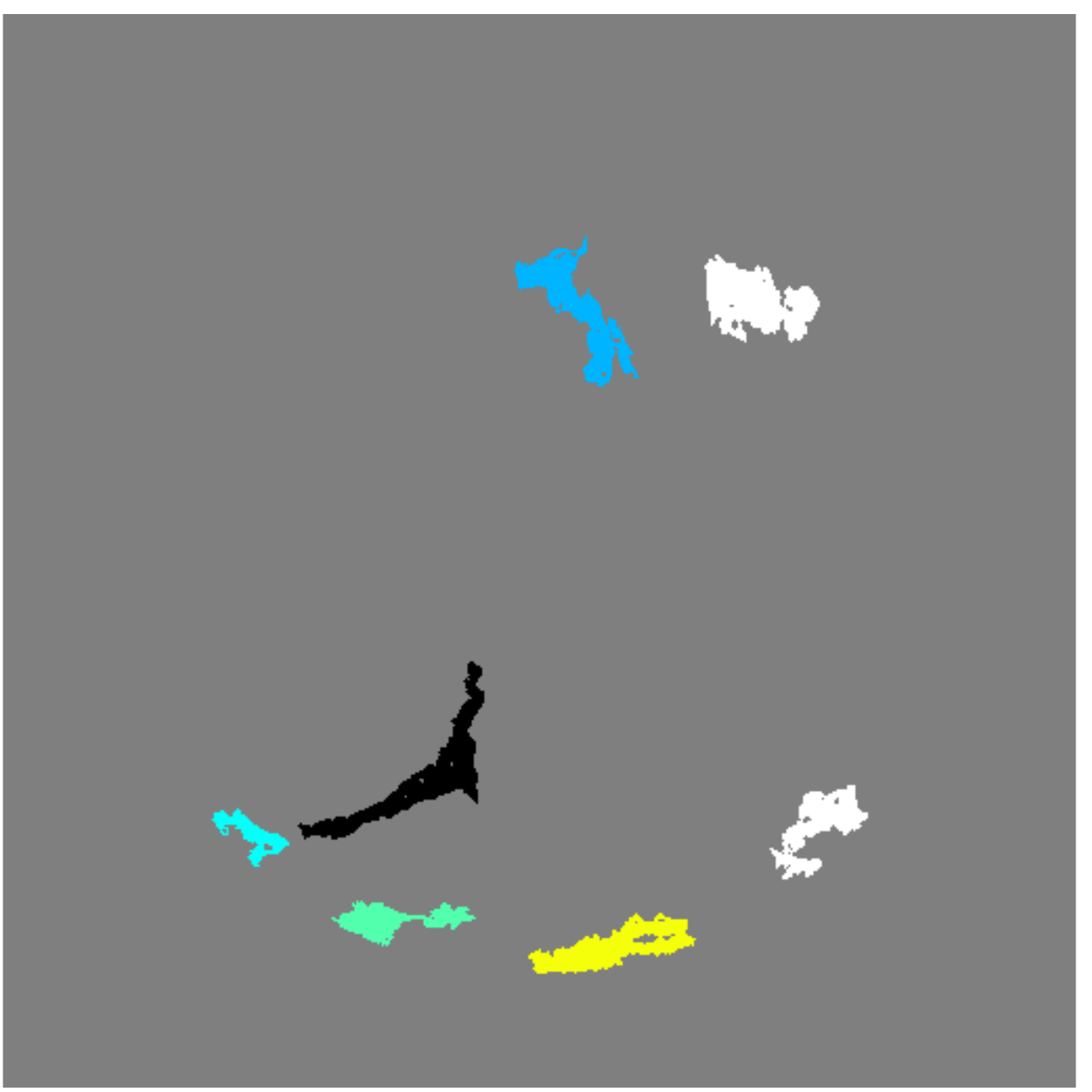}}\\[-2ex]\\
   \subfloat[15 Feb. 2013]{\includegraphics[width=0.13\textwidth]{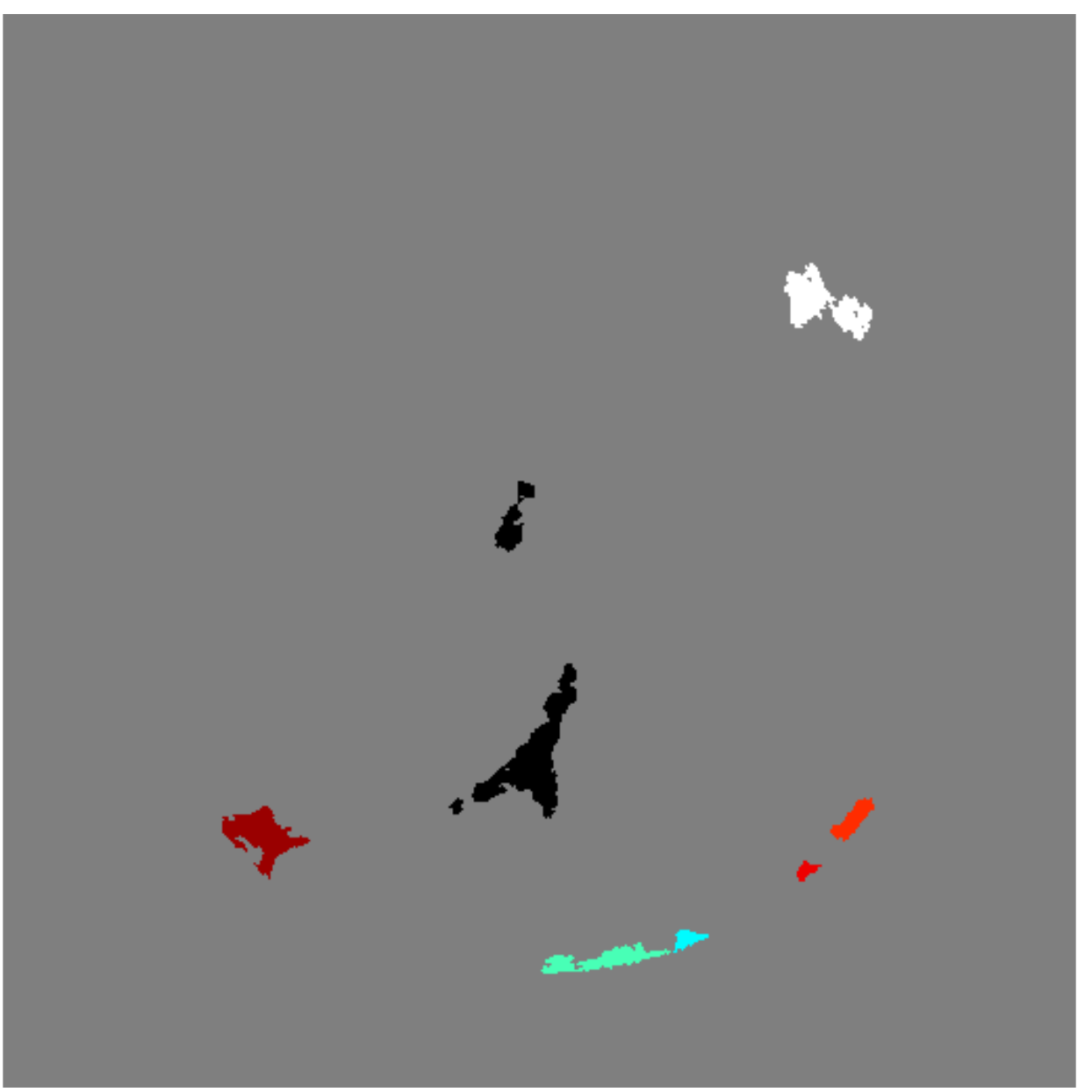}}~
   \subfloat[16 Feb. 2013]{\includegraphics[width=0.13\textwidth]{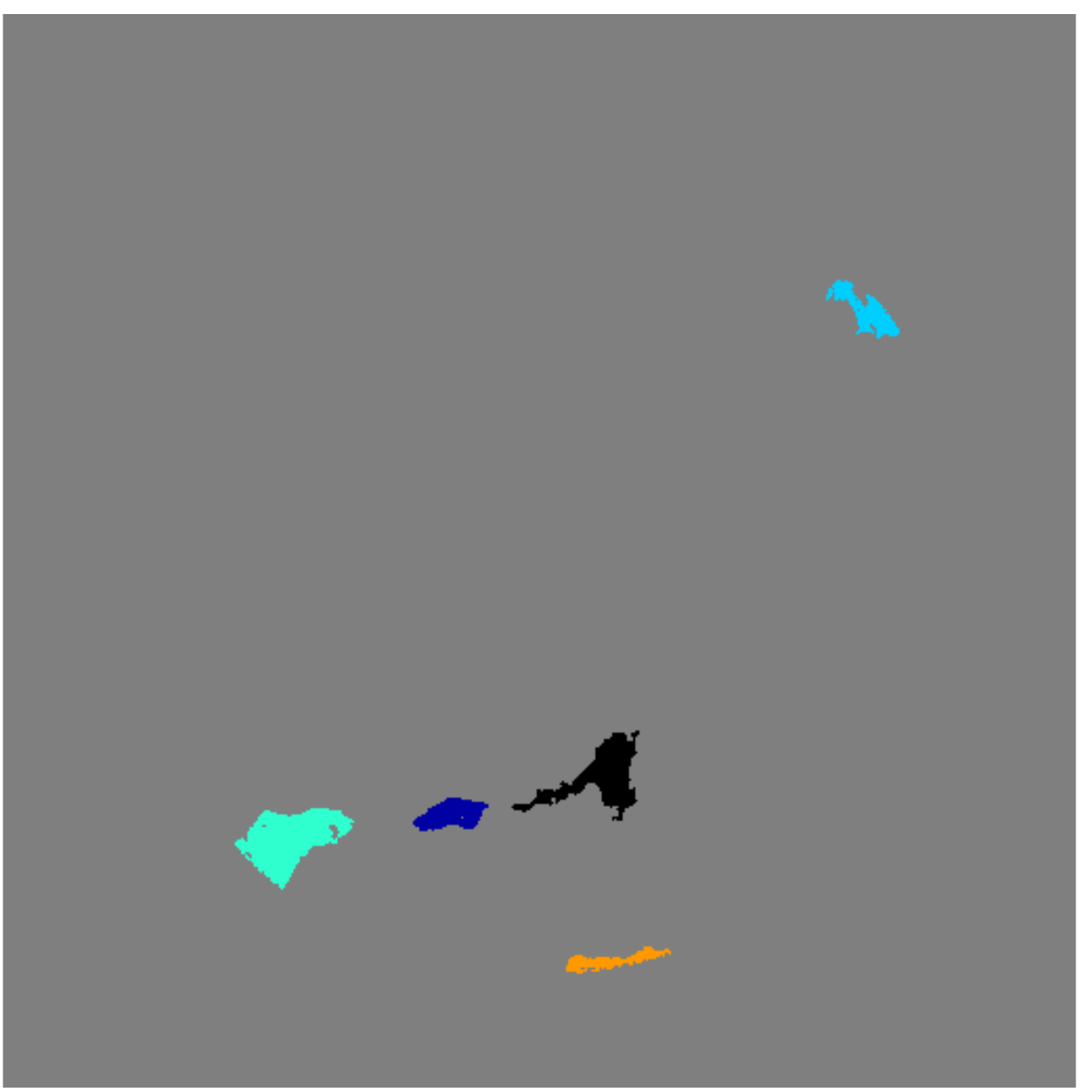}}~
   \subfloat[17 Feb. 2013]{\includegraphics[width=0.13\textwidth]{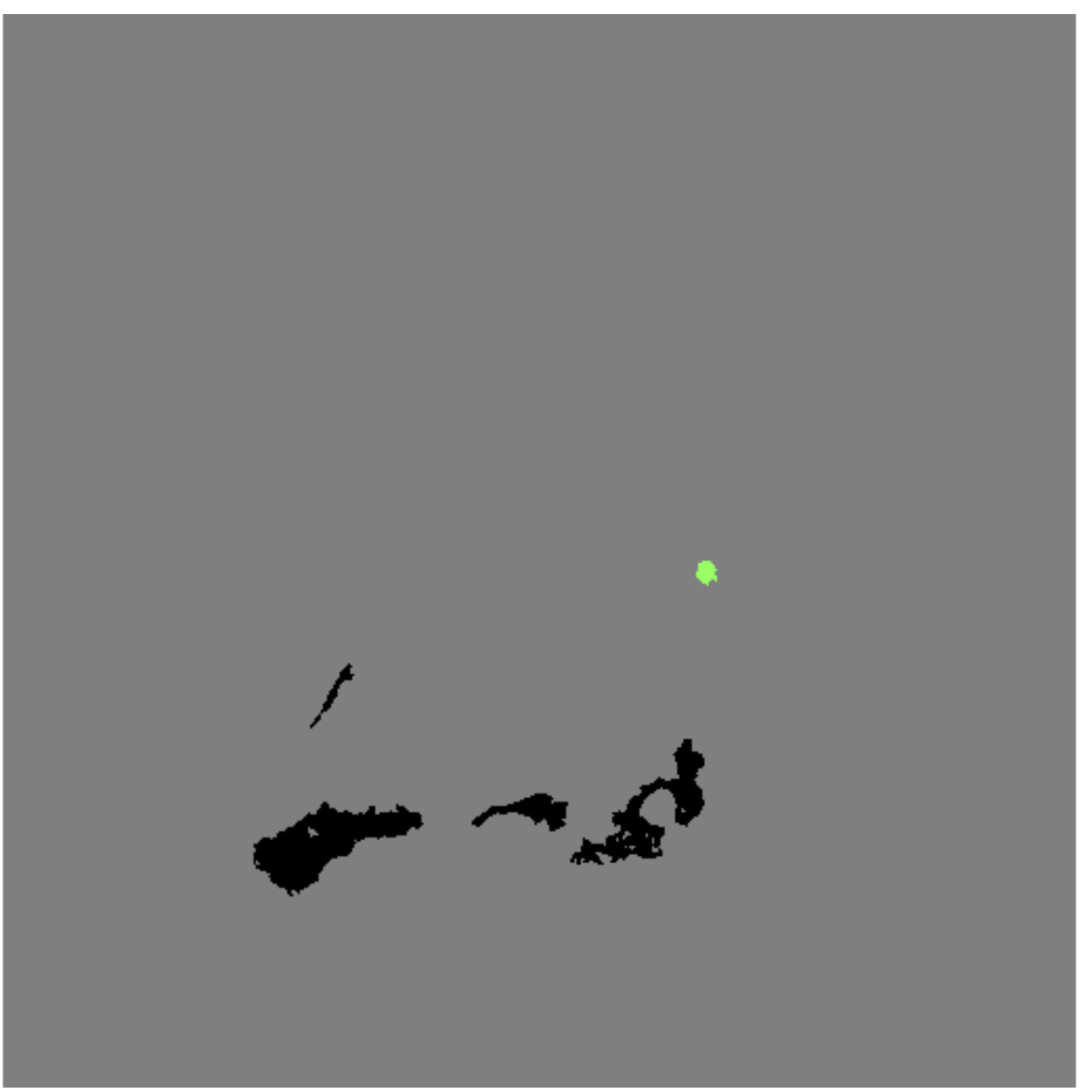}}~
   \subfloat[18 Feb. 2013]{\includegraphics[width=0.13\textwidth]{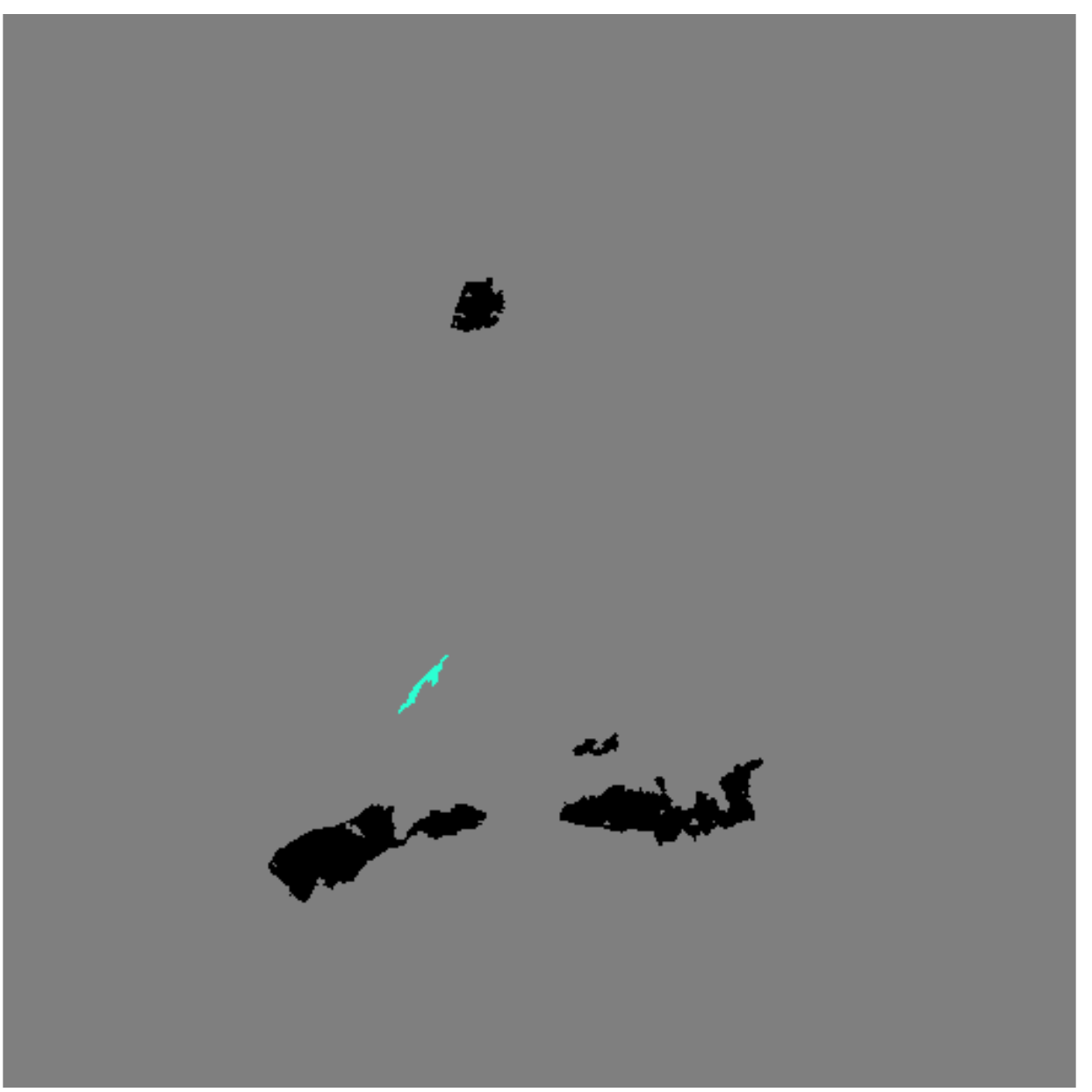}}~
   \subfloat[19 Feb. 2013]{\includegraphics[width=0.13\textwidth]{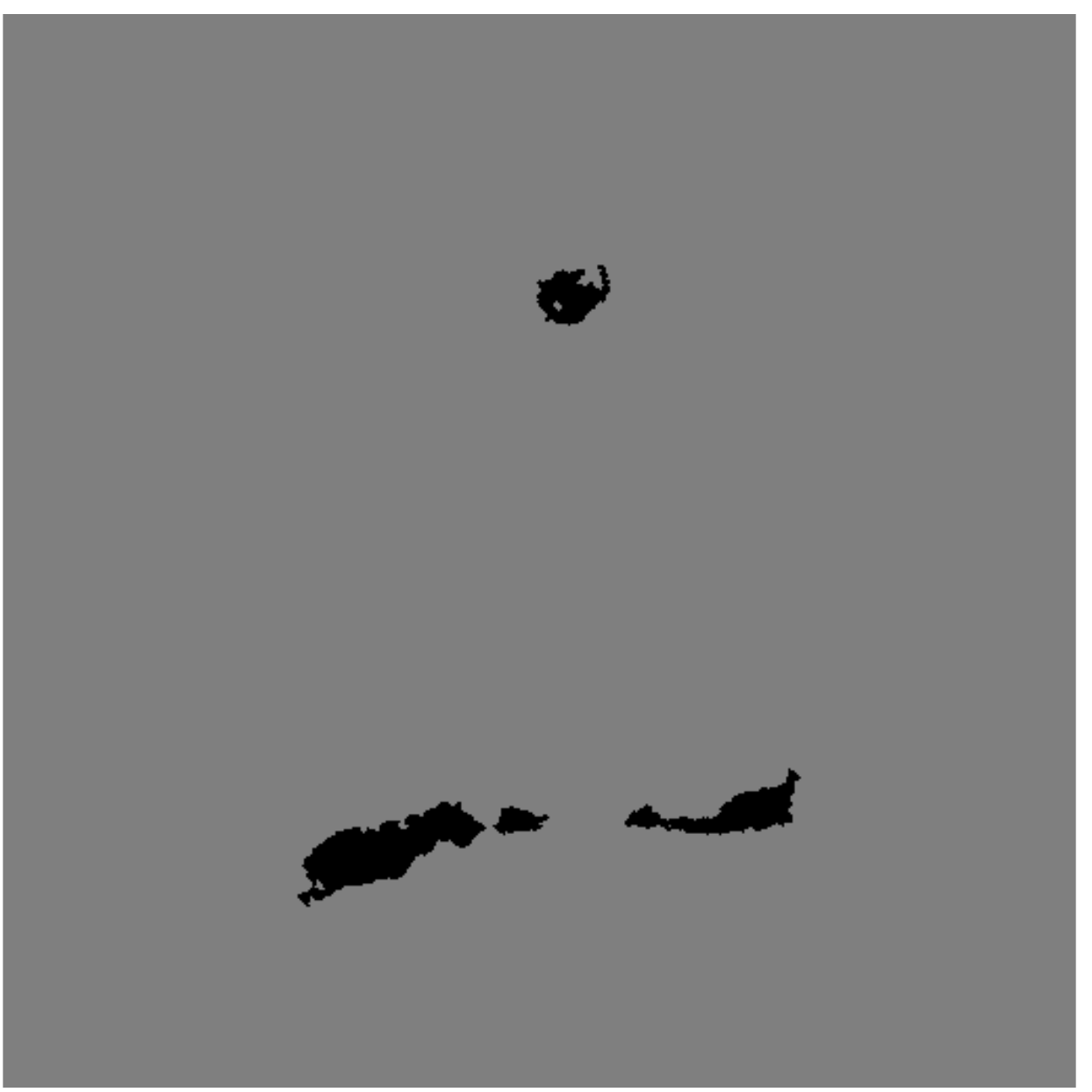}}~
   \subfloat[20 Feb. 2013]{\includegraphics[width=0.13\textwidth]{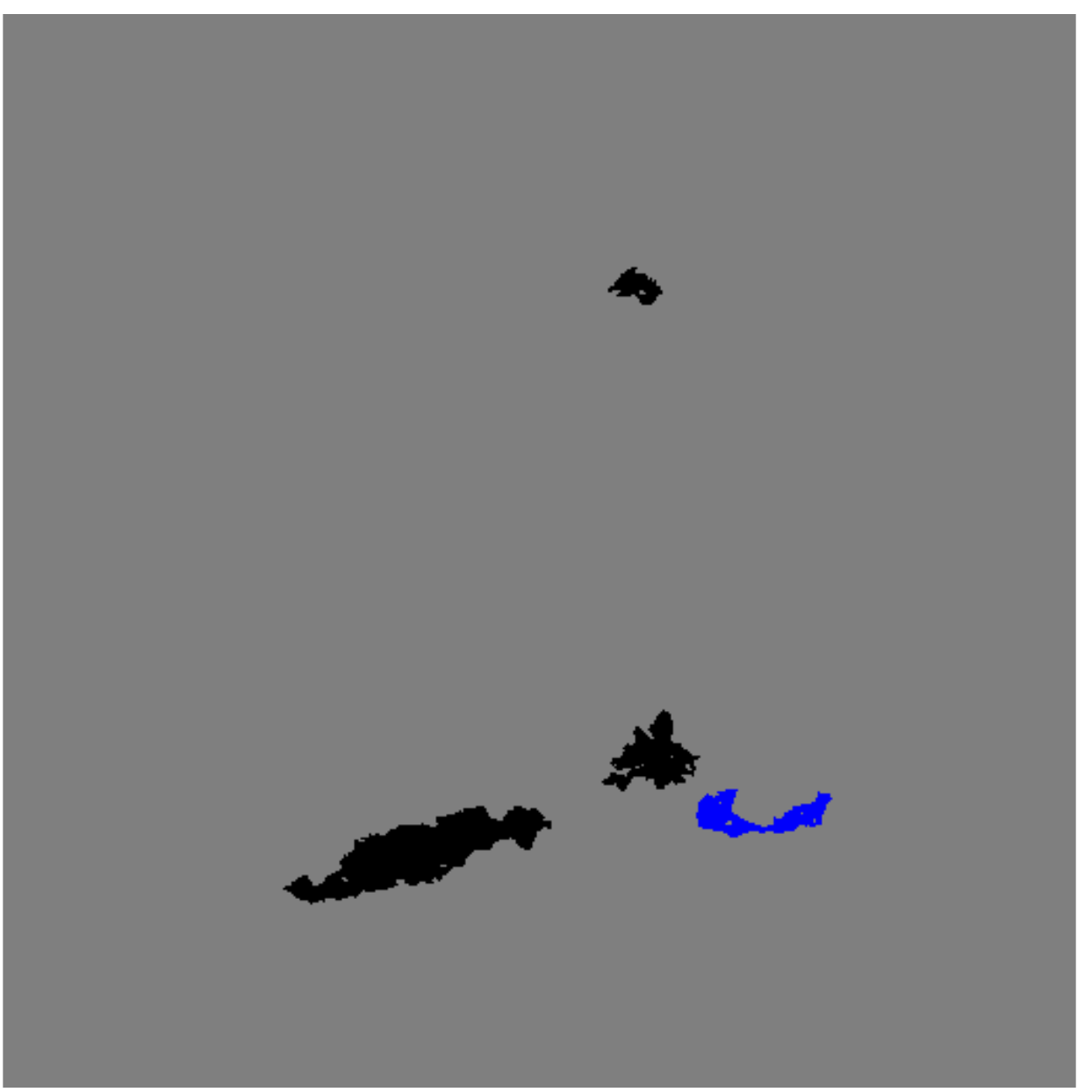}}~
   \subfloat[21 Feb. 2013]{\includegraphics[width=0.13\textwidth]{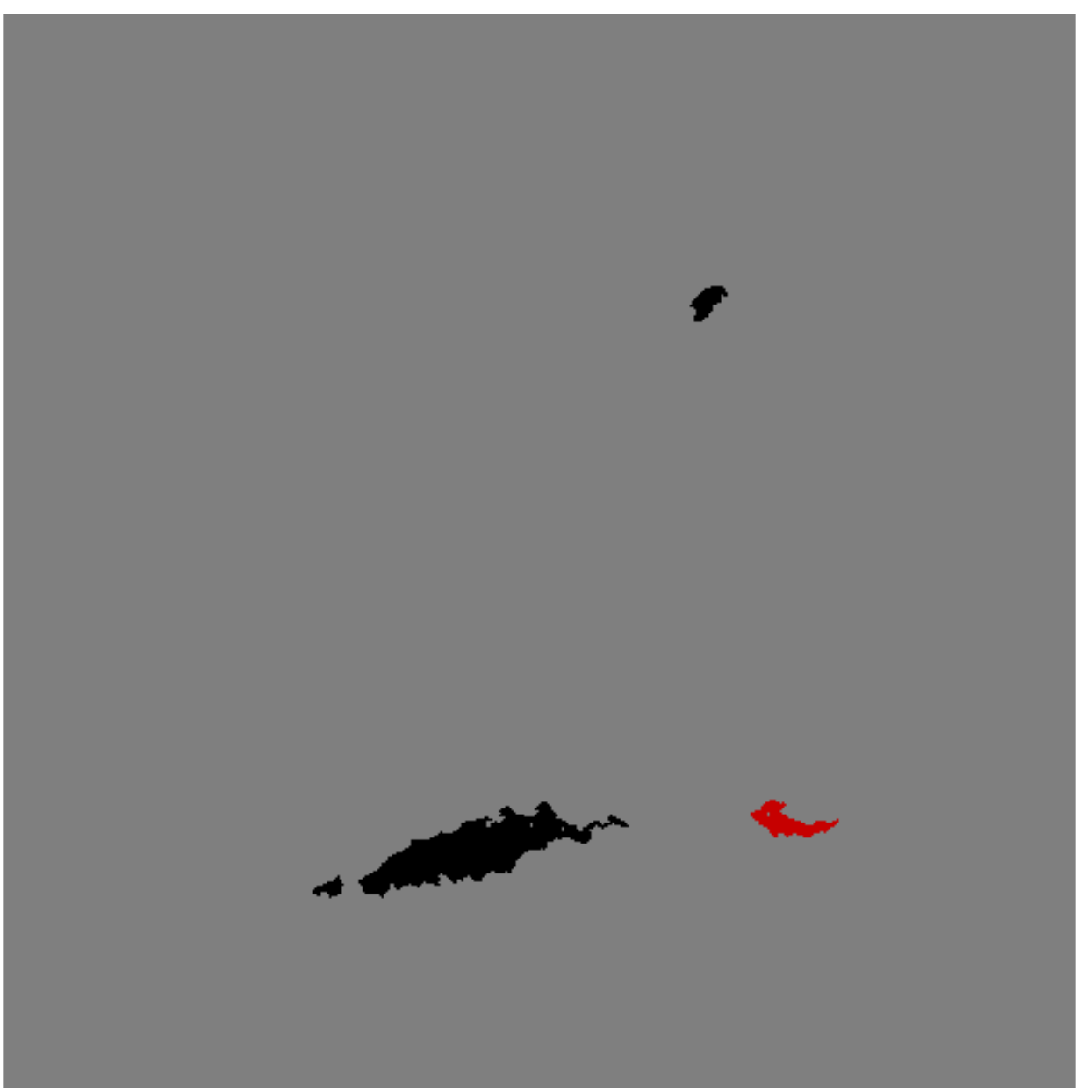}}\\[-2ex]
  \end{minipage}
  \begin{minipage}[r]{0.075\textwidth}
   \subfloat{\includegraphics[width=0.5in]{figures/skewness_colorbar}}
  \end{minipage}
  \caption{Carrington rotation 2133 skewness images for $\lambda_i/\lambda_o=50$ and $\alpha=0.3$.  The colorbar is shown to the right; skewness values $\ge2.5$ are set to white and skewness values $\le-2.5$ are set to black.}
  \label{fig:CR2133_skewness}
\end{sidewaysfigure}

Measurement of the unipolarity of the magnetic field underlying CHs serves as additional evidence in support of an accurately detected CH boundary. It should be noted, however, that while the correspondence between CHs and open magnetic field lines has been confirmed (e.g.,~\citep{krieger1973,wang1996,hassler1999,schwadron2003,antonucci2004}), CHs are observationally defined in terms of diminished EUV intensity.  However, the study of the underlying magnetic field is an important sanity check for CH segmentation, especially as the magnetogram data are relatively independent from the EUV data.  The HMI magnetogram closest to the AIA observation time is overplotted with the ACWE segmented CH boundaries and flux imbalance is measured by skewness of the flux distribution within the CH.  The skewness of the field strength distribution is visualized in Figures~\ref{fig:CR2099_skewness} and~\ref{fig:CR2133_skewness} where CH segmentations are color coded by skewness values.  In these figures, any skewness $\ge2.5$ is clipped to white and any skewness $\le-2.5$ is clipped to black.  

We note a few observations in regards to the skewness images in Figures~\ref{fig:CR2099_skewness} and~\ref{fig:CR2133_skewness}.  First, we see that the majority of the low latitude CHs have absolute skewness values $\ge2.5$.  Second, we note that the polar CHs are more likely to demonstrate an absolute skewness $<2.5$.  We expect that this is largely due to projection effects in the LOS magnetograms near the disk edge.  Third, we note that the majority of northern latitude CHs have a negative skewness and the majority of southern latitude CHs have a positive skewness, consistent with the dominant polarity of the hemispheres.  Fourth, we note that some of the regions with smaller areas and smaller skewness values may belong to filaments or other bipolar structures and could be removed with some combination of morphological or magnetic criteria.  

\subsection{Relation to High Speed Solar Wind}
\captionsetup[subfigure]{labelformat=parens}
\begin{figure}[t]
  \centering
  \subfloat[ACE proton speed.]{\includegraphics[width=0.4\textwidth]{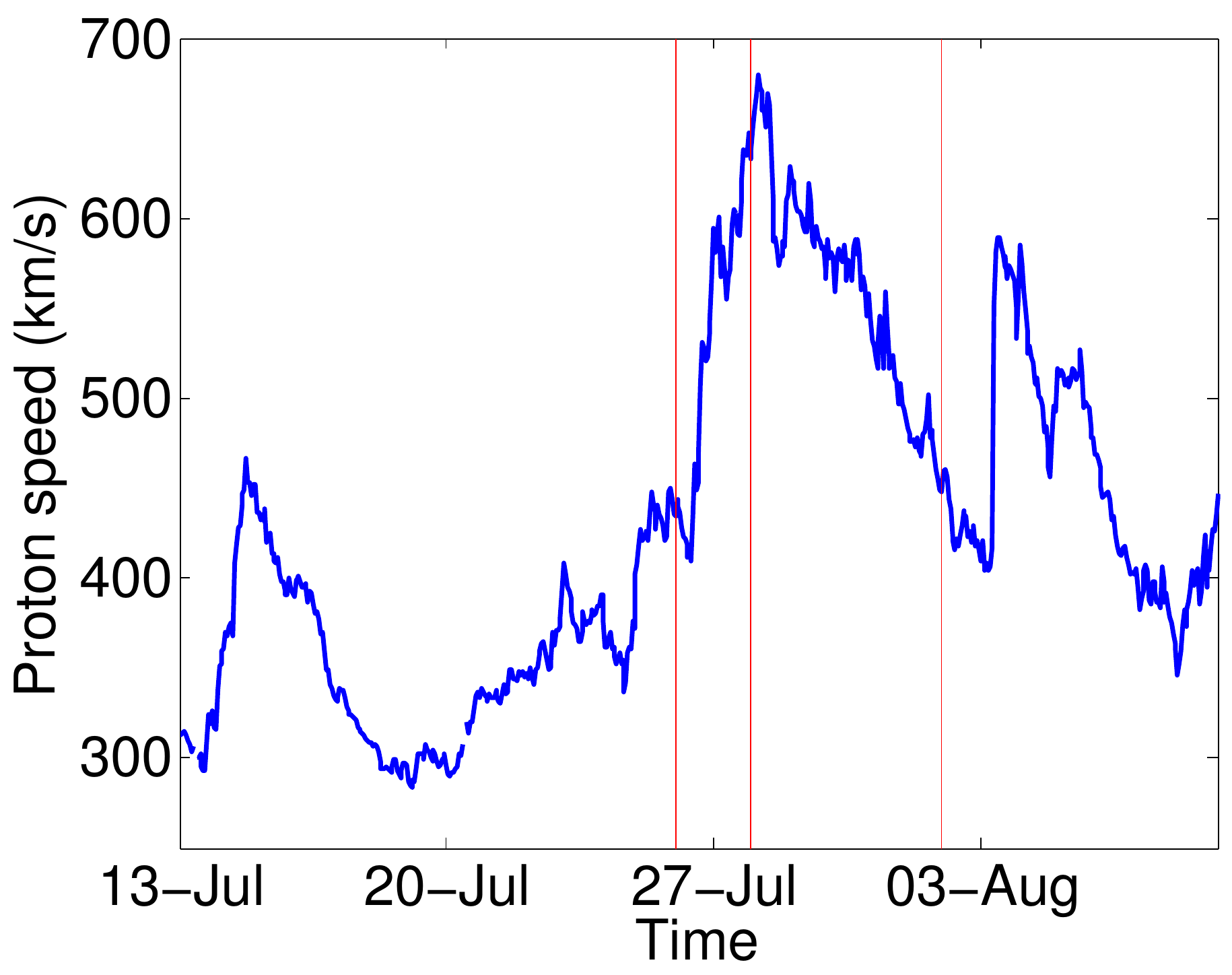}}\\
  \subfloat[ACWE 24 Jul.]{\includegraphics[width=0.2\textwidth]{figures/CR2099_file11_seg}}~~ 
  \subfloat[ACWE 26 Jul.]{\includegraphics[width=0.2\textwidth]{figures/CR2099_file13_seg}}~~ 
  \subfloat[ACWE 31 Jul.]{\includegraphics[width=0.2\textwidth]{figures/CR2099_file18_seg}}\\
  \subfloat[SPoCA 24 Jul.]{\includegraphics[width=0.2\textwidth]{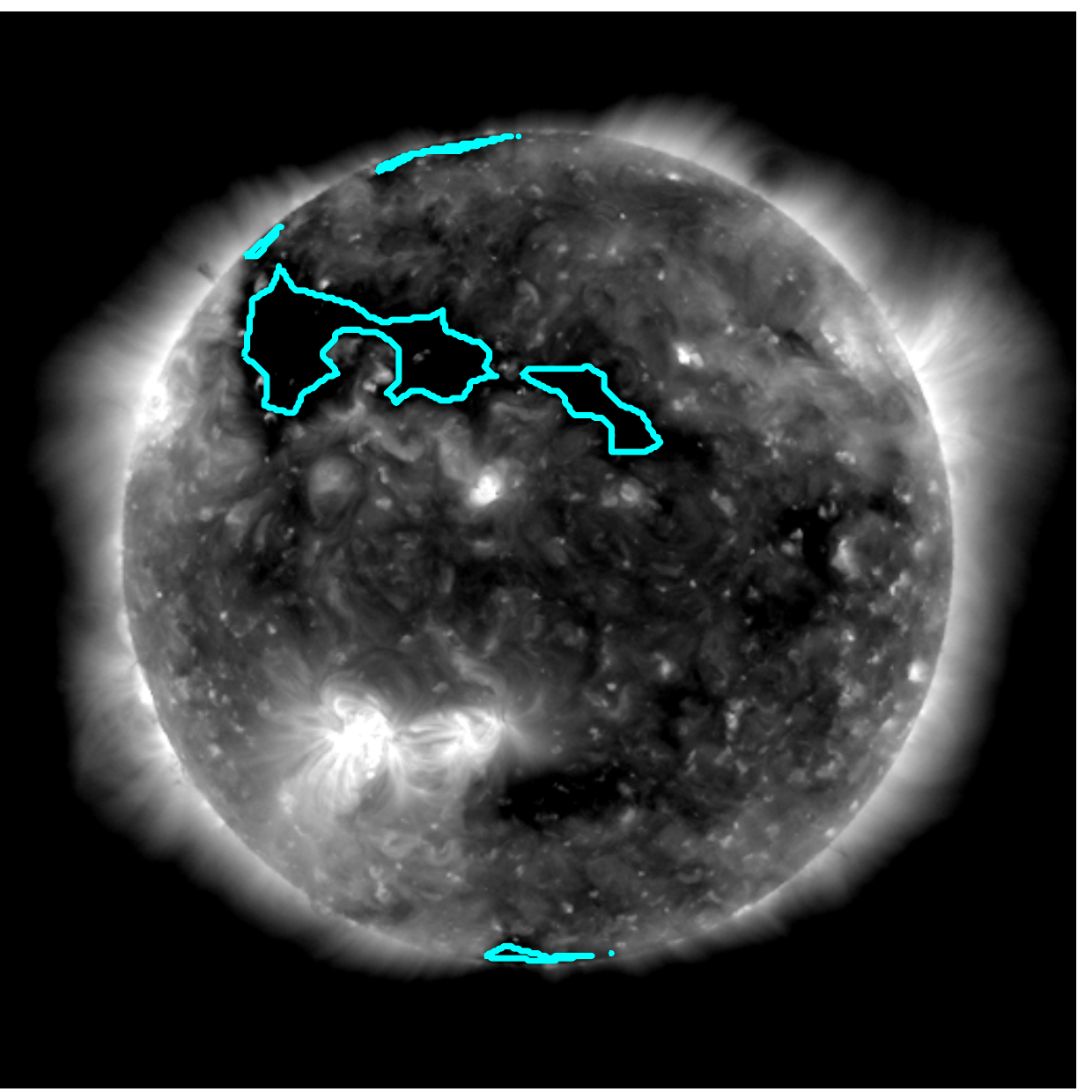}}~~ 
  \subfloat[SPoCA 26 Jul.]{\includegraphics[width=0.2\textwidth]{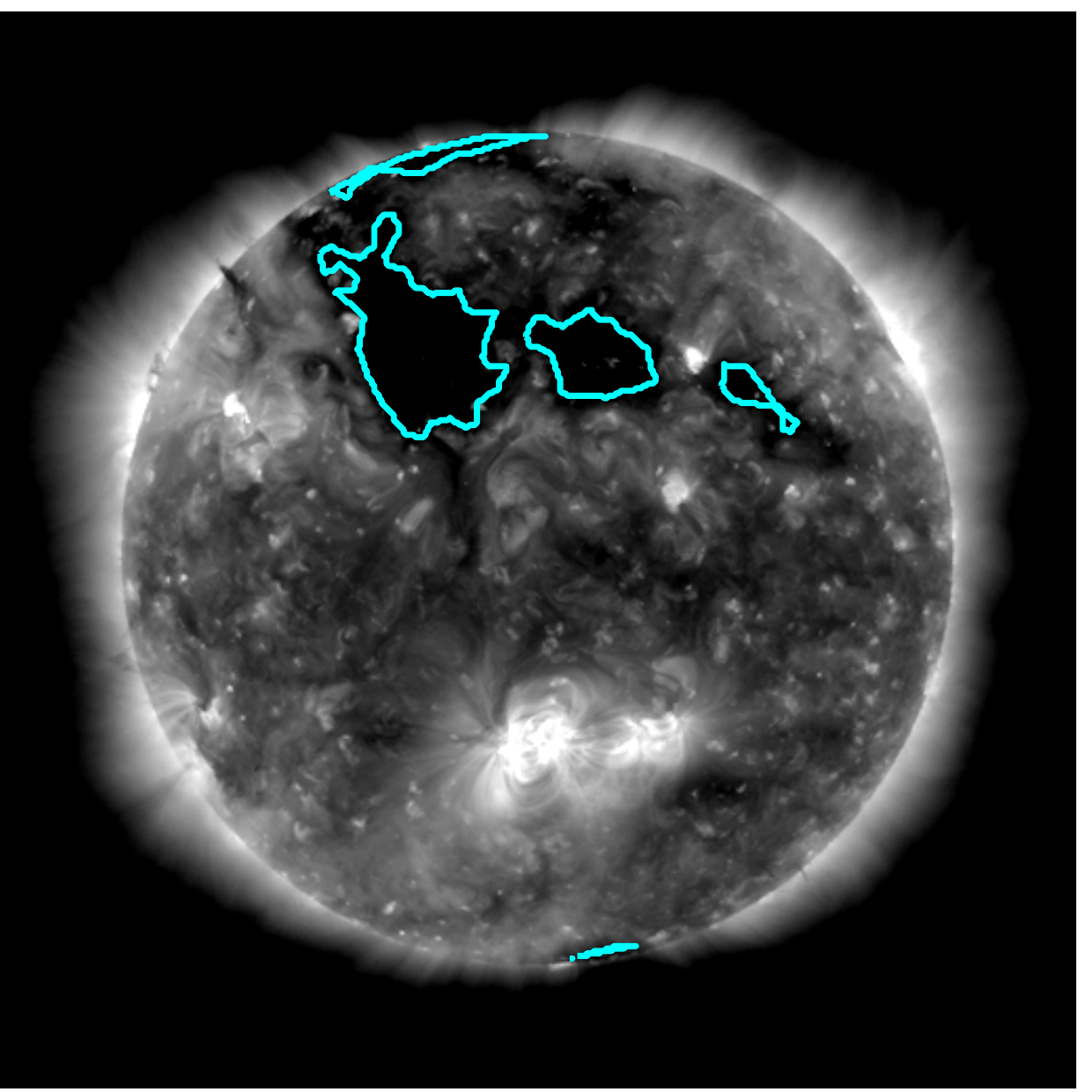}}~~ 
  \subfloat[SPoCA 31 Jul.]{\includegraphics[width=0.2\textwidth]{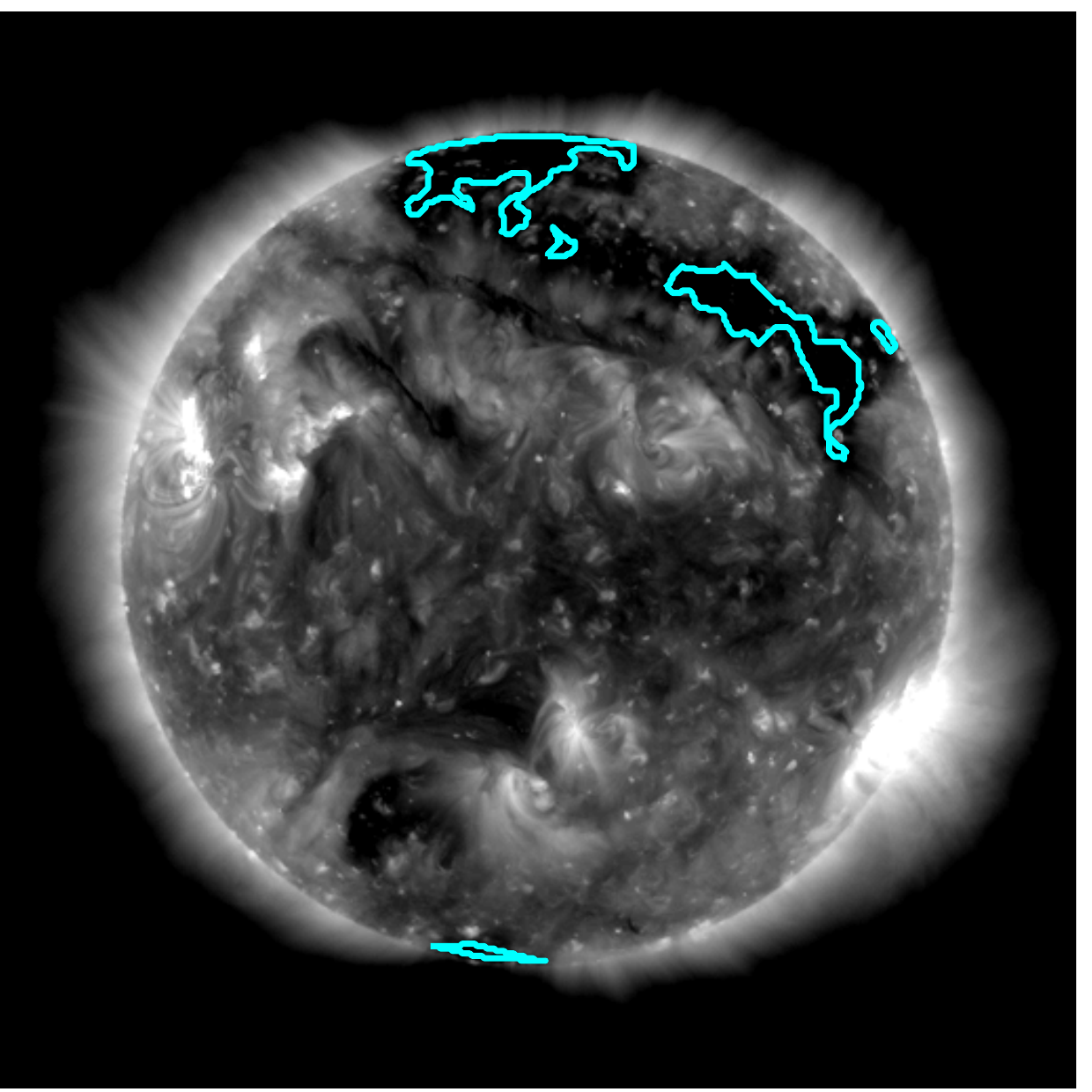}}
  \caption{Relation between (a) solar wind measurements (proton speed from ACE) and (b)-(d) location of CH boundaries corresponding to two days prior to the vertical red lines in (a).}
  \label{fig:ACE_CR2099}
\end{figure}
As coronal holes are sources of high speed solar wind, the segmented CHs are compared with solar wind stream measurements from ACE. Figure~\ref{fig:ACE_CR2099}(a) shows the proton speed measurements for the time period coinciding with CR 2099. The CH segmentations for two days prior to the the beginning (Figure~\ref{fig:ACE_CR2099}(b)), peak (Figure~\ref{fig:ACE_CR2099}(c)), and end (Figure~\ref{fig:ACE_CR2099}(d)) of a period of high-speed solar wind are also provided.  CH locations 2 days prior to the ACE data is based on the assumption of a transit time for high-speed solar wind to earth of approximately 2 days, as also supported by the results in~\citet{krista2009}. We find a large CH is beginning to rotate into the western hemisphere at the beginning of the high-speed solar wind and is rotating off the west limb by the end of the high-speed solar wind.  The comparison to ACE solar wind data is provided here as a sanity check on the likely validity of the detections. However we note that even simple detection \textbf{algorithms} will perform well in such a test. A complete study of the comparison of coronal hole location to solar wind parameters should include a cross-validation of many detection algorithms over many solar rotations.

\subsection{Comparison to the SPoCA Algorithm}
\label{sec:compare}
As a final validation of our CH segmentation method, we qualitatively compare ACWE segmentations to that of the Spatial Possibilistic Clustering Algorithm (SPoCA)~\citep{verbeeck2014} for a few select images.  Shown in Figures~\ref{fig:ACE_CR2099}(e), \ref{fig:ACE_CR2099}(f), and \ref{fig:ACE_CR2099}(g) are the SPoCA CH segmentations for the same days as the ACWE CH segmentations in Figures~\ref{fig:ACE_CR2099}(b), \ref{fig:ACE_CR2099}(c), and \ref{fig:ACE_CR2099}(d).  For all three of these images, we note a good degree of qualitative similarity between the ACWE and SPoCA segmentations.  The SPoCA segmented regions contain fewer small regions and are more convex, most likely due to the morphological erosion and closing applied in postprocessing~\citep{verbeeck2014}. 

While there is a large degree of qualitative similarity between our ACWE segmentations and those of SPoCA, the comparisons here are limited in scope.  In this paper, we have focused on a introduction to the application of ACWE for CH segmentation rather than a comprehensive comparison to other algorithms.  We strongly advocate, however, for a comprehensive and quantitative comparison of CH segmentation algorithms to be conducted in the near future, similar to the goals of the various SHINE workshops\footnote{\url{http://shinecon.org/PastMeetings/Meetings_past.php}}.  Currently, there are several factors which complicate comparison of CH segmentations.  First, ongoing algorithmic modifications for many of the commonly used methods make it difficult for any single investigator to be certain they are using the most up-to-date algorithm.  Second, implementations for algorithms are not \textbf{always} publicly available \textbf{and may not be in a freely-available language.  We do note the availability of the EZSEG algorithm of~\citet{caplan2016}\footnote{\url{http://www.predsci.com/chd/chd_software.html}} and the related magnetic feature tracking algorithm SWAMIS from~\citet{deforest2007}\footnote{\url{http://www.boulder.swri.edu/swamis/}}.  We encourage researchers to make their algorithm code available to the research community and, when possible, to implement algorithms in freely-available languages; we are currently working on a python implementation of the ACWE algorithm described here, which will be available from \url{https://wordpress.nmsu.edu/lboucher/}}.  Third, the segmentation outputs are often presented in different formats (e.g., images~\citep{krista2009} or chain codes~\citep{verbeeck2014}) and coordinates (e.g., equal-area projected~\citep{krista2009,caplan2016} or \textbf{original pixel coordinates~\citep{verbeeck2014}}). Fourth, as there is no accepted ground truth for CH segmentations, it is difficult to define validation metrics with which to compare algorithms.  None-the-less it will be critical in future study of CHs and their effect on the space weather environment to robustly and consistently test existing and future algorithms.

\section{Conclusions and Future Work}
\label{sec:conclusions}
We proposed the application of active contours without edges (ACWE) to the problem of automated CH segmentation.  As opposed to previous methods which rely on threshold based techniques, the method presented here uses image energies based on intensity homogeneity to obtain the final CH boundaries with only a seeding based on thresholding.  ACWE weights were studied for specific application to CH segmentation, and a stopping criterion based on percentage of newly changed pixels was implemented.  The complementary nature of the ACWE algorithm as compared to threshold-based algorithms will facilitate further study of CH properties, including their effect on solar wind and space weather.

The consistency of ACWE segmentation results was studied in three ways.  First, the skewness of the magnetic field underlying the CH was computed as a measure of the unipolarity and the majority of low-latitude CHs were found to have absolute skewness values $\ge2.5$.  Second, location of the segmented CHs were compared to measurements of the high speed solar wind at earth.  We found a correlation between segmented CH boundaries and solar wind proton speed measurements from the ACE satellite.  Third, a qualitative comparison of our segmented CH to the standard SPoCA method~\citep{verbeeck2014} was provided.  

There are several considerations that might lead to a better and more significant automation of CH segmentation in future studies. First, energy terms related to the magnetic field skewness could be incorporated directly into the ACWE algorithm.  Second, a quantitative comparison of our method to other methods motivated by considerations other than intensity homogeneity may provide insight into the potential complementary nature of different algorithms.  Third, while skewness values do not indicate a problem, it would be beneficial to consider the implementation of a method to discriminate between CHs and filaments and to implement other postprocessing consistent with other algorithms.  Fourth, consideration of the consistency of the CH segmentation across higher cadence images may provide insight into parameter selection.

The ultimate scientific question related to this work is the relationship of CH properties to high speed solar wind properties.  In particular, we are interested in the relationship of CH characteristics such as size, shape, 3D structure (by analysis of multiple AIA wavelengths), and temporal evolution to characteristics of the high speed solar wind and their effects on earth.  Future work will consider such relationships.  We strongly advocate for a comprehensive and quantitative analysis and comparison of CH algorithms to be undertaken in the near future.  Such a study should cross-validate multiple automated and manual algorithms as well as link the automated segmentations to solar wind and space weather models.
                                  
\begin{acks}
 The authors gratefully acknowledge NASA PAARE grant AST-0849986, NASA EPSCoR grant NNX09AP76A, and NSF CAREER grant 1255024 which helped support this work.  The authors also thank Michael Kirk for productive discussions regarding this work and comments on the manuscript.
\end{acks}


  
\bibliographystyle{spr-mp-sola}

\tracingmacros=2
\bibliography{main}  

\begin{thebibliography}{30}
\ifx\bisbn     \undefined \def\bisbn  #1{ISBN #1}\fi
\ifx\binits    \undefined \def\binits#1{#1}\fi
\ifx\bauthor   \undefined \def\bauthor#1{#1}\fi
\ifx\batitle   \undefined \def\batitle#1{#1}\fi
\ifx\bjtitle   \undefined \def\bjtitle#1{\textit{#1}}\fi
\ifx\bvolume   \undefined \def\bvolume#1{\textbf{#1}}\fi
\ifx\byear     \undefined \def\byear#1{#1}\fi
\ifx\bissue    \undefined \def\bissue#1{#1}\fi
\ifx\bfpage    \undefined \def\bfpage#1{#1}\fi
\ifx\blpage    \undefined \def\blpage #1{#1}\fi
\ifx\burl      \undefined \def\burl#1{\textsf{#1}}\fi
\ifx\href      \undefined \def\href#1#2{\textsf{#2}}\fi
\ifx\betal     \undefined \def\betal{\textit{et al.}}\fi
\ifx\bctitle   \undefined \def\bctitle#1{#1}\fi
\ifx\beditor   \undefined \def\beditor#1{#1}\fi
\ifx\bbtitle   \undefined \def\bbtitle#1{\textit{#1}}\fi
\ifx\bedition  \undefined \def\bedition#1{#1}\fi
\ifx\bseriesno \undefined \def\bseriesno#1{\textbf{#1}}\fi
\ifx\blocation \undefined \def\blocation#1{#1}\fi
\ifx\bsertitle \undefined \def\bsertitle#1{\textit{#1}}\fi
\ifx\bsnm      \undefined \def\bsnm#1{#1}\fi
\ifx\bsuffix   \undefined \def\bsuffix#1{#1}\fi
\ifx\bparticle \undefined \def\bparticle#1{#1}\fi
\ifx\barticle  \undefined \def\barticle#1{}\fi
\ifx\binstitute  \undefined \def\binstitute#1{#1}\fi
\ifx\bpublisher  \undefined \def\bpublisher#1{#1}\fi
\ifx\doiurl    \undefined
  \def\doiurl#1{\href{http://dx.doi.org/#1}{\textsf{DOI}}}\fi
\ifx\arxivurl  \undefined
  \def\arxivurl#1{\href{http://arxiv.org/abs/#1}{\textsf{arXiv}}}\fi
\ifx\adsurl    \undefined
  \def\adsurl#1{\href{http://adsabs.harvard.edu/abs/#1}{\textsf{ADS}}}\fi
\ifx\botherref \undefined \def\botherref#1{}\fi
\ifx\url       \undefined \def\url#1{\textsf{#1}}\fi
\ifx\bchapter  \undefined \def\bchapter#1{}\fi
\ifx\bbook     \undefined \def\bbook#1{}\fi
\ifx\bcomment  \undefined \def\bcomment#1{#1}\fi
\ifx\oauthor   \undefined \def\oauthor#1{#1}\fi
\ifx\citeauthoryear \undefined\def \citeauthoryear#1{#1}\fi
\def\endbibitem {}
\ifx\bconflocation  \undefined \def\bconflocation#1{#1} \fi

\bibitem[\protect\citeauthoryear{Altschuler, Trotter, and
  Orrall}{1972}]{altschuler1972}
\begin{barticle}
\bauthor{\bsnm{Altschuler}, \binits{M.D.}},
\bauthor{\bsnm{Trotter}, \binits{D.E.}},
\bauthor{\bsnm{Orrall}, \binits{F.Q.}}:
\byear{1972},
\batitle{Coronal holes}.
\bjtitle{\solphys}
\bvolume{26}(\bissue{2}),
\bfpage{354}.
\doiurl{10.1007/BF00165276}.
\end{barticle}
\endbibitem

\bibitem[\protect\citeauthoryear{Antonucci
  \textit{et~al.}}{2004}]{antonucci2004}
\begin{barticle}
\bauthor{\bsnm{Antonucci}, \binits{E.}},
\bauthor{\bsnm{Dodero}, \binits{M.A.}},
\bauthor{\bsnm{Giordano}, \binits{S.}},
\bauthor{\bsnm{Krishnakumar}, \binits{V.}},
\bauthor{\bsnm{Noci}, \binits{G.}}:
\byear{2004},
\batitle{Spectroscopic measurement of the plasma electron density and outflow
  velocity in a polar coronal hole}.
\bjtitle{\aap}
\bvolume{416}(\bissue{2}),
\bfpage{749}.
\doiurl{10.1051/0004-6361:20031650}.
\end{barticle}
\endbibitem

\bibitem[\protect\citeauthoryear{Caplan, Downs, and Linker}{2016}]{caplan2016}
\begin{barticle}
\bauthor{\bsnm{Caplan}, \binits{R.M.}},
\bauthor{\bsnm{Downs}, \binits{C.}},
\bauthor{\bsnm{Linker}, \binits{J.A.}}:
\byear{2016},
\batitle{Sychronic coronal hole mapping using multi-instrument {EUV} images:
  {D}ata preparation and detection method}.
\bjtitle{\apj}
\bvolume{823},
\bfpage{53}.
\doiurl{10.3847/0004/637X/823/1/53}.
\end{barticle}
\endbibitem

\bibitem[\protect\citeauthoryear{Chan and Vese}{2001}]{chan2001}
\begin{barticle}
\bauthor{\bsnm{Chan}, \binits{T.F.}},
\bauthor{\bsnm{Vese}, \binits{L.A.}}:
\byear{2001},
\batitle{Active contours without edges}.
\bjtitle{IEEE Transactions on Image Processing}
\bvolume{10}(\bissue{2}),
\bfpage{266}.
\doiurl{10.1109/83.902291}.
\end{barticle}
\endbibitem

\bibitem[\protect\citeauthoryear{Chiu \textit{et~al.}}{1998}]{chiu1998}
\begin{barticle}
\bauthor{\bsnm{Chiu}, \binits{M.C.}},
\bauthor{\bparticle{nd} \bsnm{C.~E.~Willey}, \binits{U.I.V.-M.}},
\bauthor{\bsnm{Betenbaugh}, \binits{T.M.}},
\bauthor{\bsnm{Maynard}, \binits{J.J.}},
\bauthor{\bsnm{Krein}, \binits{J.A.}},
\bauthor{\bsnm{Conde}, \binits{R.F.}},
\bauthor{\bsnm{Gray}, \binits{W.T.}},
\bauthor{\bsnm{{Hunt, Jr.}}, \binits{J.W.}},
\bauthor{\bsnm{Mosher}, \binits{L.E.}},
\bauthor{\bsnm{McCullough}, \binits{M.G.}},
\bauthor{\bsnm{Panneton}, \binits{P.E.}},
\bauthor{\bsnm{Staiger}, \binits{J.P.}},
\bauthor{\bsnm{Rodberg}, \binits{E.H.}}:
\byear{1998},
\batitle{{ACE} spacecraft}.
\bjtitle{\ssr}
\bvolume{86}(\bissue{1}),
\bfpage{257}.
\doiurl{10.1023/A:1005002013459}.
\end{barticle}
\endbibitem

\bibitem[\protect\citeauthoryear{Colak and Qahwaji}{2013}]{colak2013}
\begin{barticle}
\bauthor{\bsnm{Colak}, \binits{T.}},
\bauthor{\bsnm{Qahwaji}, \binits{R.}}:
\byear{2013},
\batitle{Prediction of \emph{{E}xtreme {U}ltraviolet {V}ariability
  {E}xperiment} ({EVE})/\emph{{E}xtreme {U}ltraviolet {S}pectro-{P}hotometer}
  ({ESP}) irradiance from \emph{{S}olar {D}ynamics {O}bservatory}
  ({SDO})/\emph{{A}tmospheric {I}maging {A}ssembly} ({AIA}) images using fuzzy
  image processing and machine learning}.
\bjtitle{\solphys}
\bvolume{283}(\bissue{1}),
\bfpage{143}.
\doiurl{10.1007/s11207-011-9880-9}.
\end{barticle}
\endbibitem

\bibitem[\protect\citeauthoryear{{de Toma}}{2011}]{detoma2011}
\begin{barticle}
\bauthor{\bsnm{{de Toma}}, \binits{G.}}:
\byear{2011},
\batitle{Evolution of coronal holes and implications for high-speed solar wind
  during the minimum between cycles 23 and 24}.
\bjtitle{\solphys}
\bvolume{274}(\bissue{1--2}),
\bfpage{195}.
\doiurl{10.1007/s11207-010-9677-2}.
\end{barticle}
\endbibitem

\bibitem[\protect\citeauthoryear{DeForest \textit{et~al.}}{2007}]{deforest2007}
\begin{barticle}
\bauthor{\bsnm{DeForest}, \binits{C.E.}},
\bauthor{\bsnm{Hagenaar}, \binits{H.J.}},
\bauthor{\bsnm{Lamb}, \binits{D.A.}},
\bauthor{\bsnm{Parnell}, \binits{C.E.}},
\bauthor{\bsnm{Welsch}, \binits{B.T.}}:
\byear{2007},
\batitle{Solar magnetic tracking. {I}. {S}oftware comparison and recommended
  practices}.
\bjtitle{\apj}
\bvolume{666}(\bissue{1}),
\bfpage{576}.
\doiurl{10.1086/518994}.
\end{barticle}
\endbibitem

\bibitem[\protect\citeauthoryear{{Dudok de Wit}}{2006}]{dudokdewit2006}
\begin{barticle}
\bauthor{\bsnm{{Dudok de Wit}}, \binits{T.}}:
\byear{2006},
\batitle{Fast segmentation of solar extreme ultraviolet images}.
\bjtitle{\solphys}
\bvolume{239}(\bissue{1--2}),
\bfpage{519}.
\doiurl{10.1007/s11207-006-0140-3}.
\end{barticle}
\endbibitem

\bibitem[\protect\citeauthoryear{Gopalswamy
  \textit{et~al.}}{2009}]{gopalswamy2009}
\begin{barticle}
\bauthor{\bsnm{Gopalswamy}, \binits{N.}},
\bauthor{\bsnm{M\"{a}kel\"{a}}, \binits{P.}},
\bauthor{\bsnm{Xie}, \binits{H.}},
\bauthor{\bsnm{Akiyama}, \binits{S.}},
\bauthor{\bsnm{Yashiro}, \binits{S.}}:
\byear{2009},
\batitle{{CME} interactions with coronal holes and their interplanetary
  consequences}.
\bjtitle{\jgr}
\bvolume{114}(\bissue{A3}),
\bfpage{A00A22}.
\doiurl{10.1029/2008JA013686}.
\end{barticle}
\endbibitem

\bibitem[\protect\citeauthoryear{Harvey and Recely}{2002}]{harvey2002}
\begin{barticle}
\bauthor{\bsnm{Harvey}, \binits{K.L.}},
\bauthor{\bsnm{Recely}, \binits{F.}}:
\byear{2002},
\batitle{Polar coronal holes during cycles 22 and 23}.
\bjtitle{\solphys}
\bvolume{211}(\bissue{1--2}),
\bfpage{31}.
\doiurl{10.1023/A:1022469023581}.
\end{barticle}
\endbibitem

\bibitem[\protect\citeauthoryear{Hassler \textit{et~al.}}{1999}]{hassler1999}
\begin{barticle}
\bauthor{\bsnm{Hassler}, \binits{D.M.}},
\bauthor{\bsnm{Dammasch}, \binits{I.E.}},
\bauthor{\bsnm{Lemaire}, \binits{P.}},
\bauthor{\bsnm{Brekke}, \binits{P.}},
\bauthor{\bsnm{Curdt}, \binits{W.}},
\bauthor{\bsnm{Mason}, \binits{H.E.}},
\bauthor{\bsnm{Vial}, \binits{J.-C.}},
\bauthor{\bsnm{Wilhelm}, \binits{K.}}:
\byear{1999},
\batitle{Solar wind outflow and the chromospheric magnetic network}.
\bjtitle{Science}
\bvolume{283}(\bissue{5403}),
\bfpage{810}.
\doiurl{10.1126/science.283.5403.810}.
\end{barticle}
\endbibitem

\bibitem[\protect\citeauthoryear{Henney and Harvey}{2005}]{henney2005}
\begin{bchapter}
\bauthor{\bsnm{Henney}, \binits{C.J.}},
\bauthor{\bsnm{Harvey}, \binits{J.W.}}:
\byear{2005},
\bctitle{Automated coronal hole detection using {He I 1083 nm}
  spectroheliograms and photospheric magnetograms}.
In: \bbtitle{ASP Conf Series}
\bseriesno{346},
\bfpage{261}.
\end{bchapter}
\endbibitem

\bibitem[\protect\citeauthoryear{Kass, Witkin, and
  Terzopoulos}{1988}]{kass1988}
\begin{barticle}
\bauthor{\bsnm{Kass}, \binits{M.}},
\bauthor{\bsnm{Witkin}, \binits{A.}},
\bauthor{\bsnm{Terzopoulos}, \binits{D.}}:
\byear{1988},
\batitle{Snakes: Active contour models}.
\bjtitle{International Journal of Computer Vision}
\bvolume{1}(\bissue{4}),
\bfpage{321}.
\doiurl{10.1007/BF00133570}.
\end{barticle}
\endbibitem

\bibitem[\protect\citeauthoryear{Kirk \textit{et~al.}}{2009}]{kirk2009}
\begin{barticle}
\bauthor{\bsnm{Kirk}, \binits{M.S.}},
\bauthor{\bsnm{Pesnell}, \binits{W.D.}},
\bauthor{\bsnm{Young}, \binits{C.A.}},
\bauthor{\bsnm{{Hess Webber}}, \binits{S.A.}}:
\byear{2009},
\batitle{Automated detection of {EUV} polar coronal holes during solar cycle
  23}.
\bjtitle{\solphys}
\bvolume{257}(\bissue{1}),
\bfpage{99}.
\doiurl{10.1007/s11207-009-9369-y}.
\end{barticle}
\endbibitem

\bibitem[\protect\citeauthoryear{Krieger, Timothy, and
  Roelof}{1973}]{krieger1973}
\begin{barticle}
\bauthor{\bsnm{Krieger}, \binits{A.S.}},
\bauthor{\bsnm{Timothy}, \binits{A.F.}},
\bauthor{\bsnm{Roelof}, \binits{E.C.}}:
\byear{1973},
\batitle{A coronal hole and its identification as the source of a high velocity
  solar wind stream}.
\bjtitle{\solphys}
\bvolume{29}(\bissue{2}),
\bfpage{505}.
\doiurl{10.1007/BF00150828}.
\end{barticle}
\endbibitem

\bibitem[\protect\citeauthoryear{Krista and Gallagher}{2009}]{krista2009}
\begin{barticle}
\bauthor{\bsnm{Krista}, \binits{L.D.}},
\bauthor{\bsnm{Gallagher}, \binits{P.T.}}:
\byear{2009},
\batitle{Automated coronal hole detection using local intensity thresholding
  techniques}.
\bjtitle{\solphys}
\bvolume{256}(\bissue{1--2}),
\bfpage{87}.
\doiurl{10.1007/s11207-009-9357-2}.
\end{barticle}
\endbibitem

\bibitem[\protect\citeauthoryear{Lemen \textit{et~al.}}{2012}]{lemen2012}
\begin{barticle}
\bauthor{\bsnm{Lemen}, \binits{J.R.}},
\bauthor{\bsnm{Title}, \binits{A.M.}},
\bauthor{\bsnm{Akin}, \binits{D.J.}},
\bauthor{\bsnm{Boerner}, \binits{P.E.}},
\bauthor{\bsnm{Chou}, \binits{C.}},
\bauthor{\bsnm{Drake}, \binits{J.F.}},
\bauthor{\bsnm{Duncan}, \binits{D.W.}},
\bauthor{\bsnm{Edwards}, \binits{C.G.}},
\bauthor{\bsnm{Fridlaender}, \binits{F.M.}},
\bauthor{\bsnm{Heyman}, \binits{G.F.}},
\bauthor{\bsnm{Hurlburt}, \binits{N.E.}},
\bauthor{\bsnm{Katz}, \binits{N.L.}},
\bauthor{\bsnm{Kushner}, \binits{G.D.}},
\bauthor{\bsnm{Levay}, \binits{M.}},
\bauthor{\bsnm{Lindgren}, \binits{R.W.}},
\bauthor{\bsnm{Mathur}, \binits{D.P.}},
\bauthor{\bsnm{McFeaters}, \binits{E.L.}},
\bauthor{\bsnm{Mitchell}, \binits{S.}},
\bauthor{\bsnm{Rehse}, \binits{R.A.}},
\bauthor{\bsnm{Schrijver}, \binits{C.J.}},
\bauthor{\bsnm{Wolfram}, \binits{C.J.}},
\bauthor{\bsnm{Yanari}, \binits{C.}},
\bauthor{\bsnm{Bookbinder}, \binits{J.A.}},
\bauthor{\bsnm{Cheimets}, \binits{P.N.}},
\bauthor{\bsnm{Caldwell}, \binits{D.}},
\bauthor{\bsnm{Deluca}, \binits{E.E.}},
\bauthor{\bsnm{Gates}, \binits{R.}},
\bauthor{\bsnm{Golub}, \binits{L.}},
\bauthor{\bsnm{Park}, \binits{S.}},
\bauthor{\bsnm{Podgorski}, \binits{W.A.}},
\bauthor{\bsnm{Bush}, \binits{R.I.}},
\bauthor{\bsnm{Scherrer}, \binits{P.H.}},
\bauthor{\bsnm{Gummin}, \binits{M.A.}},
\bauthor{\bsnm{Smith}, \binits{P.}},
\bauthor{\bsnm{Auker}, \binits{G.}},
\bauthor{\bsnm{Jerram}, \binits{P.}},
\bauthor{\bsnm{Pool}, \binits{P.}},
\bauthor{\bsnm{Soufli}, \binits{R.}},
\bauthor{\bsnm{Windt}, \binits{D.L.}},
\bauthor{\bsnm{Beardsley}, \binits{S.}},
\bauthor{\bsnm{Clapp}, \binits{M.}},
\bauthor{\bsnm{Lang}, \binits{J.}},
\bauthor{\bsnm{Waltham}, \binits{N.}}:
\byear{2012},
\batitle{The \emph{Atmospheric Imaging Assembly} ({AIA}) on the \emph{Solar
  Dynamics Observatory} ({SDO})}.
\bjtitle{\solphys}
\bvolume{275},
\bfpage{17}.
\doiurl{10.1007/s11207-011-9776-8}.
\end{barticle}
\endbibitem

\bibitem[\protect\citeauthoryear{Lowder \textit{et~al.}}{2014}]{lowder2014}
\begin{barticle}
\bauthor{\bsnm{Lowder}, \binits{C.}},
\bauthor{\bsnm{Qiu}, \binits{J.}},
\bauthor{\bsnm{Leamon}, \binits{R.}},
\bauthor{\bsnm{Liu}, \binits{Y.}}:
\byear{2014},
\batitle{Measurements of {EUV} coronal holes and open magnetic flux}.
\bjtitle{\apj}
\bvolume{783},
\bfpage{142}.
\doiurl{10.1088/0004-637X/783/2/142}.
\end{barticle}
\endbibitem

\bibitem[\protect\citeauthoryear{Malanushenko and
  Jones}{2005}]{malanushenko2005}
\begin{barticle}
\bauthor{\bsnm{Malanushenko}, \binits{O.V.}},
\bauthor{\bsnm{Jones}, \binits{H.P.}}:
\byear{2005},
\batitle{Differentiating coronal holes from the quiet sun by {He 1083 nm}
  imaging spectroscopy}.
\bjtitle{\solphys}
\bvolume{226}(\bissue{1}),
\bfpage{3}.
\doiurl{10.1007/s11207-005-4972-z}.
\end{barticle}
\endbibitem

\bibitem[\protect\citeauthoryear{Martens \textit{et~al.}}{2011}]{martens2011}
\begin{barticle}
\bauthor{\bsnm{Martens}, \binits{P.C.H.}},
\bauthor{\bsnm{Atrrill}, \binits{G.D.R.}},
\bauthor{\bsnm{Davey}, \binits{A.R.}},
\bauthor{\bsnm{Engell}, \binits{A.}},
\bauthor{\bsnm{Farid}, \binits{S.}},
\bauthor{\bsnm{Grigis}, \binits{P.C.}},
\bauthor{\bsnm{Kasper}, \binits{J.}},
\bauthor{\bsnm{Korreck}, \binits{K.}},
\bauthor{\bsnm{Saar}, \binits{S.H.}},
\bauthor{\bsnm{Savcheva}, \binits{A.}},
\bauthor{\bsnm{Su}, \binits{Y.}},
\bauthor{\bsnm{Testa}, \binits{P.}},
\bauthor{\bsnm{Wills-Davey}, \binits{M.}},
\bauthor{\bsnm{Bernasconi}, \binits{P.N.}},
\bauthor{\bsnm{Raouafi}, \binits{N.-E.}},
\bauthor{\bsnm{Delouille}, \binits{V.A.}},
\bauthor{\bsnm{Hochedez}, \binits{J.F.}},
\bauthor{\bsnm{Cirtain}, \binits{J.W.}},
\bauthor{\bsnm{DeForest}, \binits{C.E.}},
\bauthor{\bsnm{Angryk}, \binits{R.A.}},
\bauthor{\bsnm{{De Moortel}}, \binits{I.}},
\bauthor{\bsnm{Wiegelmann}, \binits{T.}},
\bauthor{\bsnm{Georgoulis}, \binits{M.K.}},
\bauthor{\bsnm{McAteer}, \binits{R.T.J.}},
\bauthor{\bsnm{Timmons}, \binits{R.P.}}:
\byear{2011},
\batitle{Computer vision for the \textit{{S}olar {D}ynamics {O}bservatory}
  ({SDO})}.
\bjtitle{\solphys}
\bvolume{275}(\bissue{1--2}),
\bfpage{79}.
\doiurl{10.1007/s11207-101-9697-y}.
\end{barticle}
\endbibitem

\bibitem[\protect\citeauthoryear{Mumford and Shah}{1989}]{mumford1989}
\begin{barticle}
\bauthor{\bsnm{Mumford}, \binits{D.}},
\bauthor{\bsnm{Shah}, \binits{J.}}:
\byear{1989},
\batitle{Optimal approximation by piecewise smooth functions and associated
  variational problems}.
\bjtitle{Communications on Pure and Applied Math}
\bvolume{42}(\bissue{5}),
\bfpage{577}.
\doiurl{10.1002/cpa.3160420503}.
\end{barticle}
\endbibitem

\bibitem[\protect\citeauthoryear{Robbins, Henney, and
  Harvey}{2006}]{robbins2006}
\begin{barticle}
\bauthor{\bsnm{Robbins}, \binits{S.}},
\bauthor{\bsnm{Henney}, \binits{C.J.}},
\bauthor{\bsnm{Harvey}, \binits{J.W.}}:
\byear{2006},
\batitle{Solar wind forecasting with coronal holes}.
\bjtitle{\solphys}
\bvolume{233}(\bissue{2}),
\bfpage{265}.
\doiurl{10.1007/s11207-006-0064-y}.
\end{barticle}
\endbibitem

\bibitem[\protect\citeauthoryear{Rotter \textit{et~al.}}{2015}]{rotter2015}
\begin{barticle}
\bauthor{\bsnm{Rotter}, \binits{T.}},
\bauthor{\bsnm{Veronig}, \binits{A.M.}},
\bauthor{\bsnm{Temmer}, \binits{M.}},
\bauthor{\bsnm{Vr\v{s}nak}, \binits{B.}}:
\byear{2015},
\batitle{Real-time solar wind prediction based on {SDO}/{AIA} coronal hole
  data}.
\bjtitle{\solphys}
\bvolume{290}(\bissue{5}),
\bfpage{1355}.
\doiurl{10.1007/s11207-015-0680-5}.
\end{barticle}
\endbibitem

\bibitem[\protect\citeauthoryear{Scherrer \textit{et~al.}}{2012}]{scherrer2012}
\begin{barticle}
\bauthor{\bsnm{Scherrer}, \binits{P.H.}},
\bauthor{\bsnm{Schou}, \binits{J.}},
\bauthor{\bsnm{Bush}, \binits{R.I.}},
\bauthor{\bsnm{Kosovichev}, \binits{A.G.}},
\bauthor{\bsnm{Bogart}, \binits{R.S.}},
\bauthor{\bsnm{Hoeksema}, \binits{J.T.}},
\bauthor{\bsnm{Liu}, \binits{Y.}},
\bauthor{\bsnm{T.~L.~Duvall}, \binits{J.}},
\bauthor{\bsnm{Zhao}, \binits{J.}},
\bauthor{\bsnm{Title}, \binits{A.M.}},
\bauthor{\bsnm{Schrijver}, \binits{C.J.}},
\bauthor{\bsnm{Tarbell}, \binits{T.D.}},
\bauthor{\bsnm{Tomczyk}, \binits{S.}}:
\byear{2012},
\batitle{The \emph{Helioseismic and Magnetic Imager} ({HMI}) investigation for
  the \emph{Solar Dynamics Observatory} ({SDO})}.
\bjtitle{\solphys}
\bvolume{275},
\bfpage{207}.
\doiurl{10.1007/s11207-011-9834-2}.
\end{barticle}
\endbibitem

\bibitem[\protect\citeauthoryear{Scholl and Habbal}{2008}]{scholl2008}
\begin{barticle}
\bauthor{\bsnm{Scholl}, \binits{I.F.}},
\bauthor{\bsnm{Habbal}, \binits{S.R.}}:
\byear{2008},
\batitle{Automatic detection and classification of coronal holes and filaments
  based on {EUV} and magnetogram observations of the solar disk}.
\bjtitle{\solphys}
\bvolume{248}(\bissue{2}),
\bfpage{425}.
\doiurl{10.1007/s11207-007-9075-6}.
\end{barticle}
\endbibitem

\bibitem[\protect\citeauthoryear{Schwadron and McComas}{2003}]{schwadron2003}
\begin{barticle}
\bauthor{\bsnm{Schwadron}, \binits{N.A.}},
\bauthor{\bsnm{McComas}, \binits{D.J.}}:
\byear{2003},
\batitle{Solar wind scaling law}.
\bjtitle{\apj}
\bvolume{599}(\bissue{2}),
\bfpage{1395}.
\doiurl{10.1086/379541}.
\end{barticle}
\endbibitem

\bibitem[\protect\citeauthoryear{Verbeeck \textit{et~al.}}{2014}]{verbeeck2014}
\begin{barticle}
\bauthor{\bsnm{Verbeeck}, \binits{C.}},
\bauthor{\bsnm{Delouille}, \binits{V.}},
\bauthor{\bsnm{Mampaey}, \binits{B.}},
\bauthor{\bsnm{{De Visscher}}, \binits{R.}}:
\byear{2014},
\batitle{The {SPoCA}-suite: {S}oftware for extraction, characterization, and
  tracking of active regions and coronal holes on {EUV} images}.
\bjtitle{\aap}
\bvolume{561},
\bfpage{A64}.
\doiurl{10.1051/0004-6361/201321243}.
\end{barticle}
\endbibitem

\bibitem[\protect\citeauthoryear{Vr\v{s}nak, Temmer, and
  Veronig}{2007}]{vrsnak2007}
\begin{barticle}
\bauthor{\bsnm{Vr\v{s}nak}, \binits{B.}},
\bauthor{\bsnm{Temmer}, \binits{M.}},
\bauthor{\bsnm{Veronig}, \binits{A.M.}}:
\byear{2007},
\batitle{Coronal holes and solar wind high-speed streams: {I.} {F}orecasting
  the solar wind parameters}.
\bjtitle{\solphys}
\bvolume{240}(\bissue{2}),
\bfpage{315}.
\doiurl{10.1007/s11207-007-0285-8}.
\end{barticle}
\endbibitem

\bibitem[\protect\citeauthoryear{Wang, Hawley, and {Sheeley,
  Jr.}}{1996}]{wang1996}
\begin{barticle}
\bauthor{\bsnm{Wang}, \binits{Y.-M.}},
\bauthor{\bsnm{Hawley}, \binits{S.H.}},
\bauthor{\bsnm{{Sheeley, Jr.}}, \binits{N.R.}}:
\byear{1996},
\batitle{The magnetic nature of coronal holes}.
\bjtitle{Science}
\bvolume{271}(\bissue{5248}),
\bfpage{464}.
\doiurl{10.1126/science.271.5248.464}.
\end{barticle}
\endbibitem

\end{thebibliography}

\end{article} 

\end{document}